\documentclass[twocolumn, twocolappendix]{aastex63}

\usepackage{amsmath}

\newcommand{\fermi}{\emph{Fermi}}
\newcommand{\Tbobs}{$T_{b,\text{obs}}$}
\newcommand{\gammaray}{$\gamma$-ray}

\shorttitle{Kinematics of Parsec-Scale Jets of Blazars at 43 GHz}
\shortauthors{Weaver et al.}

\watermark{Draft}
\setwatermarkfontsize{3in}

\begin{document}

\title{Kinematics of Parsec-Scale Jets of Gamma-Ray Blazars at 43 GHz during Ten Years of the VLBA-BU-BLAZAR Program}
\correspondingauthor{Zachary R. Weaver}
\email{zweaver@bu.edu}

%-------------------------------------------------------------------------------
%---   AUTHOR LIST   -----------------------------------------------------------
%-------------------------------------------------------------------------------

\author[0000-0001-6314-0690]{Zachary R. Weaver}
\affiliation{Institute for Astrophysical Research, Boston University, 725 Commonwealth Avenue, Boston, MA 02215, USA}

\author[0000-0001-6158-1708]{Svetlana G. Jorstad}
\affiliation{Institute for Astrophysical Research, Boston University, 725 Commonwealth Avenue, Boston, MA 02215, USA}
\affiliation{Astronomical Institute, St. Petersburg State University, Universitetskij, Pr. 28, Petrodvorets, St. Petersburg 198504, Russia}

\author[0000-0001-7396-3332]{Alan P. Marscher}
\affiliation{Institute for Astrophysical Research, Boston University, 725 Commonwealth Avenue, Boston, MA 02215, USA}

\author[0000-0002-9407-7804]{Daria A. Morozova}
\affiliation{Astronomical Institute, St. Petersburg State University, Universitetskij, Pr. 28, Petrodvorets, St. Petersburg 198504, Russia}

\author[0000-0002-4218-0148]{Ivan S. Troitsky}
\affiliation{Astronomical Institute, St. Petersburg State University, Universitetskij, Pr. 28, Petrodvorets, St. Petersburg 198504, Russia}

\author{Iv\'an Agudo}
\affiliation{Instituto de Astrof\'isica de Andaluc\'ia (IAA), CSIC, Glorieta de la Astronom\'ia s/n, E-18008, Granada, Spain}

\author{Jos\'e L. G\'omez}
\affiliation{Instituto de Astrof\'isica de Andaluc\'ia (IAA), CSIC, Apartado 3004, E-18080, Granada, Spain}

\author{Anne L\"ahteenm\"aki}
\affiliation{Aalto University Mets\"ahovi Radio Observatory, Mets\"ahovintie 114, FI-02540 Kylm\"al\"a, Finland}
\affiliation{Aalto University Department of Electronics and Nanoengineering, P.O. Box 15500, FI-00076, AALTO, Finland}

\author{Joni Tammi}
\affiliation{Aalto University Mets\"ahovi Radio Observatory, Mets\"ahovintie 114, FI-02540 Kylm\"al\"a, Finland}

\author{Merja Tornikoski}
\affiliation{Aalto University Mets\"ahovi Radio Observatory, Mets\"ahovintie 114, FI-02540 Kylm\"al\"a, Finland}

%-------------------------------------------------------------------------------
%---   ABSTRACT   --------------------------------------------------------------
%-------------------------------------------------------------------------------

\begin{abstract}
    We analyze the parsec-scale jet kinematics from 2007 June to 2018 December of a sample of $\gamma$-ray bright blazars monitored roughly monthly with the Very Long Baseline Array at 43 GHz under the VLBA-BU-BLAZAR program. We implement a novel piece-wise linear fitting method to derive the kinematics of 521 distinct emission knots from a total of 3705 total intensity images in 22 quasars, 13 BL Lacertae objects, and 3 radio galaxies.  Apparent speeds of these components range from $0.01c$ to $78c$, and 18.6\% of knots (other than the ``core'') are quasi-stationary. One-fifth of moving knots exhibit non-ballistic motion, with acceleration along the jet within 5 pc of the core (projected) and deceleration farther out. These accelerations occur mainly at locations coincident with quasi-stationary features. We calculate the physical parameters of 273 knots with statistically significant motion, including their Doppler factors, Lorentz factors, and viewing angles. We determine the typical values of these parameters for each jet and the average for each subclass of active galactic nuclei. We investigate the variability of the position angle of each jet over the ten years of monitoring. The fluctuations in position of the quasi-stationary components in radio galaxies tend to be parallel to the jet, while no directional preference is seen in the components of quasars and BL Lacertae objects. We find a connection between \gammaray\ states of blazars and their parsec-scale jet properties, with blazars with brighter 43 GHz cores typically reaching higher $\gamma$-ray maxima during flares.
\end{abstract}


%-------------------------------------------------------------------------------
%---   INTRODUCTION   ----------------------------------------------------------
%-------------------------------------------------------------------------------

\section{Introduction}
\label{sec:Introduction}

The ultra-relativistic flow speeds of the jets of some active galactic nuclei (AGNs) strongly affect their observed characteristics through beaming, aberration, and light-travel delays. These physical effects cause the illusion of apparent superluminal motion, greatly enhanced flux densities, and reduced timescales of variability of flux and polarization \citep[e.g.,][]{Blandford1977b}. Blazars \citep{Angel1980}, with their jet directions oriented nearly along our line of sight \cite[e.g.,][]{Lister2016}, exhibit the most extreme characteristics of all AGNs. The extraordinary resolution afforded by very-long baseline interferometry (VLBI) enables detailed studies of the inner jets of AGNs on distance scales from $<1$ to $>100$ pc from the central supermassive black hole. Monitoring programs over at least a decade are able to (1) systematically characterize the long-term variability of the parsec-scale jet, including knot motions, jet wobbling, and precession; (2) reveal the connection between multi-wavelength outbursts and multi-messenger events in the jet; and (3) provide an observational basis for theoretical models of jet formation, acceleration, and collimation.

Most blazar jets feature ``knots" of emission that travel away from a (presumed stationary) ``core" at apparent speeds that range from $\lesssim1c$ to $\gtrsim50c$ \citep[e.g.,][]{Lister2016, Jorstad2017}.
In a number of blazars,
the jets are observed (in projection) to undergo rapid swings of position angle in their innermost regions, but remain collimated on kiloparsec scales \citep[for a review of several sources, see][]{Agudo2009}. The projected jet position angles can vary considerably \citep[by up to 150$\degr$ over a 12-16 yr time span;][]{Lister2013}.
Monotonic swings with no apparent periodicity (``wobbles") have been reported in some objects \citep[e.g.,][]{Jorstad2004,Agudo2007, Agudo2012}, while precession (sinusoidal variations) has been claimed in others \citep[e.g.,][]{Stirling2003, Bach2005, Lobanov2005, Savolainen2006, MartiVidal2013, Britzen2018, Britzen2019}.  The origin of the wobbles and precession is not well understood, but the presence of such changes in only the innermost (parsec-scale) regions of the jet implies that the causal mechanisms are related to the origin of the jet, e.g., accretion disk precession \citep[e.g.,][]{McHardy1990, Conway1995, Villata1999, Britzen2005, Perucho2012, Fromm2013, Larionov2013, Britzen2017, Raiteri2017, Raiteri2021}, instabilities in the jet flow \citep[e.g.,][]{Nakamura2001, Moll2008, Mignone2010, Liska2018}, or interactions in a binary supermassive black hole system \citep[e.g.,][]{Britzen2018, Qian2019}.

In order to understand the physical processes in blazars, it is important to relate changes observed in the radio jet to multi-wavelength variability and outbursts. Connections between the high-energy emission and radio jets of AGNs have been theorized for $\sim$40 years \citep[e.g.,][]{Konigl1981, Marscher1987}, with observational evidence provided by joint radio and \gammaray\ studies, the latter based on data from the \emph{Compton Gamma Ray Observatory} \citep{Hartman1999}. The results have highlighted the dichotomy between blazar subclasses. In the traditional AGN unification scheme \citep{Urry1995}, flat-spectrum radio quasars (FSRQs) and BL Lacertae objects (BLs) are differentiated by their optical properties and radio morphology \citep[e.g.,][]{Wardle1984}. During the \emph{Compton} era, FSRQs were found to have higher \gammaray\ fluxes when they also exhibited high optical polarization and a high-radio-frequency outburst \citep{Valtaoja1995}. There was no similarly clear correlation between the \gammaray\ emission in BLs and their radio variations \citep{Lahteenmaki2003}. For both subtypes, however, there existed a correlation between the brightness of the radio ``core" and \gammaray\ flux \citep{Jorstad2001}, as well as a connection between the ejections of superluminal radio knots and \gammaray\ outbursts \citep{Jorstad2001b}.

The \fermi\ Large Area Telescope (LAT) has ushered in a new era of \gammaray-radio studies. Individual sources show a wide variety of behavior during major \gammaray\ events. The typical duration of a high \gammaray\ state in blazars is several months \citep{Williamson2014}, although shorter-timescale fluctuations are often observed \citep[e.g.,][]{Agudo2011,Weaver2019, Weaver2020}. The majority of these \gammaray\ and very-high-energy (VHE, $\geq100$ GeV) outbursts can be connected with the propagation of a knot through the core or other quasi-stationary component in the parsec-scale radio jet \citep[e.g.,][]{Marscher2010, Agudo2011, Casadio2015, Liodakis2020, HESS2021, MAGIC2021}. Two major issues exist with this model for blazar flares: the presence of ``orphan" \gammaray\ flares \citep[e.g.,][]{Krawczynski2004, MacDonald2015} with no counterparts at longer wavelengths, and the need for an intense external photon field at these locations in the jet \citep[e.g.,][]{Aleksic2011, Joshi2014}, the source of which is unknown. Emission-line flares with short time delays relative to the \gammaray\ outburst have also been observed in several blazars \citep{LeonTavares2013, Larionov2020, Hallum2021}, which could provide additional seed photons \citep{Isler2015, LeonTavares2015, Larionov2020}. Detection of a 290 TeV neutrino with a direction of origin coincident with the flaring blazar TXS 0506+056 \citep[][]{IceCube2018b, IceCube2018a}, suggests an association between flaring blazars and high-energy neutrinos \citep[e.g.,][]{Rodrigues2018, Murase2018}. Thus, blazar jets may produce high-energy hadronic cosmic-rays \citep[e.g.,][]{Keivani2018, Cerruti2019} alongside electromagnetic outbursts and changes in the parsec-scale structure \citep[for discussion of the parsec-scale jet of TXS 0506+056, see][]{Li2020}.

VLBI observations can also provide constraints on theoretical models of AGN jet launching, acceleration, and collimation, which are challenging to model. The launching mechanism in AGNs can be ascribed to one (or both) of two processes. The Blandford-Znajek mechanism \citep{Blandford1977} utilizes the rotational energy of the black hole, with the spin of the black hole winding up the magnetic field lines anchored to the event horizon, resulting in a relativistic Poynting flux dominated jet \citep[e.g.,][]{Beskin2000, Komissarov2001, Komissarov2005, Tchekhovskoy2010}. Alternatively, magnetic field lines anchored to the accretion disk can launch mildly relativistic mass-dominated winds \citep{Blandford1982}. One of the primary goals of VLBI monitoring programs of blazar jets is to provide an observational framework for comparing numerical simulations of jets \citep[e.g.,][]{Vlahakis2003, Komissarov2005, Lyubarsky2009, Meier2012, Pu2015, Chatterjee2019} with the actual parsec-scale behavior.

A number of observing programs follow time changes in the jets of a large number of blazars. The largest has monitored several hundreds of the brightest AGNs in the northern sky, first under the 2 cm VLBA survey \citep{Kellermann1998}, and later as part of the MOJAVE\footnote{\url{https://www.physics.purdue.edu/MOJAVE/}} program. \citet{Lister2019} find that $60\%$ of the 409 monitored sources exhibit acceleration of emission features in the jet. Other than a few outliers, the 15 GHz emission features have a  distribution of apparent velocities that fall within $\pm 40\%$ of the median speed of a given source. No jet features are reported to have a bulk Lorentz factor $\Gamma > 50$. Through Monte-Carlo modeling of the parent population, the authors show that the data are consistent with a jet population that has a simple unbeamed power-law luminosity function incorporating pure luminosity evolution, and a power-law Lorentz factor distribution ranging from 1.25 to 50, with a slope $-1.4 \pm 0.2$.

The Boston University group and its collaborators have been monitoring blazar jets with the VLBA at 43 GHz since 1993. The research focuses on the connection between jet properties and multi-wavelength behavior. This includes analysis of events in the jets of \gammaray\ blazars during the \emph{Compton} era \citep{Jorstad2001}, accretion disk-jet associations \citep{Marscher2002, Ritaban2011},
and relations between optical and millimeter-wave linear polarization \citep{Jorstad2005}. Over the past 10 years, we have been monitoring the jet kinematics of a sample of 38 blazars at a roughly monthly cadence at 43 GHz under the VLBA-BU-BLAZAR program\footnote{\url{http://www.bu.edu/blazars/BEAM-ME.html}}. For comparison with the longer-wavelength data, we obtain well-sampled \gammaray\ light curves from the \fermi\ LAT. The higher observing frequency relative to other VLBA monitoring programs allows emission features to be tracked closer to the ``core" of the sources. Also, the more frequent monitoring allows faster, shorter-lived features to be observed. Based on the first 5 years of data, we found that nearly one-third of moving knots show evidence of acceleration in the jet \citep{Jorstad2017}. Furthermore, knots in FSRQs tend to have higher Doppler and Lorentz factors, with smaller viewing and opening angles, than the knots in BLs and radio galaxies (RGs). A number of radio jet - \gammaray\ event associations have been detected using these data \citep{Marscher2010, Agudo2011, Agudo2011a, Wehrle2012, Jorstad2013b, Casadio2015, Jorstad2016, Larionov2020, HESS2021, MAGIC2021}.

In this paper, we provide the data and detailed analyses of the parsec-scale jet behavior of a sample of \gammaray\ bright AGNs over our decade-long monitoring program, VLBA-BU-BLAZAR. We implement a new piece-wise linear fitting method to determine the jet kinematics. For consistency, we reanalyze previously-published data to produce a uniform data set that covers the first 10.5 years of VLBA monitoring associated with \fermi\ LAT observations, 2007 June to 2018 December. The paper is structured as follows. In $\S$\ref{sec:ObsandData}, we detail the observations, their calibration, and analysis from 2013 January through 2018 December, and describe the piece-wise fitting of the core-knot separations versus time that we apply to all of the data. The data for individual knots and sources are available in Tables~\ref{tab:GaussianComponents} and \ref{tab:JetStructure}. The parsec-scale jet structure and the behavior of the jet position angle with time (Table~\ref{tab:JetPAs}) of each source are discussed in $\S$\ref{sec:ParsecScaleStructure}.
Section~\ref{sec:Speeds} describes the moving/stationary features (Tables~\ref{tab:JetSpeeds} and \ref{tab:Stationary}) and accelerations (Table~\ref{tab:JetAccel}) of all knots in our sample. We determine the statistically robust physical parameters of knots in our sample in $\S$\ref{sec:JetPhysicalParams} (Table~\ref{tab:PhysParams}), and provide estimates for the ``typical" physical parameters of each jet (Table~\ref{tab:AvePhysParams}). We
briefly discuss connections between millimeter-wave and \gammaray\ states of the AGNs in $\S$\ref{sec:Discussion}. A summary of our findings is given in $\S$\ref{sec:Summary}.
Throughout this work, we assume a standard $\Lambda$CDM cosmology with rounded values of cosmological parameters based on those obtained by the Planck collaboration \citep{Planck2020}, the results from SNIa studies \citep{Riess2016}, and VLT-KMOS data \citep{KMOS2021}: $H_\circ = 70$ km s$^{-1}$ Mpc$^{-1}$, $\Omega_{\text{m}} = 0.3$, and $\Omega_{\Lambda} = 0.7$.

%-------------------------------------------------------------------------------
%---   OBSERVATIONS AND DATA ANALYSIS   ----------------------------------------
%-------------------------------------------------------------------------------

\section{Observations and Data Analysis}
\label{sec:ObsandData}

\begin{deluxetable*}{lllDccc}
  \tablecaption{The VLBA-BU-BLAZAR Sample.\label{tab:SourceList}}
  \tablewidth{0pt}
  \tablehead{
  \colhead{Source} & \colhead{Name} & \colhead{Type} & \multicolumn2c{$z$} & \colhead{$\langle S_{37} \rangle$} & \colhead{$\langle S_\gamma \rangle$} & \colhead{$\langle S_R \rangle$} \\
  \colhead{} & \colhead{} & \colhead{} & \multicolumn2c{} & \colhead{[Jy]} & \colhead{[$10^{-8}$ ph cm$^{-2}$ s$^{-1}$]} & \colhead{[mag]}
  }
  \decimalcolnumbers
  \startdata
  0219+428\tablenotemark{a} & 3C 66A       & BL   & 0.444\tablenotemark{f}  & $\phn 0.71 \pm 0.24$ & $\phn     10.81 \pm \phn     12.86$ & $14.40 \pm 0.42$ \\
  0235+164                  & A0 0235+16   & BL   & 0.940\tablenotemark{f}  & $\phn 1.99 \pm 1.16$ & $\phn     11.77 \pm \phn     14.91$ & $17.45 \pm 0.98$ \\
  0316+413\tablenotemark{b} & 3C 84        & RG   & 0.0176                  & $26.11     \pm 4.57$ & $\phn     35.54 \pm \phn     20.40$ & $12.82 \pm 0.14$ \\
  0336--019                 & CTA 26       & FSRQ & 0.852                   & $\phn 2.24 \pm 0.56$ & $\phn     12.34 \pm \phn     10.13$ & $16.86 \pm 0.52$ \\
  0415+379                  & 3C 111       & RG   & 0.0485                  & $\phn 4.06 \pm 2.13$ & $\phn \phn 7.43 \pm \phn \phn 4.68$ & $17.38 \pm 0.22$ \\
  0420--014                 & OA 129       & FSRQ & 0.916                   & $\phn 4.11 \pm 2.14$ & $\phn \phn 7.68 \pm \phn \phn 5.18$ & $17.50 \pm 0.78$ \\
  0430+052\tablenotemark{c} & 3C 120       & RG   & 0.033                   & $\phn 3.40 \pm 1.64$ & $\phn \phn 5.43 \pm \phn \phn 4.84$ & $13.90 \pm 0.13$ \\
  0528+134                  & PKS 0528+13  & FSRQ & 2.060                   & $\phn 1.84 \pm 1.20$ & $\phn \phn 8.85 \pm \phn \phn 9.93$ & $19.38 \pm 0.27$ \\
  0716+714                  & S5 0716+71   & BL   & 0.3\tablenotemark{f}    & $\phn 2.44 \pm 1.19$ & $\phn     20.78 \pm \phn     14.67$ & $13.19 \pm 0.63$ \\
  0735+178                  & PKS 0735+17  & BL   & 0.424                   & $\phn 0.77 \pm 0.23$ & $\phn \phn 6.21 \pm \phn     11.77$ & $15.98 \pm 0.47$ \\
  0827+243                  & OJ 248       & FSRQ & 0.939                   & $\phn 1.23 \pm 0.61$ & $\phn \phn 6.87 \pm \phn     11.26$ & $16.91 \pm 0.39$ \\
  0829+046                  & OJ 049       & BL   & 0.182                   & $\phn 0.83 \pm 0.37$ & $\phn \phn 6.53 \pm \phn \phn 3.93$ & $15.52 \pm 0.38$ \\
  0836+710                  & 4C+71.07     & FSRQ & 2.172                   & $\phn 2.12 \pm 0.69$ & $\phn     13.91 \pm \phn     19.40$ & $16.60 \pm 0.09$ \\
  0851+202                  & OJ 287       & BL   & 0.306                   & $\phn 5.94 \pm 2.20$ & $\phn \phn 9.08 \pm \phn \phn 7.46$ & $14.37 \pm 0.49$ \\
  0954+658                  & S4 0954+65   & BL   & 0.368\tablenotemark{f}  & $\phn 1.19 \pm 0.40$ & $\phn     10.87 \pm \phn     20.24$ & $15.58 \pm 0.96$ \\
  1055+018\tablenotemark{a} & 4C+01.28     & FSRQ\tablenotemark{g} & 0.890  & $\phn 5.22 \pm 1.36$ & $\phn \phn 9.91 \pm \phn \phn 5.72$ & $15.92 \pm 0.83$ \\
  1101+384\tablenotemark{b} & Mkn 421      & BL   & 0.030                   & $\phn 0.49 \pm 0.16$ & $\phn     17.83 \pm \phn \phn 9.50$ & $12.41 \pm 0.39$ \\
  1127--145                 & PKS 1127--14 & FSRQ & 1.184                   & $\ldots$             & $\phn \phn 6.80 \pm \phn \phn 5.05$ & $16.74 \pm 0.18$ \\
  1156+295                  & 4C+29.45     & FSRQ & 0.729                   & $\phn 2.08 \pm 0.83$ & $\phn     16.51 \pm \phn     21.60$ & $15.71 \pm 1.00$ \\
  1219+285                  & WCom         & BL   & 0.102                   & $\phn 0.41 \pm 0.16$ & $\phn \phn 6.29 \pm \phn \phn 2.65$ & $15.26 \pm 0.37$ \\
  1222+216                  & 4C+21.35     & FSRQ & 0.435                   & $\phn 1.98 \pm 0.61$ & $\phn     33.62 \pm \phn     55.61$ & $15.10 \pm 0.43$ \\
  1226+023                  & 3C 273       & FSRQ & 0.158                   & $17.76     \pm 4.58$ & $\phn     29.03 \pm \phn     40.00$ & $12.63 \pm 0.13$ \\
  1253--055                 & 3C 279       & FSRQ & 0.538                   & $19.00     \pm 5.31$ & $\phn     66.30 \pm         117.31$ & $15.02 \pm 0.66$ \\
  1308+326                  & B2 1308+32   & FSRQ & 0.998                   & $\phn 1.74 \pm 0.72$ & $\phn \phn 5.25 \pm \phn \phn 4.24$ & $18.03 \pm 0.83$ \\
  1406--076                 & PKS 1406--07 & FRSQ & 1.494                   & $\phn 0.94 \pm 0.26$ & $\phn \phn 6.54 \pm \phn     10.19$ & $18.28 \pm 0.42$ \\
  1510--089                 & PKS 1510--08 & FSRQ & 0.361                   & $\phn 3.31 \pm 1.25$ & $\phn     96.60 \pm         120.47$ & $15.88 \pm 0.53$ \\
  1611+343                  & DA 406       & FSRQ & 1.40                    & $\phn 3.11 \pm 0.71$ & $\phn \phn 5.25 \pm \phn \phn 2.44$ & $17.31 \pm 0.21$ \\
  1622--297\tablenotemark{h}                 & PKS 1622--29 & FSRQ & 0.815                   & $\ldots$             & $\phn     12.09 \pm \phn \phn 7.57$ & $18.18 \pm 0.25$ \\
  1633+382                  & 4C+38.41     & FSRQ & 1.814                   & $\phn 3.78 \pm 1.39$ & $\phn     30.31 \pm \phn     30.63$ & $16.68 \pm 0.63$ \\
  1641+399                  & 3C 345       & FSRQ & 0.595                   & $\phn 5.83 \pm 1.16$ & $\phn \phn 9.22 \pm \phn \phn 5.83$ & $17.15 \pm 0.46$ \\
  1652+398\tablenotemark{d} & Mkn 501      & BL   & 0.034                   & $\phn 0.93 \pm 0.17$ & $\phn \phn 5.27 \pm \phn \phn 2.38$ & $13.26 \pm 0.09$ \\
  1730--130                 & NRAO 530     & FSRQ & 0.902                   & $\phn 3.47 \pm 0.94$ & $\phn     11.96 \pm \phn \phn 9.99$ & $17.63 \pm 0.38$ \\
  1749+096\tablenotemark{a} & OT 081       & BL   & 0.322                   & $\phn 3.82 \pm 1.21$ & $\phn \phn 8.89 \pm \phn     12.39$ & $16.49 \pm 0.52$ \\
  1959+650\tablenotemark{e} & 1ES 1959+65  & BL   & 0.047                   & $\phn 0.32 \pm 0.14$ & $\phn \phn 5.13 \pm \phn \phn 3.29$ & $14.24 \pm 0.25$ \\
  2200+420                  & BL Lac       & BL   & 0.069                   & $\phn 4.46 \pm 2.47$ & $\phn     40.48 \pm \phn     25.08$ & $13.64 \pm 0.56$ \\
  2223--052                 & 3C 446       & FSRQ & 1.404                   & $\phn 3.80 \pm 2.00$ & $\phn \phn 5.02 \pm \phn \phn 2.89$ & $18.09 \pm 0.35$ \\
  2230+114                  & CTA 102      & FSRQ & 1.037                   & $\phn 3.50 \pm 0.89$ & $        118.81 \pm         185.75$ & $15.65 \pm 1.04$ \\
  2251+158                  & 3C 454.3     & FSRQ & 0.859                   & $17.01     \pm 7.72$ & $        206.62 \pm         337.67$ & $15.02 \pm 0.79$ \\
  \enddata
  \tablenotetext{a}{Source added to the sample in 2009.}
  \tablenotetext{b}{Source added to the sample in 2010.}
  \tablenotetext{c}{Source added to the sample in 2012.}
  \tablenotetext{d}{Source added to the sample in 2014.}
  \tablenotetext{e}{Source added to the sample in 2017.}
  \tablenotetext{f}{Redshift has not been confirmed.}
  \tablenotetext{g}{Classified as a BL object in \citet{Jorstad2017}.}
  \tablenotetext{h}{Removed from the sample after 2017 July.}
\end{deluxetable*}

As part of the VLBA-BU-BLAZAR monitoring program, we have obtained roughly monthly observations with the VLBA at 43 GHz (7 mm wavelength) of a sample of AGNs detected as \gammaray\ sources. The results of the observations from 2007 June to 2013 January have been presented in \citet{Jorstad2017}. In this work, we extend the analyzed period with observations obtained from 2013 January through 2018 December, bringing the total monitoring period to 10.5 years. In order to interpret the results of our study more consistently, we apply a new formalism (described below) to the entire data set to calculate the knot motions and kinematic parameters.

The sample studied here consists of a total of 38 sources, of which 22 are FSRQs, 13 are BLs, and 3 are RGs. Each source has been detected at \gammaray\ energies by the \fermi\ LAT \citep{Abdo2010, Acero2015}, with an average flux density at 43 GHz exceeding 0.5 Jy, a declination north of $-30\degr$, and an optical magnitude in \emph{R}-band brighter than $18.5^{\text{m}}$.
The original sample is described in \citet{Jorstad2017}.
Several additional sources that have been revealed to have occasionally strong \gammaray\ flux (as observed with the \fermi-LAT) and sufficiently high optical flux to meet the selection criteria have been included in the current sample.

The observations of the sample were performed roughly monthly via dynamical scheduling, over a 24-hr period at each epoch, using all available VLBA antennas. (At some epochs 1-2 antennas could not collect data owing to technical problems or weather conditions.) A total of 32 or 33 objects were observed during each epoch, with 40-45 min integration time per source (8-9 scans of $\sim5$-min duration), spread over the 6-10 hr time span with the optimal visibility of the object. Eight sources were observed during alternating sessions; these objects are in crowded right ascension ranges and exhibit relatively long timescales of variability of the jet structure at 43 GHz. Although the observations and data processing were carried out in a similar manner as in \citet{Jorstad2017}, with the latter using the Astronomical Image Processing System \citep[AIPS;][]{vanMoorsel1996}, some details were different. A new data recording system (Mark5C) was installed, allowing a data recording rate of 2048 Mb per second, as opposed to the previous rate of 512 Mb per second. The resulting increase in data quality allows a sufficient signal-to-noise ratio to be achieved without averaging intermediate frequency bands (IFs). Thus, since 2014 February the final calibrated data available on the project website include visibilities at 4 IFs (centered at 43.0075, 43.0875, 43.1515, and 43.2155 GHz). We have calibrated the instrumental polarization separately for each IF.

We have continued to monitor the optical \emph{BVRI} photometric brightness and \emph{R}-band linear polarization of the sample at the 1.83-m Perkins Telescope (PTO, Boston University\footnote{Operated by Lowell Observatory prior to 2019}, Flagstaff, Arizona) over $\sim1$ week time intervals during dark/gray lunar phases. One such session occurred during most months.

Table~\ref{tab:SourceList} gives a list of the monitored sources, along with their characteristics:
1---name based on B1950 coordinates;
2---common name;
3---subclass of blazar;
4---redshift, $z$, according to the NASA Extragalactic Database\footnote{\url{https://ned.ipac.caltech.edu/}};
5---average total intensity at 37 GHz, $\langle S_{37} \rangle$, and its standard deviation as observed with the Mets\"ahovi Radio Observatory of Aalto University, Finland, between 2007 January and 2018 December;
6---average \gammaray\ photon flux at 0.1-200 GeV, $\langle S_\gamma \rangle$, and its standard deviation based on the \gammaray\ light curve of each source, calculated from the photon and spacecraft data provided by the \fermi\ Space Science Center \citep[for details, see][]{Williamson2014} between 2007 January and 2018 December;
7---average \emph{R}-band magnitude, $\langle S_R \rangle$ and its standard deviation according to our observations at the Perkins Telescope between 2007 January and 2018 December.

\subsection{Total Intensity Image Modeling}
\label{subsec:ModelingTotalIntensityImages}

The parsec-scale jet morphology of blazars features a ``core" \citep[e.g.,][]{Jorstad2001, Kellermann2004, Jorstad2005, Lister2013, Jorstad2017, Lister2018}, an assumed stationary feature located at the upstream end of the jet seen in the VLBI images. For most sources, the core is usually the brightest feature in the jet \citep{Jorstad2017}, but sometimes in some blazars an emission component downstream of the core can be more prominent \citep[e.g., 2251+158;][]{Jorstad2001}. The exact location of this core relative to the central black hole is frequency dependent as the core represents the region where the jet becomes optically thick at a given frequency, with VLBI at higher frequencies being able to image closer to the black hole \citep[see, e.g.,][]{OSullivan2009, EHT2019I}. One or more ``knots'' of emission are often observed downstream of the core \citep[e.g.,][]{Jorstad2001}. Some of these are quasi-stationary, without systematic motions relative to the core, while others move down the jet, in most cases at apparent superluminal speeds \citep[e.g.,][]{Cohen1971, Cohen1977, Jorstad2001,Jorstad2017,Lister2018}.

In order to represent the total intensity structure of each source at each epoch, we model the frequency-averaged visibility data with a series of circular Gaussian components that best fit the data. The model fitting involves iterations of the \texttt{MODELFIT} task within the Difmap \citep{Shepherd1997} software package. We use the term ``knot" to refer to these Gaussian components, which correspond to a (usually) compact feature of enhanced brightness in the jet. Initially, a single circular Gaussian component approximating the brightness distribution of the core is used, then knots are added at the approximate locations of bright features identified in the image. Each addition of a knot is followed by hundreds of iterations with \texttt{MODELFIT} to determine the parameters of the knot that yield the best agreement between the model and $uv$ data, according to a $\chi^2$ test. The knot parameters that we fit are:
$S$---flux density;
$R_{\text{obs}}$---distance with respect to the core;
$\Theta_{\text{obs}}$---relative position angle (P.A.) with respect to the core, measured north through east;
and $a$---the angular size of the knot, corresponding to the FWHM of the circular Gaussian component.
The iterative process of adding knots is ended when the addition of a new knot does not significantly improve the $\chi^2$ value of the model. Often, the model of a previous epoch is used as the starting model for the next epoch for a given source, as we assume that the jet does not drastically change structure between our roughly monthly observations. We utilize the previously-published Difmap models of each source at each epoch from 2007 June to 2012 December \citep{Jorstad2017}. In this work we provide models for each source from 2013 January to 2018 December.

\begin{deluxetable*}{cccccccccr}
  \tablecaption{Modeling of Jets by Gaussian Components 2013--2018\label{tab:GaussianComponents}}
  \tablewidth{0pt}
  \tablehead{
  \colhead{Source} & \colhead{$\chi^2$} & \colhead{RMS} & \colhead{Epoch} & \colhead{MJD} & \colhead{$S \pm \sigma_S$} & \colhead{$R \pm \sigma_R$} & \colhead{$\Theta \pm \sigma_\Theta$} & \colhead{$a \pm \sigma_a$} & \colhead{$T_{b,\text{obs}}$} \\
  \colhead{} & \colhead{} & \colhead{} & \colhead{[yr]} & \colhead{} & \colhead{[Jy]} & \colhead{[mas]} & \colhead{[deg]} & \colhead{[mas]} & \colhead{[$\times 10^8$ K]}
  }
  \colnumbers
  \startdata
  0219+428 & 1.131 & 0.050 & 2013.038 & 56307 & $0.129 \pm 0.009$ & $0.000 \phantom{\pm} \phn \phd \phn \phn \phn$ & $\phantom{-} \phd \phd 0.0 \phantom{\pm} \phn \phd \phn \phn \phn$ & $0.020 \pm 0.009$ & 2418.8L \\
           &       &       & 2013.038 & 56307 & $0.071 \pm 0.010$ & $0.084 \pm 0.030$ & $-165.0 \pm \phn 5.8$ & $0.081 \pm 0.022$ & 81.2 \\
           &       &       & 2013.038 & 56307 & $0.044 \pm 0.012$ & $0.391 \pm 0.099$ & $-157.1 \pm \phn 4.6$ & $0.183 \pm 0.037$ & 9.9 \\
           &       &       & 2013.038 & 56307 & $0.025 \pm 0.014$ & $0.737 \pm 0.296$ & $-176.0 \pm \phn 5.8$ & $0.329 \pm 0.057$ & 1.7 \\
           &       &       & 2013.038 & 56307 & $0.031 \pm 0.015$ & $2.396 \pm 0.443$ & $-177.8 \pm \phn 2.7$ & $0.512 \pm 0.067$ & 0.9 \\
  0219+428 & 0.792 & 0.035 & 2013.288 & 56398 & $0.203 \pm 0.013$ & $0.000 \phantom{\pm} \phn \phd \phn \phn \phn$ & $\phantom{-} \phd \phd 0.0 \phantom{\pm} \phn \phd \phn \phn \phn$ & $0.053 \pm 0.013$ & 542.0 \\
           &       &       & 2013.288 & 56398 & $0.047 \pm 0.011$ & $0.149 \pm 0.060$ & $-161.5 \pm \phn 6.8$ & $0.122 \pm 0.029$ & 23.7 \\
           &       &       & 2013.288 & 56398 & $0.030 \pm 0.011$ & $0.448 \pm 0.098$ & $-160.4 \pm \phn 3.8$ & $0.148 \pm 0.036$ & 10.3 \\
           &       &       & 2013.288 & 56398 & $0.021 \pm 0.015$ & $0.843 \pm 0.613$ & $-169.4 \pm     11.1$ & $0.558 \pm 0.077$ & 0.5 \\
           &       &       & 2013.288 & 56398 & $0.025 \pm 0.014$ & $2.474 \pm 0.372$ & $-177.3 \pm \phn 2.2$ & $0.398 \pm 0.062$ & 1.2 \\
  \enddata
\tablecomments{Values of $T_{\text{b,obs}}$ denoted by the letter ``L" represent the lower limits to the brightness temperature; see $\S$\ref{subsec:ModelingTotalIntensityImages} for details.\newline (Table~\ref{tab:GaussianComponents} (10192 lines) is published in its entirety in the machine-readable format. A portion is shown here for guidance regarding its form and content.)}
\end{deluxetable*}

The uncertainties of the model parameters were calculated using the formalism described in \citet{Jorstad2017}, which is based on an empirical relation between the uncertainties and the brightness temperatures of knots, $T_{b,\text{obs}} = 7.5 \times 10^8 S/a^2$ K \citep[e.g.,][]{Jorstad2005, Casadio2015}. Estimates of the errors in each model parameter are as follows:
$\sigma_{\text{X}} \approx 1.3 \times 10^4\ T_{b,\text{obs}}^{-0.6}$, $\sigma_{\text{Y}} \approx 2\sigma_{\text{X}}$, $\sigma_S \approx 0.09\ T_{b,\text{obs}}^{-0.1}$, and $\sigma_a \approx 6.5\ T_{b,\text{obs}}^{-0.25}$. Here, $\sigma_{\text{X}}$ and $\sigma_{\text{Y}}$ are the $1\sigma$ uncertainties in right ascension and declination in mas, $\sigma_S$ the uncertainty in the flux density in Jy, and $\sigma_a$ the uncertainty in component size in mas. Note that the typical north-south synthesized beam is $\sim2\times$ the typical east-west beam size, which increases the uncertainty in declination. As in \citet{Jorstad2017}, we have also added a minimum positional error of 0.005 mas (related to the resolution of the observations) and a typical amplitude calibration error of 5\% to these uncertainties.

Table~\ref{tab:GaussianComponents} gives the total intensity jet models and brightness temperature values for all components in the jet of each source from 2013 January to 2018 December as follows:
1---source name according to its B1950 coordinates;
2---$\chi^2$ value of the model fit;
3---root-mean-square difference between the observed and model visibilities;
4---epoch of the start of the observation;
5---MJD of the start of the observation;
6---the flux density, $S$, and its uncertainty $\sigma_S$, in Jy;
7---distance with respect to the core, $R_{\text{obs}}$, and its uncertainty $\sigma_{R}$, in mas;
8---Position angle of the knot with respect to the core, $\Theta_{\text{obs}}$, and its uncertainty $\sigma_{\Theta}$, measured north through east in degrees;
9---angular size of the knot, $a$, and its uncertainty $\sigma_a$, in mas; and
10---observed brightness temperature \Tbobs, in units of $10^{8}$ K.
The core at each epoch is located at position $(X,Y) = (0,0)$, where $X$ is the relative right ascension and $Y$ is the relative declination, and $X$ and $Y$ are related to $R_{\text{obs}}$ and $\Theta_{\text{obs}}$ through $R_{\text{obs}} = \sqrt{X^2 + Y^2}$ and $\tan{\Theta_{\text{obs}}} = X / Y$. For 9.8\% of knots, the model fit yielded a size less than 0.02 mas. This angular size is too small to be resolved on the longest VLBA baselines. Thus, lower limits to \Tbobs\ for these knots are calculated using $a = 0.02$ ($\sim20\%$ of the  resolution of the longest baselines) and denoted by ``L" in column 10 of Table~\ref{tab:GaussianComponents}. The total intensity jet models and brightness temperature values for all components in the jet of each source from 2007 June to 2012 December can be found in Table 2 of \citet{Jorstad2017}.

\subsection{Identification of Components and Calculation of Apparent Speed}
\label{subsec:KnotIdentitySpeed}

As discussed above, the VLBI ``core" is defined as the bright, compact (either unresolved or partially resolved) emission feature at the upstream end of the jet, relative to which the positions of all other components are measured. To identify components, we make use of the fact that all four model parameters $(S, R_{\text{obs}}, \Theta_{\text{obs}}$, and $a$) should not change abruptly with time given the regularity of our observations. If a knot is identified at $N \geq 4$ epochs, it is assigned a kinematic classification and identification number (ID).

Previous studies \citep[e.g.,][]{Homan2015,Jorstad2017} have found that a given knot can exhibit both acceleration and deceleration as it propagates down the jet. In order to describe the locations along the jet of such regions of changing velocity, we have created a new formalism for the calculation of kinematic parameters of a knot, utilizing piece-wise linear fits to separation (from the core) versus time data instead of polynomial fits as used in \citet{Jorstad2017}. This simplifies the calculation of kinematic parameters by describing the motion in terms of the mean velocity of each segment of time within which the velocity is roughly constant and significantly different from that of other segments. This allows us to estimate the distance down the jet at which acceleration occurs. The new procedure consists of the following steps:

\begin{enumerate}
    \item \emph{Piece-wise Linear Fits to $X$ and $Y$:} We start with the assumption that knots detected at $N \geq 4$ epochs can have their $X$ and $Y$ positions fit by a linear trend of the form

    \begin{equation}
        X(t_i) = a_0 + a_1 (t_i - t_{\text{mid}})
        \label{eqn:1}
    \end{equation}
    \noindent and
    \begin{equation}
        Y(t_i) = b_0 + b_1 (t_i - t_{\text{mid}}),
        \label{eqn:2}
    \end{equation}
    \noindent where $t_i$ is the epoch of observation, $i = 1, \ldots, N$, and $t_{\text{mid}} \equiv (t_N + t_1) / 2$. We use the \texttt{statsmodels} package\footnote{\url{https://www.statsmodels.org/stable/index.html}} \citep{Seabold2010} weighted least-squares program to calculate the best-fit parameters and reduced $\chi^2$ value of the fit. These fit $\chi^2$ values are then compared to the critical $\chi^2_{\text{crit}}$, corresponding to a significance level of $\zeta = 0.05$ for $f = N - 2$ degrees of freedom %\citep[where the two comes from the linear fit minus one,][]
    \citep{Bowker1972}. If $\chi^2 < \chi^2_{\text{crit}}$, our assumption was valid. However, if the opposite is true, the knot may be experiencing acceleration in the jet and a more complicated fit is necessary.
    In such cases, we use the \texttt{pwlf} package\footnote{\url{https://jekel.me/piecewise_linear_fit_py/index.html}} to perform a weighted least-squares piece-wise linear fit to the data \citep{Jekel2019}. The advantage of such a fit over higher-order polynomial fits is that it relies on simple linear fits for segments of the data, so that there are no concerns about over-fitting the data or estimating kinematic parameters from higher-order fits.

    We define the piece-wise fits in this work in terms of the number of line segments present for each coordinate of motion $X$ and $Y$. A single-segment fit describes the simple linear case as defined above. A two-segment fit indicates that the motion can be described by two different speeds with a single acceleration region at a particular break point, $t_b$, and a three-segment fit indicates three speeds and two acceleration regions and breakpoints. In a two-segment fit, there are 5 free parameters, while in a three-segment fit there are 8 free parameters. A piece-wise fit with the fewest number of segments for which $\chi^2 \leq \chi^2_{\text{crit}}$ is used to fit the data, where $\chi^2_{\text{crit}}$ is the $\chi^2$ value corresponding to a significance level of $\zeta=0.05$ for $f = N - \ell$, where $\ell$ is the number of fit parameters. However, a two-segment fit is only used if $N > 8$, and a three-segment fit is only used if $N > 15$, given the number of degrees of freedom. In cases for which the $\chi^2$ criterion is not reached for even a three-segment piece-wise fit, we use the simplest fit that adequately describes the data ($\chi^2$ closest to $\chi^2_{\text{crit}}$). There are only several such cases. In some cases the $X$ and $Y$ motion breakpoints can overlap in time, generally defined as $|t_{b,X} - t_{b,Y}| \leq 0.25$ yr. In these situations we average $t_{b,X}$ and $t_{b,Y}$ and use this average time as a single break-point. The model fit is then re-run with the break times frozen to the averaged values. The overlap value of 0.25 yr was chosen because in any time interval $\Delta t = 0.25$ yr, our roughly monthly observations will lead to only 2 or 3 observations within the time interval so that a calculation of the speed of such a segment is unreliable. The uncertainties of all best-fit parameters are calculated from the diagonals of the co-variance matrices of the fits.

    \item \emph{Calculation of the Epoch of Ejection, $t_\circ$:} The epoch of ejection is defined as the time when the centroid of a knot coincides with the centroid of the core, with the knot motion extrapolated back to $(X,Y) = (0,0)$. In general, all observations of the motion of a knot are important for calculating its kinematic parameters. However, for determining $t_\circ$, the most important observations are when the knot is closest to the core, before any potential acceleration. We thus extrapolate the linear fits to $X$ and $Y$ back to $(0,0)$ from the first identified break point (or use the first ten observations of a knot in the case of no acceleration). This extrapolation provides the time of ejection along each coordinate, $t_{x\circ}$ and $t_{y\circ}$. We then calculate the true time of ejection as:

    \begin{equation}
        t_\circ = \frac{t_{\mathrm{x\circ}}/\sigma_{t_{\mathrm{x\circ}}}^2 + t_{\mathrm{y\circ}} / \sigma_{t_{\mathrm{y\circ}}}^2}{1/\sigma_{t_{\mathrm{x\circ}}}^2 + 1/\sigma_{t_{\mathrm{y\circ}}}^2}
        \label{eq:Tcirc}
    \end{equation}

    \noindent Here, the $1\sigma$ propagated uncertainties in $t_{x\circ}$ and $t_{y\circ}$ are defined as $\sigma_{t_{x\circ}}$ and $\sigma_{t_{y\circ}}$, respectively. The uncertainty in $t_\circ$ is then calculated as

    \begin{equation}
        \sigma_{t_\circ} = \sqrt{\frac{(t_\circ - t_{\mathrm{x\circ}})^2/\sigma_{t_{\mathrm{x\circ}}}^2 + (t_\circ - t_{\mathrm{y\circ}})^2 / \sigma_{t_{\mathrm{y\circ}}}^2}{1/\sigma_{t_{\mathrm{x\circ}}}^2 + 1/\sigma_{t_{\mathrm{y\circ}}}^2}}
         \label{eq:TcircErr}
    \end{equation}

    For a few cases, weighting $t_{x\circ}$ and $t_{y\circ}$ by uncertainties does not provide a robust solution since they are significantly different from each other, a situation that we define as $|t_{x\circ} - t_{y\circ}| \geq 0.5$ yr. This often occurs in the case of knot motion directly along the $X$ or $Y$ axis or for knots whose ejections occurred before our monitoring. In such cases, we construct a best-fit linear trend to the $R$ coordinate (either the first ten epochs of observation in the case of purely linear motion or the first segment of motion for knots showing acceleration) and find $t_\circ$ as its root, with its corresponding uncertainty.

    To each uncertainty in $t_\circ$, regardless of calculation method, we add a minimum uncertainty of $0.005 / \mu$ yr, where 0.005 mas is the size of half a pixel in the VLBA images and $\mu$ is the proper motion of the component (defined below), in mas yr$^{-1}$. This minimum error is typically $\sim3$-$4$ days and is generally much less than the error computed as indicated above.

    \item \emph{Calculation of Apparent Speeds:} Based on the piece-wise linear fits that best fit the $X$ and $Y$ data, we calculate the proper motion $\mu$, direction of motion $\Phi$, and apparent speed $\beta_{\mathrm{app}}$ for each time segment. We continue to define the speed and direction of the motion as in \citet{Jorstad2005}, with $\langle \mu_j \rangle = \sqrt{\langle \mu_{x,j} \rangle^2 + \langle \mu_{y,j} \rangle^2}$ and $\langle \Phi_j \rangle = \tan^{-1}(\langle\mu_{x,j} \rangle / \langle \mu_{y,j} \rangle)$, where the subscript $j$ indicates the number of the line segment. As the motion in each coordinate $X$ and $Y$ can be described independently by up to 3 linear segments, the maximum range of $j = 1, 2, \ldots, 5$. We arrange the break-points chronologically.%, so that in any given linear segment the speeds along the \emph{X} and \emph{Y} axes are constant, and at a break-point only the speed in either \emph{X} or \emph{Y} is changing at a given time. {\bf(THIS IS NOT TRUE IF $t_{b,X}=t_{b,Y}$)}

    \item \emph{Calculation of Apparent Accelerations:} If $j \neq 1$, there is a change in the speed and direction of motion of the knot over time as it moves down the jet. In the case of two or more line segments, we define a region in time between the mid-points of the line segments as the ``acceleration region," $\Delta t$. While the acceleration may occur over a shorter or longer time period than this mid-point definition, using the mid-points of a segment to determine $\Delta t$ characterizes the acceleration without \emph{a priori} knowledge of the cause of acceleration. The acceleration along each coordinate $X$ and $Y$ is defined as $\dot{\mu}_x = \Delta \mu_x / \Delta t$ and $\dot{\mu}_y = \Delta \mu_y / \Delta t$. Then the total acceleration parallel, $\dot{\mu}_\parallel$, and perpendicular, $\dot{\mu}_\perp$, to the jet axis can be calculated through $\dot{\mu}_\parallel = \dot{\mu}_x \sin{\langle \Theta_{\text{jet}} \rangle} + \dot{\mu}_y \cos{\langle \Theta_{\text{jet}} \rangle}$ and $\dot{\mu}_\perp = \dot{\mu}_x \cos{\langle \Theta_{\text{jet}} \rangle} - \dot{\mu}_y \sin{\langle \Theta_{\text{jet}} \rangle}$, where $\langle \Theta_{\text{jet}} \rangle$ is the average jet direction. How we calculate the average jet direction for each source is discussed in $\S$\ref{subsec:JetPAs}.
\end{enumerate}

\begin{figure*}[t]
    \figurenum{1}
    \begin{center}
        \includegraphics[width=0.65\textwidth]{{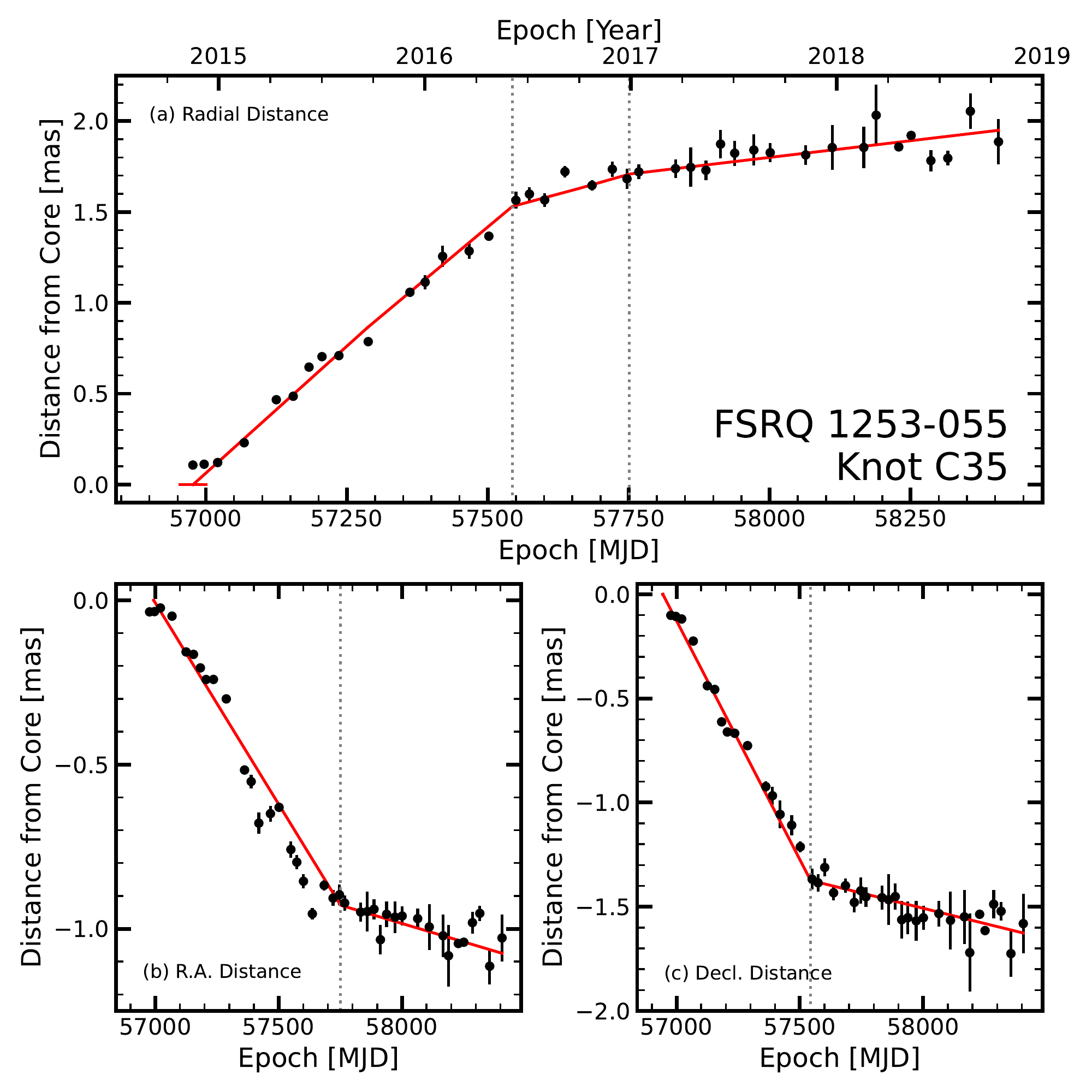}}
        \caption{An example of the piece-wise linear fitting procedure described in $\S$\ref{subsec:KnotIdentitySpeed} for the knot \emph{C35} of the FSRQ 1253--055. The three panels represent the total distances from the core along the (a) radial \emph{R}-axis, (b) \emph{X}-axis (relative right ascension), and (c) \emph{Y}-axis (relative declination). The red lines represent the piece-wise linear fits to the data, while gray dotted lines indicate break-point dates. The fit line in (a) is constructed from the combination of the fits in (b) and (c). The estimated epoch of ejection, $t_\circ$, is shown in (a) as a horizontal red line. See text for details.\label{fig:ExamplePiecewise}}
    \end{center}
\end{figure*}

To interpret the results of the jet kinematics in a uniform manner across ten years of observation, we have applied the above method to all of the knots presented in \citet{Jorstad2017}, in addition to the new knots observed since 2012 December. As an example of this fitting procedure, Figure~\ref{fig:ExamplePiecewise} shows the result for knot \emph{C35} of the FSRQ 1253$-$055. Panel (b) and (c) presents the linear fits to the separation of the knot versus time along the $X$ and $Y$ dimensions, respectively.
In this example, 2 segments are needed to represent both the $X$ and $Y$ motions accurately. Panel (a) shows the motion of the knot relative to the core along the radial $R$ dimension, combining the $X$ and $Y$ models.
For this example, there are three measurements of the speed of the knot and two estimates for the acceleration.

%-------------------------------------------------------------------------------
%---   PARSEC-SCALE JET STRUCTURE   --------------------------------------------
%-------------------------------------------------------------------------------

\section{Parsec-Scale Jet Structure}
\label{sec:ParsecScaleStructure}

Given the above determination of the kinematic parameters of the jet components, we have separated the knots into different types according to their properties. Images of all sources at all epochs possess a core (described above), which we label as \emph{A0}. All remaining components are classified based on their proper motion. Those with significant proper motion ($\mu \geq 2\sigma_\mu$) are designated as moving features of type \emph{M}. Components without this proper motion are termed ``quasi-stationary" (often shortened to ``stationary'' below) features of type \emph{St}. We continue the naming scheme introduced in \citet{Jorstad2017} for all previously analyzed sources. For the two sources added to the sample later (the BLs 1652+398 and 1959+650), designations start at 1 for both moving and quasi-stationary features.

We have calculated the average parameters of all features in all of the blazars, and list the results in Table~\ref{tab:JetStructure} in the following format:
1---B1950 name of the source;
2---ID of the knot;
3---the number of epochs, $N$, at which the knot is detected;
4---mean flux density, $\langle S \rangle$, of the knot and its standard deviation, in Jy;
5---average distance of the knot, $\langle R \rangle$, and its standard deviation, in mas\footnote{This average represents the typical distance at which the knot was observed and is not necessarily the mid-point of its trajectory.};
6---average position angle of the knot with respect to the core, $\langle \Theta \rangle$, and its standard deviation, in degrees. For the core, $\langle \Theta \rangle$ is calculated using the weighted average of the $\Theta$ of the all knots in the jet over all epochs at which $R \leq 0.7$ mas, and thus represents the average projected direction of the inner jet at a given epoch. Counter-jet features are not included in this calculation;
7---mean size of the knot, $\langle a \rangle$, and its standard deviation, in mas; and
8---kinematic type of the knot (see above).

\begin{deluxetable*}{lcrllllr}
  \tablecaption{Jet Structure\label{tab:JetStructure}}
  \tablewidth{0pt}
  \tablehead{
  \colhead{Source} & \colhead{Knot} & \colhead{$N$} & \colhead{$\langle S \rangle$} & \colhead{$\langle R \rangle$} & \colhead{$\langle \Theta \rangle$} & \colhead{$\langle a \rangle$} & \colhead{Type} \\
  \colhead{} & \colhead{} & \colhead{} & \colhead{[Jy]} & \colhead{[mas]} & \colhead{[deg]} & \colhead{[mas]} & \colhead{}
  }
  \colnumbers
  \startdata
  0219+428 & A0 & 62 & $0.24 \pm 0.09$ & $0.00 \pm 0.02$ & $-166.0 \pm 10.0$     & $0.04 \pm 0.02$ & Core\\
           & A1 & 52 & $0.08 \pm 0.04$ & $0.17 \pm 0.04$ & $-167.9 \pm 10.9$     & $0.09 \pm 0.03$ & St\\
           & A2 & 36 & $0.04 \pm 0.02$ & $0.37 \pm 0.06$ & $-168.5 \pm \phn 9.5$ & $0.14 \pm 0.04$ & St\\
           & A3 & 50 & $0.03 \pm 0.01$ & $0.65 \pm 0.09$ & $-173.2 \pm \phn 5.8$ & $0.21 \pm 0.09$ & St\\
           & A4 & 56 & $0.03 \pm 0.01$ & $2.31 \pm 0.09$ & $-177.1 \pm \phn 1.4$ & $0.46 \pm 0.14$ & St\\
           & B1 & 6  & $0.02 \pm 0.01$ & $1.03 \pm 0.19$ & $-175.7 \pm \phn 1.3$ & $0.27 \pm 0.06$ & M\\
           & B2 & 10 & $0.02 \pm 0.01$ & $1.04 \pm 0.22$ & $-172.7 \pm \phn 6.0$ & $0.40 \pm 0.10$ & M\\
           & B3 & 6  & $0.04 \pm 0.03$ & $0.48 \pm 0.26$ & $-160.7 \pm 18.3$     & $0.15 \pm 0.11$ & M\\
           & B4 & 7  & $0.01 \pm 0.01$ & $1.35 \pm 0.45$ & $-168.6 \pm \phn 1.3$ & $0.41 \pm 0.10$ & M\\
           & B5 & 11 & $0.02 \pm 0.02$ & $0.94 \pm 0.56$ & $-169.5 \pm \phn 3.2$ & $0.28 \pm 0.14$ & M\\
           & B6 & 4  & $0.01 \pm 0.01$ & $1.40 \pm 0.24$ & $-177.5 \pm \phn 3.2$ & $0.43 \pm 0.06$ & M\\
  0235+164\tablenotemark{a} & A0 & 126 & $1.30 \pm 0.83$ & $0.00 \pm 0.02$ & $\phantom{-000.0}\ldots$         & $0.06 \pm 0.03$ & Core\\
           & B1 & 6  & $0.16 \pm 0.23$ & $0.40 \pm 0.17$ & $-\phn 17.7 \pm \phn 5.8$ & $0.26 \pm 0.13$ & M \\
           & B2 & 15 & $1.03 \pm 0.88$ & $0.22 \pm 0.08$ & $\phantom{-}163.1 \pm 12.2$ & $0.15 \pm 0.06$ & M \\
           & B3 & 40 & $0.25 \pm 0.14$ & $0.21 \pm 0.07$ & $\phantom{-}164.9 \pm \phn 6.4$ & $0.13 \pm 0.05$ & M\\
           & B4 & 16 & $0.15 \pm 0.11$ & $0.19 \pm 0.06$ & $-\phn 63.4 \pm 21.0$ & $0.13 \pm 0.10$ & M\\
           & B5 & 17 & $0.41 \pm 0.29$ & $0.28 \pm 0.10$ & $\phantom{-0} 59.1 \pm \phn 4.0$ & $0.20 \pm 0.07$ & M\\
           & B6 & 19 & $0.19 \pm 0.16$ & $0.27 \pm 0.11$ & $\phantom{-0} 51.2 \pm \phn 6.2$ & $0.23 \pm 0.14$ & M\\
           & B7 & 6  & $0.04 \pm 0.01$ & $0.26 \pm 0.03$ & $\phantom{-00} 9.6 \pm \phn 5.5$ & $0.11 \pm 0.05$ & M\\
  \enddata
  \tablenotetext{a}{The direction of the inner jet for 0235+164 is uncertain, since features detected in the jet have a range of P.A. exceeding 180$\degr$.}
  \tablecomments{(Table~\ref{tab:JetStructure} (559 lines) is published in its entirety in the machine-readable format. A portion is shown here for guidance regarding its form and content.)}
\end{deluxetable*}

Figure~\ref{fig:TotalIntensityImages} presents an image of each source at a single epoch between 2013 January and 2018 December, when the most prominent features in the jet were visible. The black circles indicate the sizes and positions of the features according to the model parameters for that epoch. These images do not contain labels for all observed features in the jet over the 10 years of observations. We list the important parameters for each image in Table~\ref{tab:TotalIntensityImageTable} as follows:
1---B1950 name of the source;
2---epoch of observation;
3---total intensity peak of the map, $I_{\text{peak}}$, in mJy beam$^{-1}$;
4---lowest shown contour, $I_{\text{low}}^{\text{cnt}}$, in mJy beam$^{-1}$;
5---size of the restoring beam, used to produce images of similar fidelity across all epochs, in units of mas $\times$ mas, with all restoring beam orientation angles set to $-10\degr$;
6---antennas with no data for the source at that particular observation epoch
and
7---size and orientation angle of the synthesized beam corresponding to the $uv$-coverage
for the epoch of observation, in units of mas $\times$ mas and degrees.

\begin{figure*}
  \figurenum{2}
  \begin{center}
    \includegraphics[width=0.95\textwidth]{{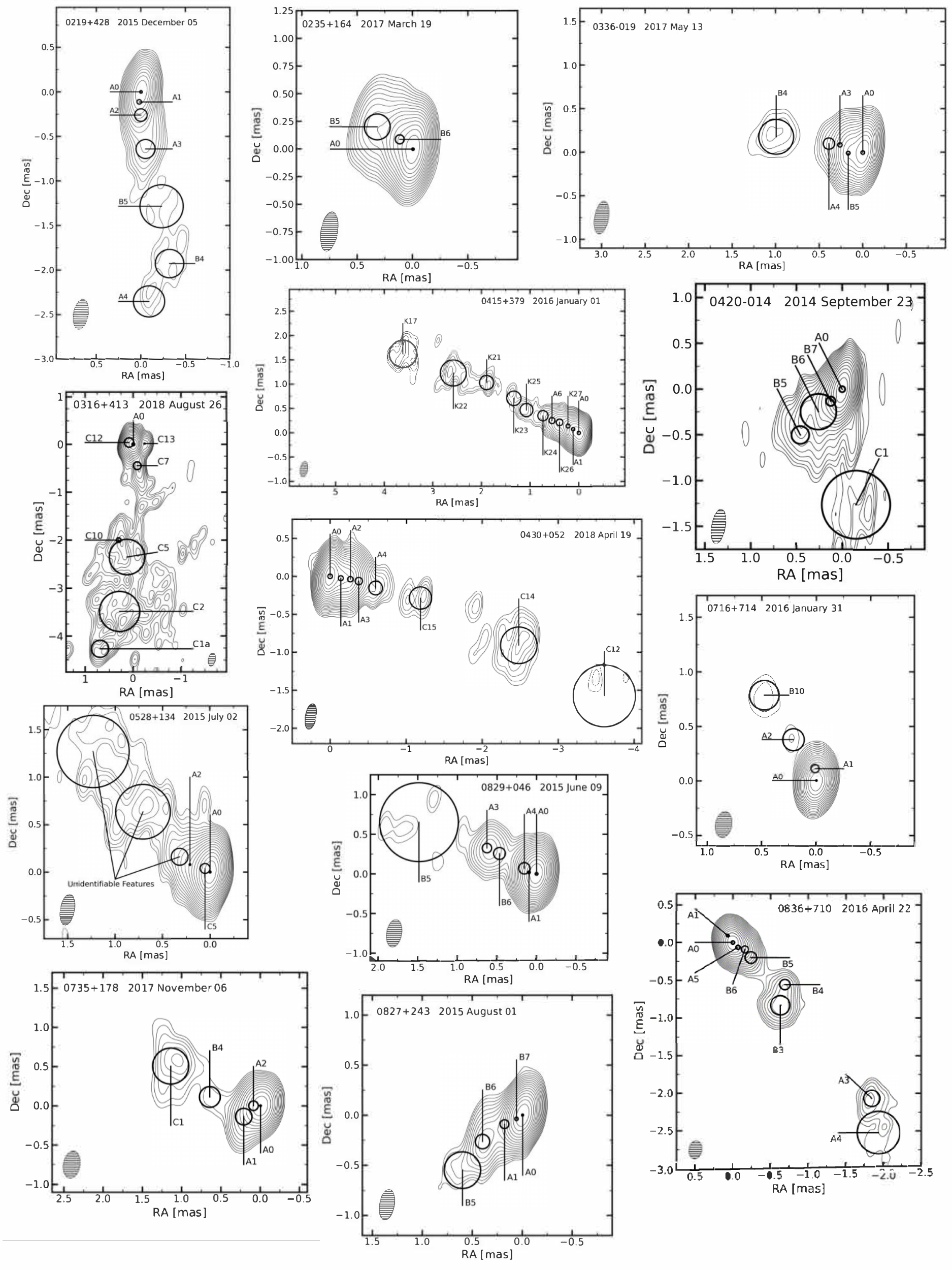}}
    \caption{Total intensity images of the VLBA-BU-BLAZAR sample at 43 GHz with uniform weighting. Parameters of all images are listed in Table~\ref{tab:TotalIntensityImageTable}. Contours decrease by a factor of $\sqrt{2}$ from the peak flux density. Black circles on the images indicate the jet features according to the model fits. The shaded ellipse in the bottom left corner represents the restoring beam. \label{fig:TotalIntensityImages}}
  \end{center}
\end{figure*}

\begin{figure*}
  \figurenum{2}
  \begin{center}
    \includegraphics[width=0.95\textwidth]{{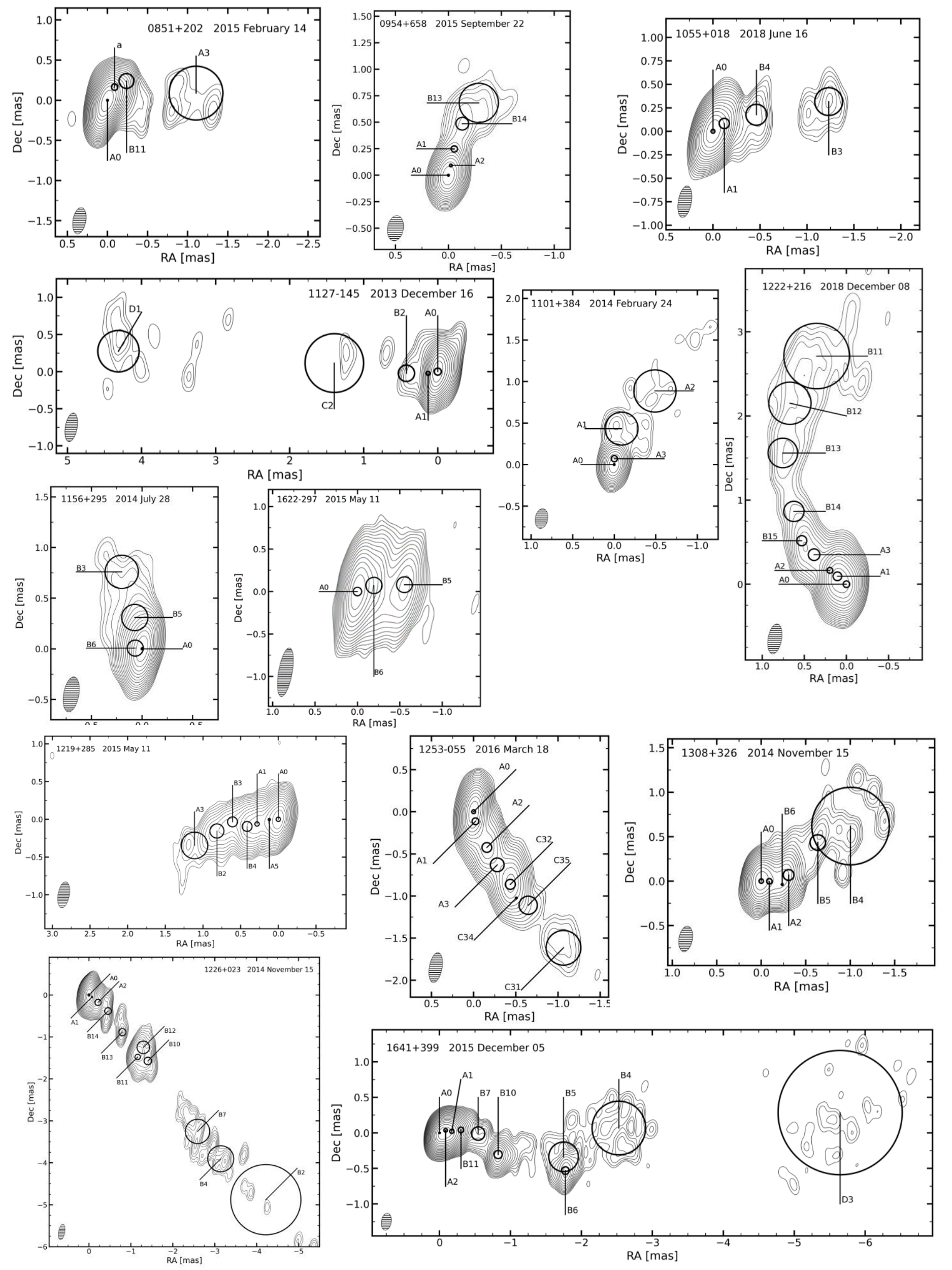}}
    \caption{(Continued.)\label{fig:TotalIntensityImagesb}}
  \end{center}
\end{figure*}

\begin{figure*}
  \figurenum{2}
  \begin{center}
    \includegraphics[width=0.95\textwidth]{{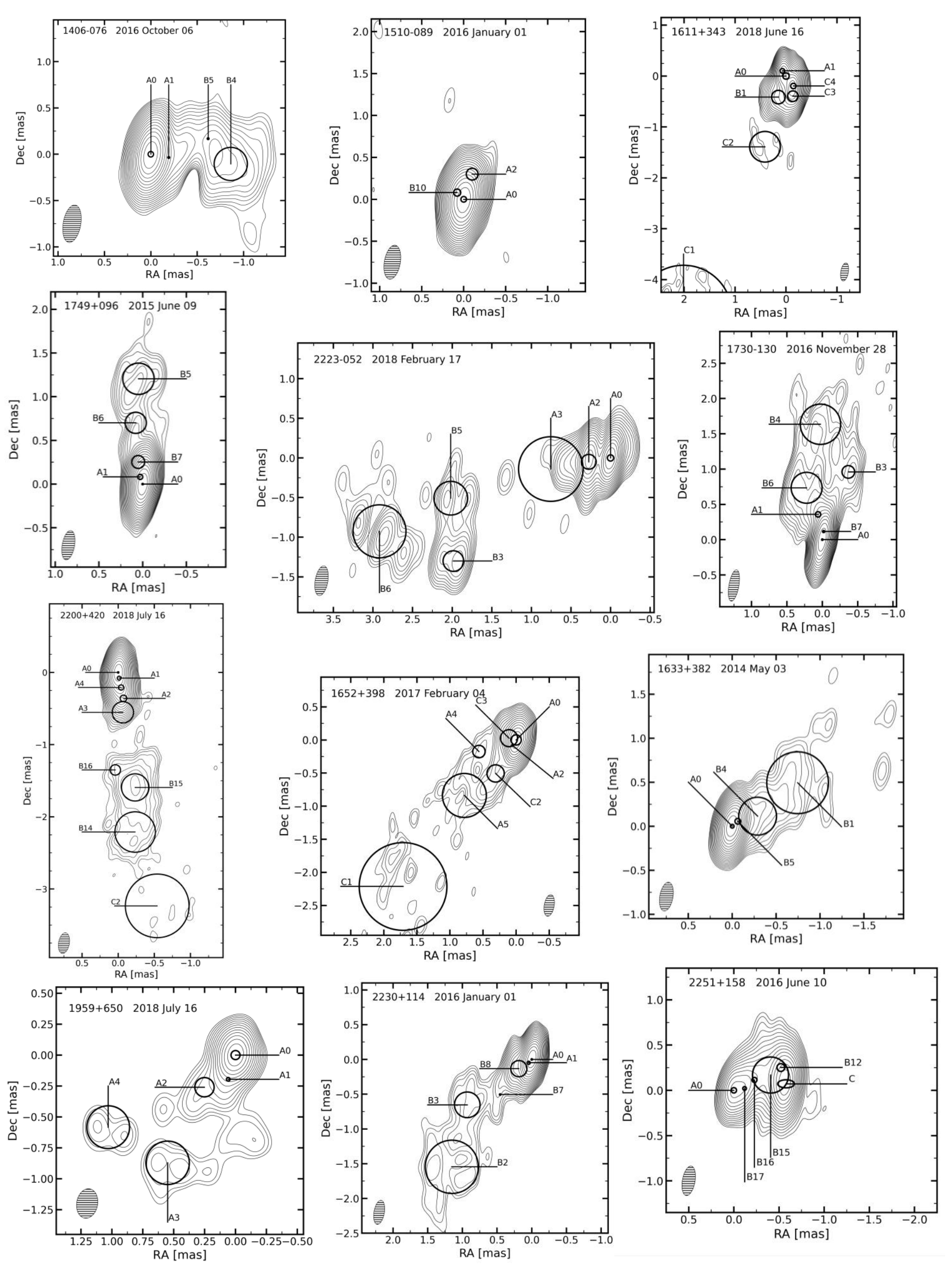}}
    \caption{(Continued.)\label{fig:TotalIntensityImagesc}}
  \end{center}
\end{figure*}

\begin{deluxetable*}{llrrrcr}
  \tablecaption{Parameters of VLBA 43 GHz Images in Figure~\ref{fig:TotalIntensityImages}.\label{tab:TotalIntensityImageTable}}
  \tablewidth{0pt}
  \tablehead{
  \colhead{Source} & \colhead{Epoch} & \colhead{$I_{\text{peak}}$} & \colhead{$I_{\text{low}}^{\text{cnt}}$} & \colhead{Beam Size} & \colhead{Unavailable} & \colhead{Synthesized Beam Size} \\
  \colhead{} & \colhead{} &  \colhead{} & \colhead{} & \colhead{} & \colhead{Antennas\tablenotemark{a}} & \colhead{} \\
  \colhead{} & \colhead{} & \colhead{[mJy beam$^{-1}$]} & \colhead{[mJy beam$^{-1}$]} & \colhead{[mas $\times$ mas]} & \colhead{} & \colhead{[mas $\times$ mas, deg]}
  }
  \colnumbers
  \startdata
  0219+428   & 2015 Dec 05 &  233 &  1.82 & $0.17 \times 0.34$ & $\ldots$ & $0.16 \times 0.28,\ -\phn 4.1$ \\
  0235+164   & 2017 Mar 19 &  866 &  3.38 & $0.15 \times 0.36$ & $\ldots$ & $0.16 \times 0.33,\ -\phn 3.4$ \\
  0316+413   & 2018 Aug 26 & 1572 & 12.28 & $0.15 \times 0.28$ & $\ldots$ & $0.16 \times 0.30,\ \phantom{-}10.0$ \\
  0336--019  & 2017 May 13 & 1214 &  6.71 & $0.16 \times 0.38$ & $\ldots$ & $0.16 \times 0.37,\ -\phn 2.8$ \\
  0415+379   & 2016 Jan 01 &  578 &  2.26 & $0.16 \times 0.32$ & $\ldots$ & $0.16 \times 0.33,\ -\phn 2.5$ \\
  0420--014  & 2014 Sep 23 & 1207 &  4.71 & $0.15 \times 0.38$ & HN       & $0.16 \times 0.39,\ -\phn 2.9$ \\
  0430+052   & 2018 Apr 19 & 1117 &  2.18 & $0.14 \times 0.34$ & $\ldots$ & $0.15 \times 0.36,\ -\phn 4.4$ \\
  0528+134   & 2015 Jul 02 &  366 &  0.72 & $0.15 \times 0.33$ & $\ldots$ & $0.16 \times 0.37,\ -\phn 0.9$ \\
  0716+714   & 2016 Jan 31 & 1498 &  8.28 & $0.15 \times 0.24$ & $\ldots$ & $0.16 \times 0.28,\ -\phn 7.1$ \\
  0735+178   & 2017 Nov 06 &  204 &  2.25 & $0.21 \times 0.35$ & BR, SC   & $0.16 \times 0.73,\ -20.0$     \\
  0827+243   & 2015 Aug 01 &  322 &  1.26 & $0.15 \times 0.31$ & $\ldots$ & $0.16 \times 0.31,\ -\phn 3.1$ \\
  0829+046   & 2015 Jun 09 &  723 &  2.83 & $0.19 \times 0.35$ & $\ldots$ & $0.17 \times 0.39,\ -\phn 7.8$ \\
  0836+710   & 2016 Apr 22 &  854 &  3.33 & $0.17 \times 0.24$ & $\ldots$ & $0.16 \times 0.24,\ - 16.0$    \\
  0851+202   & 2015 Feb 14 & 4807 &  9.39 & $0.16 \times 0.33$ & FD       & $0.20 \times 0.57,\ \phantom{-}26.0$ \\
  0954+658   & 2015 Sep 22 &  723 &  3.99 & $0.15 \times 0.24$ & $\ldots$ & $0.17 \times 0.24,\ \phantom{-} \phn 8.5$\\
  1055+018   & 2018 Jun 16 & 2195 &  8.57 & $0.14 \times 0.34$ & SC       & $0.15 \times 0.68,\ - 18.0$ \\
  1101+384   & 2014 Feb 24 &  281 &  3.10 & $0.15 \times 0.24$ & FD       & $0.16 \times 0.28,\ - 21.0$ \\
  1127--145  & 2013 Dec 16 & 1112 &  4.34 & $0.16 \times 0.39$ & KP       & $0.17 \times 0.46,\ -\phn 8.2$\\
  1156+295   & 2014 Jul 28 & 1087 &  4.25 & $0.15 \times 0.35$ & $\ldots$ & $0.16 \times 0.31,\ \phantom{-} \phn 0.5$\\
  1219+285   & 2015 May 11 &  205 &  0.80 & $0.15 \times 0.35$ & $\ldots$ & $0.17 \times 0.30,\ -\phn 8.3$\\
  1222+216   & 2018 Dec 08 &  473 &  1.85 & $0.16 \times 0.36$ & $\ldots$ & $0.16 \times 0.33,\ -10.0$ \\
  1226+023   & 2014 Nov 15 & 3612 &  7.05 & $0.15 \times 0.38$ & $\ldots$ & $0.18 \times 0.44,\ -\phn 5.6$ \\
  1253--055  & 2016 Mar 18 & 6266 & 24.47 & $0.15 \times 0.36$ & $\ldots$ & $0.17 \times 0.40,\ -\phn 4.6$\\
  1308+326   & 2014 Nov 15 &  571 &  2.23 & $0.15 \times 0.29$ & $\ldots$ & $0.17 \times 0.35,\ - 15.0$\\
  1406--076  & 2016 Oct 06 &  654 &  1.28 & $0.19 \times 0.41$ & HN, MK   & $0.20 \times 0.82,\ \phantom{-}23.0$\\
  1510--089  & 2016 Jan 01 & 4277 &  5.91 & $0.19 \times 0.41$ & $\ldots$ & $0.14 \times 0.45,\ -10.0$\\
  1611+343   & 2018 Jun 16 &  616 &  1.70 & $0.15 \times 0.36$ & SC       & $0.16 \times 0.37,\ -27.0$\\
  1622--297  & 2015 May 11 &  589 &  3.25 & $0.16 \times 0.58$ & HN       & $0.13 \times 0.40,\ -\phn 4.8$\\
  1633+382   & 2014 May 03 & 2230 &  4.36 & $0.15 \times 0.33$ & NL       & $0.15 \times 0.30,\ -\phn 9.2$\\
  1641+399   & 2015 Dec 05 & 1800 &  1.76 & $0.14 \times 0.24$ & $\ldots$ & $0.15 \times 0.28,\ - 17.0$\\
  1652+398   & 2017 Feb 04 &  175 &  0.97 & $0.15 \times 0.33$ & BR       & $0.15 \times 0.31,\ - 20.0$\\
  1730--130  & 2016 Nov 28 & 1812 &  3.54 & $0.14 \times 0.45$ & $\ldots$ & $0.14 \times 0.38,\ -\phn 4.5$\\
  1749+096   & 2015 Jun 09 & 3310 &  3.23 & $0.14 \times 0.34$ & $\ldots$ & $0.15 \times 0.36,\ -\phn 7.0$ \\
  1959+650   & 2018 Jul 16 &  126 &  1.39 & $0.17 \times 0.24$ & HN, PT, SC & $0.20 \times 0.34,\ -\phn 7.0$ \\
  2200+420   & 2018 Jul 16 & 1745 &  1.70 & $0.15 \times 0.29$ & $\ldots$ & $0.16 \times 0.25,\ -\phn 8.1$\\
  2223--052  & 2018 Feb 17 &  486 &  1.90 & $0.16 \times 0.38$ & MK, SC   & $0.27 \times 0.81,\ -\phn 5.0$\\
  2230+114   & 2016 Jan 01 & 1840 &  3.59 & $0.15 \times 0.35$ & $\ldots$ & $0.14 \times 0.40,\ -\phn 7.9$\\
  2251+158   & 2016 Jun 10 & 5115 & 28.26 & $0.14 \times 0.33$ & $\ldots$ & $0.13 \times 0.33,\ -\phn 6.1$\\
  \enddata
  \tablenotetext{a}{Antenna abbreviations can be found at the NRAO VLBA website (\url{https://science.nrao.edu/facilities/vlba/docs/manuals/oss/sites}). Individual antennas are sometimes unavailable due to weather, maintenance, or undetected fringes (in the case of the weakest sources, e.g., 1959+650).}
\end{deluxetable*}

\subsection{Observed Brightness Temperatures}
\label{subsec:ObsBrightnessTemps}

The observed brightness temperatures of all jet features from 2007 June to 2012 December are listed in Table 2 of \citet{Jorstad2017}, while those from 2013 January to 2018 December are listed in Table~\ref{tab:GaussianComponents} of this paper. Combined, there are a total of 19,644 estimates for the brightness temperatures of components. Of these, 10.1\% of the $T_{\text{b,obs}}$ values are lower limits (see $\S$\ref{subsec:ModelingTotalIntensityImages}).

\begin{figure*}
    \figurenum{3}
    \begin{center}
        \includegraphics[width=\textwidth]{{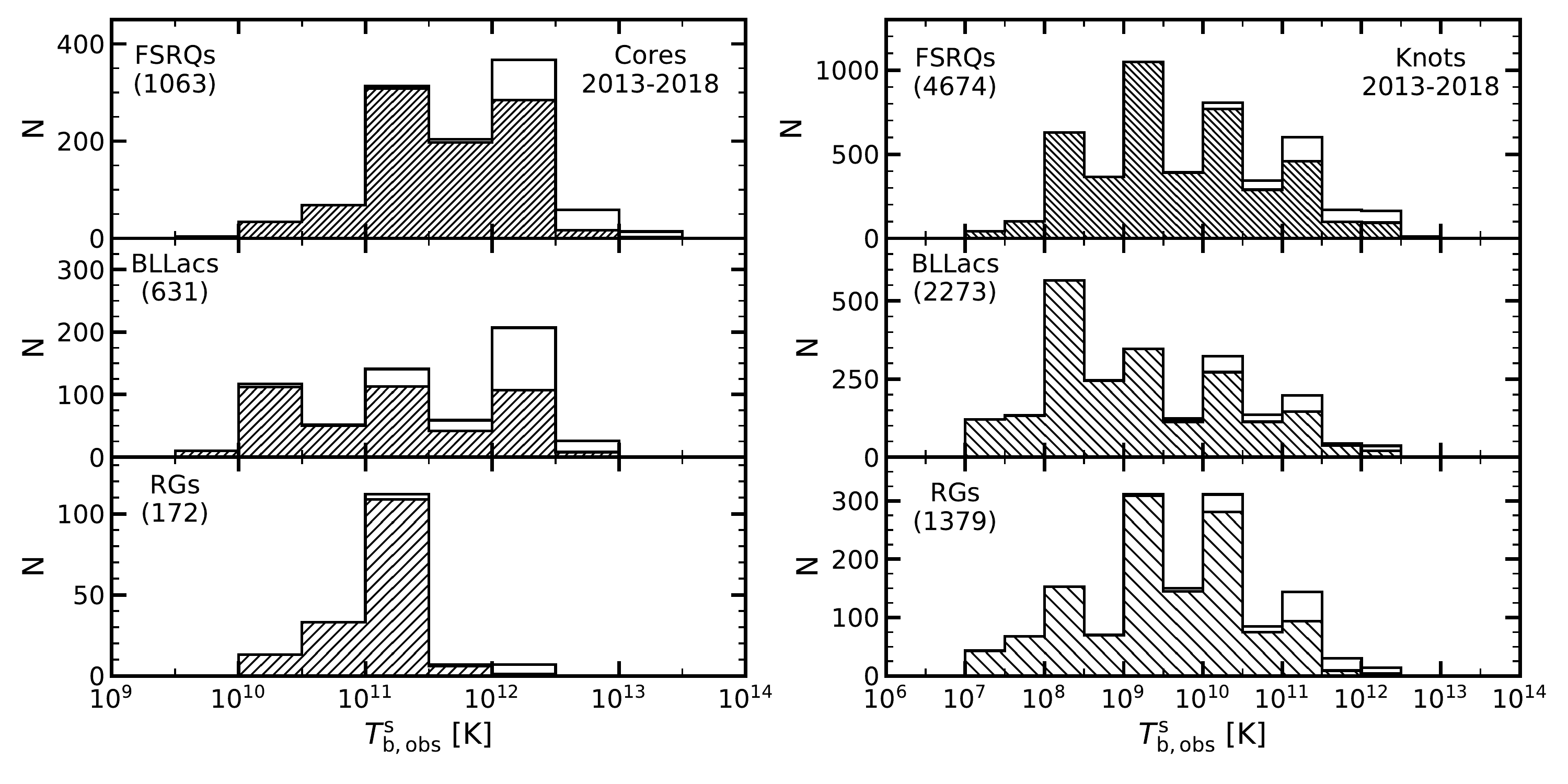}}
        \caption{Host-galaxy frame brightness temperature distributions of the core (left) and non-core (knots, right) features between 2013 January and 2018 December. Unshaded regions of the distributions represent components with lower limits to $T_{\mathrm{b,obs}}^{\mathrm{s}}$. The bin sizes are neither linear nor logarithmic. Values in parentheses are the number of $T_{\text{b,obs}}$ measurements of components for each subclass, including lower limits. \label{fig:JorstadTbs}}
    \end{center}
\end{figure*}

For a direct comparison with the brightness temperatures analyzed in \citet{Jorstad2017}, Figure~\ref{fig:JorstadTbs} (left) displays the brightness temperatures of all cores in the host galaxy frame ($T_{\mathrm{b,obs}}^{\mathrm{s}} = T_{\mathrm{b,obs}}(1+z)$) from 2013 January to 2018 December for each subclass of blazar. The RG distribution peaks at $(1-5) \times 10^{11}$ K. Only a few cores have epochs when $T_{\mathrm{b,obs}}^{\mathrm{s}} \geq 5 \times 10^{11}$ K. The FSRQ distribution is bimodal, with peaks at both $(1-5) \times 10^{11}$ K and $(1-5) \times 10^{12}$ K. The highest peak of the BL distribution is also at $(1-5) \times 10^{12}$ K, but includes a large number of lower limits. The BL distribution also has the most significant number of cores with low brightness temperatures $(T_{\mathrm{b,obs}}^{\mathrm{s}} < 10^{11}$ K). The distributions of the brightness temperatures for knots, other than the cores, between 2013 January and 2018 December are shown in Figure~\ref{fig:JorstadTbs} (right). In general, these knots have lower brightness temperatures than the cores, but with much wider dispersion about the mean. Both distributions are similar in shape to those presented in \citet{Jorstad2017}, indicating no apparent time variability in the populations of brightness temperatures of components as a whole for the sources in our sample.

A direct comparison of the distributions for each subclass presented in Figure~\ref{fig:JorstadTbs} (as with a Kolmogorov-Smirnov (KS) Test) would ignore two crucial aspects of the data: (1) that the brightness temperature values for the cores and the knots of a particular object in our sample are correlated by virtue of coming from the same source, rather than being an independent measurement from a random blazar, and (2) the relatively large number (in some cases, $\sim50\%$, e.g., the core of the BL 1959+650) of lower limits present in the sample. In order to compare the brightness temperatures of the cores and knots in our sample between each subclass, we derive ``typical" values for the brightness temperatures of components for each source by estimating the survival function of the brightness temperature of the cores and knots to get the median value (see Appendix~\ref{app:survival}).

Figure~\ref{fig:CoreHists} (left) shows the median brightness temperatures of core components of blazars in our sample between 2013 January and 2018 December, while Fig.~\ref{fig:CoreHists} (right) shows the same for the entire observing period of the program (2007 June - 2018 December). The three RGs in the sample have very similar median core brightness temperatures, $\sim5\times10^{11}$ K. The distribution of the median $T_{\mathrm{b,obs}}^{\mathrm{s}}$ for BLs is very wide-spread, ranging from $10^{10}$ K to $\sim10^{13}$ K. The cores of FSRQs are more narrowly distributed around a central brightness temperature, $\sim 10^{12}$ K. As the process of calculating the median brightness temperature has already taken into account the presence of lower-limits, we can now meaningfully compare the distributions for each subclass through a KS test. Values of the test-statistic, $\mathcal{D}$, and significance, $p$, values for the test between each subclass distribution are given in Table~\ref{tab:KSCores}. The differences between the FSRQ and RG core distributions are statistically significant at a $\sim3\sigma$ level, with $p = 0.009$, once all cores from 2007 June to 2018 December have been taken into account. The FSRQ and BL cores also appear to be different, but at a less significant level ($p = 0.086$). There is no statistical difference between the BL and RG distributions, but it is important to note that the number of sources is small (13 for BLs and 3 for RGs).

\begin{figure*}
    \figurenum{4}
    \begin{center}
        \includegraphics[width=0.75\textwidth]{{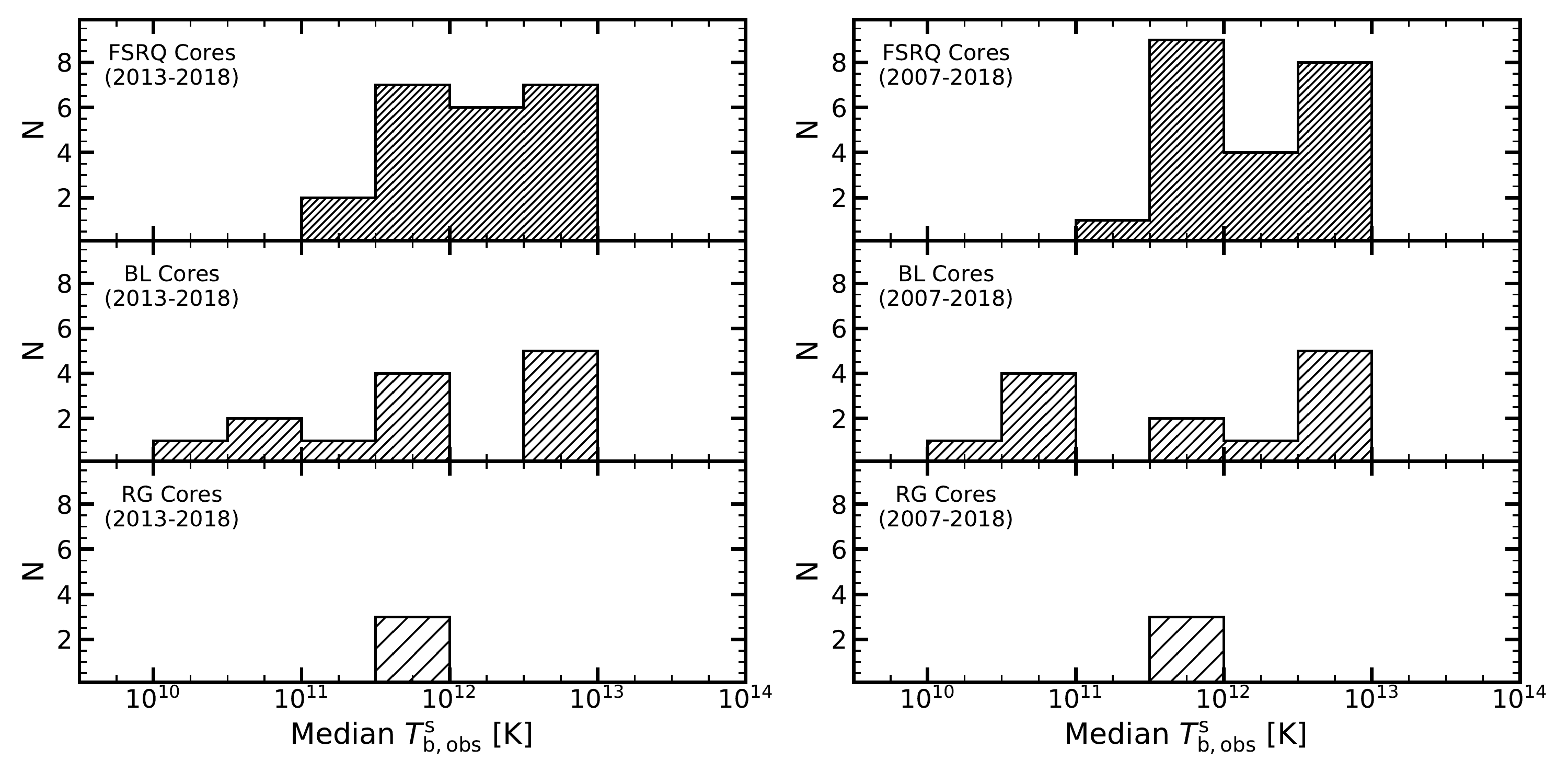}}
        \caption{Median core component host-galaxy frame brightness temperature distributions of blazars in the VLBA-BU-BLAZAR sample, as determined through survival analysis. The bin sizes are neither linear nor logarithmic. \label{fig:CoreHists}}
    \end{center}
\end{figure*}

\begin{deluxetable}{llcc}
    \tablecaption{KS test statistic, $\mathcal{D}$-, and $p$-values for the median brightness temperatures of jet cores.\label{tab:KSCores}}
    \tablewidth{0pt}
    \tablehead{
            \colhead{Subclass 1} & \colhead{Subclass 2} & \colhead{$\mathcal{D}$} & \colhead{$p$}
    }
    \startdata
    FSRQ & BL & 0.388 & 0.129 \\
    FSRQ & RG & 0.818 & 0.030 \\
    BL   & RG & 0.615 & 0.229 \\
    \tableline
    FSRQ & BL & 0.416 & 0.086 \\
    FSRQ & RG & 0.909 & 0.009 \\
    BL   & RG & 0.538 & 0.375 \\
    \enddata
    \tablecomments{The data for section 1 of this table consist of observations taken from 2013 January through 2018 December. The data for section 2 were taken from 2007 June through 2018 December.}
\end{deluxetable}

\begin{figure*}
    \figurenum{5}
    \begin{center}
        \includegraphics[width=0.75\textwidth]{{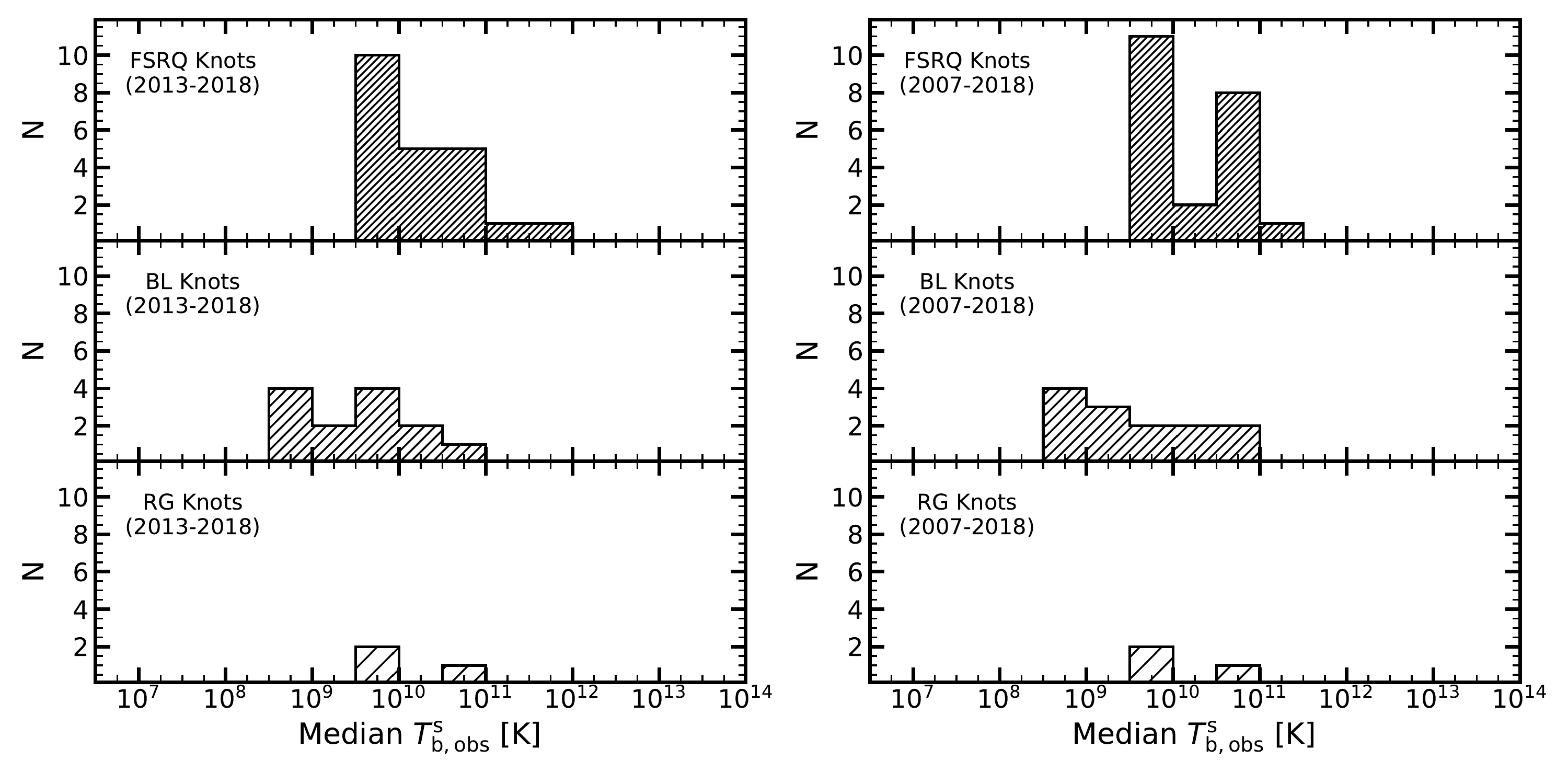}}
        \caption{Median knot component host-galaxy frame brightness temperature distributions of blazars in the VLBA-BU-BLAZAR sample, as determined through survival analysis. The bin sizes are neither linear nor logarithmic. \label{fig:KnotHists}}
    \end{center}
\end{figure*}

Similarly to the core distributions, in Figure~\ref{fig:KnotHists} we show the median knot brightness temperature distributions for each subclass, including knots observed from 2013 January to 2018 December (left) and 2007 June to 2018 December (right). In general, the median knot brightness temperature distribution for each subclass is centered on lower values than that of the core components. The knots of FSRQs are more concentrated around the mean value than the knots of BLs. The knots of the RG 0316+428 have a higher median $T_{\mathrm{b,obs}}^{\mathrm{s}}$ than the other two RG in the sample. The $\mathcal{D}$- and $p$-values for a KS test between the knot distributions are given in Table~\ref{tab:KSKnots}. The only distributions that show a statistically significant difference are the knots of FSRQs compared with BLs, at a significance $p = 0.002$ for the entire observed time period.

\begin{deluxetable}{llcc}
    \tablecaption{KS test statistic, $\mathcal{D}$-, and $p$-values for the median brightness temperatures of jet knots.\label{tab:KSKnots}}
    \tablewidth{0pt}
    \tablehead{
            \colhead{Subclass 1} & \colhead{Subclass 2} & \colhead{$\mathcal{D}$} & \colhead{$p$}
    }
    \startdata
    FSRQ & BL & 0.538 & 0.010 \\
    FSRQ & RG & 0.363 & 0.750 \\
    BL   & RG & 0.615 & 0.229 \\
    \tableline
    FSRQ & BL & 0.615 & 0.002 \\
    FSRQ & RG & 0.439 & 0.574 \\
    BL   & RG & 0.615 & 0.229 \\
    \enddata
    \tablecomments{The data for section 1 of this table consist of observations taken from 2013 January through 2018 December. The data for section 2 were taken from 2007 June through 2018 December.}
\end{deluxetable}

Despite the low significance levels according to a KS test (likely due to the number of sources in our sample), we see a general trend across the median core and knot brightness temperature distributions: \emph{FSRQs and BLs have more intense cores than those of RGs, but the knots in the extended jets of BLs are significantly less intense than the knots of RGs and FSRQs.} A small fraction of analyzed knots are weak diffuse features with $T_{\text{b,obs}}^{\text{s}} < 10^8$ K. The highest brightness temperatures, near $10^{13}$ K, are discussed in $\S$\ref{subsec:TbandDelta}.

\subsection{Jet Position Angles}
\label{subsec:JetPAs}

With over 10 years of observations, we can constrain the bulk characteristics of the jet, such as the average position angle, $\langle \Theta_{\text{jet}} \rangle$, and its temporal evolution. Such constraints are important for determining whether there are sudden changes in the direction of the jet, or if precession is common in blazars. In addition, since the calculation of acceleration along or perpendicular to the jet axis depends on the position angle, such information can allow us to account for the effect of changes in $ \Theta_{\text{jet}}$ on the accelerations.

\begin{figure*}[t]
    \figurenum{6}
    \begin{center}
        \includegraphics[width=\textwidth]{{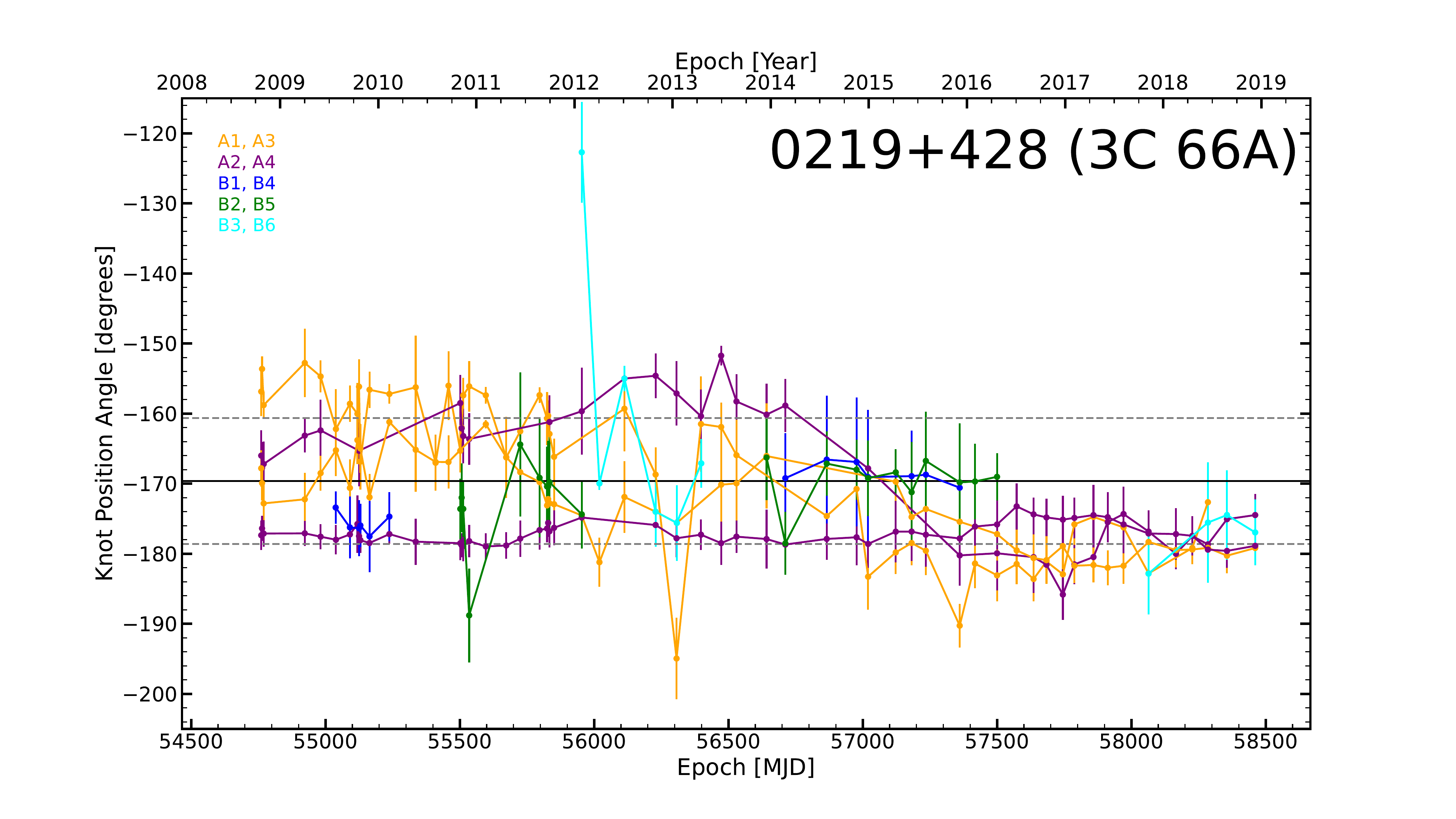}}
        \caption{Jet position angles of each knot, relative to the core, of the BL 0219+428 (3C 66A). The colors of each knot are the same as in Figure~\ref{fig:ExampleR}, and are designated in the upper left corner. We classify this source as having a constant jet position angle according to our criterion (see the text and Table~\ref{tab:JetPAs}).
        The black line indicates $\langle \Theta_{\text{jet}} \rangle$, and the two gray dashed lines show the standard deviation of the PA with respect to $\Theta_{\text{jet}}$, $\sigma_{\langle \Theta_{\text{jet}} \rangle}$, which is listed in column 7 of Table~\ref{tab:JetPAs}.
        Similar plots for all sources in the sample are available in the figure set. \newline (The complete figure set (38 images) is available online).\label{fig:ExampleTheta}}
    \end{center}
\end{figure*}

In order to determine how the jet position angles vary, we have constructed plots of the observed position angle, $\Theta$, versus time for all knots in each source. Figure~\ref{fig:ExampleTheta} shows an example of such plots for the source 0219+428. Figure SET 6 provides a similar plot for each source in the sample.

%---- Figure Set 6
\figsetstart
    \label{figset:6}
    \figsetnum{6}
    \figsettitle{Jet Position Angles}
        % Number 1
        \figsetgrpstart
        \figsetgrpnum{6.1}
        \figsetgrptitle{0219+428}
        \figsetplot{{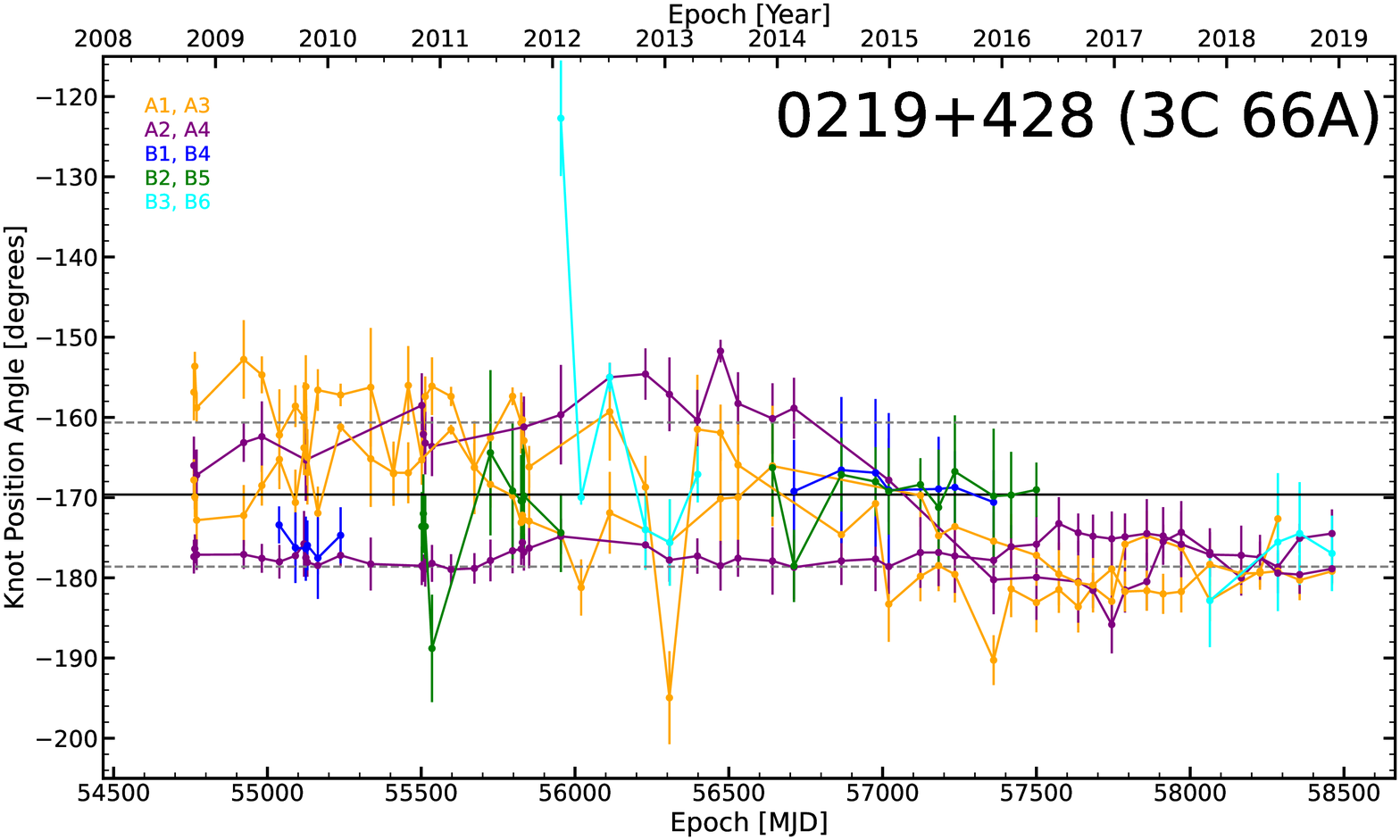}}
        \figsetgrpnote{Jet position angles of each knot, relative to the core, of the BL 0219+428.}
        \figsetgrpend
        %
        % Number 2
        \figsetgrpstart
        \figsetgrpnum{6.2}
        \figsetgrptitle{0235+164}
        \figsetplot{{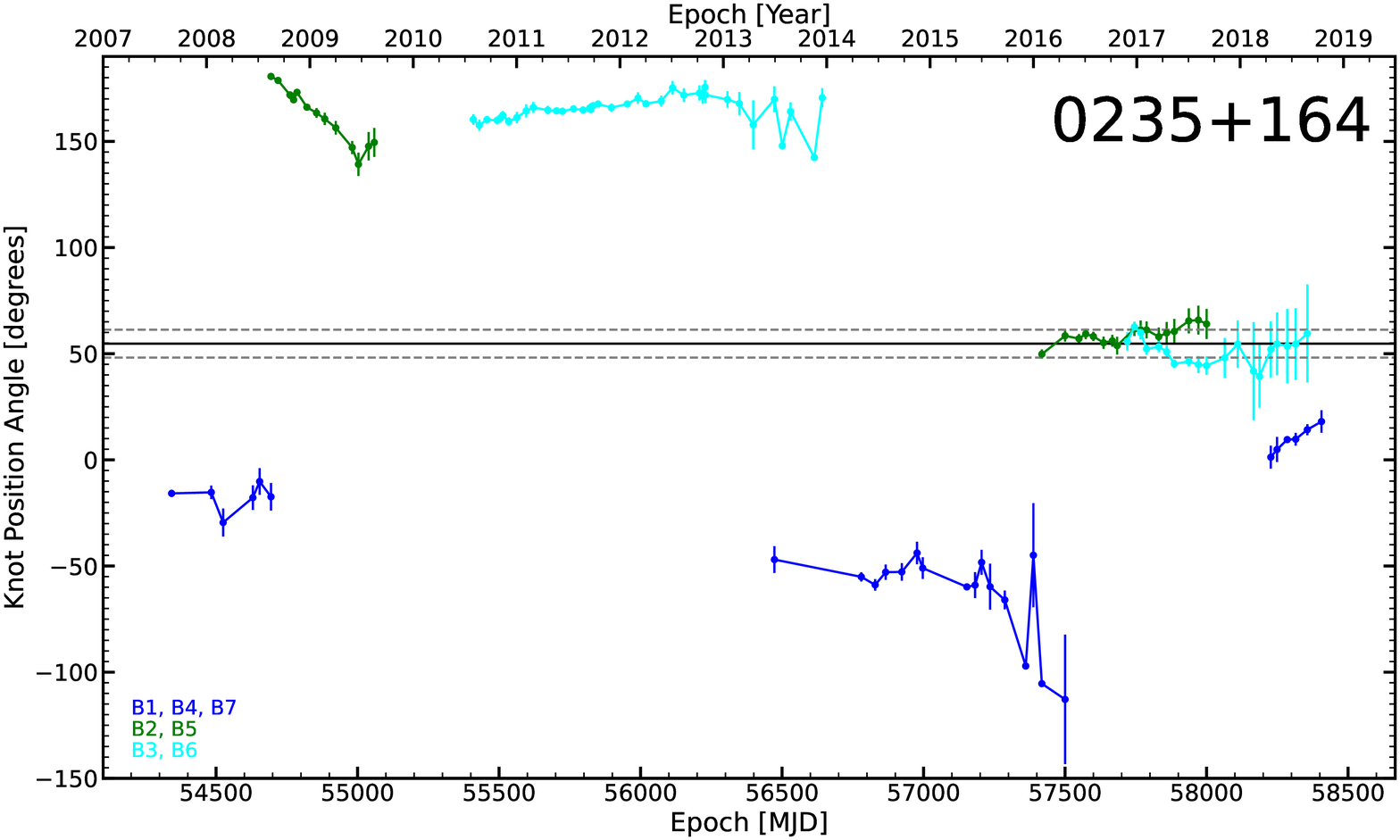}}
        \figsetgrpnote{Jet position angles of each knot, relative to the core, of the BL 0235+164.}
        \figsetgrpend
        %
        % Number 3
        \figsetgrpstart
        \figsetgrpnum{6.3}
        \figsetgrptitle{0316+413}
        \figsetplot{{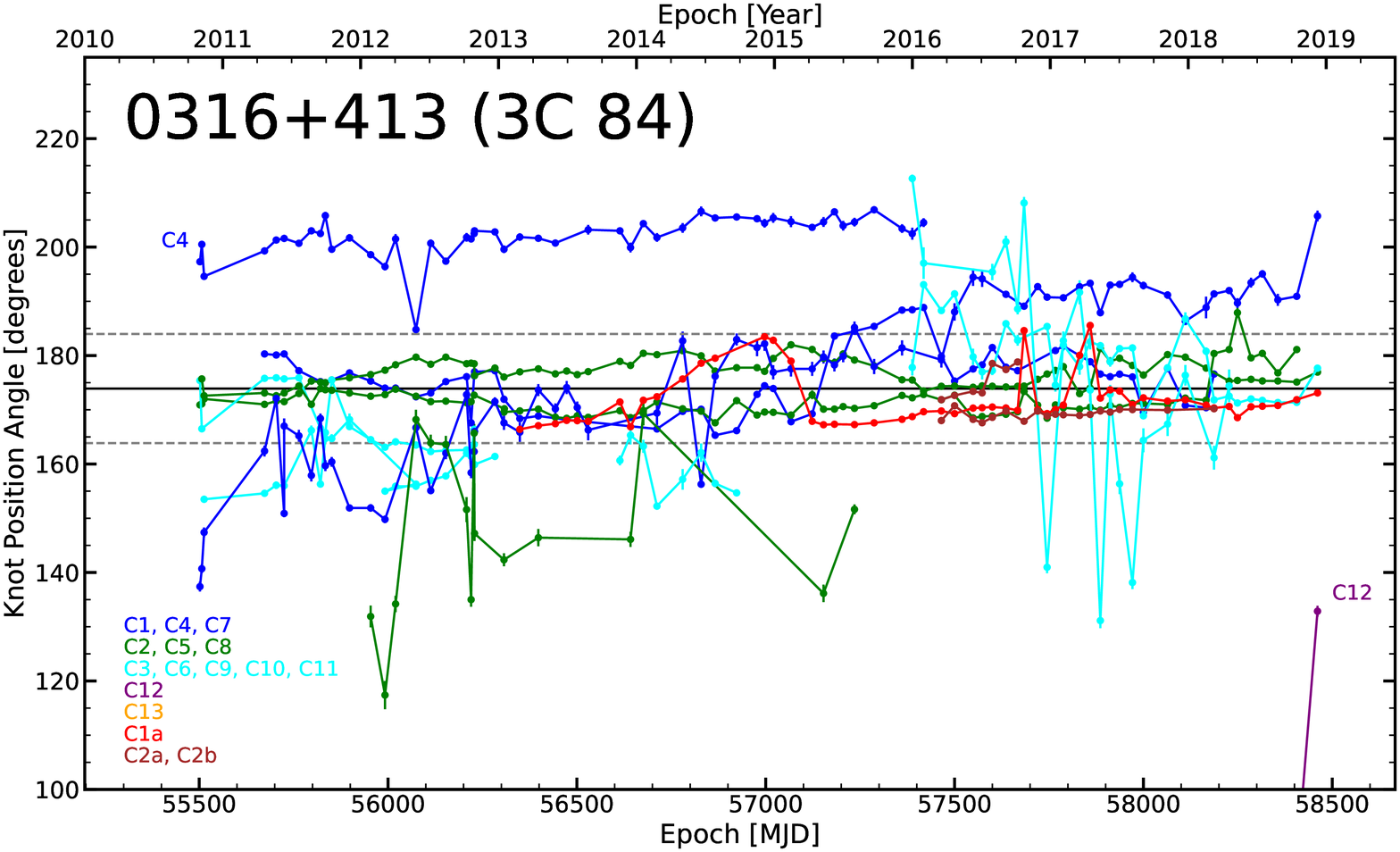}}
        \figsetgrpnote{Jet position angles of each knot, relative to the core, of the RG 0316+413.}
        \figsetgrpend
        %
        % Number 4
        \figsetgrpstart
        \figsetgrpnum{6.4}
        \figsetgrptitle{0336-019}
        \figsetplot{{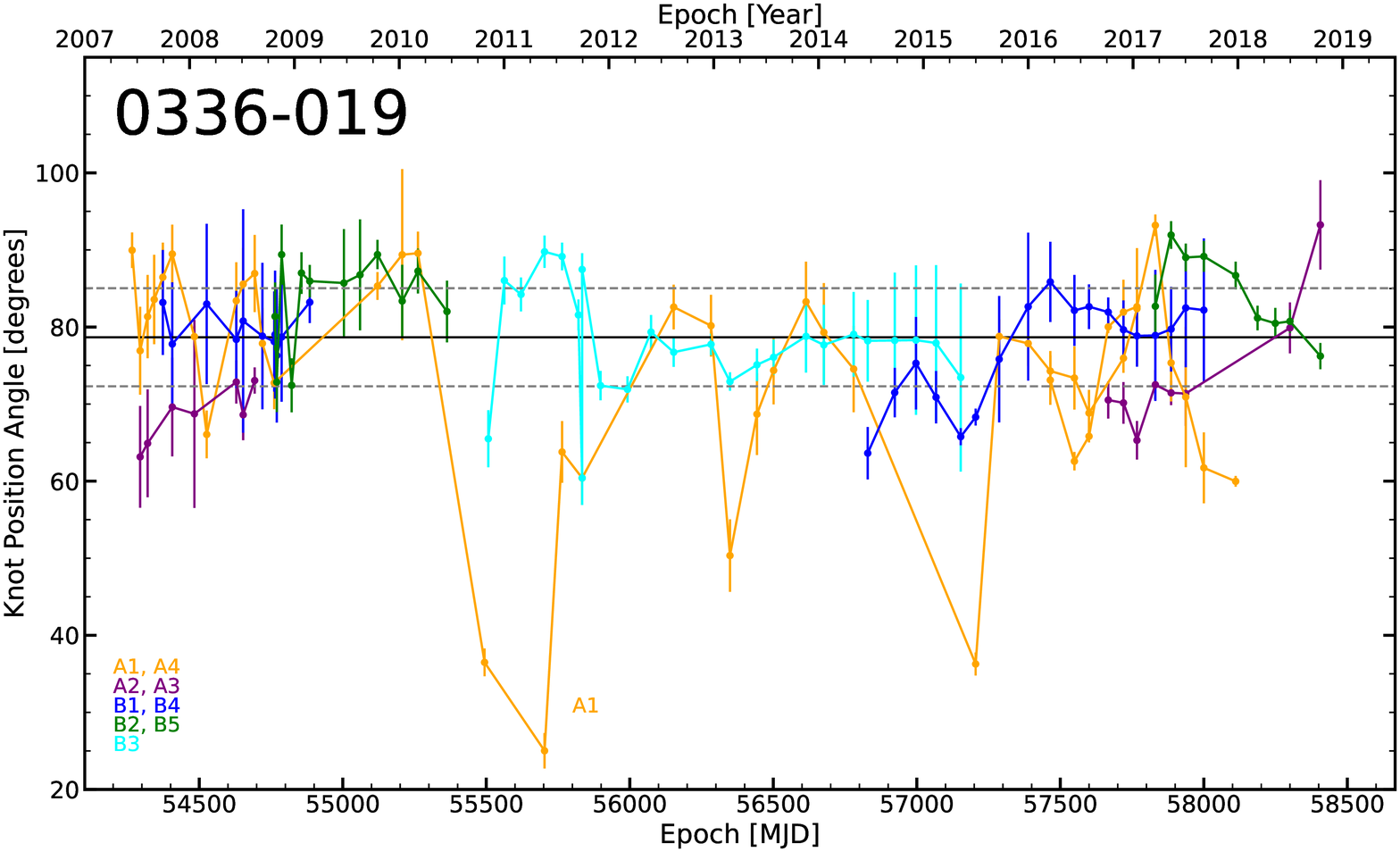}}
        \figsetgrpnote{Jet position angles of each knot, relative to the core, of the FSRQ 0336-019.}
        \figsetgrpend
        %
        % Number 5
        \figsetgrpstart
        \figsetgrpnum{6.5}
        \figsetgrptitle{0415+379}
        \figsetplot{{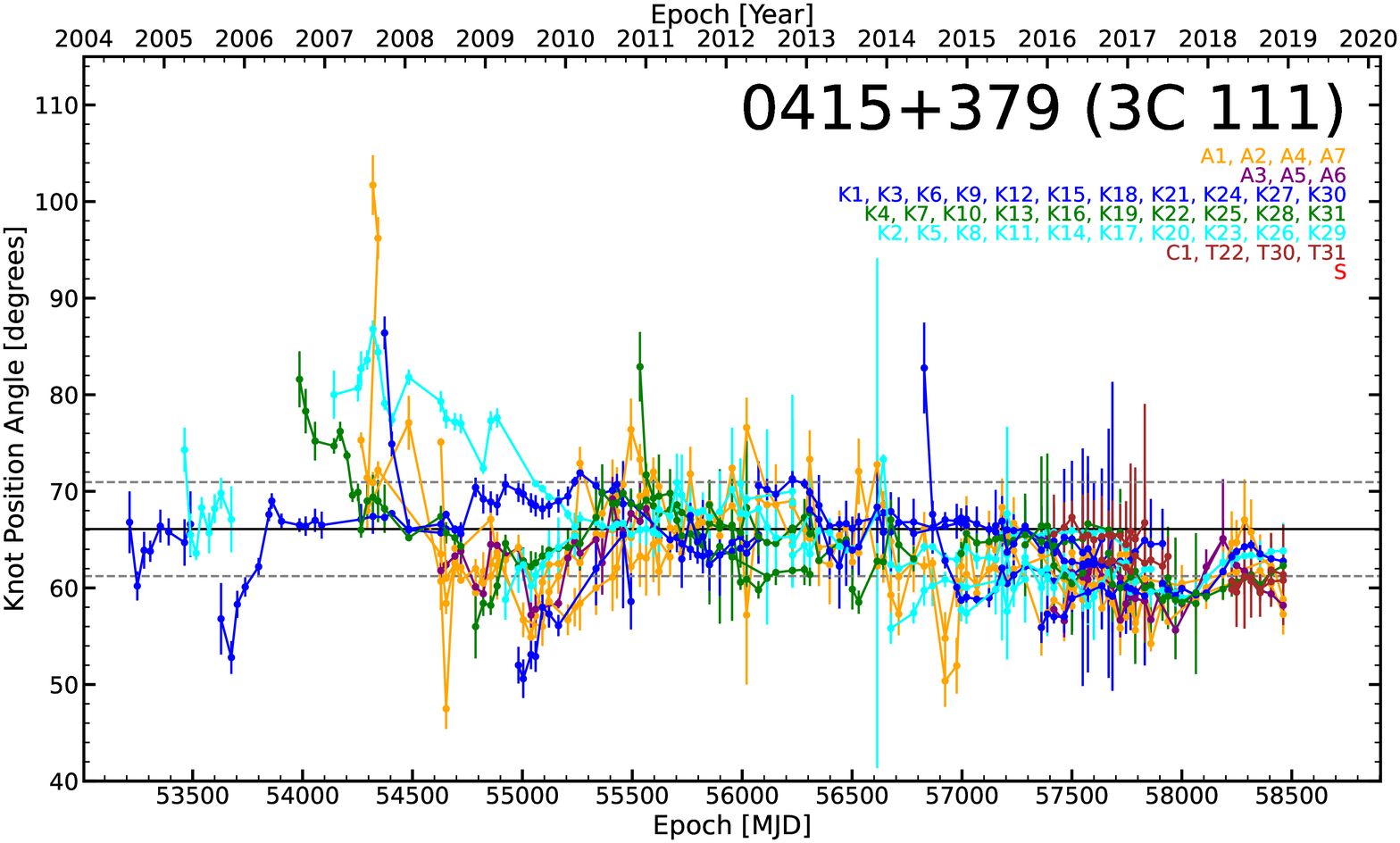}}
        \figsetgrpnote{Jet position angles of each knot, relative to the core, of the RG 0415+379.}
        \figsetgrpend
        %
        % Number 6
        \figsetgrpstart
        \figsetgrpnum{6.6}
        \figsetgrptitle{0420-014}
        \figsetplot{{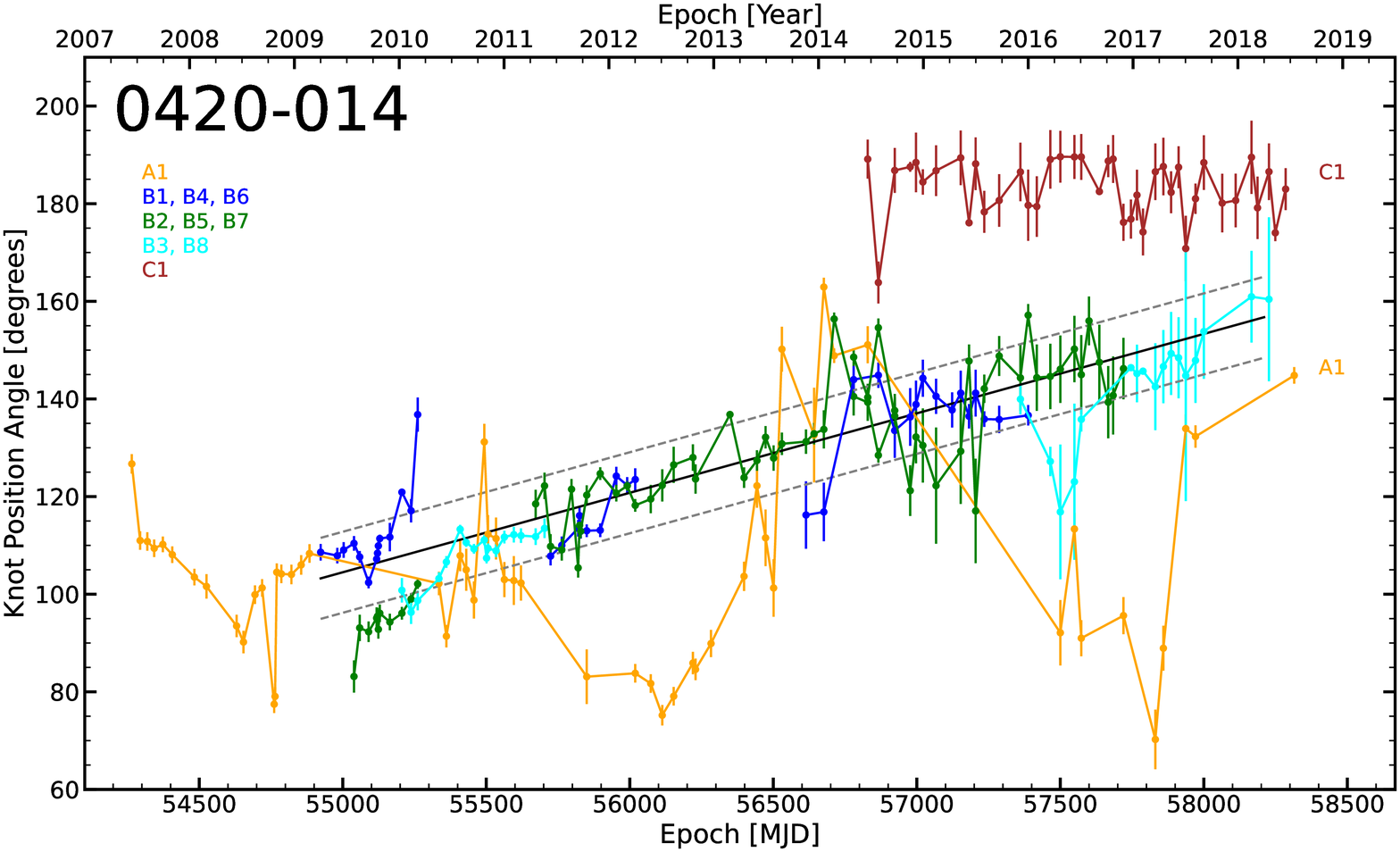}}
        \figsetgrpnote{Jet position angles of each knot, relative to the core, of the FSRQ 0420-014.}
        \figsetgrpend
        %
        % Number 7
        \figsetgrpstart
        \figsetgrpnum{6.7}
        \figsetgrptitle{0430+052}
        \figsetplot{{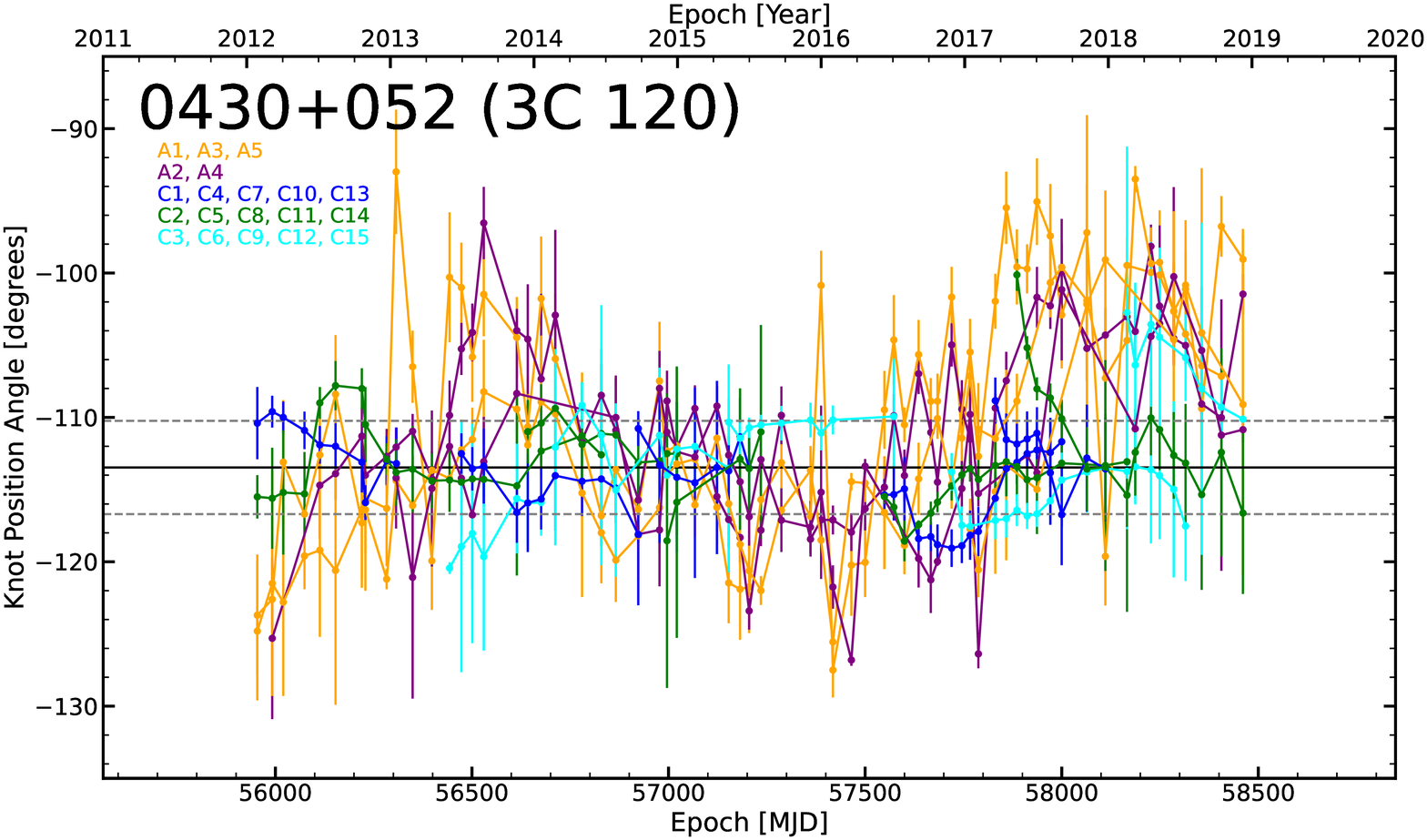}}
        \figsetgrpnote{Jet position angles of each knot, relative to the core, of the RG 0430+052.}
        \figsetgrpend
        %
        % Number 8
        \figsetgrpstart
        \figsetgrpnum{6.8}
        \figsetgrptitle{0528+134}
        \figsetplot{{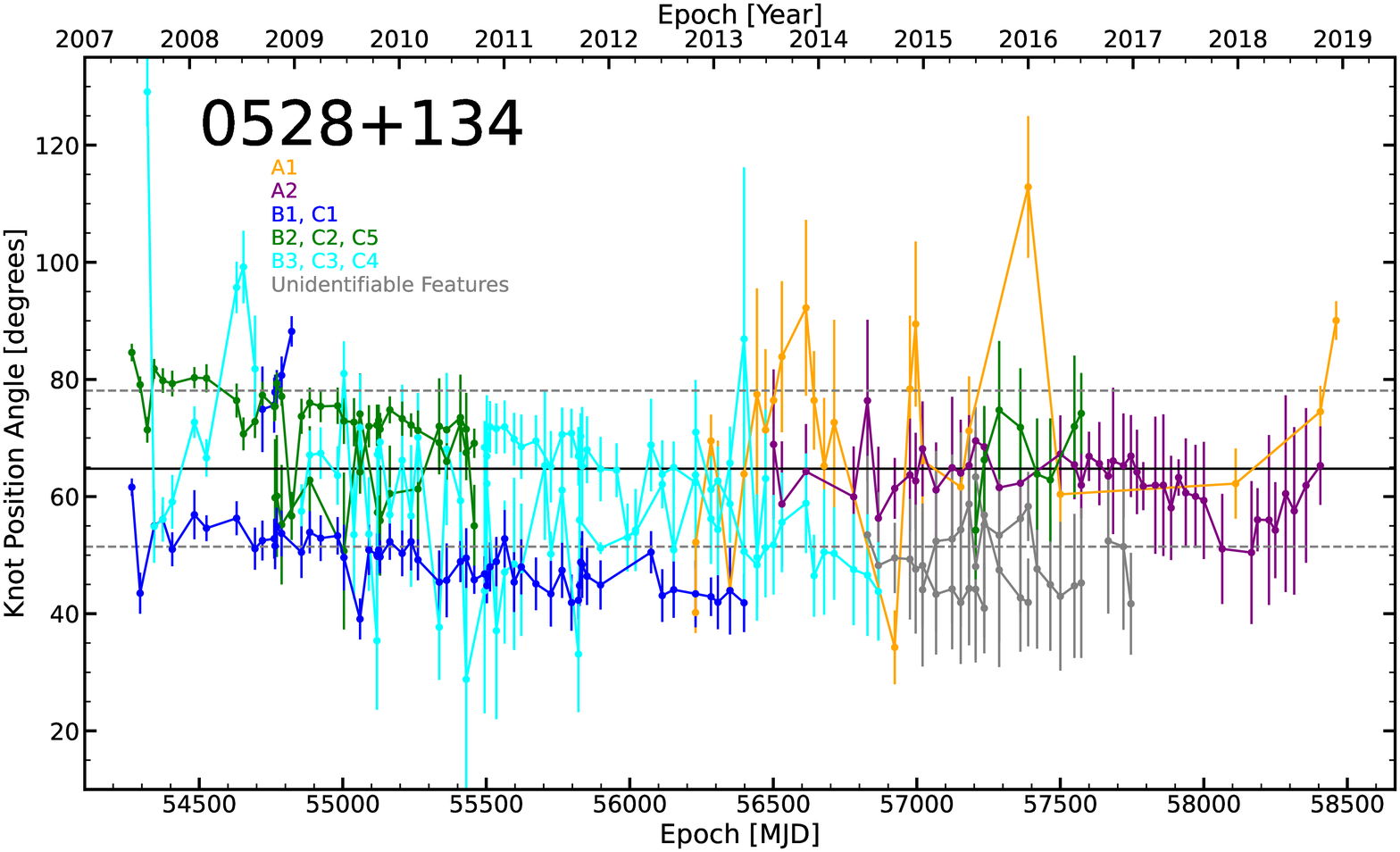}}
        \figsetgrpnote{Jet position angles of each knot, relative to the core, of the FSRQ 0528+134.}
        \figsetgrpend
        %
        % Number 9
        \figsetgrpstart
        \figsetgrpnum{6.9}
        \figsetgrptitle{0716+714}
        \figsetplot{{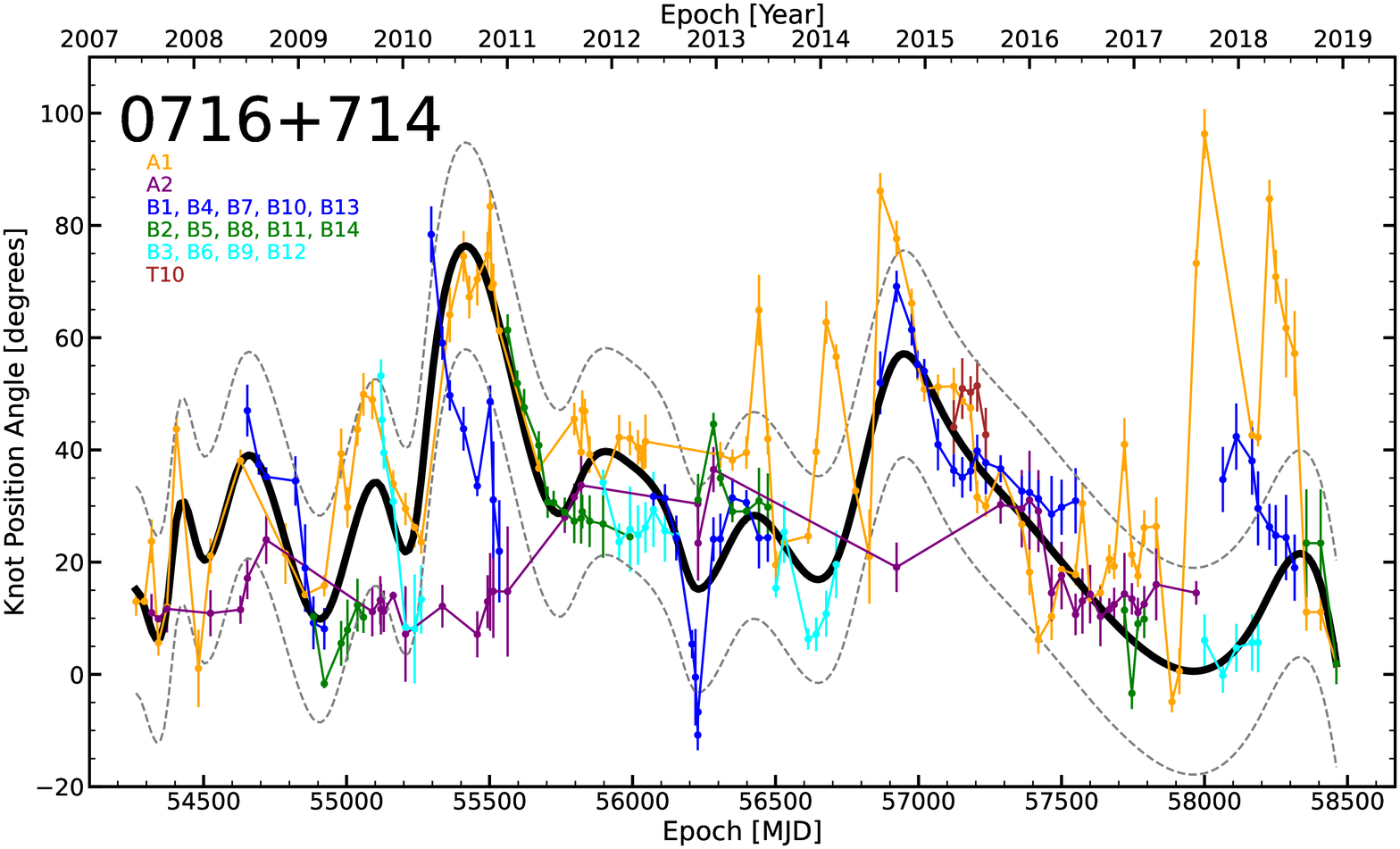}}
        \figsetgrpnote{Jet position angles of each knot, relative to the core, of the BL 0716+714.}
        \figsetgrpend
        %
        % Number 10
        \figsetgrpstart
        \figsetgrpnum{6.10}
        \figsetgrptitle{0735+178}
        \figsetplot{{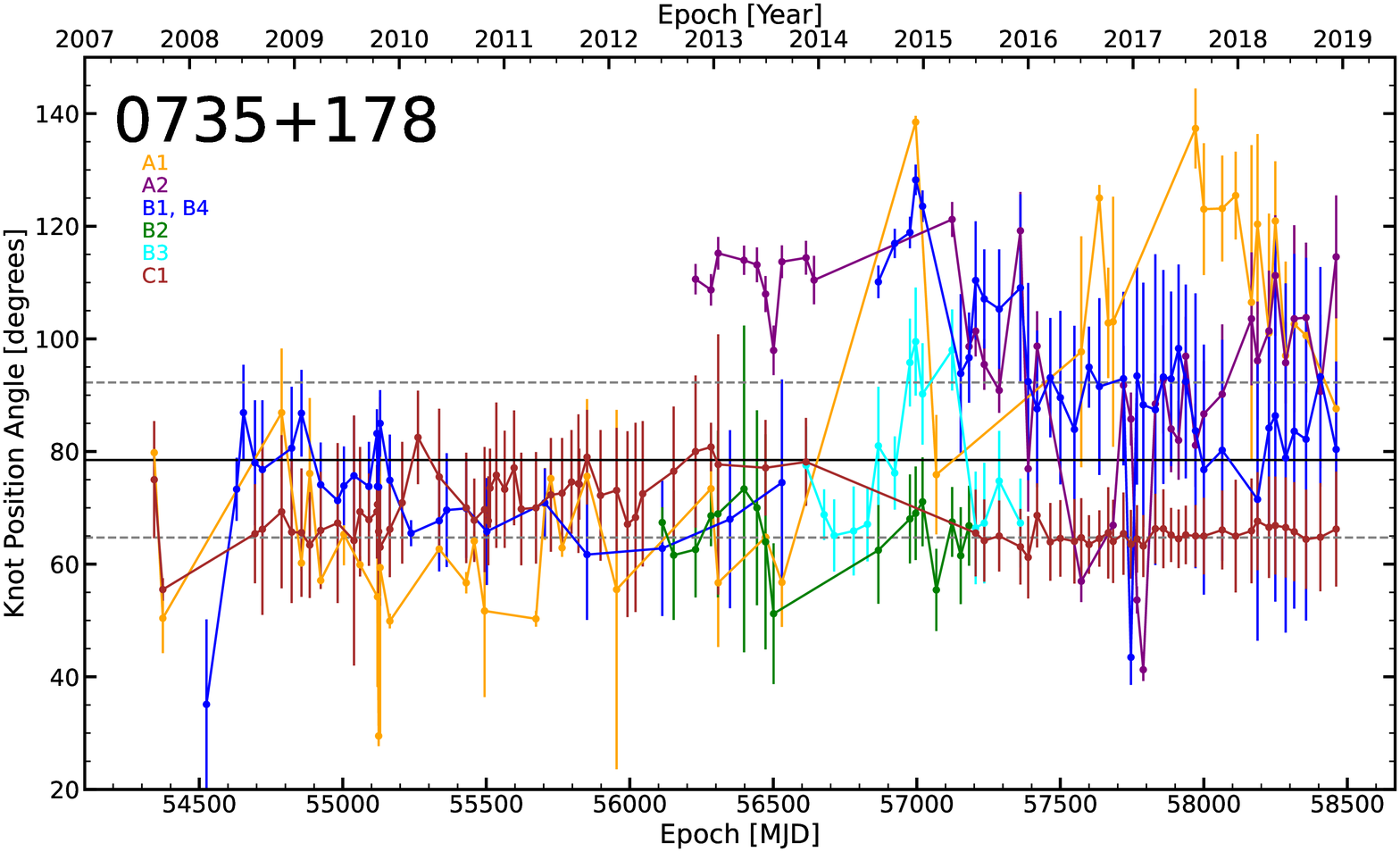}}
        \figsetgrpnote{Jet position angles of each knot, relative to the core, of the BL 0735+178.}
        \figsetgrpend
        %
        % Number 11
        \figsetgrpstart
        \figsetgrpnum{6.11}
        \figsetgrptitle{0827+243}
        \figsetplot{{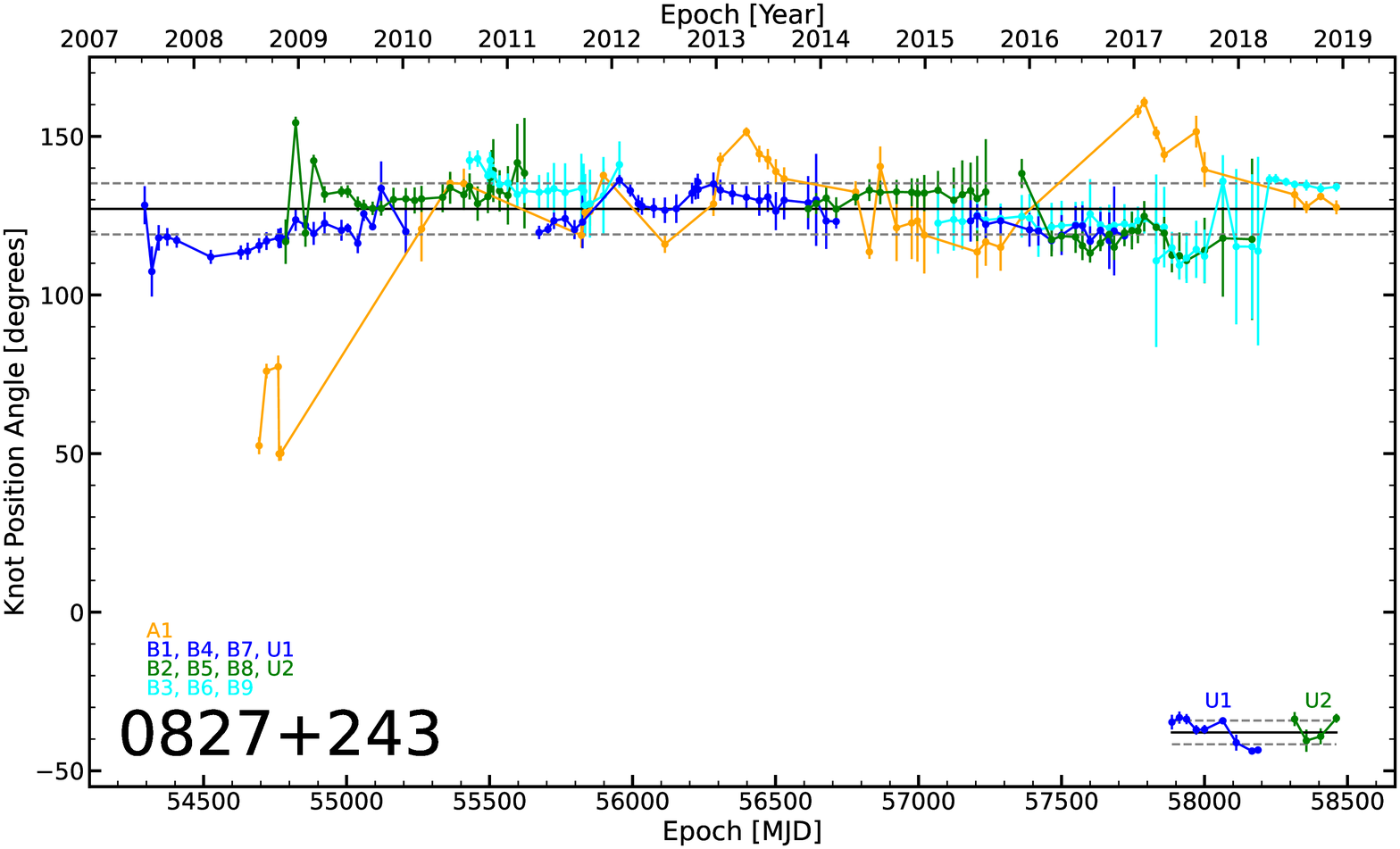}}
        \figsetgrpnote{Jet position angles of each knot, relative to the core, of the FSRQ 0827+243.}
        \figsetgrpend
        %
        % Number 12
        \figsetgrpstart
        \figsetgrpnum{6.12}
        \figsetgrptitle{0829+046}
        \figsetplot{{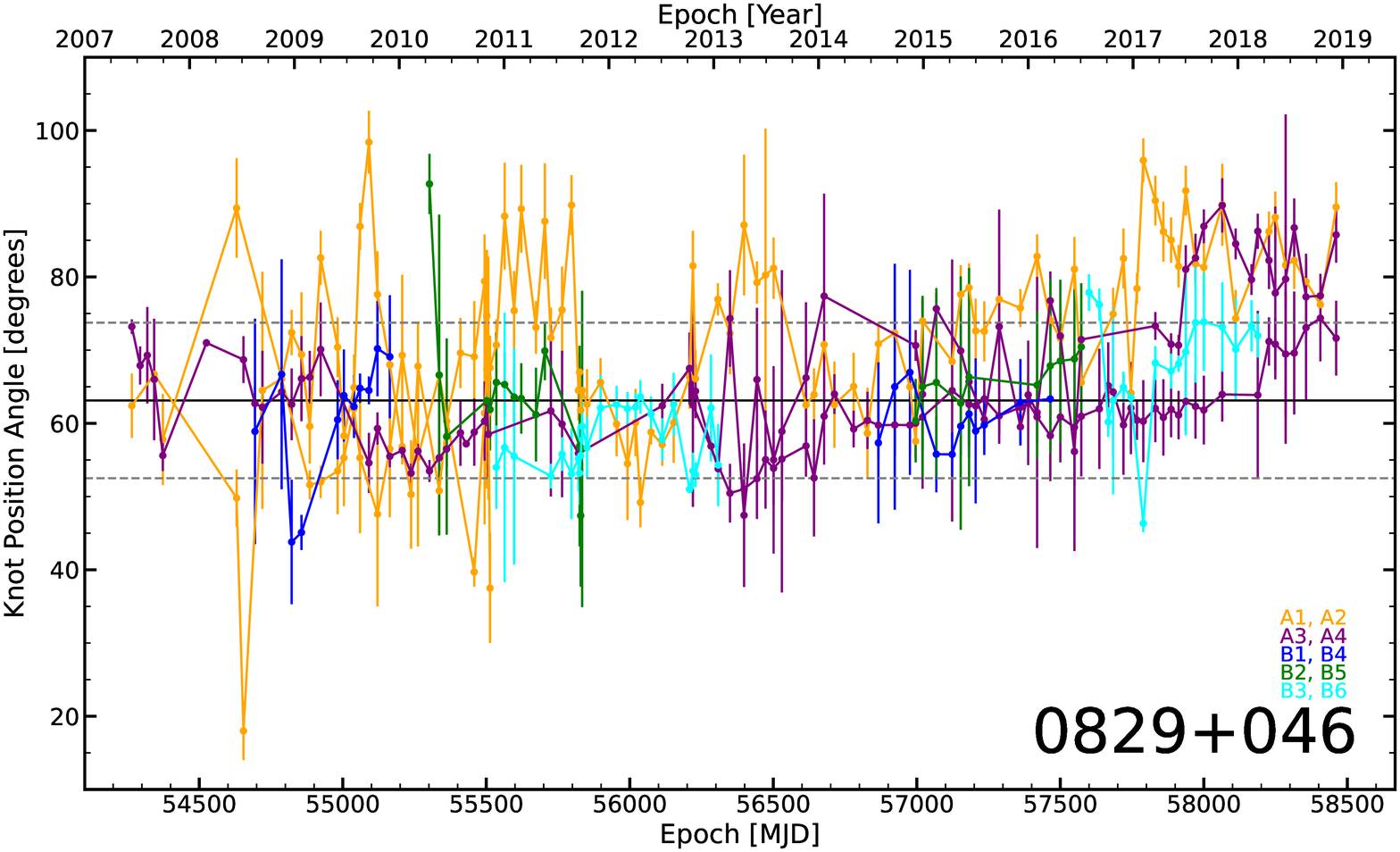}}
        \figsetgrpnote{Jet position angles of each knot, relative to the core, of the BL 0829+046.}
        \figsetgrpend
        %
        % Number 13
        \figsetgrpstart
        \figsetgrpnum{6.13}
        \figsetgrptitle{0836+710}
        \figsetplot{{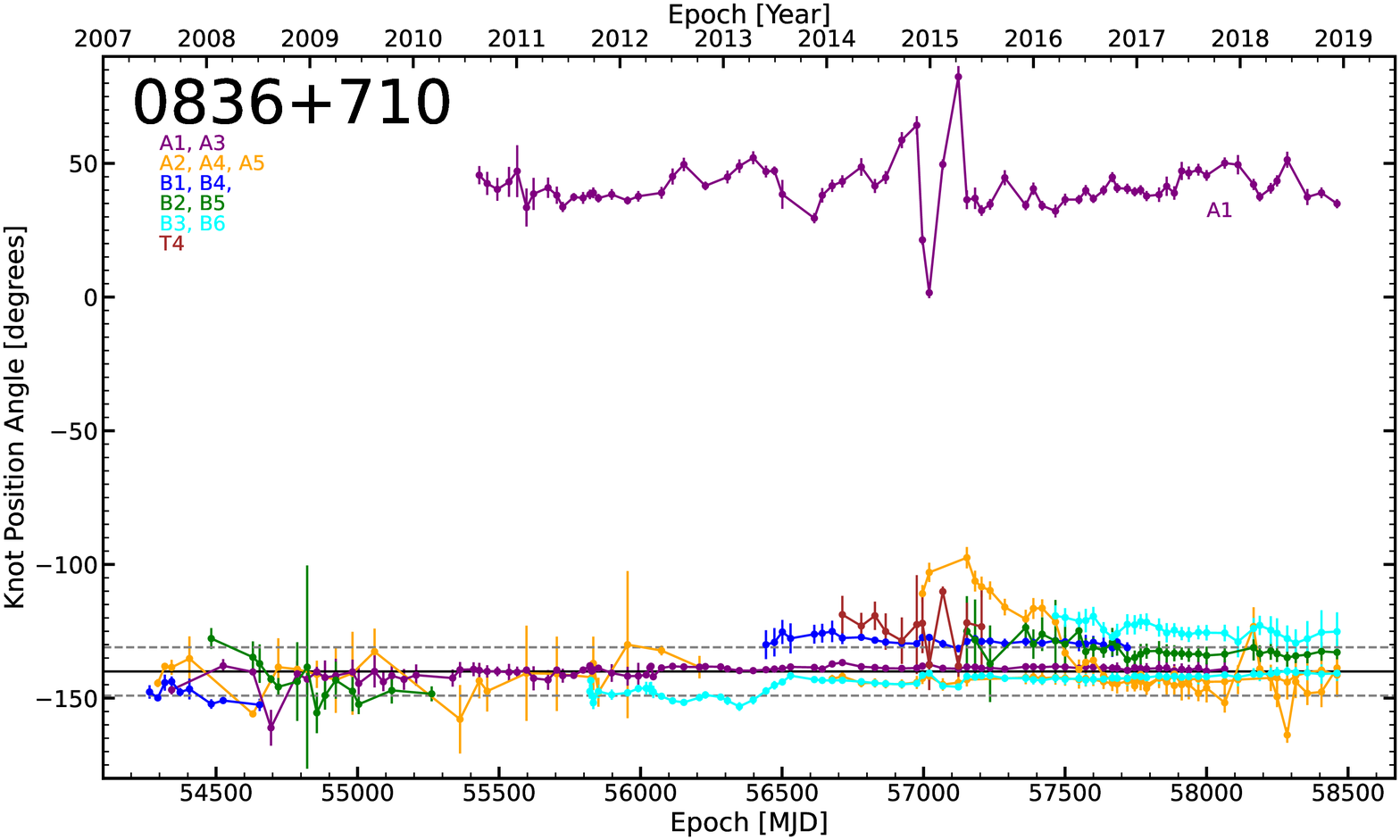}}
        \figsetgrpnote{Jet position angles of each knot, relative to the core, of the FSRQ 0836+710.}
        \figsetgrpend
        %
        % Number 14
        \figsetgrpstart
        \figsetgrpnum{6.14}
        \figsetgrptitle{0851+202}
        \figsetplot{{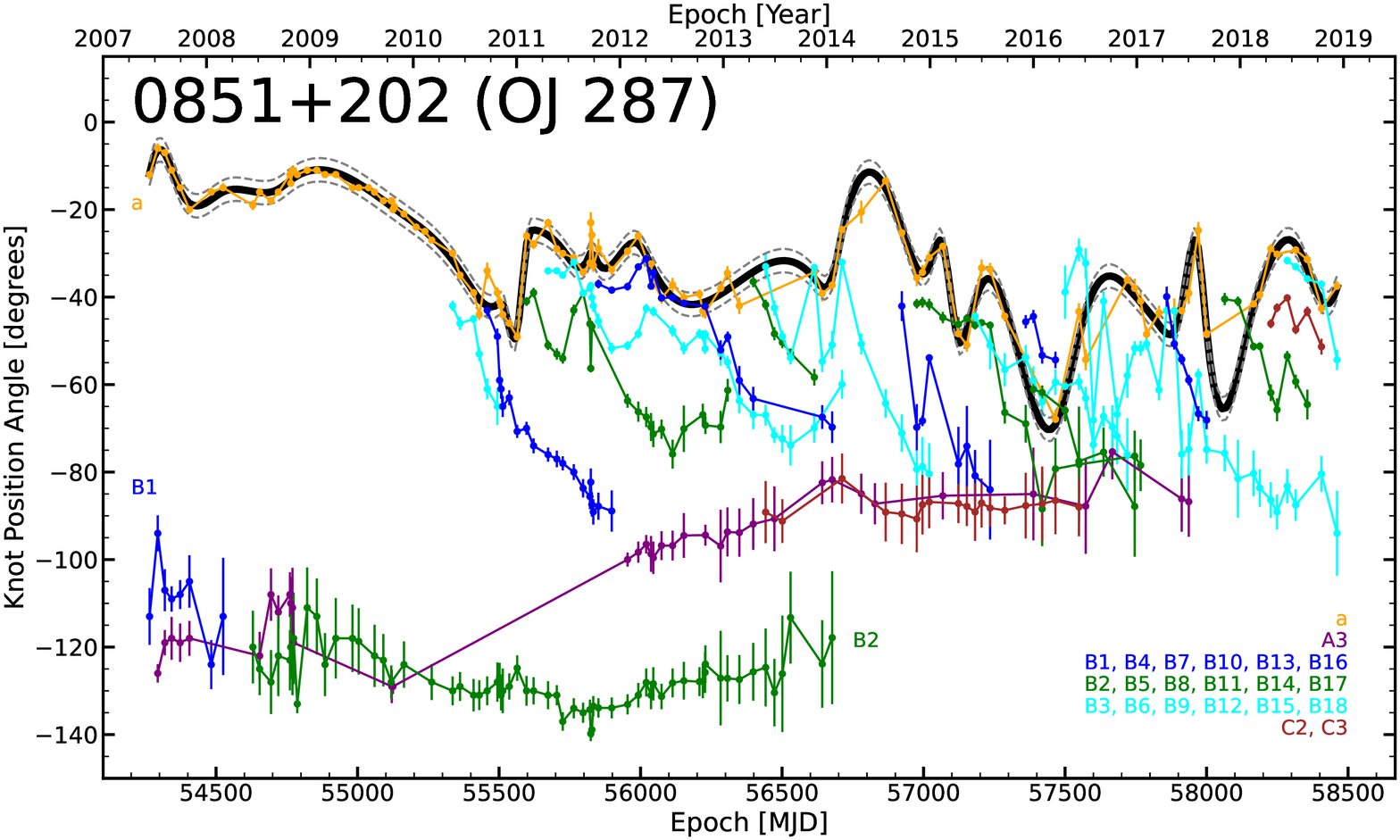}}
        \figsetgrpnote{Jet position angles of each knot, relative to the core, of the BL 0851+202.}
        \figsetgrpend
        %
        % Number 15
        \figsetgrpstart
        \figsetgrpnum{6.15}
        \figsetgrptitle{0954+658}
        \figsetplot{{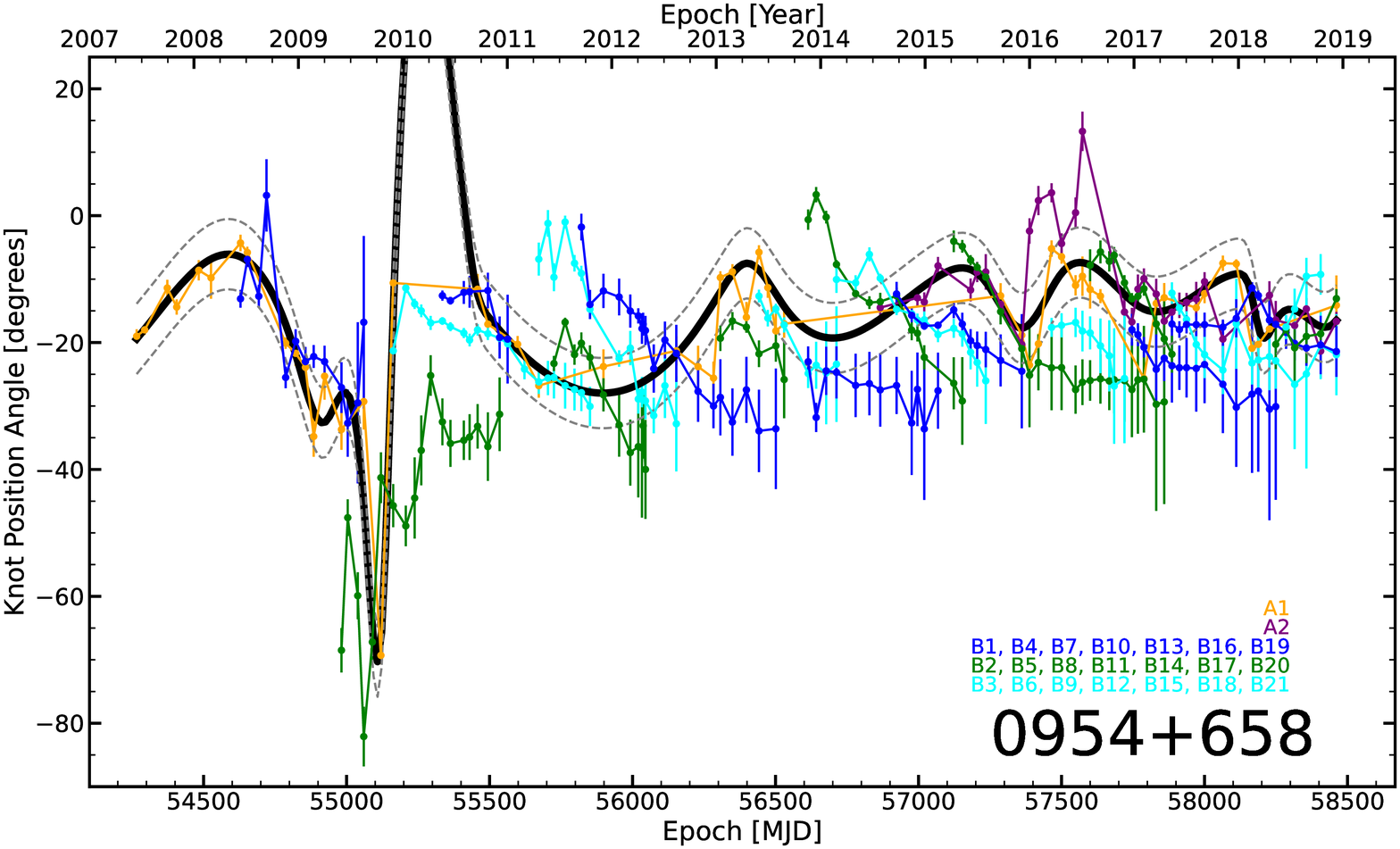}}
        \figsetgrpnote{Jet position angles of each knot, relative to the core, of the BL 0954+658.}
        \figsetgrpend
        %
        % Number 16
        \figsetgrpstart
        \figsetgrpnum{6.16}
        \figsetgrptitle{1055+018}
        \figsetplot{{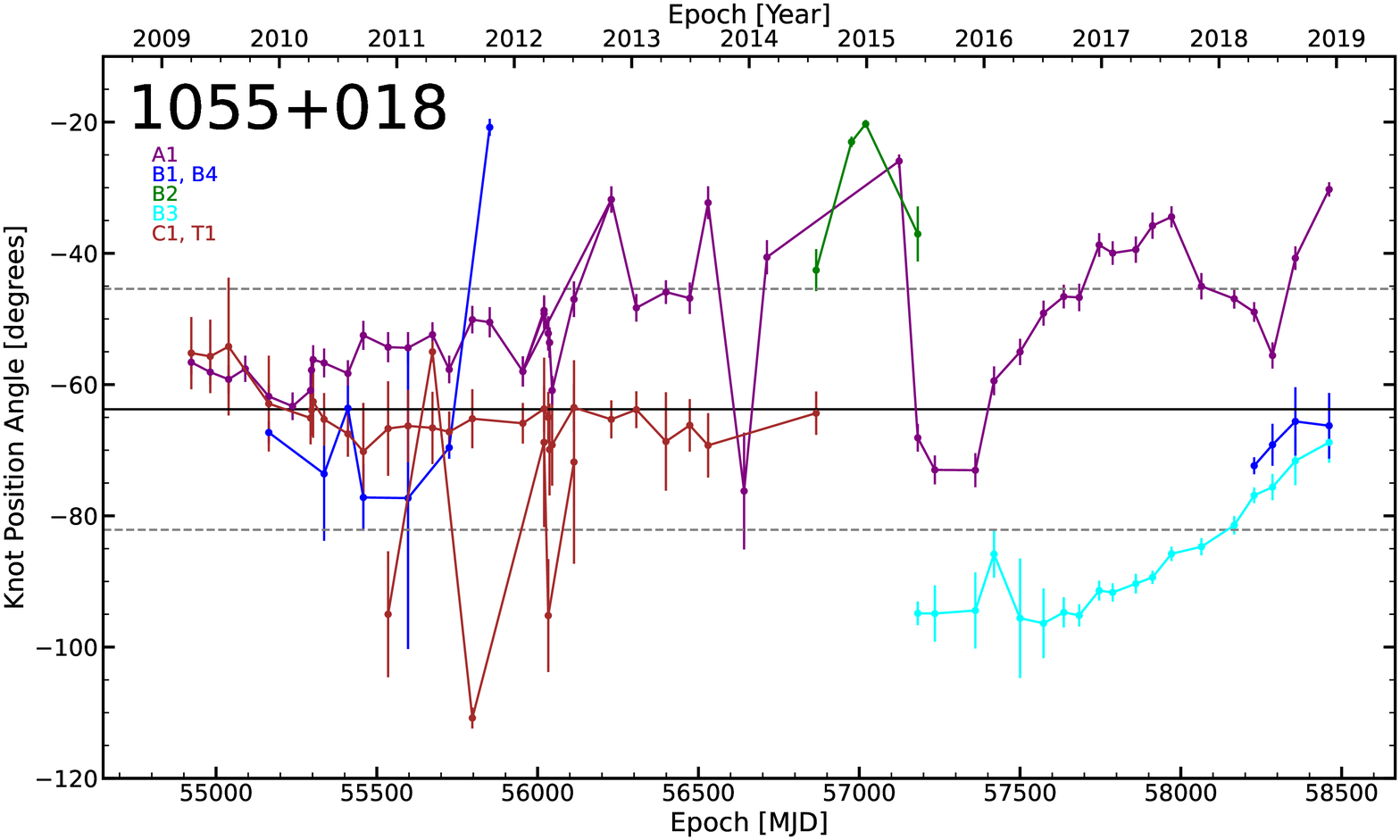}}
        \figsetgrpnote{Jet position angles of each knot, relative to the core, of the FSRQ 1055+018.}
        \figsetgrpend
        %
        % Number 17
        \figsetgrpstart
        \figsetgrpnum{6.17}
        \figsetgrptitle{1101+384}
        \figsetplot{{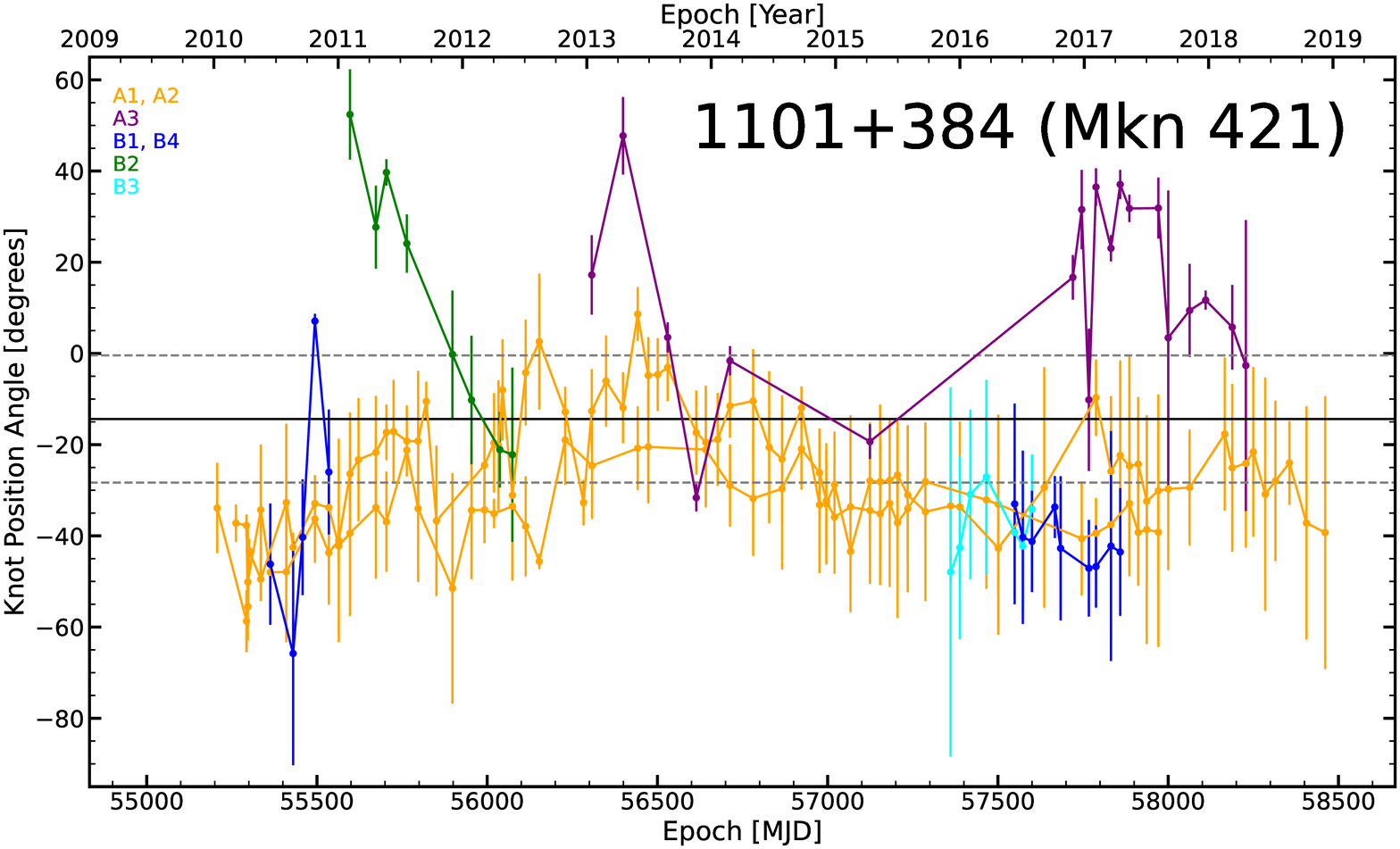}}
        \figsetgrpnote{Jet position angles of each knot, relative to the core, of the BL 1101+384.}
        \figsetgrpend
        %
        % Number 18
        \figsetgrpstart
        \figsetgrpnum{6.18}
        \figsetgrptitle{1127-145}
        \figsetplot{{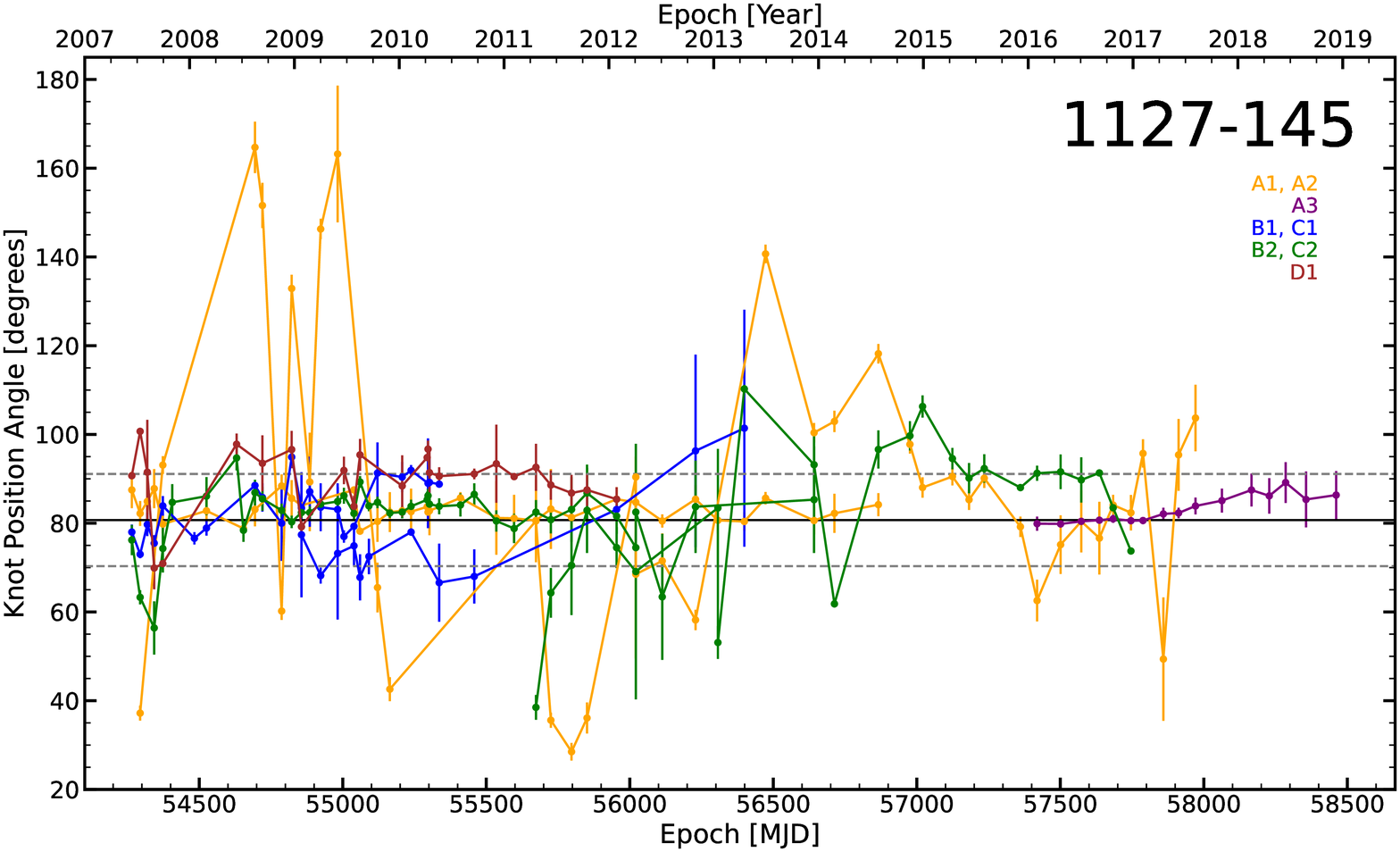}}
        \figsetgrpnote{Jet position angles of each knot, relative to the core, of the FSRQ 1127-145.}
        \figsetgrpend
        %
        % Number 19
        \figsetgrpstart
        \figsetgrpnum{6.19}
        \figsetgrptitle{1156+295}
        \figsetplot{{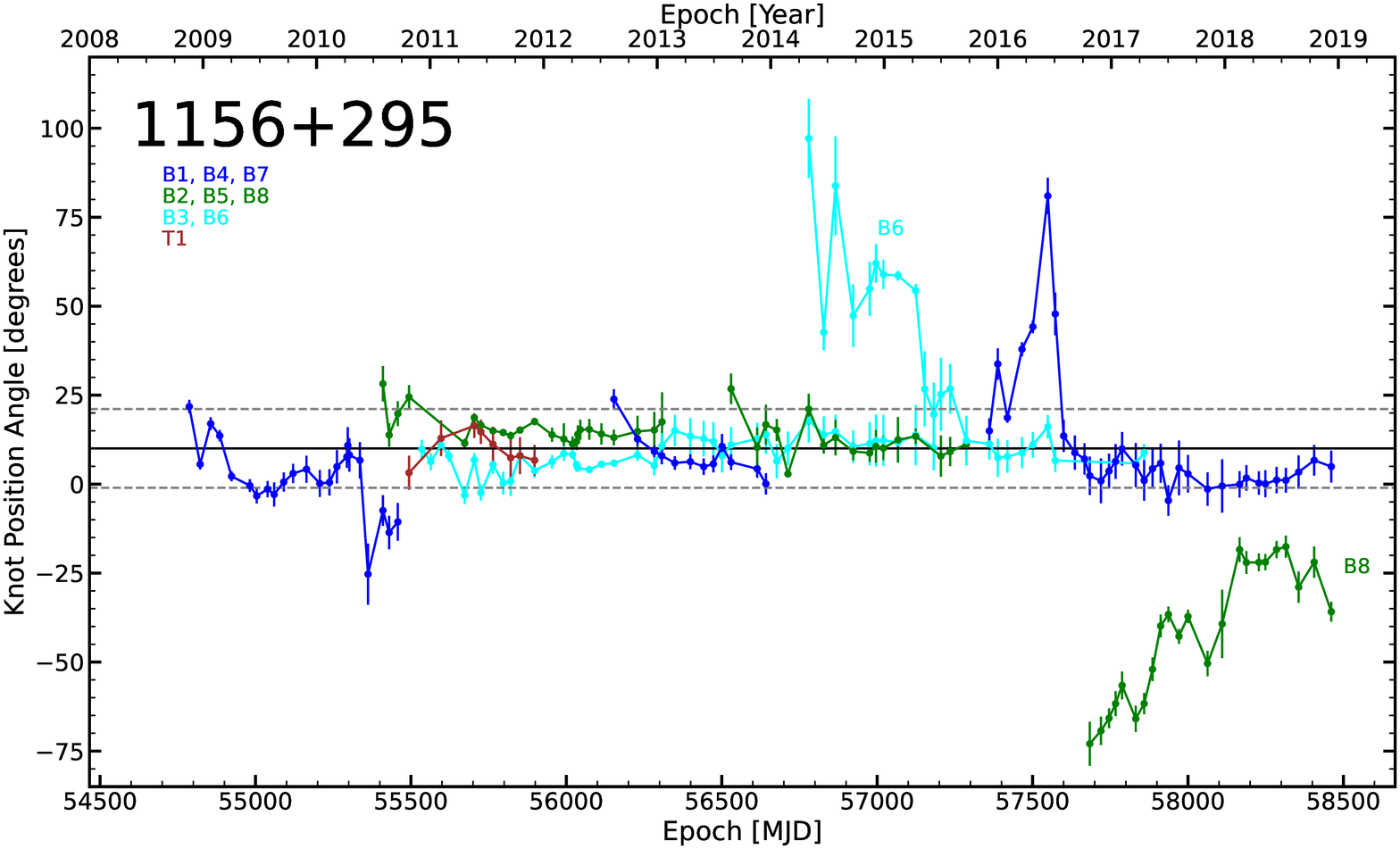}}
        \figsetgrpnote{Jet position angles of each knot, relative to the core, of the FSRQ 1156+295.}
        \figsetgrpend
        %
        % Number 20
        \figsetgrpstart
        \figsetgrpnum{6.20}
        \figsetgrptitle{1219+285}
        \figsetplot{{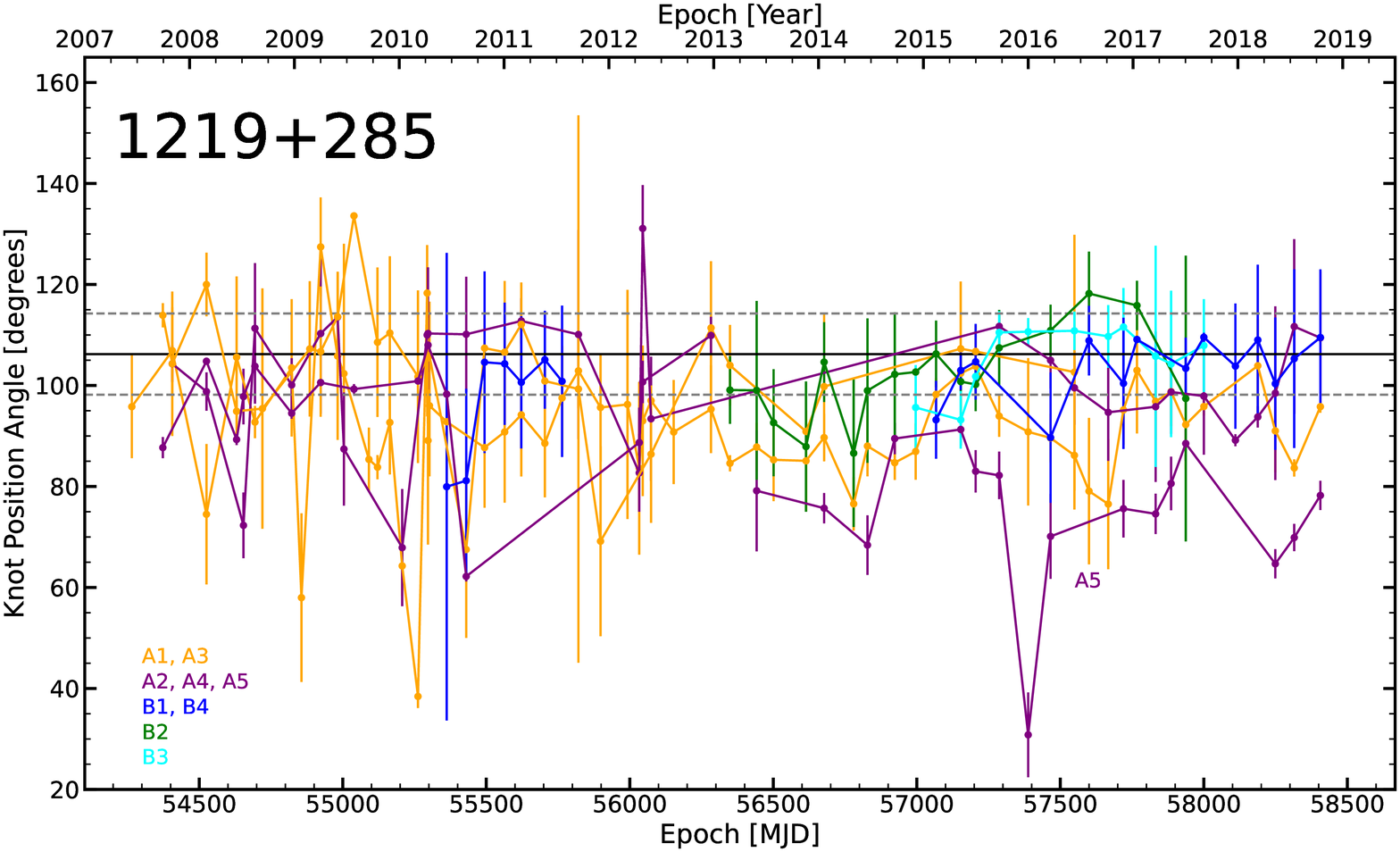}}
        \figsetgrpnote{Jet position angles of each knot, relative to the core, of the BL 1219+285.}
        \figsetgrpend
        %
        % Number 21
        \figsetgrpstart
        \figsetgrpnum{6.21}
        \figsetgrptitle{1222+216}
        \figsetplot{{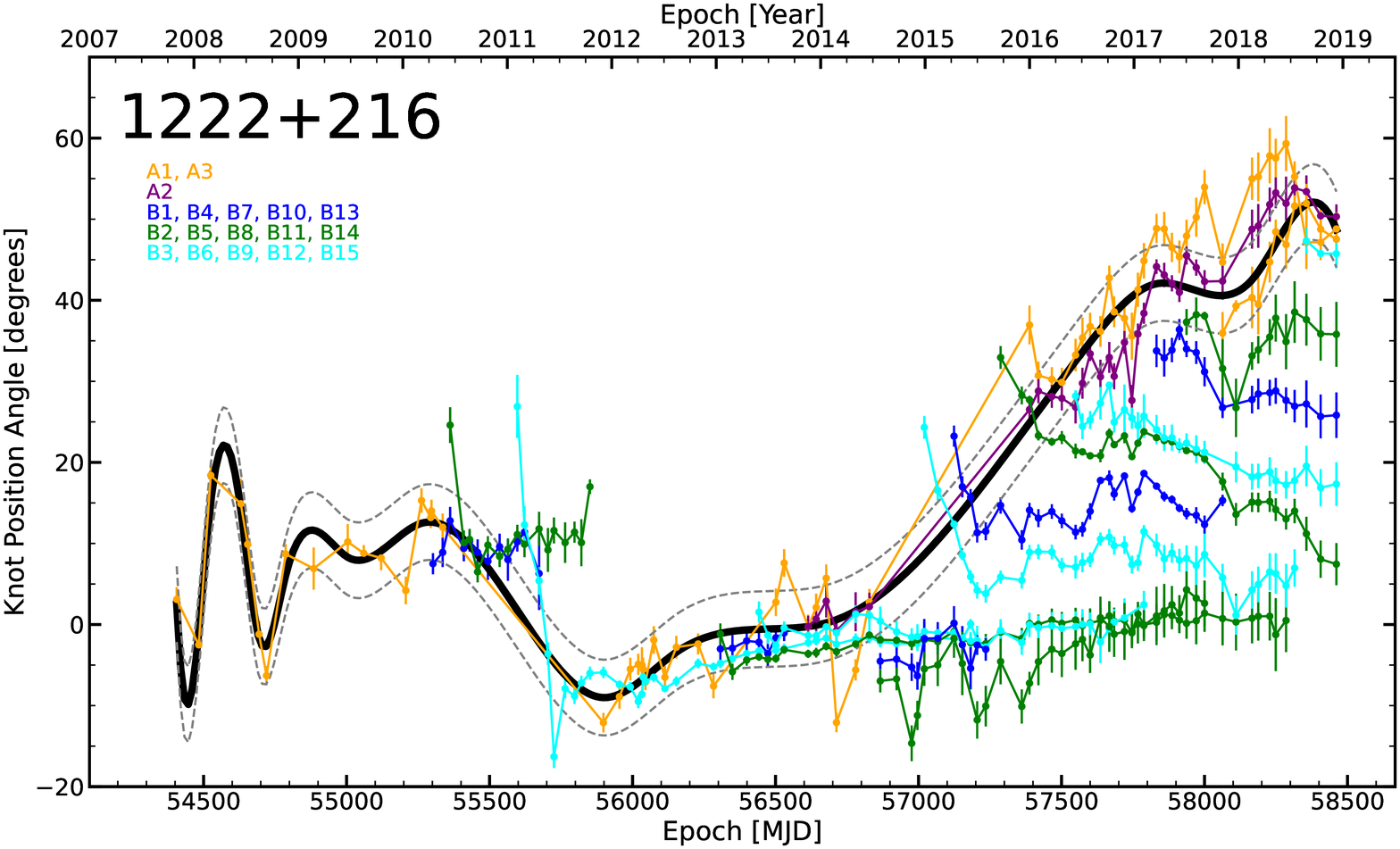}}
        \figsetgrpnote{Jet position angles of each knot, relative to the core, of the FSRQ 1222+216.}
        \figsetgrpend
        %
        % Number 22
        \figsetgrpstart
        \figsetgrpnum{6.22}
        \figsetgrptitle{1226+023}
        \figsetplot{{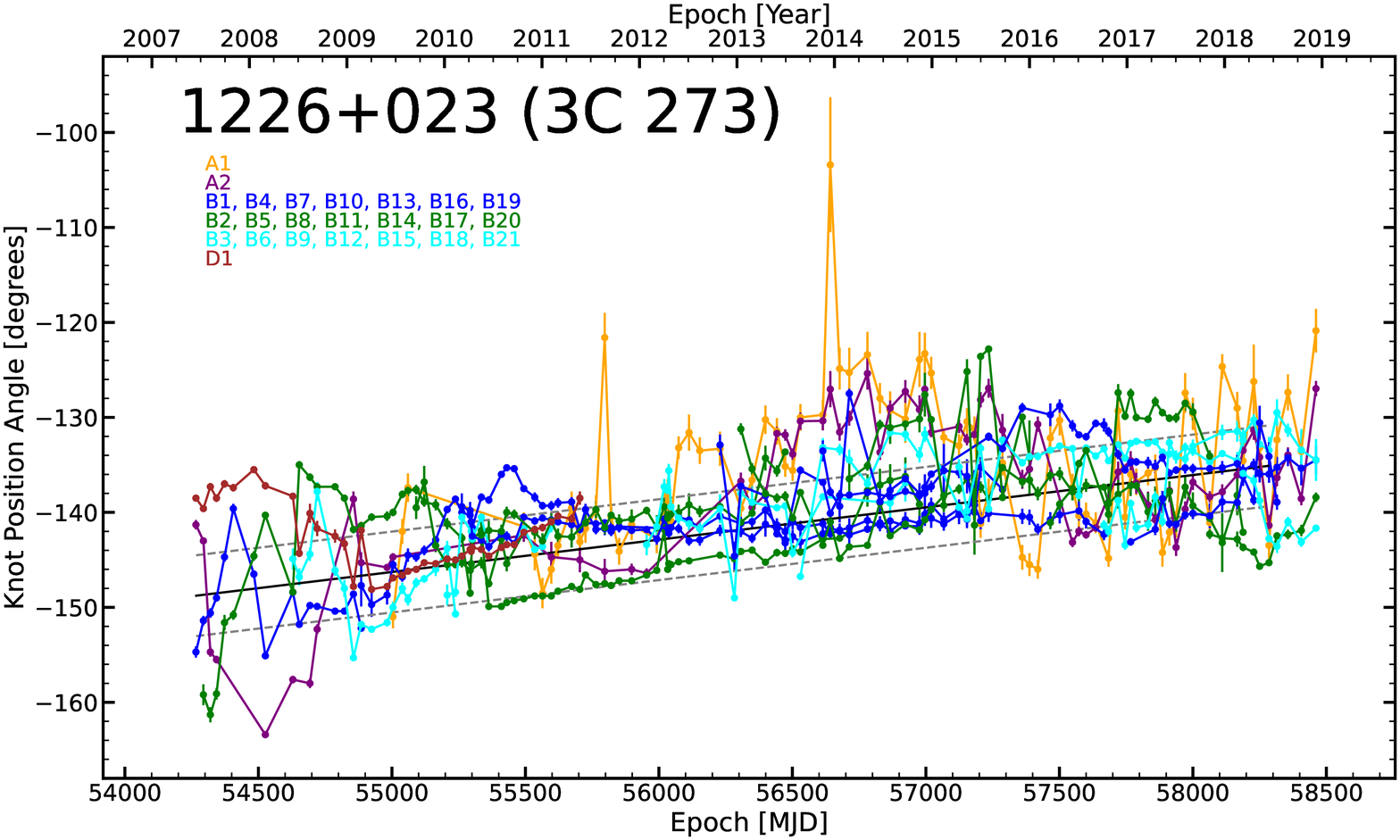}}
        \figsetgrpnote{Jet position angles of each knot, relative to the core, of the FSRQ 1226+023.}
        \figsetgrpend
        %
        % Number 23
        \figsetgrpstart
        \figsetgrpnum{6.23}
        \figsetgrptitle{1253-055}
        \figsetplot{{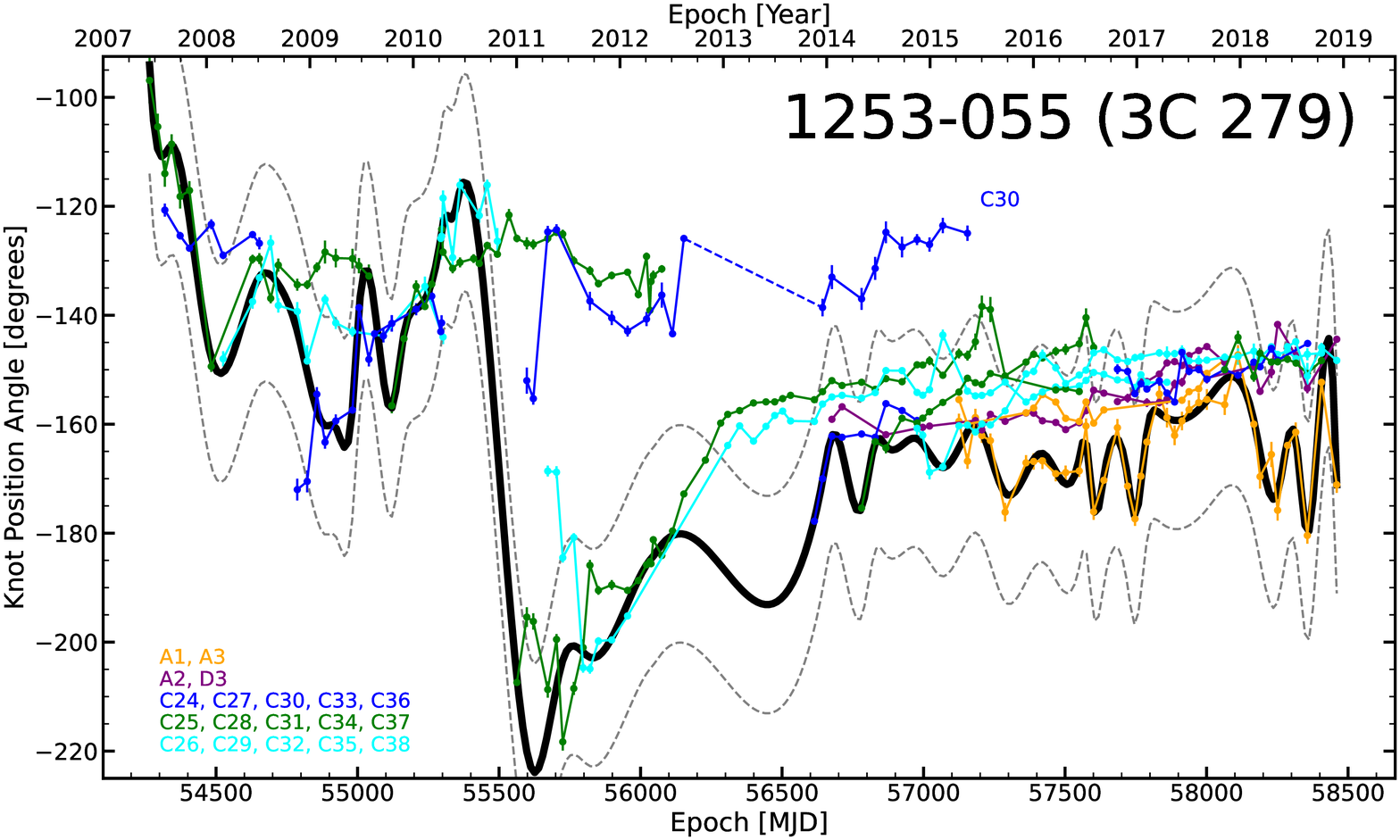}}
        \figsetgrpnote{Jet position angles of each knot, relative to the core, of the FSRQ 1253-055.}
        \figsetgrpend
        %
        % Number 24
        \figsetgrpstart
        \figsetgrpnum{6.24}
        \figsetgrptitle{1308+326}
        \figsetplot{{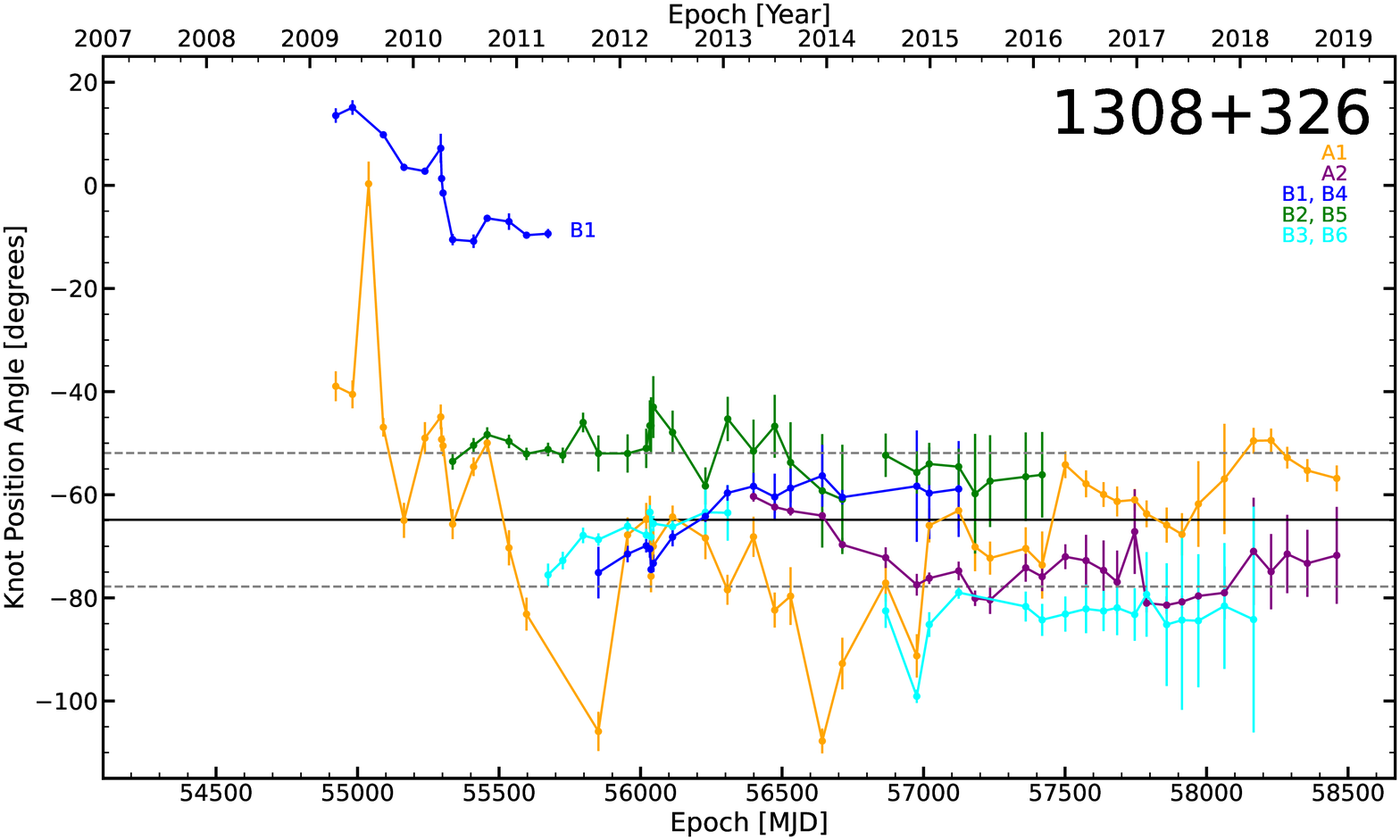}}
        \figsetgrpnote{Jet position angles of each knot, relative to the core, of the FSRQ 1308+326.}
        \figsetgrpend
        %
        % Number 25
        \figsetgrpstart
        \figsetgrpnum{6.25}
        \figsetgrptitle{1406-076}
        \figsetplot{{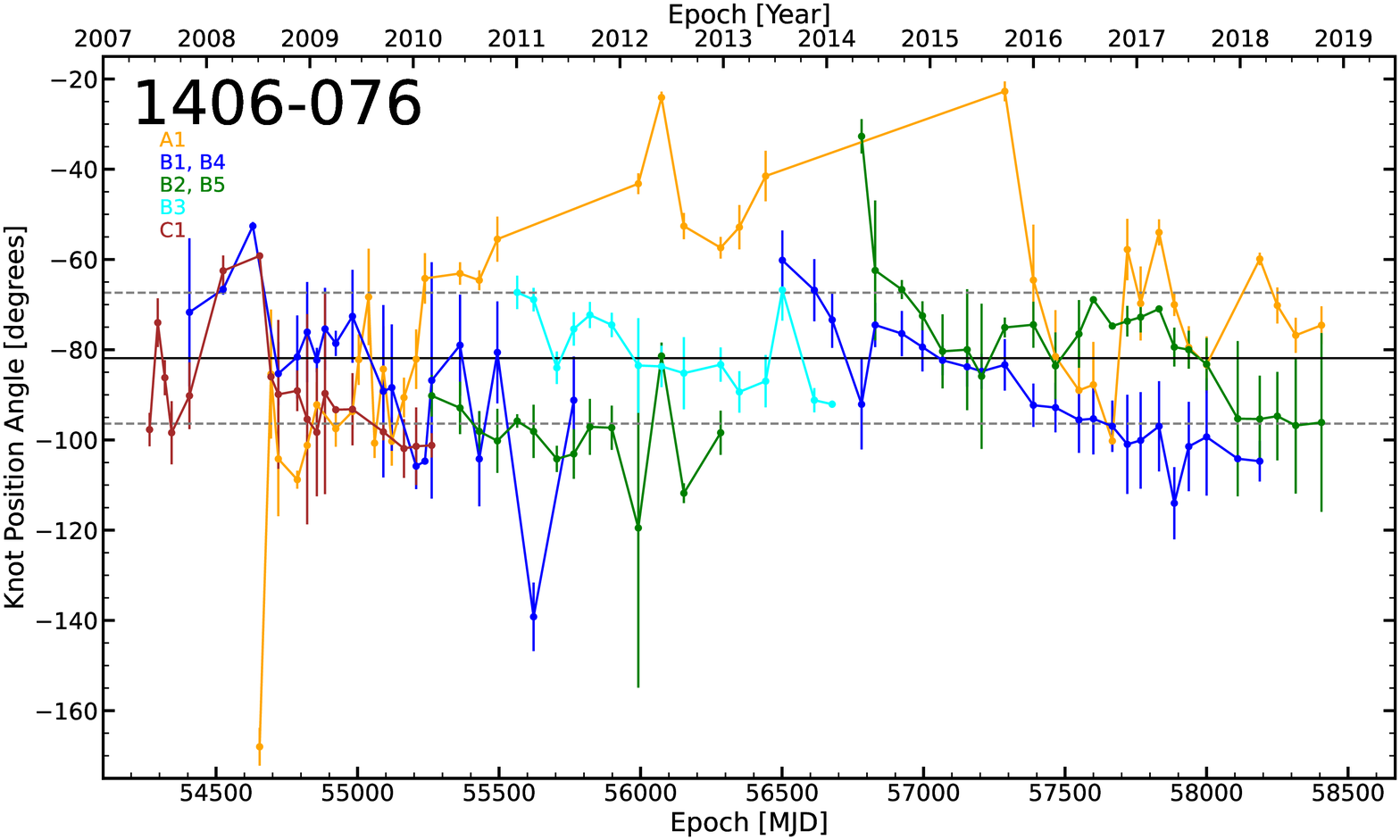}}
        \figsetgrpnote{Jet position angles of each knot, relative to the core, of the FSRQ 1406-076.}
        \figsetgrpend
        %
        % Number 26
        \figsetgrpstart
        \figsetgrpnum{6.26}
        \figsetgrptitle{1510-089}
        \figsetplot{{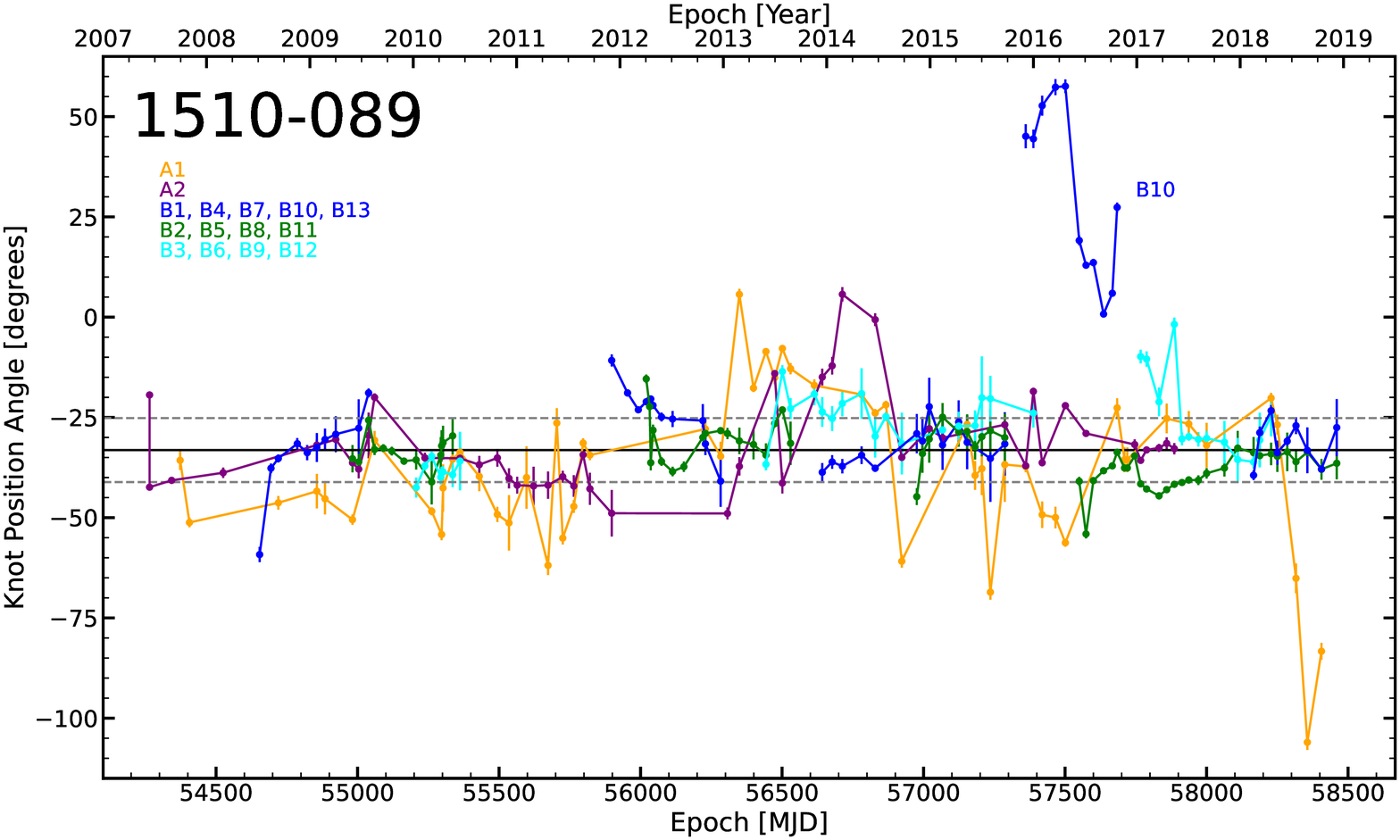}}
        \figsetgrpnote{Jet position angles of each knot, relative to the core, of the FSRQ 1510-089.}
        \figsetgrpend
        %
        % Number 27
        \figsetgrpstart
        \figsetgrpnum{6.27}
        \figsetgrptitle{1611+343}
        \figsetplot{{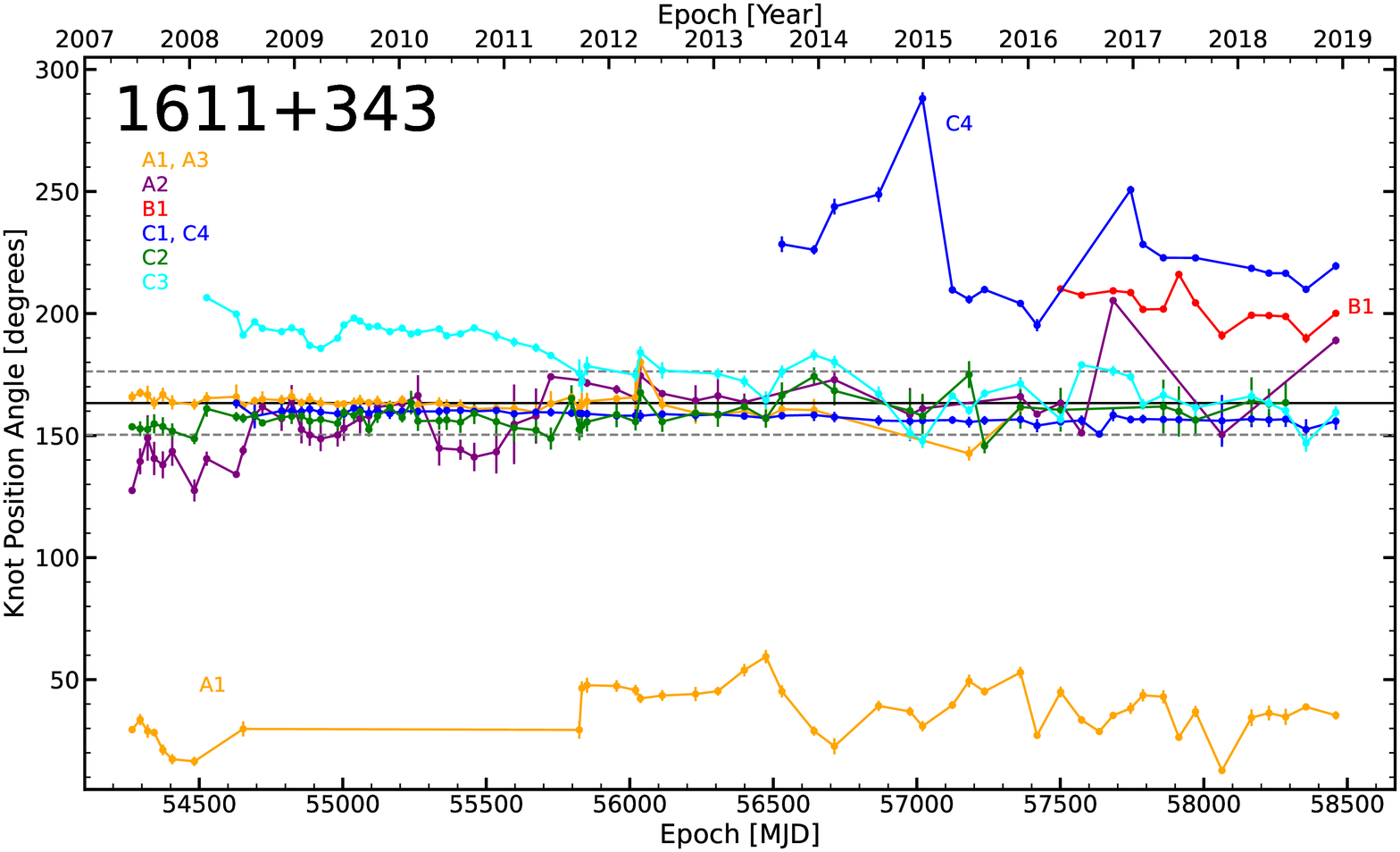}}
        \figsetgrpnote{Jet position angles of each knot, relative to the core, of the FSRQ 1611+343.}
        \figsetgrpend
        %
        % Number 28
        \figsetgrpstart
        \figsetgrpnum{6.28}
        \figsetgrptitle{1622-297}
        \figsetplot{{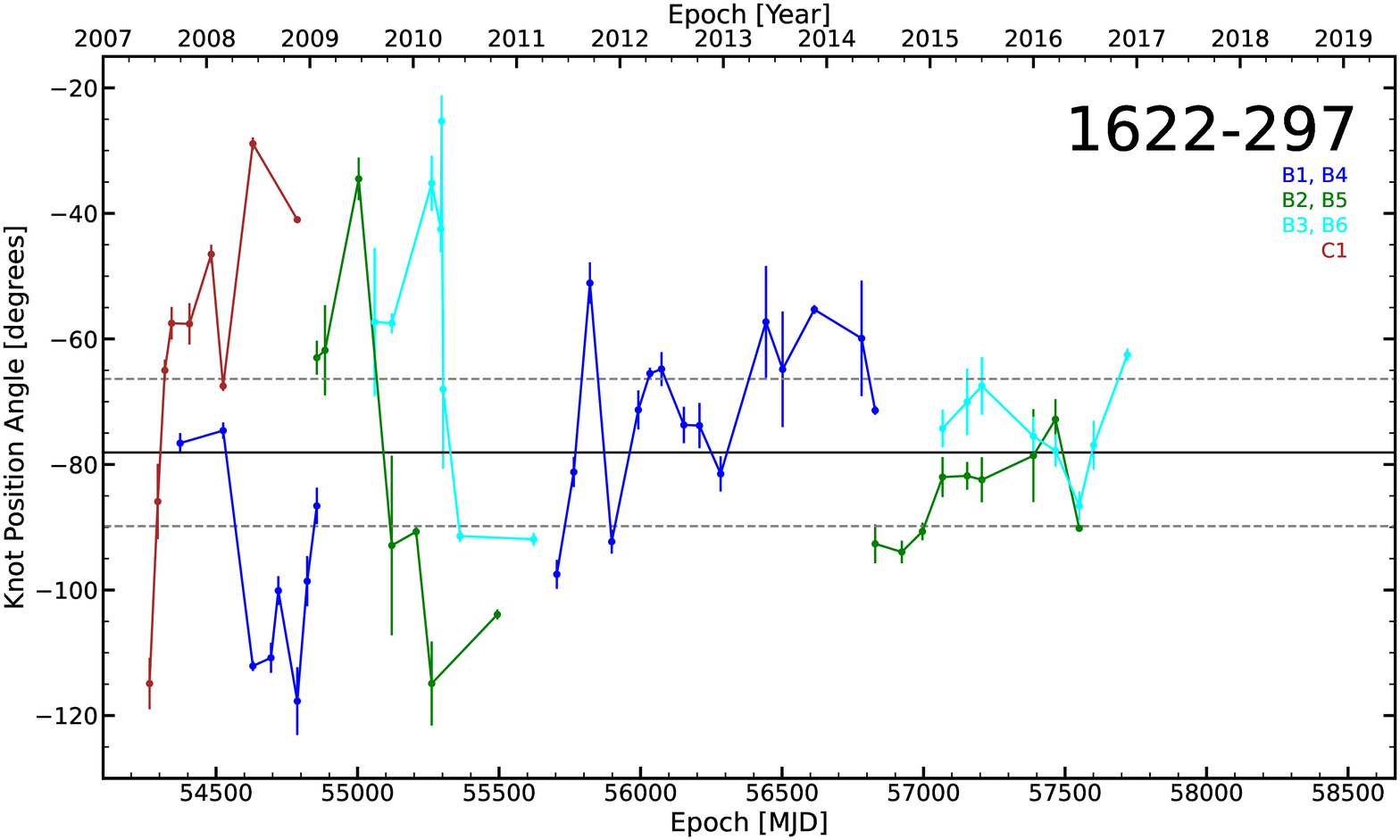}}
        \figsetgrpnote{Jet position angles of each knot, relative to the core, of the FSRQ 1622-297.}
        \figsetgrpend
        %
        % Number 29
        \figsetgrpstart
        \figsetgrpnum{6.29}
        \figsetgrptitle{1633+382}
        \figsetplot{{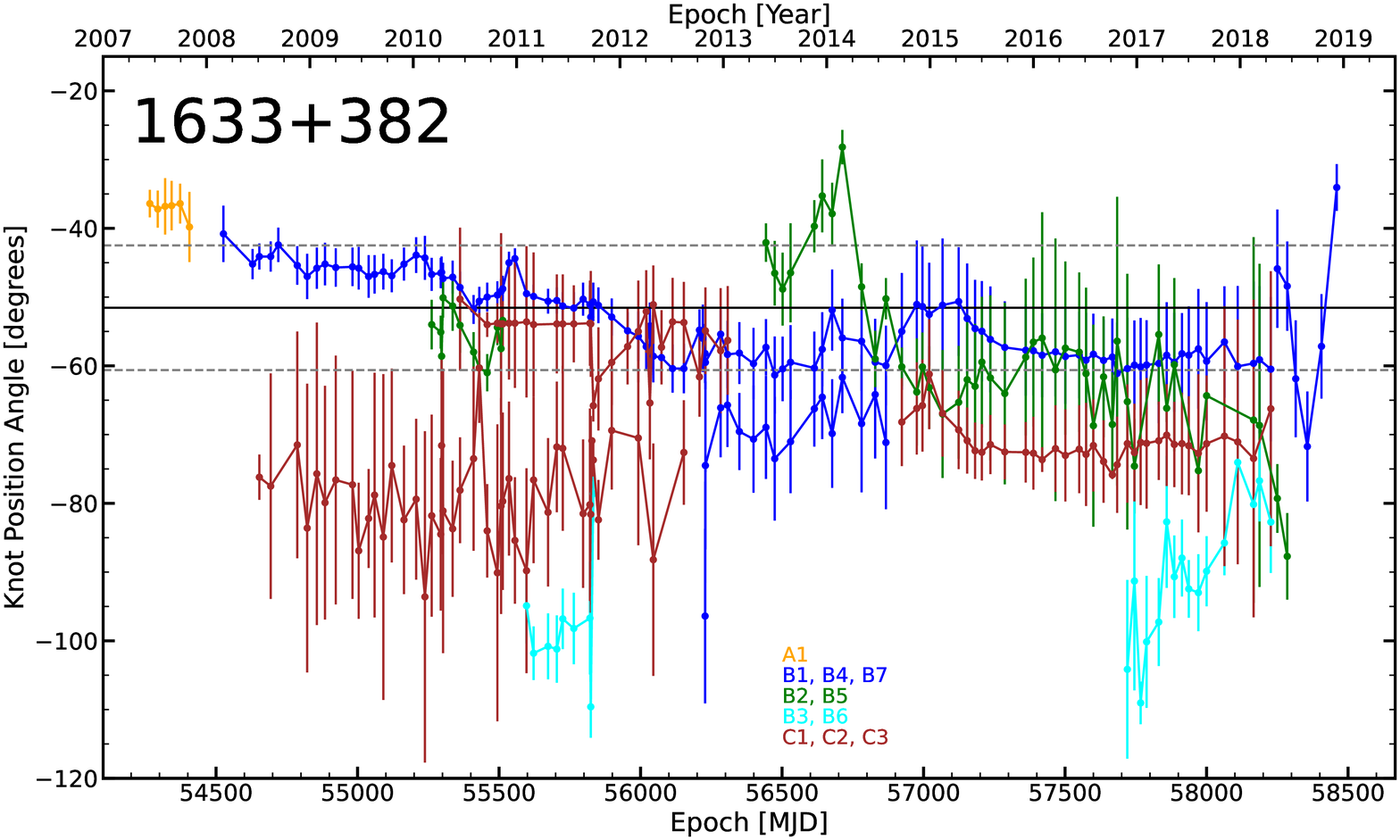}}
        \figsetgrpnote{Jet position angles of each knot, relative to the core, of the FSRQ 1633+382.}
        \figsetgrpend
        %
        % Number 30
        \figsetgrpstart
        \figsetgrpnum{6.30}
        \figsetgrptitle{1641+399}
        \figsetplot{{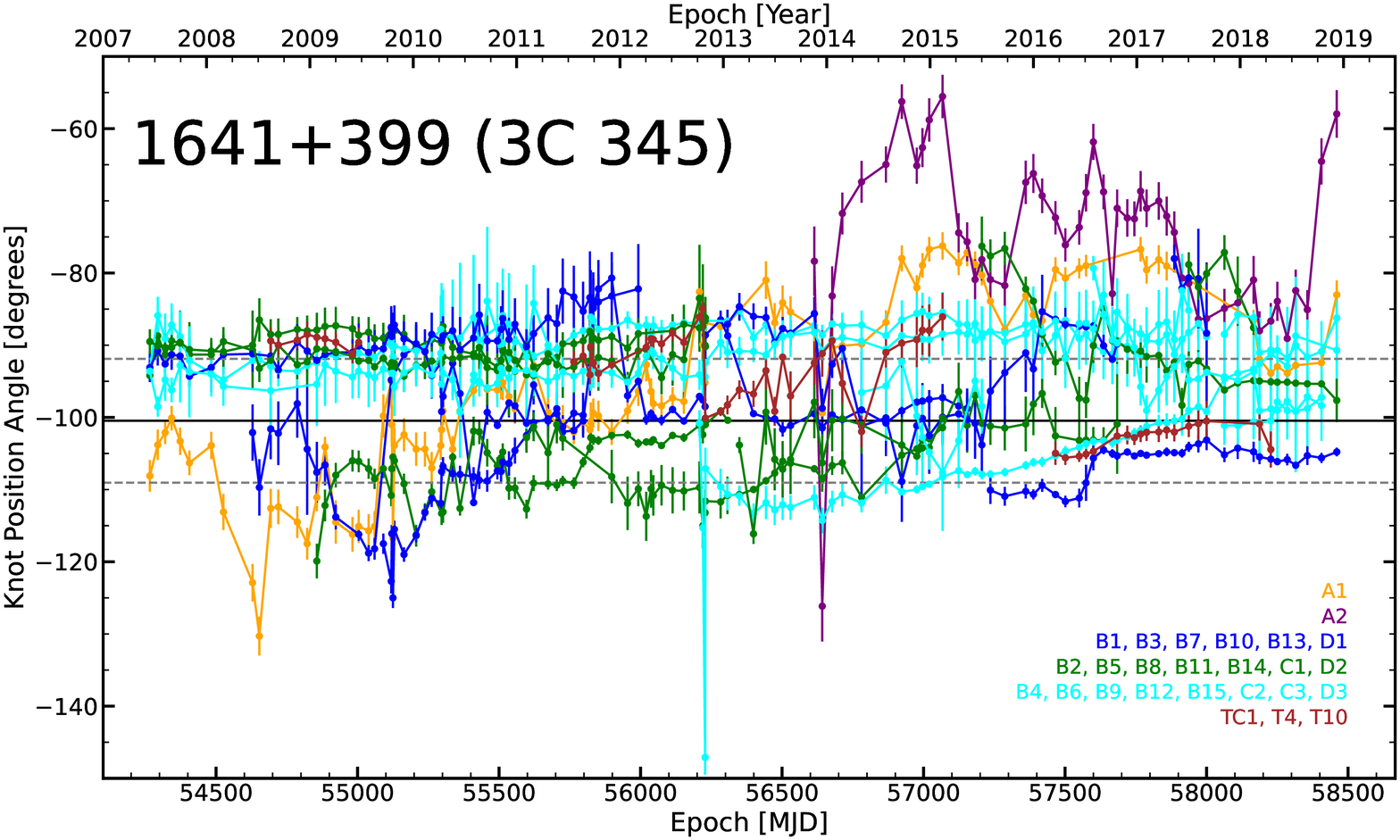}}
        \figsetgrpnote{Jet position angles of each knot, relative to the core, of the FSRQ 1641+399.}
        \figsetgrpend
        %
        % Number 31
        \figsetgrpstart
        \figsetgrpnum{6.31}
        \figsetgrptitle{1652+398}
        \figsetplot{{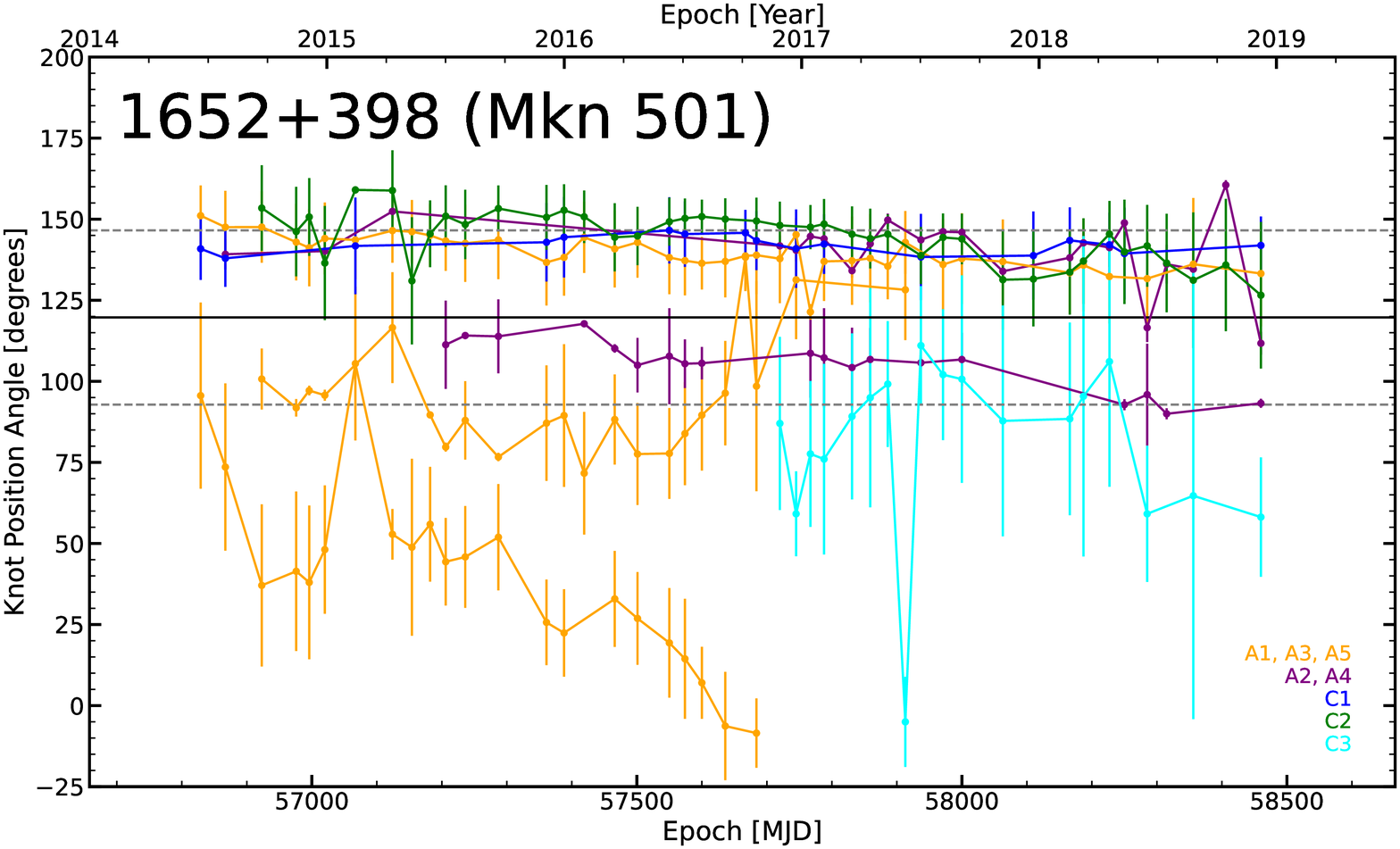}}
        \figsetgrpnote{Jet position angles of each knot, relative to the core, of the BL 1652+398.}
        \figsetgrpend
        %
        % Number 32
        \figsetgrpstart
        \figsetgrpnum{6.32}
        \figsetgrptitle{1730-130}
        \figsetplot{{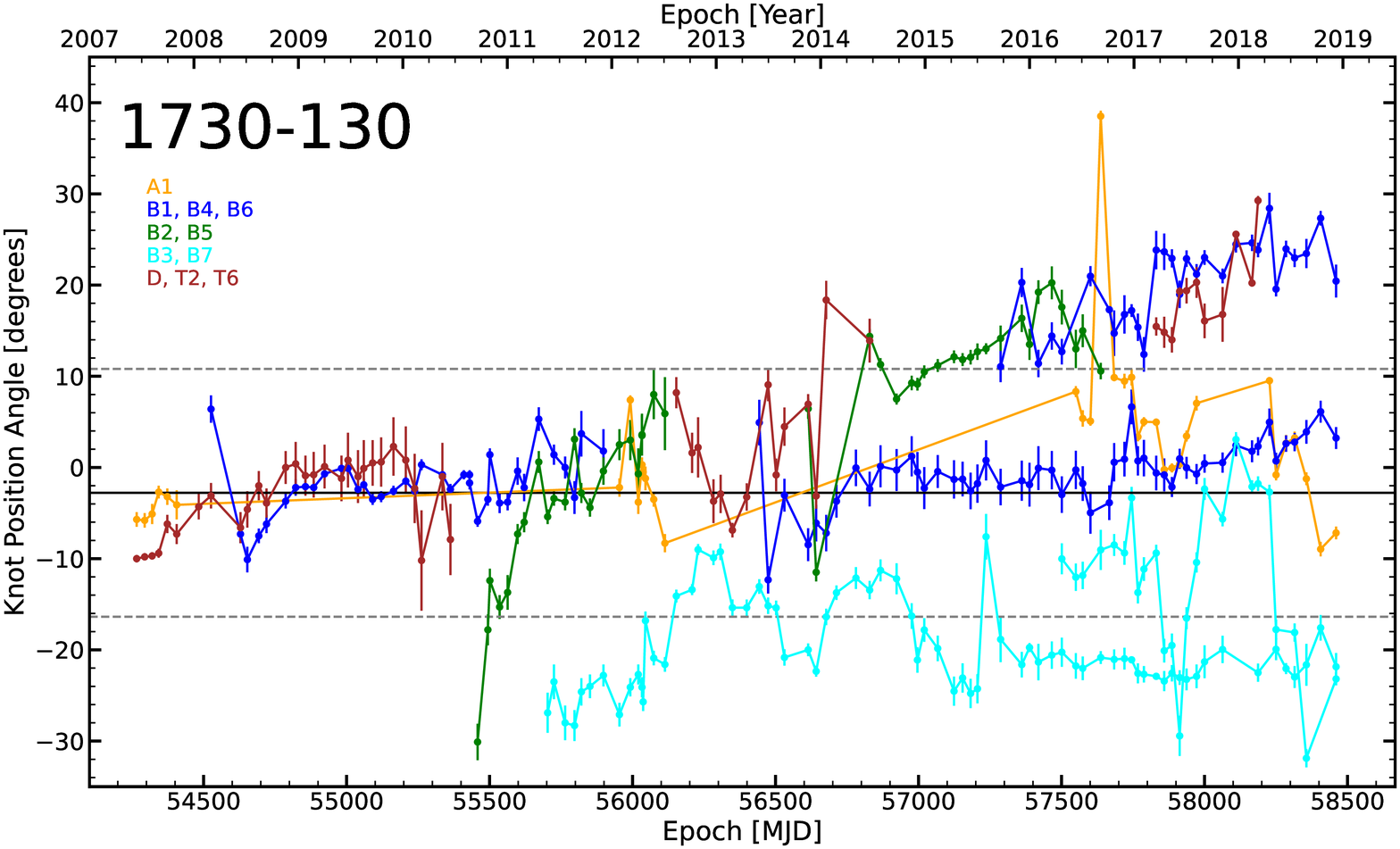}}
        \figsetgrpnote{Jet position angles of each knot, relative to the core, of the FSRQ 1730-130.}
        \figsetgrpend
        %
        % Number 33
        \figsetgrpstart
        \figsetgrpnum{6.33}
        \figsetgrptitle{1749+096}
        \figsetplot{{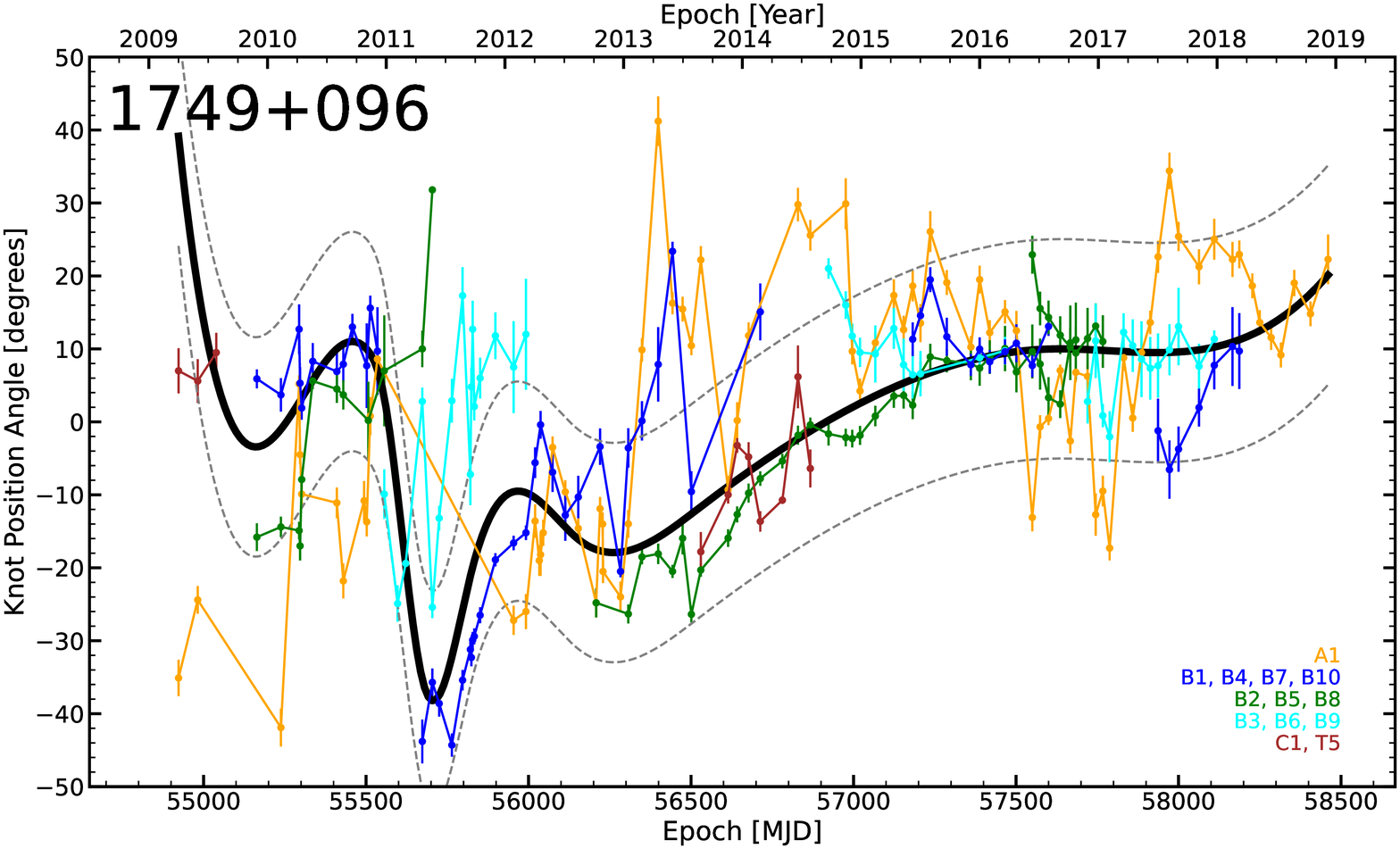}}
        \figsetgrpnote{Jet position angles of each knot, relative to the core, of the BL 1749+096.}
        \figsetgrpend
        %
        % Number 34
        \figsetgrpstart
        \figsetgrpnum{6.34}
        \figsetgrptitle{1959+650}
        \figsetplot{{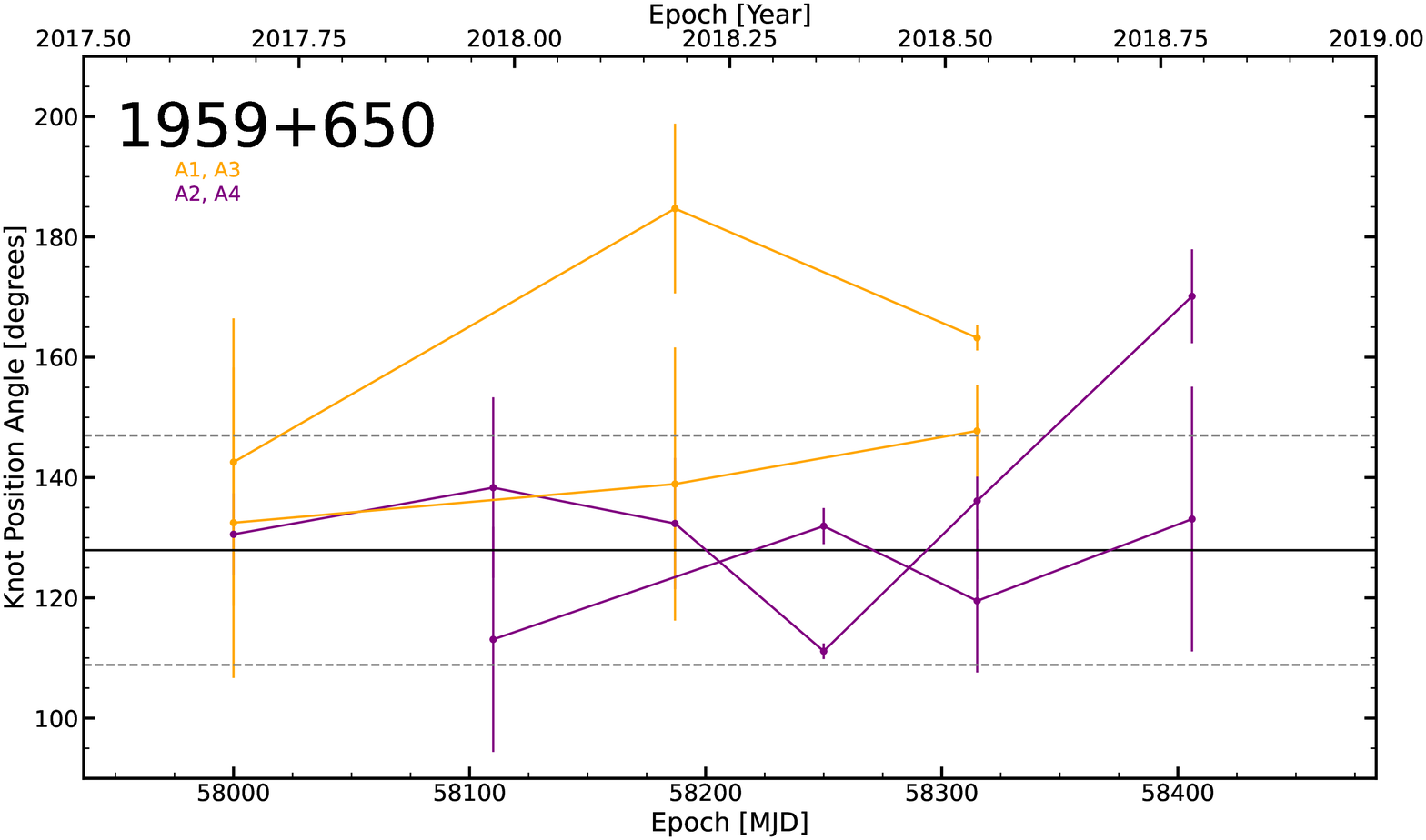}}
        \figsetgrpnote{Jet position angles of each knot, relative to the core, of the BL 1959+650.}
        \figsetgrpend
        %
        % Number 35
        \figsetgrpstart
        \figsetgrpnum{6.35}
        \figsetgrptitle{2200+420}
        \figsetplot{{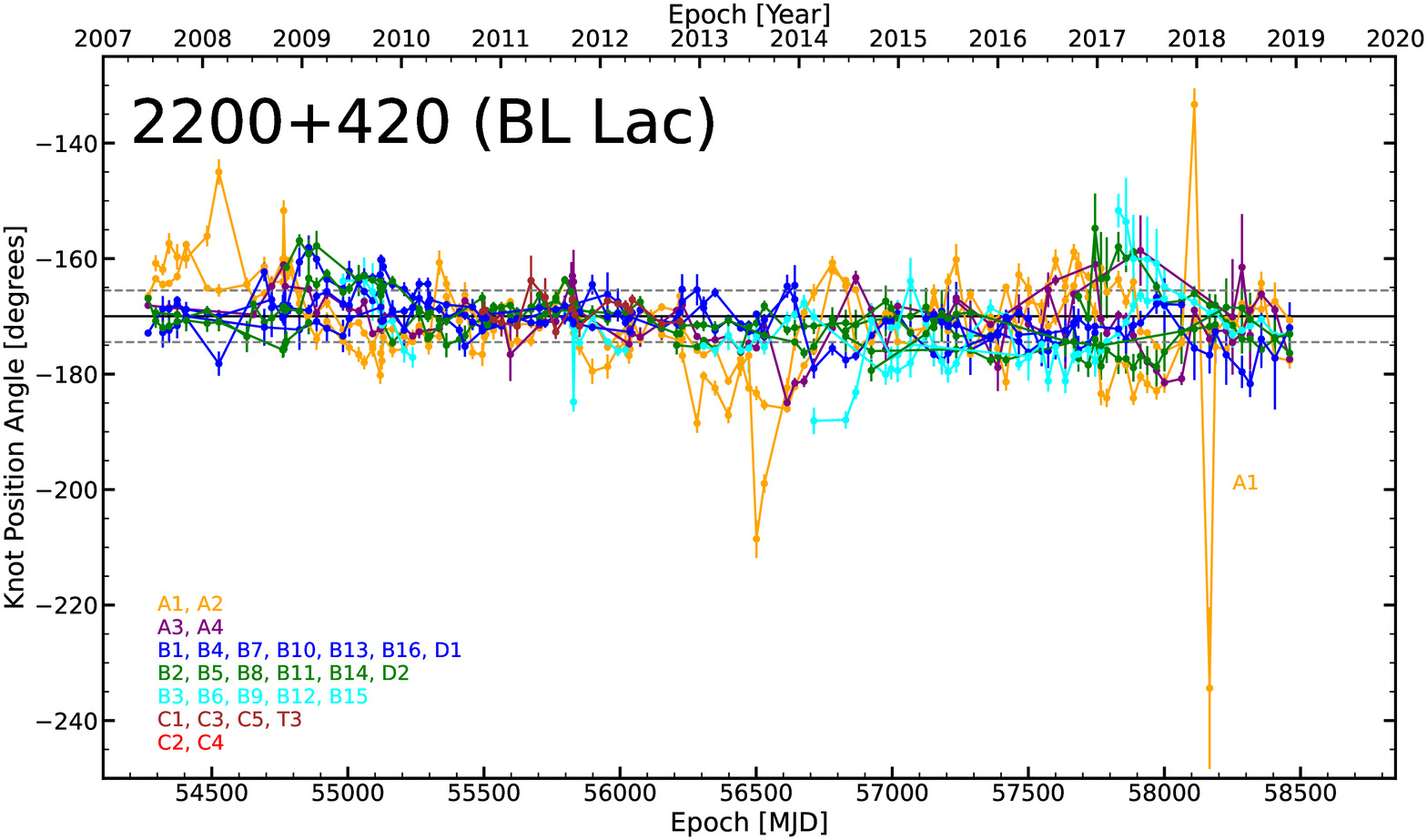}}
        \figsetgrpnote{Jet position angles of each knot, relative to the core, of the BL 2200+420.}
        \figsetgrpend
        %
        % Number 36
        \figsetgrpstart
        \figsetgrpnum{6.36}
        \figsetgrptitle{2223-052}
        \figsetplot{{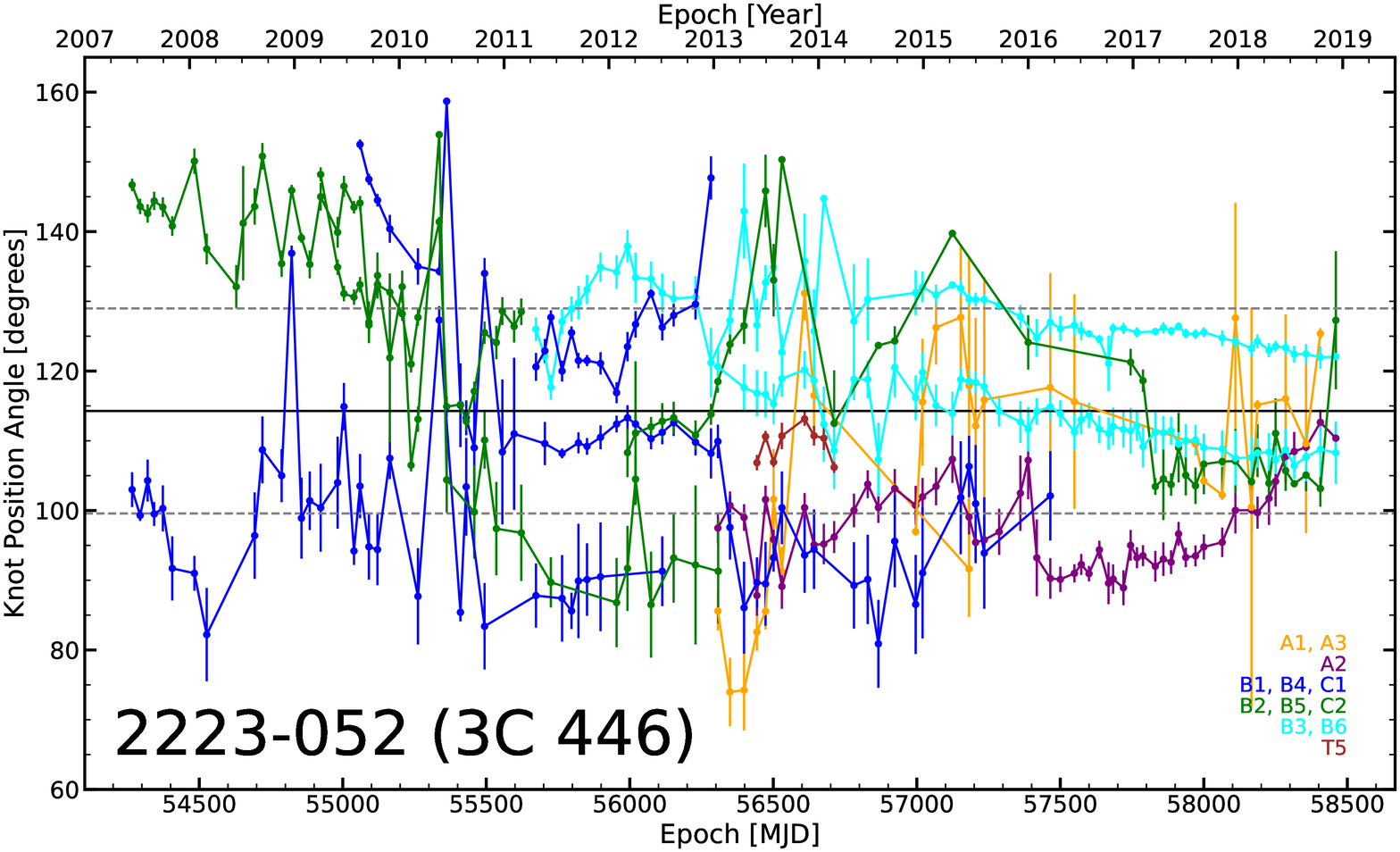}}
        \figsetgrpnote{Jet position angles of each knot, relative to the core, of the FSRQ 2223-052.}
        \figsetgrpend
        %
        % Number 37
        \figsetgrpstart
        \figsetgrpnum{6.37}
        \figsetgrptitle{2230+114}
        \figsetplot{{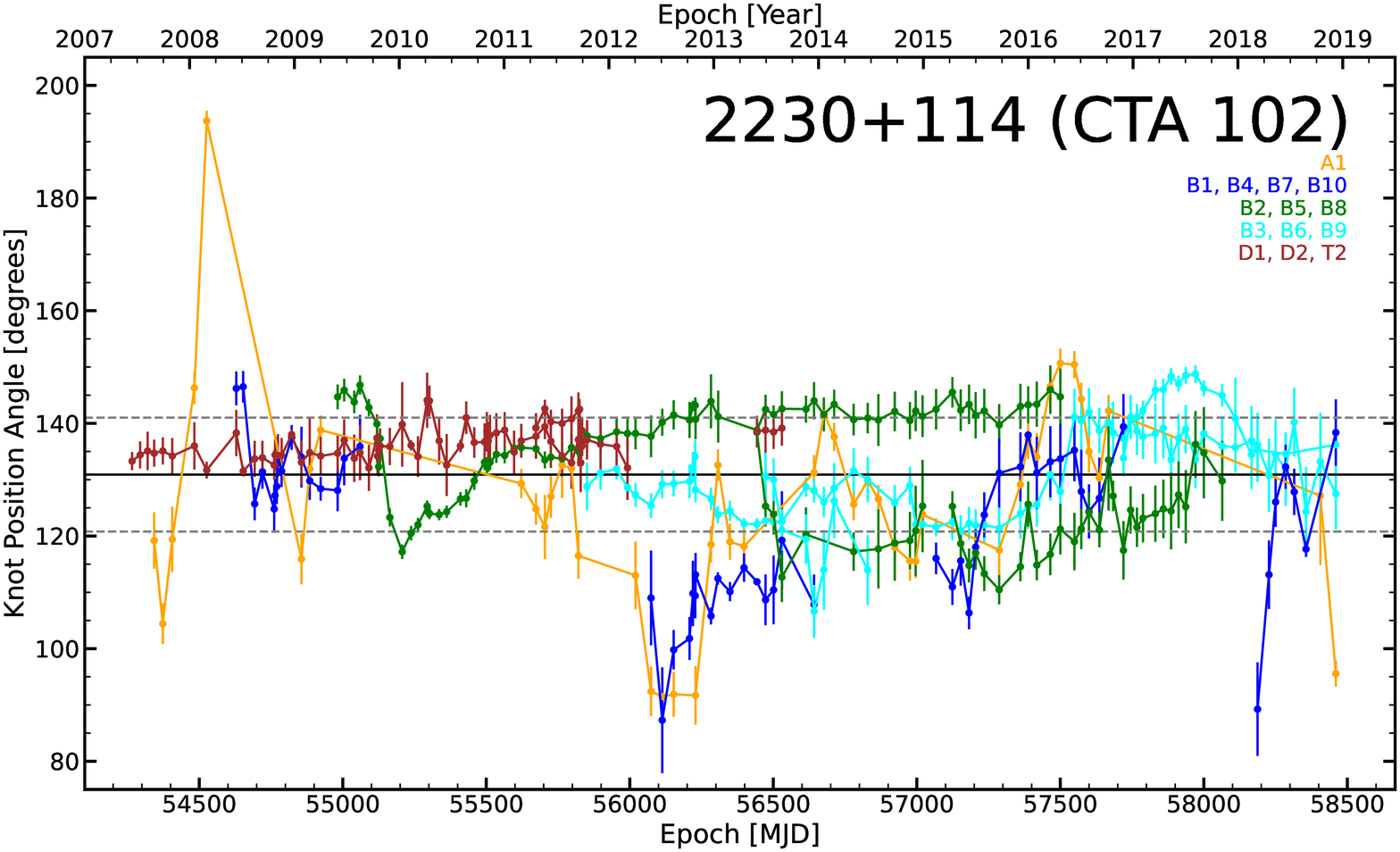}}
        \figsetgrpnote{Jet position angles of each knot, relative to the core, of the FSRQ 2230+114.}
        \figsetgrpend
        %
        % Number 38
        \figsetgrpstart
        \figsetgrpnum{6.38}
        \figsetgrptitle{2251+158}
        \figsetplot{{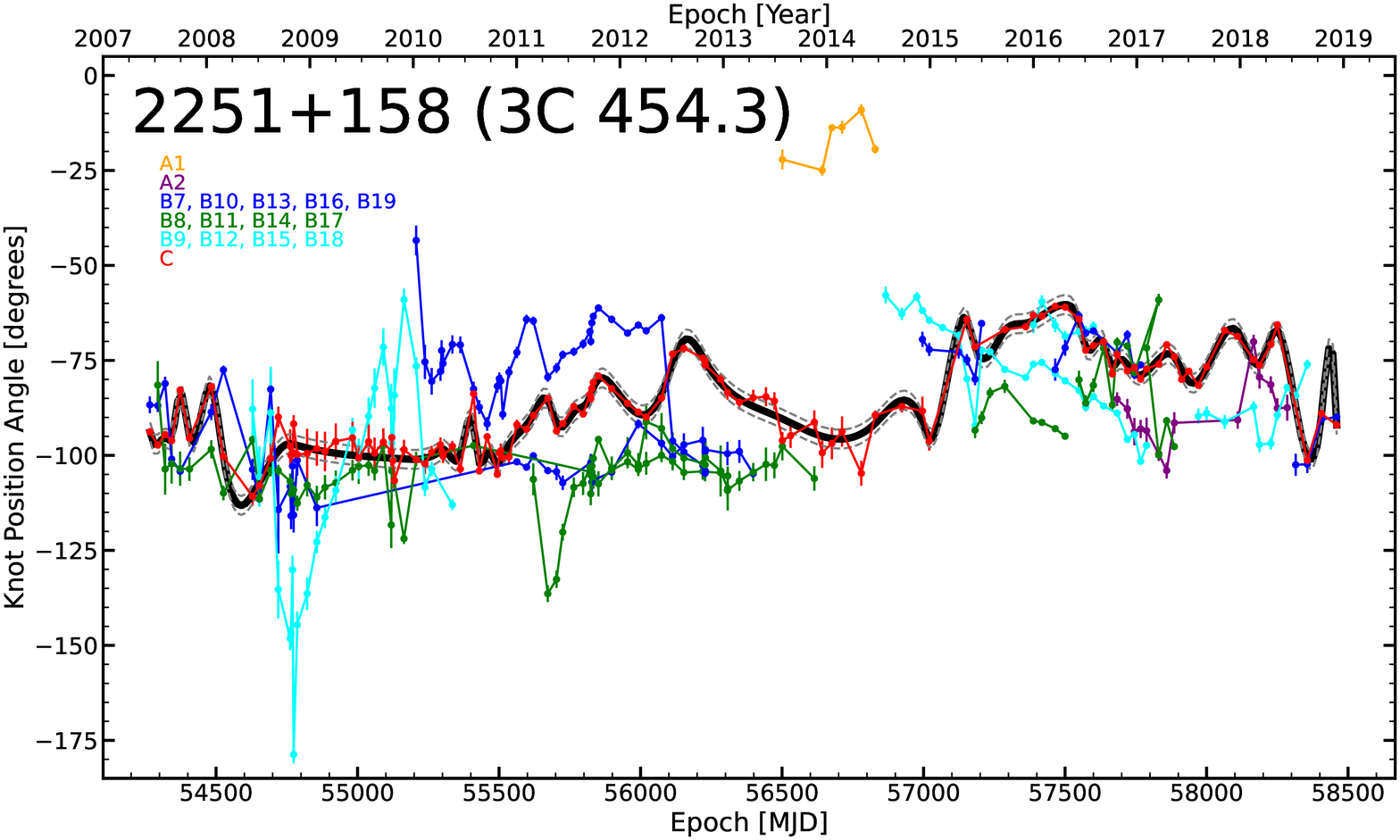}}
        \figsetgrpnote{Jet position angles of each knot, relative to the core, of the FSRQ 2251+158.}
        \figsetgrpend
\figsetend
%----

We have visually inspected the plots in Figure SET 6 to see if there are any overall trends or abrupt changes in the position angles of knots. This leads us to classify the average jet position angle behavior into three categories: 1) $\langle \Theta_{\text{jet}} (t) \rangle =$ constant (i.e., variations are insignificant), labeled as \emph{C}; 2) a linear dependence, $\langle \Theta_{\text{jet}} (t) \rangle = a_1 * (t - t_{\text{ave}}) + a_0$, where $t_{\text{ave}}$ is the midpoint of the observed period, labeled with \emph{L}; and 3) a cubic spline fit, $\langle \Theta_{\text{jet}} (t) \rangle = \text{spline}(t-t_{\text{ave}})$, labeled as \emph{S}.

In order to perform a more quantitative placement of the sources into these categories, we have calculated the average inner jet direction, $\langle \Theta_{\text{jet}} \rangle$, and standard deviation, $\sigma_{\langle\Theta_{\text{jet}} \rangle}$, by averaging $\Theta$ of all moving knots within a distance from the core of $\sim3\times$ the observing beam FWHM (i.e., the inner jet). For several jets, there are one or two knots with $\Theta$ significantly different from the others, which we exclude from this calculation.
We have compared $\sigma_{\langle\Theta_{\text{jet}} \rangle}$ with the average uncertainty of the knot position angle measurement within the inner jet, $\langle \sigma_{\Theta} \rangle$. If $\sigma_{\langle\Theta_{\text{jet}} \rangle} \leq 3\times \langle \sigma_{\Theta} \rangle$, the position angle is classified as  constant, \emph{C}.
For sources where this criterion is not met, we have employed either linear or spline fits as determined from visual inspection of the behavior of $\Theta$ for all knots.
To confirm that these fits provide a better estimate of the jet position angle, we have compared the standard deviation of the fit to the standard deviation of a constant $\langle \Theta_{\text{jet}} \rangle$. In the majority of cases, the fit reduced the standard deviation of the knot position angles to
within the $3\times\langle \sigma_{\Theta} \rangle$ criterion.
This criterion remained unmet after the fit in only four sources (0716+714, 1226+023, 1253-055, and 1749+096), but even in these cases the fits improved the standard deviation, and so were adopted. Finally, a handful of sources have a very wide dispersion, but no clear trend in the position angles of knots. For these sources, we have created a fourth category (\emph{W} for ``wide"), where we have adopted a constant value of $\langle \Theta_{\text{jet}} \rangle$, but with a standard deviation $>3\times \langle \sigma_{\Theta} \rangle$.

Table~\ref{tab:JetPAs} describes the fits for each source as follows:
1---source name;
2---constant average position angle of the knots in a jet, $\langle \Theta_{\text{jet}} \rangle$ (in degrees), regardless of the best fit;
3---average error of knot position angles, $\langle \sigma_{\Theta} \rangle$ (in degrees);
4---categorization of the jet position angle behavior, with categories as described above\footnote{Due to their complexity, spline fits are available upon request. Linear fits are provided below.};
5---midpoint in time of all the knot observations, $t_{\text{ave}}$, for the fits;
6---standard deviation of the best fit to knot position angles, $\sigma_{\langle \Theta_{\text{jet}} \rangle}$, in degrees; and
7---knots that were used to determine the average position angles and fits.

The majority of sources in our sample (23/38, 60.5\%) have knots in the jet with position angles that are relatively constant over time, i.e., have a narrow spread. For these we take the value in column 2 of Table~\ref{tab:JetPAs} to be $\langle \Theta_{\text{jet}} \rangle$ in the analyses that follow. Six other sources (15.8\% of the sample) are well described by a constant jet direction, but have very wide jets in terms of the spread of the position angles of knots. Despite the latter property, we adopt the value listed in column 2 for the subsequent analyses. Only two sources (the FSRQs 0420$-$014 and 1226+023) have clear linear trends of knot position angle variations over time. The linear fit for 0420$-$014 is $\langle \Theta_{\text{jet}} (t) \rangle = (5.9 \pm 0.2)*(t - 2013.24) + (127.0 \pm 0.7)$ degrees, while the fit for 1226+023 is $\langle \Theta_{\text{jet}} (t) \rangle = (1.2 \pm 0.1) * (t - 2013.79) - (140.9 \pm 0.2)$ degrees. The remaining seven sources (18.4\%) display very complex behavior. Some (e.g., the FSRQ 1222+216) visually resemble jet rotation (possibly precession), while others exhibit abrupt changes in the jet position angle (e.g., the FSRQ 1253$-$055). For the sources best fit with a linear trend or spline, we use the fits to calculate the jet position angle at the time of acceleration for the analysis in $\S$\ref{subsec:KnotAccels}. We use the average of the fit over the observed time period for each knot for the analysis of stationary features in $\S$\ref{subsec:QuasiStationaryFeatures}. For all sources, we use $\sigma_{\langle \Theta_{\text{jet}} \rangle}$ as the uncertainty on $\langle \Theta_{\text{jet}} \rangle$ in the subsequent analyses.

\begin{deluxetable*}{lrccrrr}
    \tablecaption{Jet Position Angles\label{tab:JetPAs}}
    \tablewidth{0pt}
    \tablehead{
    \colhead{Source} & \colhead{$\langle \Theta_{\text{jet}} \rangle$} & \colhead{$\langle \sigma_{\Theta} \rangle$} & \colhead{Type\tablenotemark{a}}  & \colhead{$t_{\text{ave}}$} & \colhead{$\sigma_{\langle \Theta_{\text{jet}} \rangle}$} & \colhead{Knots Included} \\
    \colhead{} & \colhead{[deg]} & \colhead{[deg]} & \colhead{} & \colhead{[yr]} & \colhead{[deg]} & \colhead{}
    }
    \colnumbers
    \startdata
        0219+428  & $-169.6$ & $\phn 5.5$ & C & $\ldots$  &  $9.0$ & B1---B6 \\
        0235+164  & $  54.7$ & $\phn 6.8$ & C & $\ldots$  &  $6.6$ & B5, B6 \\
        0316+413  & $ 173.3$ & $\phn 0.4$ & W & $\ldots$  &  $9.6$ & All except C4, C8, C12, C13 \\
        0336--019 & $  78.7$ & $\phn 4.6$ & C & $\ldots$  &  $6.4$ & B1---B5 \\
        0415+379  & $  66.1$ & $\phn 2.2$ & C & $\ldots$  &  $4.9$ & K1---K31 \\
        0420--014 & $ 121.0$ & $\phn 3.7$ & L & $2013.24$ &  $8.3$ & B1---B9 \\
        0430+052  & $-113.5$ & $\phn 2.9$ & C & $\ldots$  &  $3.2$ & C1---C15 \\
        0528+134  & $  64.8$ & $\phn 6.0$ & C & $\ldots$  & $13.3$ & B1---B3; C1---C5 \\
        0716+714  & $  24.3$ & $\phn 3.9$ & S & $2012.98$ & $18.4$ & A1, A2; B1---B14; T10 \\
        0735+178  & $  78.5$ & $10.7$     & C & $\ldots$  & $13.8$ & B1---B4; C1 \\
        0827+243  & $ 127.1$ & $\phn 5.8$ & C & $\ldots$  & $ 8.1$ & B1---B9 \\
        0829+046  & $  63.1$ & $\phn 5.7$ & C & $\ldots$  & $10.6$ & A1---A4; B1---B6 \\
        0836+710  & $-140.0$ & $\phn 3.3$ & C & $\ldots$  & $ 9.0$ & B1---B6 \\
        0851+202  & $-24.1 $ & $\phn 1.2$ & S & $2012.44$ & $ 2.7$ & a \\
        0954+658  & $- 13.7$ & $\phn 2.1$ & S & $2013.97$ & $ 5.5$ & A1, A2 \\
        1055+018  & $-63.8 $ & $\phn 6.5$ & C & $\ldots$  & $18.4$ & B1; B4; T1 \\
        1101+384  & $- 14.4$ & $14.2$     & C & $\ldots$  & $13.9$ & A1; B1, B3, B4 \\
        1127--145 & $  80.7$ & $\phn 4.3$ & C & $\ldots$  & $10.4$ & B1, B2; C1, C2; D1 \\
        1156+295  & $  10.0$ & $\phn 3.8$ & C & $\ldots$  & $11.1$ & B1---B7 \\
        1219+285  & $ 106.2$ & $10.6$     & C & $\ldots$  & $ 8.0$ & B1---B4 \\
        1222+216  & $  23.5$ & $\phn 1.9$ & S & $2014.11$ & $ 4.7$ & A1, A2, A3 \\
        1226+023  & $-141.3$ & $\phn 0.7$ & L & $2013.79$ & $ 4.3$ & B1---B22 \\
        1253--055 & $-152.7$ & $\phn 1.0$ & S & $2013.37$ & $20.0$ & A1; C24---C38 \\
        1308+326  & $- 64.9$ & $\phn 4.7$ & C & $\ldots$  & $12.9$ & B2---B6 \\
        1406--076 & $- 81.9$ & $\phn 7.2$ & C & $\ldots$  & $14.5$ & B1---B5; C1 \\
        1510--089 & $ -33.1$ & $\phn 2.8$ & C & $\ldots$  & $ 8.0$ & All except B10 \\
        1611+343  & $ 163.3$ & $\phn 3.3$ & W & $\ldots$  & $13.0$ & A2, A3; C1---C3 \\
        1622--297 & $- 78.1$ & $\phn 3.3$ & W & $\ldots$  & $11.7$ & B4---B6 \\
        1633+382  & $- 51.6$ & $\phn 6.1$ & C & $\ldots$  & $ 9.1$ & B1, B2, B4, B5, B7 \\
        1641+399  & $-100.5$ & $\phn 2.6$ & W & $\ldots$  & $ 8.6$ & B1---B15; C1---C3; D1---D3 \\
        1652+398  & $ 119.7$ & $11.8$     & C & $\ldots$  & $26.9$ & A2---A5; C2, C3 \\
        1730--130 & $-  2.8$ & $\phn 1.4$ & W & $\ldots$  & $13.6$ & A1; B1---B7; C1; D; T2 \\
        1749+096  & $-  1.2$ & $\phn 2.3$ & S & $2014.03$ & $15.0$ & A1; B1---B10; C1; T5 \\
        1959+650  & $ 127.9$ & $12.9$     & C & $\ldots$  & $19.1$ & A1---A4 \\
        2200+420  & $-170.0$ & $\phn 1.8$ & C & $\ldots$  & $ 4.5$ & B1---B16; C1---C5; D1, D2; T3 \\
        2223--052 & $ 114.3$ & $\phn 3.2$ & W & $\ldots$  & $14.7$ & A1---A3; B1---B6; C1, C2; T5 \\
        2230+114  & $ 130.9$ & $\phn 3.6$ & C & $\ldots$  & $10.1$ & B1---B10; D1, D2 \\
        2251+158  & $- 81.8$ & $\phn 1.6$ & S & $2012.67$ & $2.5$ & C \\
    \enddata
    \tablenotetext{a}{Final fit to the jet position angle behavior, labeled as follow: C---$\langle \Theta_{\text{jet}} (t)\rangle =$ constant, for which the value in column 2 is used in subsequent analyses in this work; W---sources for which $\langle \Theta_{\text{jet}} (t) \rangle =$ constant, but $\sigma_{\langle\Theta_{\text{jet}} \rangle} > 3\times \langle \sigma_{\Theta} \rangle$; L---linear dependence, $\langle \Theta_{\text{jet}} (t) \rangle = a_1 * (t - t_{\text{ave}}) + a_0$, where $t_{\text{ave}}$ is the midpoint of the observed period; and S---a spline fit of the 3rd order, $\langle \Theta_{\text{jet}}(t) \rangle = \text{spline}(t-t_{\text{ave}})$.}
    \tablecomments{PAs are measured north through east.}
\end{deluxetable*}

\section{Motions of Knots in the Jets}
\label{sec:Speeds}

Between 2007 and 2018 we have identified 559 distinct emission features in the parsec-scale jets of the 38 blazars in the VLBA-BU-BLAZAR sample. Of this total number, 38 are the 43 GHz cores. Among the remaining knots, 96 have been classified as quasi-stationary (18.4\% of non-core components), leaving 425 classified as moving knots. Accelerations along the jet cause 75 of the moving knots (17.6\%) to have multiple estimates for their apparent speed as they traverse different regions along their trajectory. The only sources that do not exhibit apparent superluminal motion in the parsec-scale jet are the BLs 1652+398 and 1959+650. Interestingly, we find superluminal motion in the generally subluminal jet of the RG 0316+413 (3C\ 84) in one knot, \emph{C10}, with an apparent speed of $\beta_{\text{app}} = 1.38c \pm 0.03c$. We consider this estimate of the knot speed to be reliable, with multiple observations of its motion close to the core and a well-defined epoch of ejection. Also, we detect knot motions up to $\sim2c$ in the BL 1101+384 (Mkn\ 421).

\begin{figure*}[t]
    \figurenum{7}
    \begin{center}
        \includegraphics[width=\textwidth]{{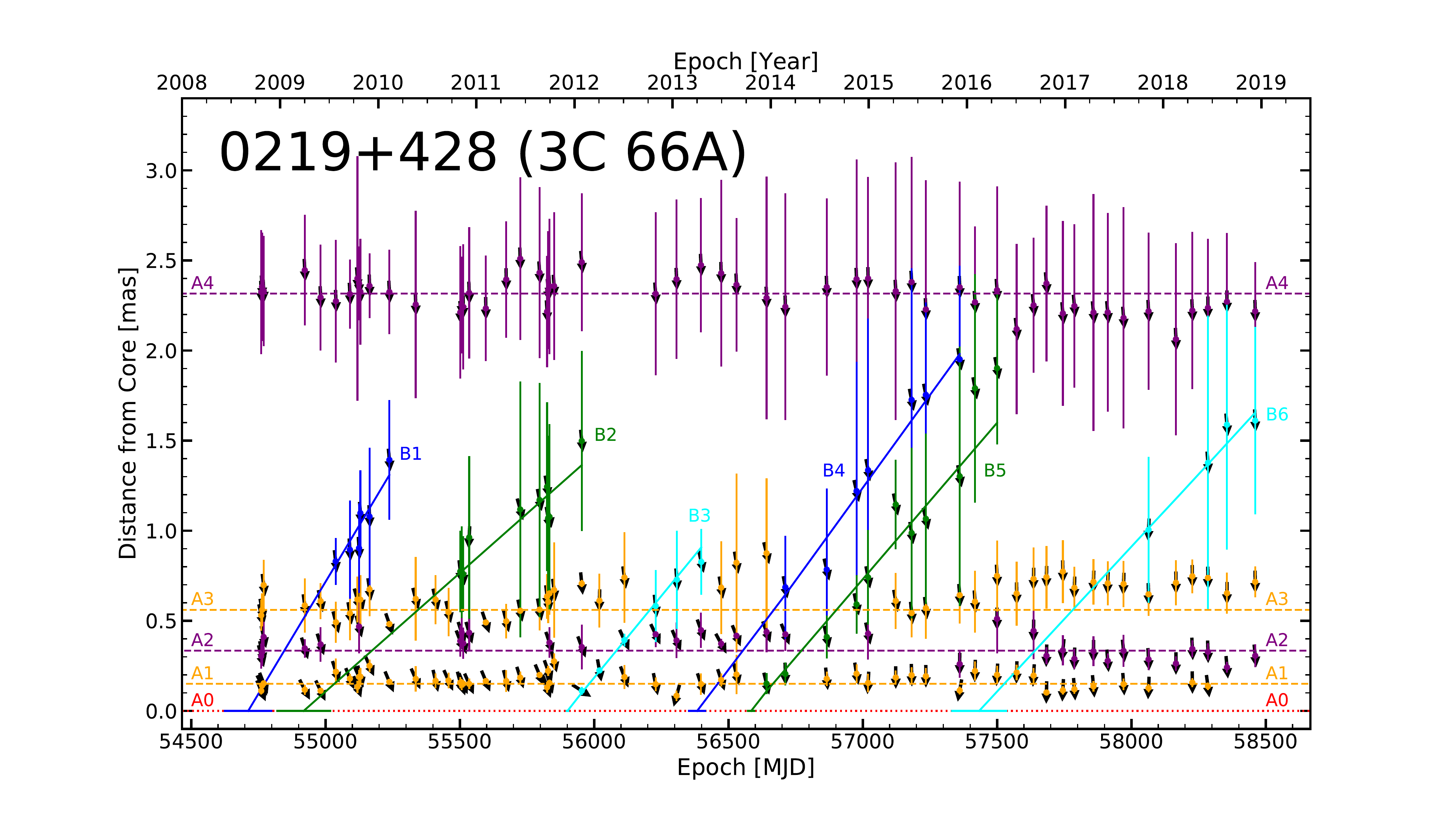}}
        \caption{Separation vs.\ time of the knots in the jet of the BL 0219+428 (3C 66A). The colors of the knots --- the same as in Figure~\ref{fig:ExampleTheta} --- are coded based on their motion designation: blue, green, and cyan represent bright moving knots, orange and purple correspond to stationary components, and brown is used for trailing knots (appearing to split off from other knots), relatively faint moving features, or to differentiate between closely spaced knots. Red dotted lines are used to show the core component \emph{A0}.
        Horizontal lines along \emph{A0} indicate the uncertainty in time of the epoch of ejection. See the text for more details. Plots for all sources in the sample are available in the figure set.\newline(The complete figure set (38 images) is available online.)\label{fig:ExampleR}}
    \end{center}
\end{figure*}

For each source, we have constructed a plot displaying the separation of all knots from the core (Figure SET 7). Each knot is color-coded in the corresponding figure in Figure SET 6 according to its motion type. In each figure, the black vectors show the PA of knots with respect to the core at the corresponding epoch. The solid lines represent the piece-wise linear fits to the knot motions, the red dotted line shows the position of the core, and dashed colored lines mark the average positions of the stationary components. Error bars on each measurement are the approximate $1\sigma$ positional uncertainties based on $T_{\text{b,obs}}$. Lines extrapolating the knot motion back to the epoch of ejection $t_\circ$ may appear slightly bent in some cases owing to (1) $R$ being a quadratic combination of $X$ and $Y$, or (2) instances where $|t_{\text{x}\circ} - t_{\text{y}\circ}| \geq 0.5$ yr so that the epoch of ejection is calculated using a linear fit to the first segment of \emph{R} data. These apparent bends have no effect on the apparent speeds.
Figure~\ref{fig:ExampleR} presents an example of these plots for the BL 0219+428 (3C\ 66A).

%---- Figure Set 7
\figsetstart
    \label{figset:7}
    \figsetnum{7}
    \figsettitle{Knot Separations from Core}
        % Number 1
        \figsetgrpstart
        \figsetgrpnum{7.1}
        \figsetgrptitle{0219+428}
        \figsetplot{{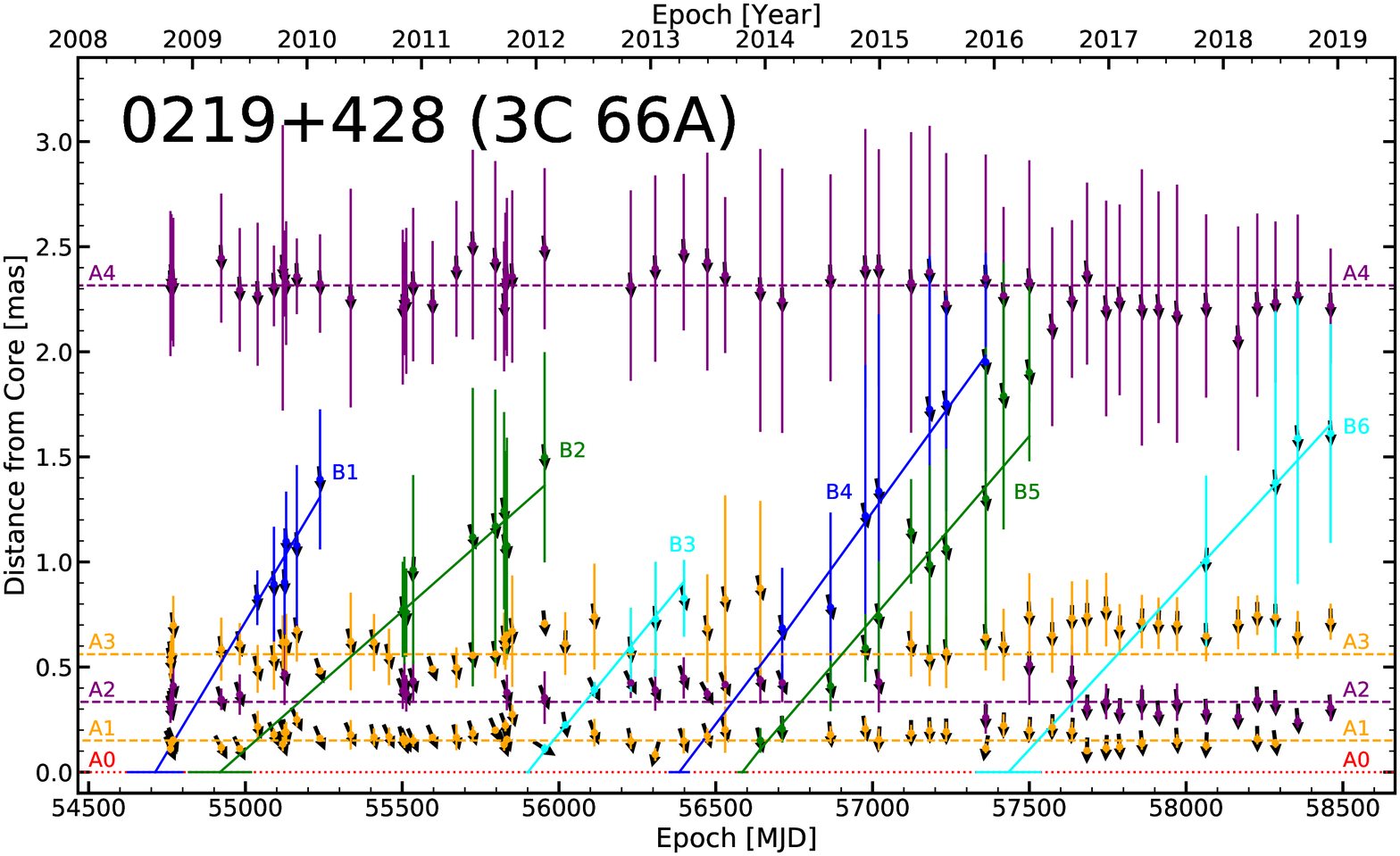}}
        \figsetgrpnote{Separation vs. time of the knots in the jet, relative to the core, of the BL 0219+428.}
        \figsetgrpend
        %
        % Number 2
        \figsetgrpstart
        \figsetgrpnum{7.2}
        \figsetgrptitle{0235+164}
        \figsetplot{{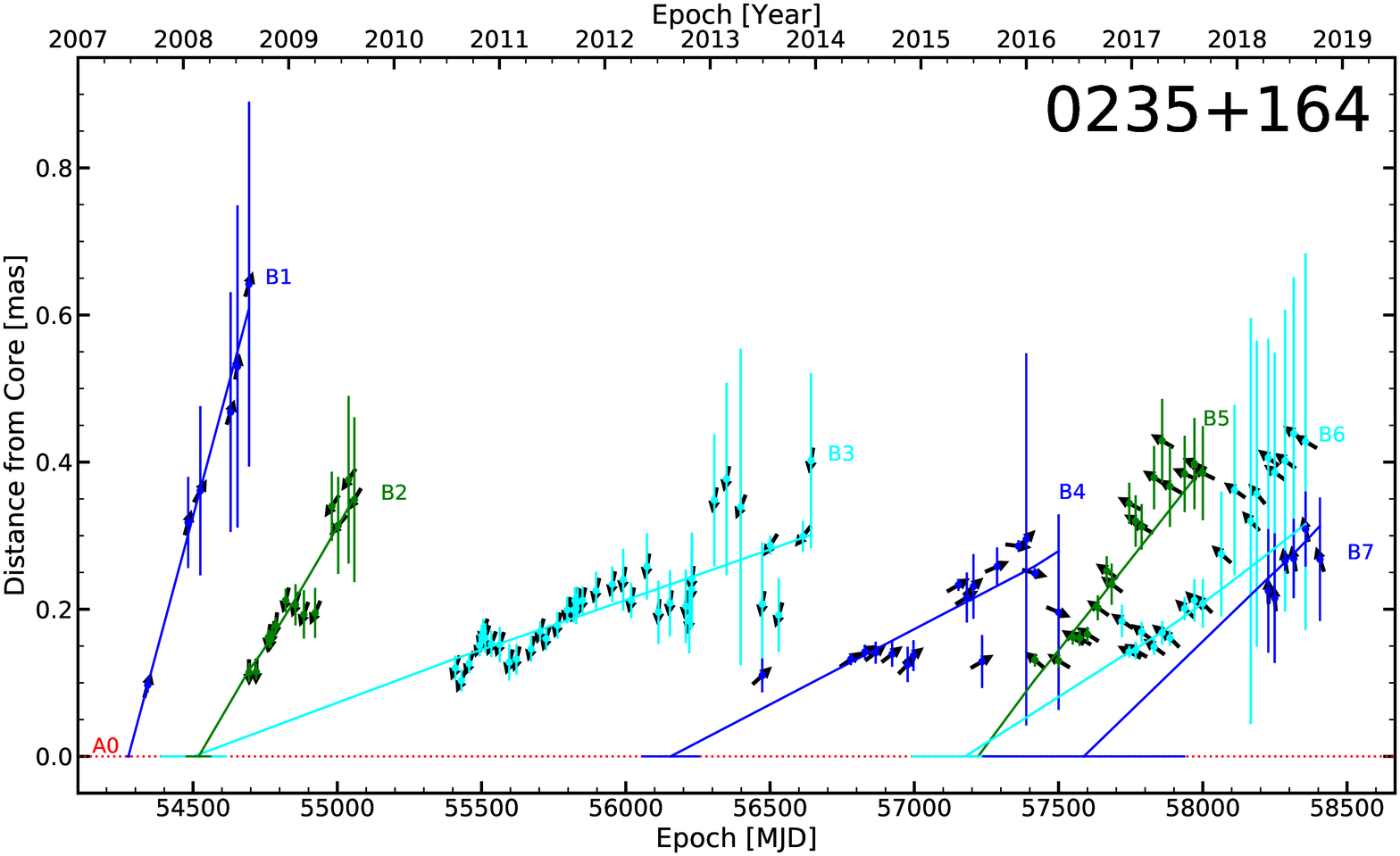}}
        \figsetgrpnote{Separation vs. time of the knots in the jet, relative to the core, of the BL 0235+164.}
        \figsetgrpend
        %
        % Number 3
        \figsetgrpstart
        \figsetgrpnum{7.3}
        \figsetgrptitle{0316+413}
        \figsetplot{{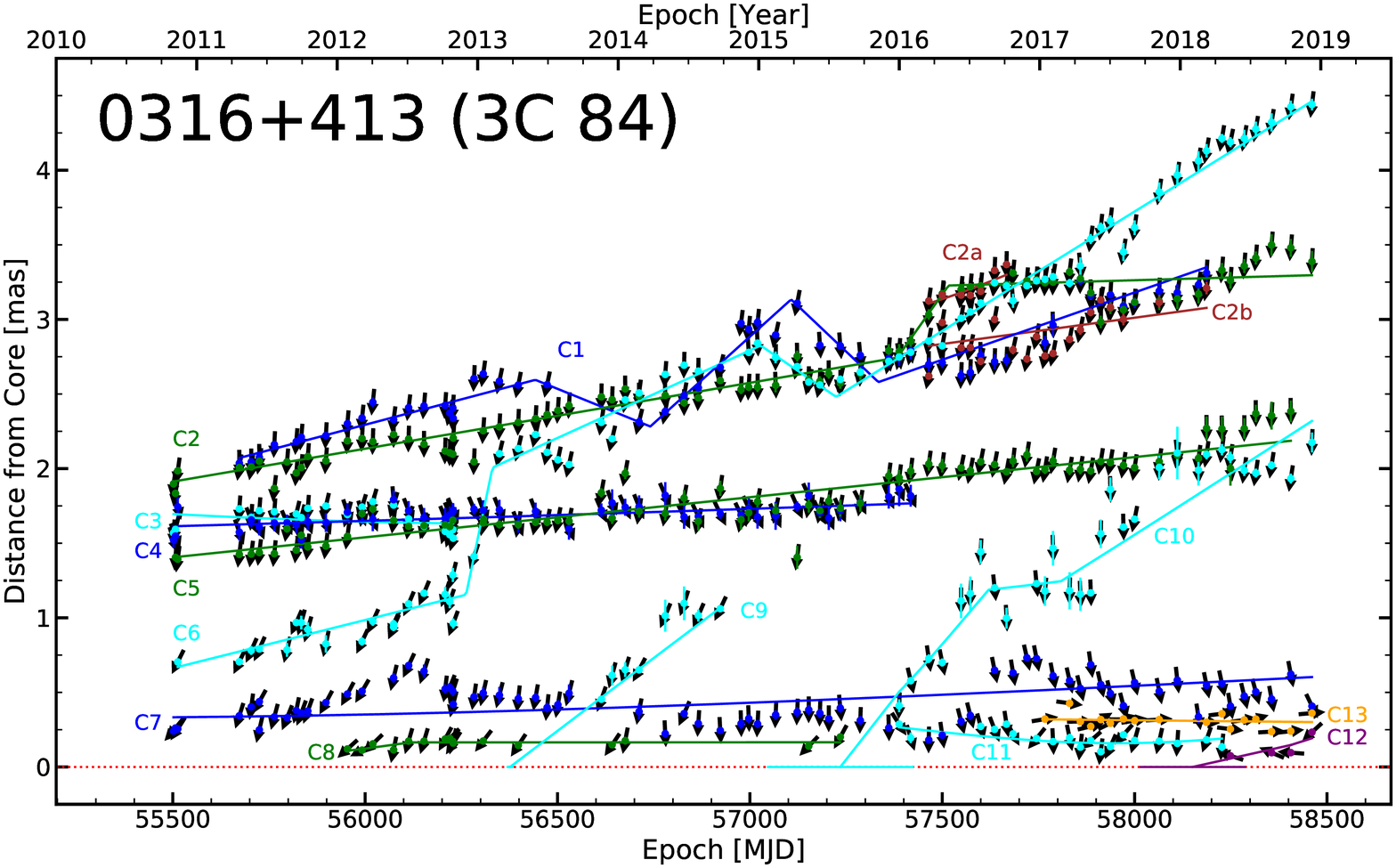}}
        \figsetgrpnote{Separation vs. time of the knots in the jet, relative to the core, of the RG 0316+413.}
        \figsetgrpend
        %
        % Number 4
        \figsetgrpstart
        \figsetgrpnum{7.4}
        \figsetgrptitle{0336-019}
        \figsetplot{{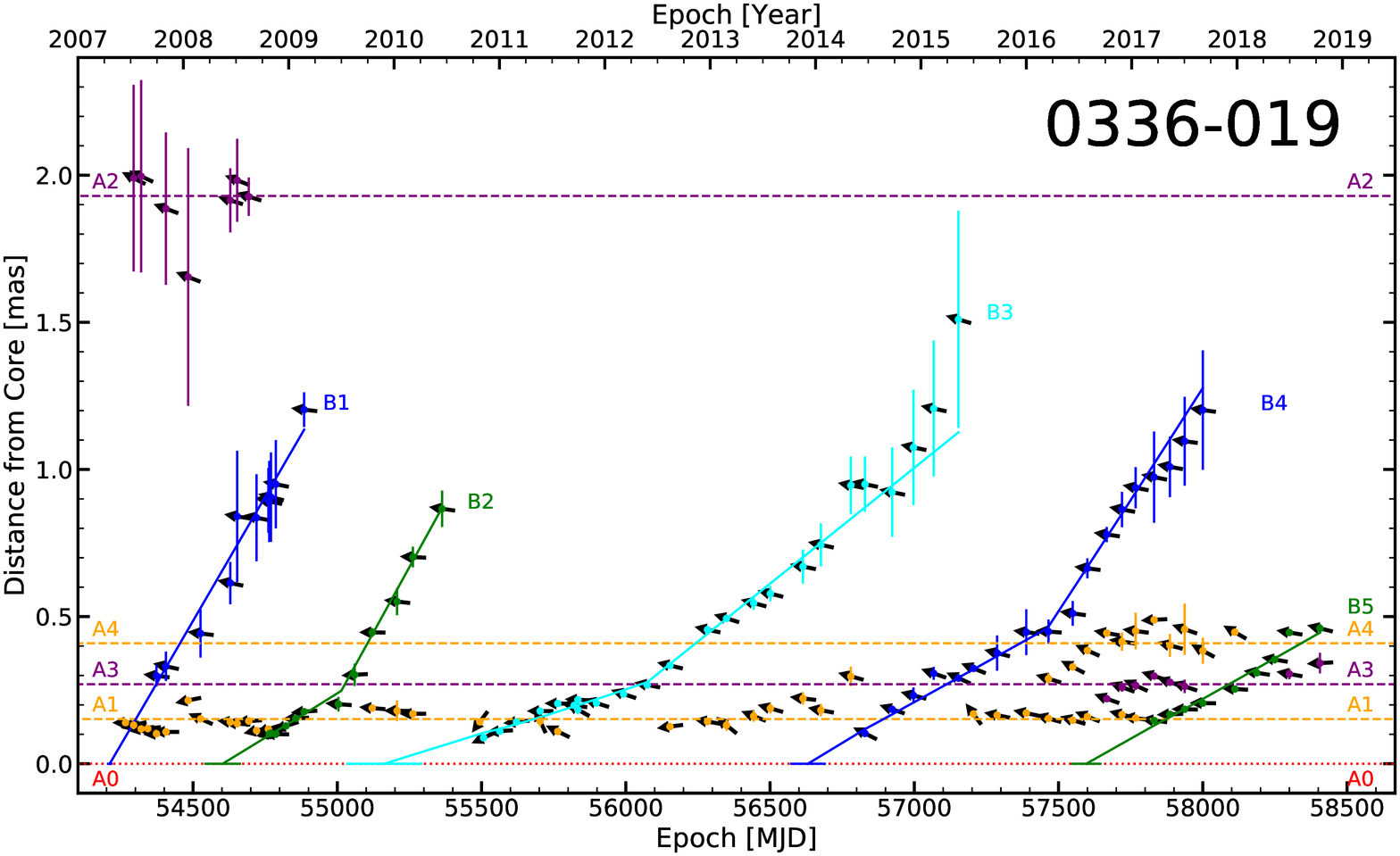}}
        \figsetgrpnote{Separation vs. time of the knots in the jet, relative to the core, of the FSRQ 0336-019.}
        \figsetgrpend
        %
        % Number 5
        \figsetgrpstart
        \figsetgrpnum{7.5}
        \figsetgrptitle{0415+379}
        \figsetplot{{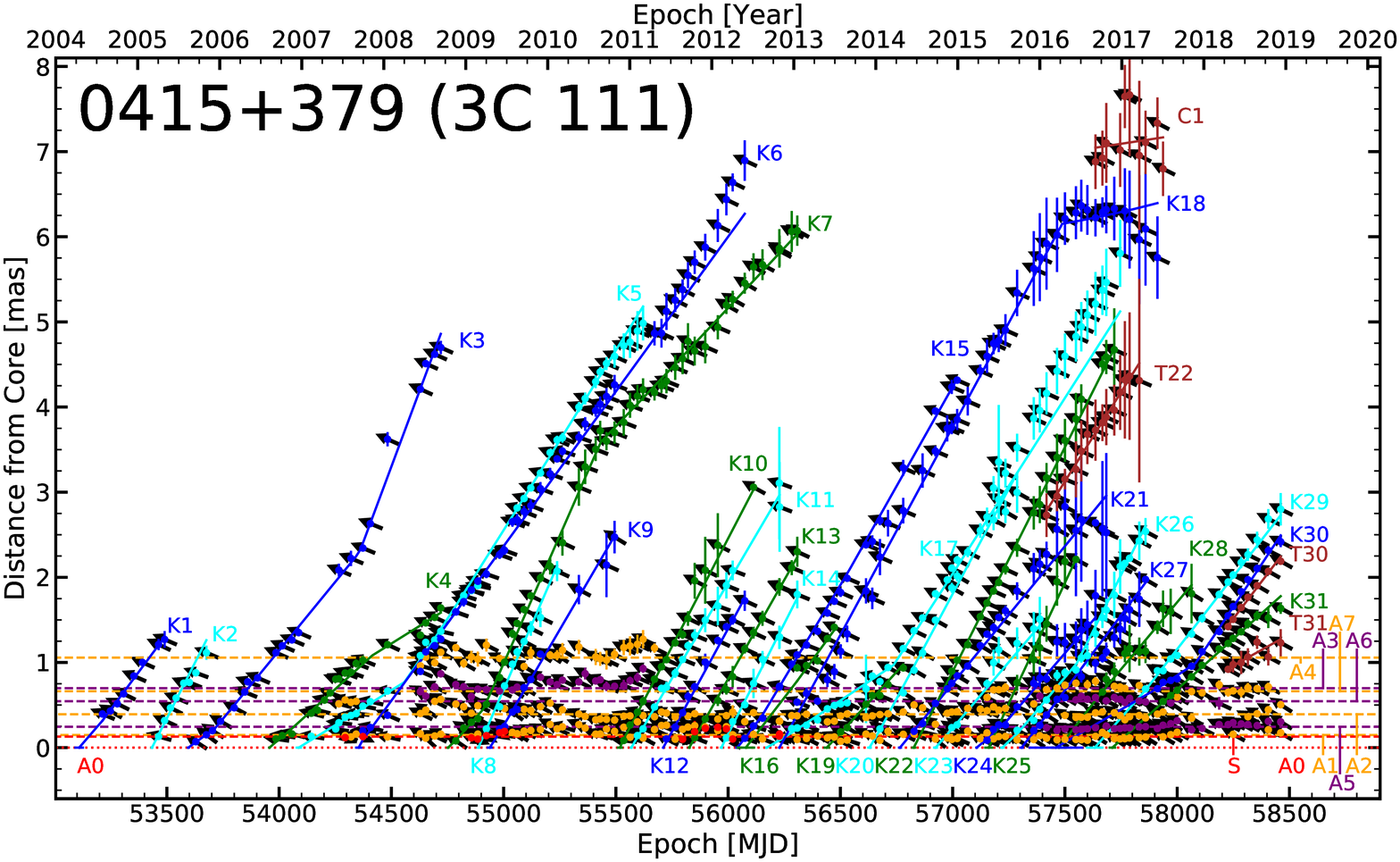}}
        \figsetgrpnote{Separation vs. time of the knots in the jet, relative to the core, of the RG 0415+379.}
        \figsetgrpend
        %
        % Number 6
        \figsetgrpstart
        \figsetgrpnum{7.6}
        \figsetgrptitle{0420-014}
        \figsetplot{{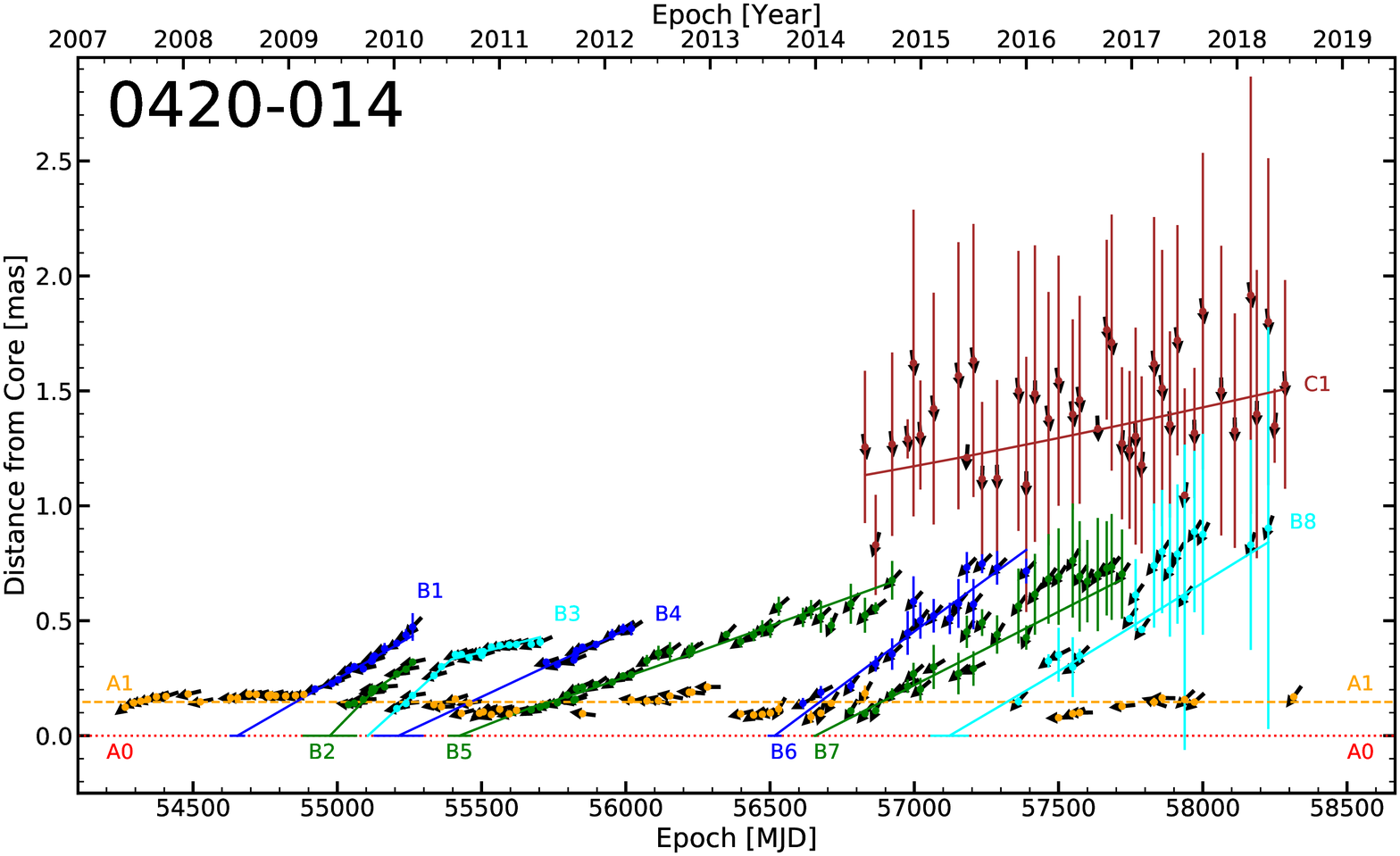}}
        \figsetgrpnote{Separation vs. time of the knots in the jet, relative to the core, of the FSRQ 0420-014.}
        \figsetgrpend
        %
        % Number 7
        \figsetgrpstart
        \figsetgrpnum{7.7}
        \figsetgrptitle{0430+052}
        \figsetplot{{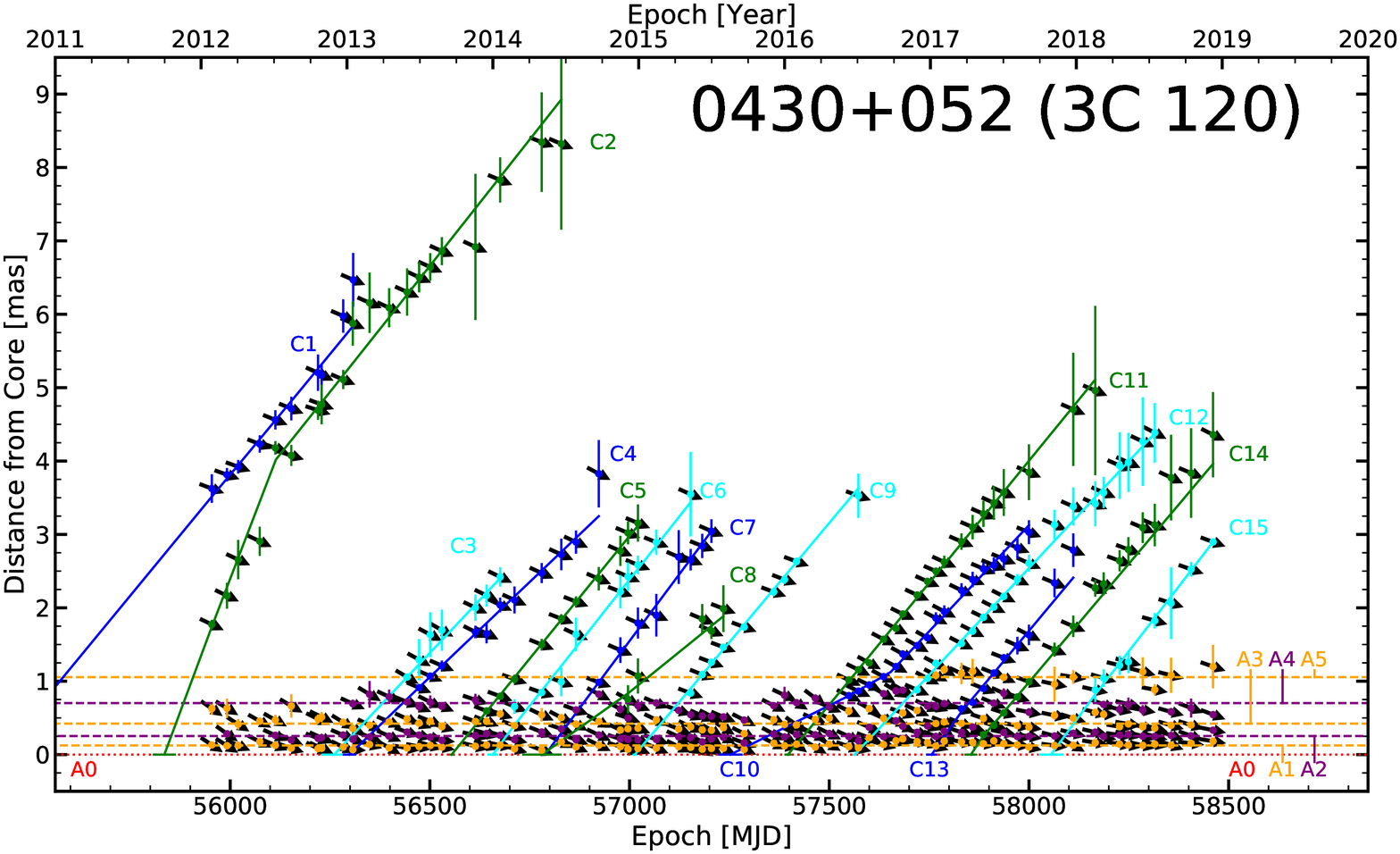}}
        \figsetgrpnote{Separation vs. time of the knots in the jet, relative to the core, of the RG 0430+052.}
        \figsetgrpend
        %
        % Number 8
        \figsetgrpstart
        \figsetgrpnum{7.8}
        \figsetgrptitle{0528+134}
        \figsetplot{{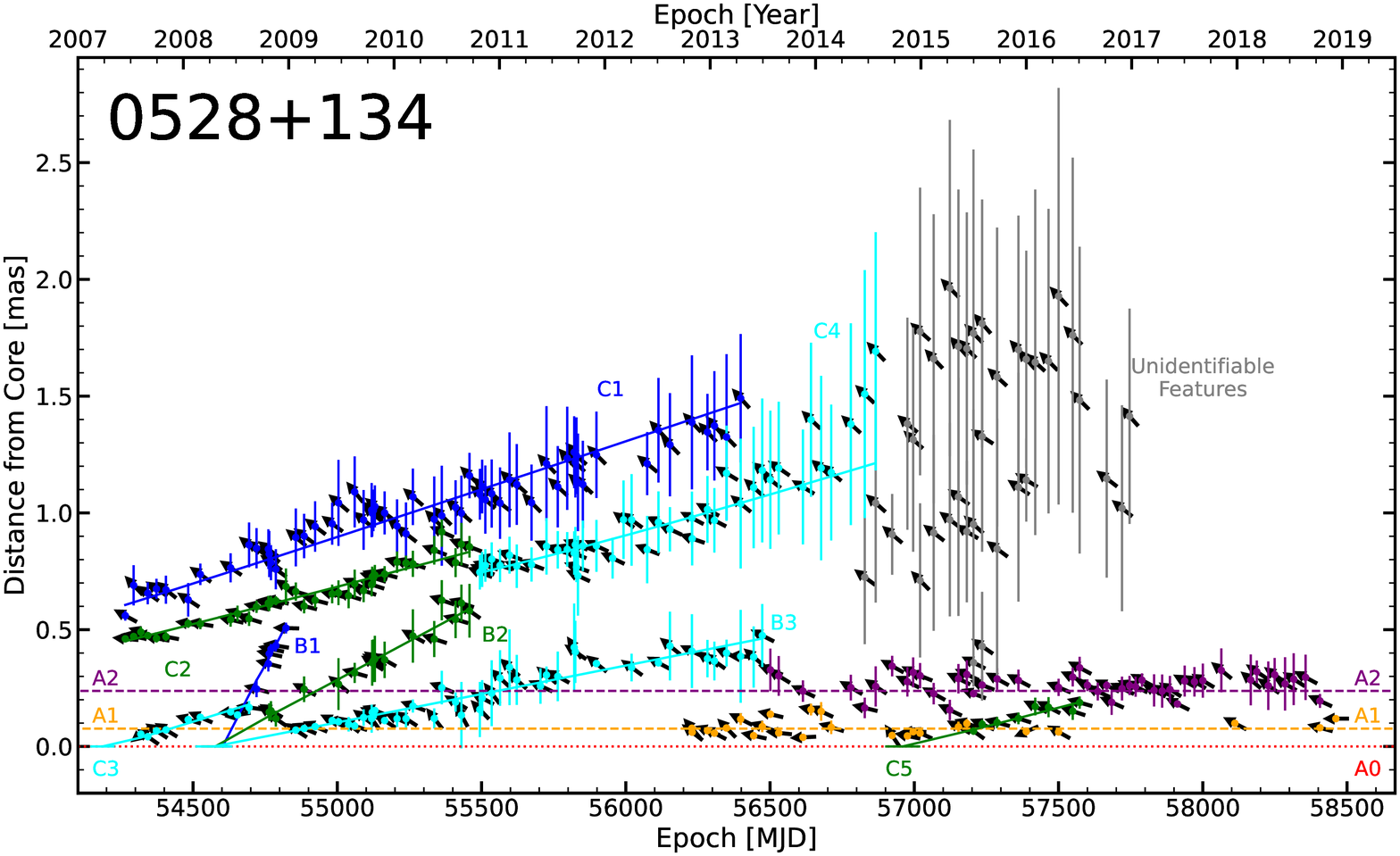}}
        \figsetgrpnote{Separation vs. time of the knots in the jet, relative to the core, of the FSRQ 0528+134.}
        \figsetgrpend
        %
        % Number 9
        \figsetgrpstart
        \figsetgrpnum{7.9}
        \figsetgrptitle{0716+714}
        \figsetplot{{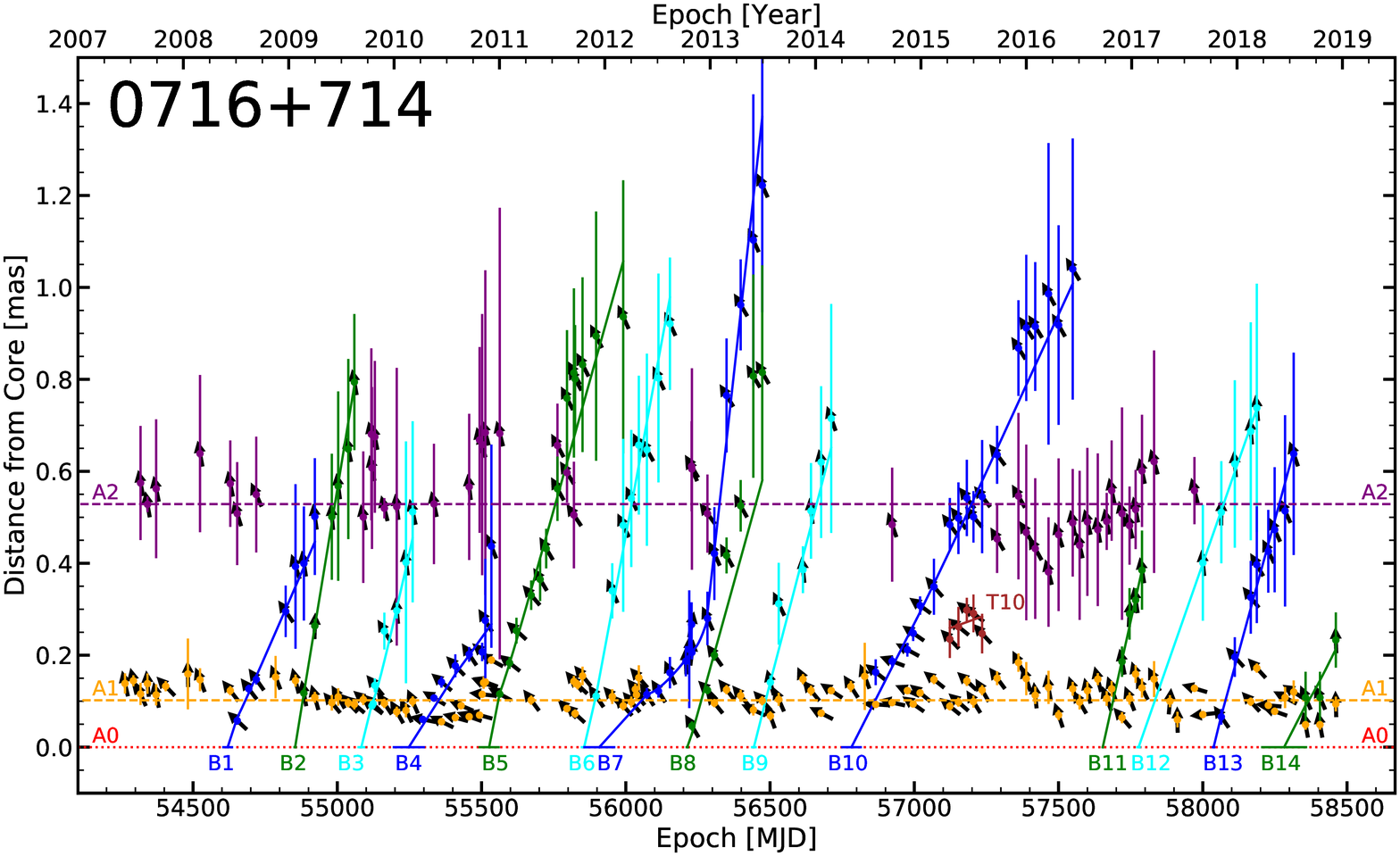}}
        \figsetgrpnote{Separation vs. time of the knots in the jet, relative to the core, of the BL 0716+714.}
        \figsetgrpend
        %
        % Number 10
        \figsetgrpstart
        \figsetgrpnum{7.10}
        \figsetgrptitle{0735+178}
        \figsetplot{{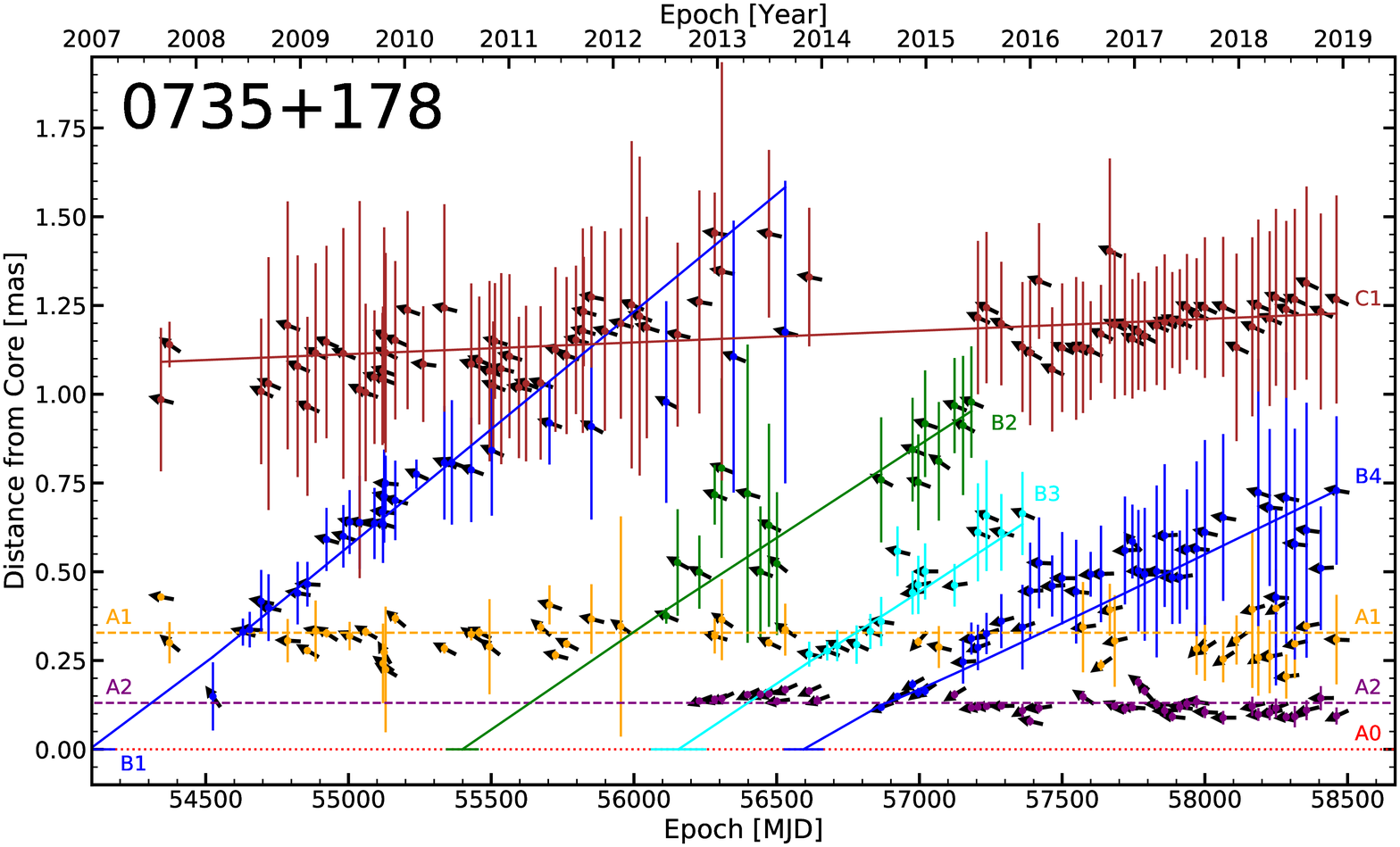}}
        \figsetgrpnote{Separation vs. time of the knots in the jet, relative to the core, of the BL 0735+178.}
        \figsetgrpend
        %
        % Number 11
        \figsetgrpstart
        \figsetgrpnum{7.11}
        \figsetgrptitle{0827+243}
        \figsetplot{{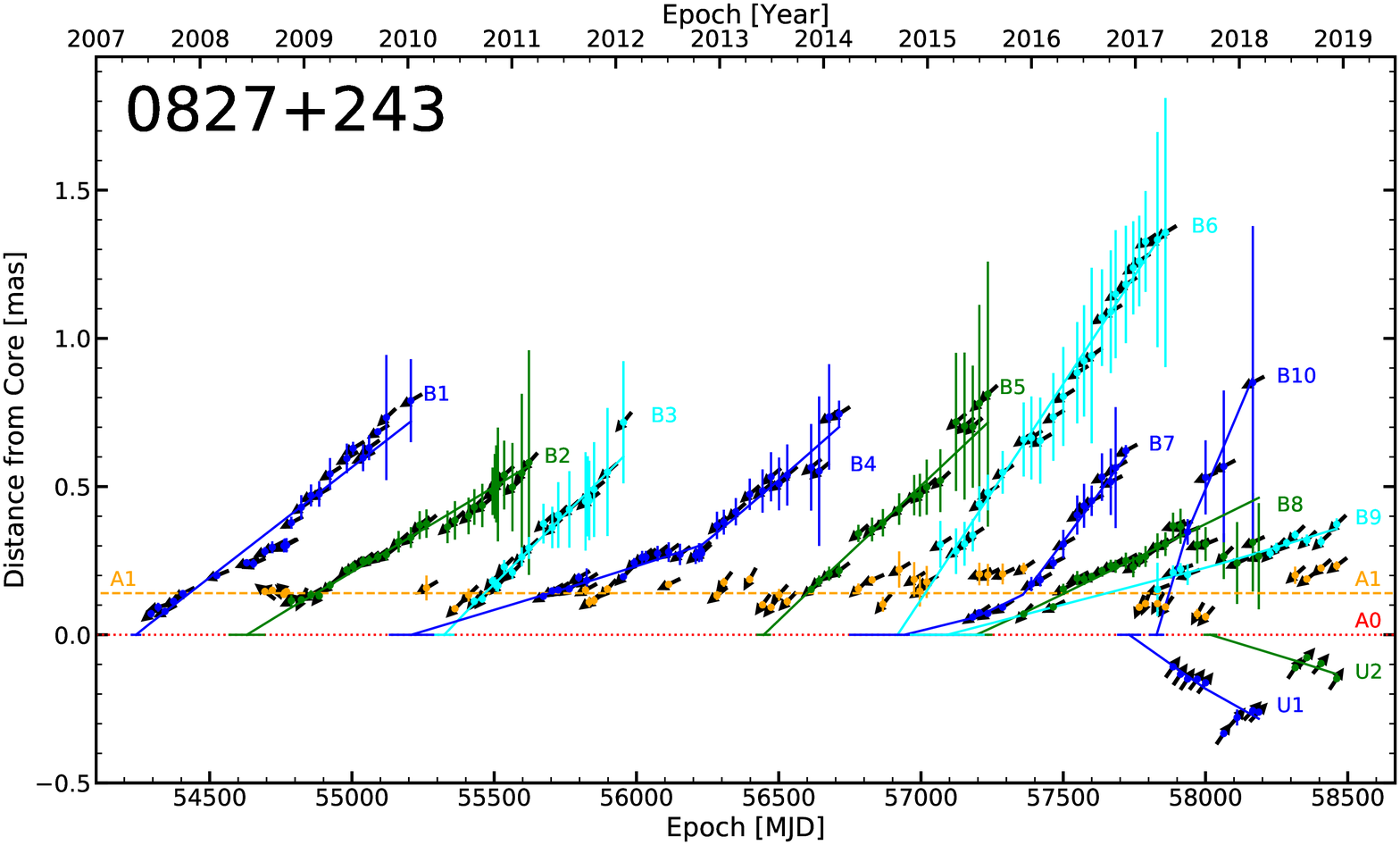}}
        \figsetgrpnote{Separation vs. time of the knots in the jet, relative to the core, of the FSRQ 0827+243.}
        \figsetgrpend
        %
        % Number 12
        \figsetgrpstart
        \figsetgrpnum{7.12}
        \figsetgrptitle{0829+046}
        \figsetplot{{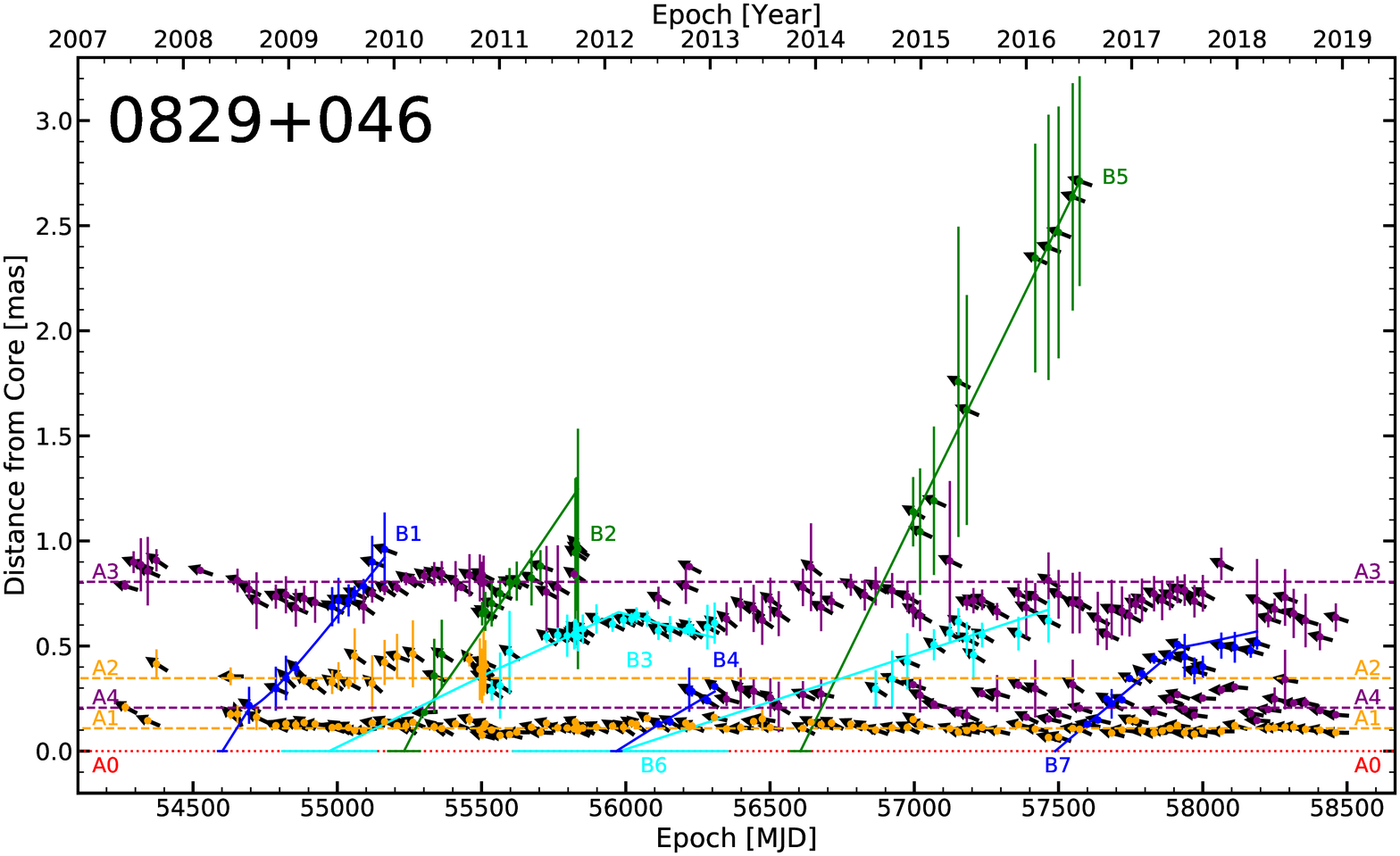}}
        \figsetgrpnote{Separation vs. time of the knots in the jet, relative to the core, of the BL 0829+046.}
        \figsetgrpend
        %
        % Number 13
        \figsetgrpstart
        \figsetgrpnum{7.13}
        \figsetgrptitle{0836+710}
        \figsetplot{{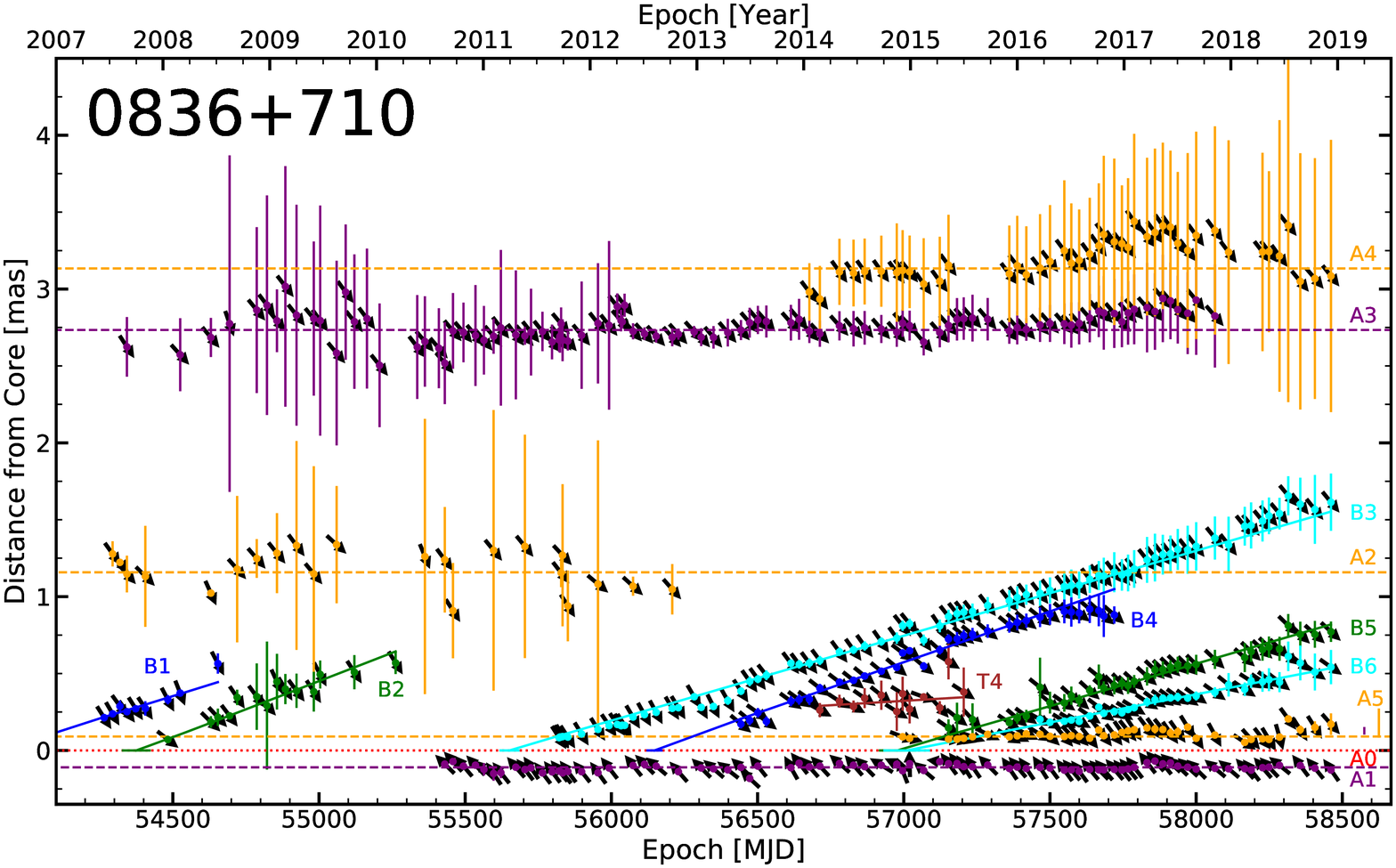}}
        \figsetgrpnote{Separation vs. time of the knots in the jet, relative to the core, of the FSRQ 0836+710.}
        \figsetgrpend
        %
        % Number 14
        \figsetgrpstart
        \figsetgrpnum{7.14}
        \figsetgrptitle{0851+202}
        \figsetplot{{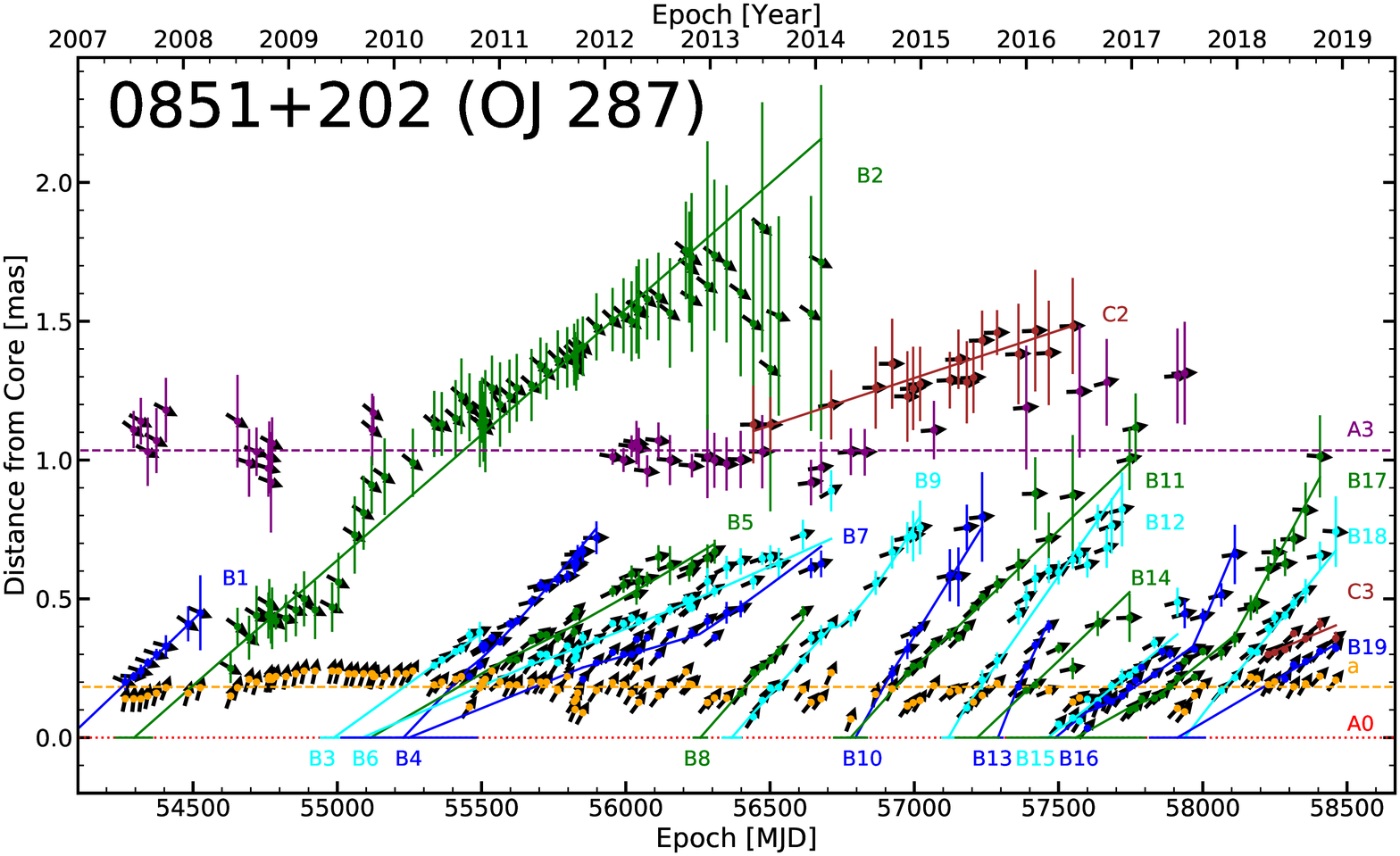}}
        \figsetgrpnote{Separation vs. time of the knots in the jet, relative to the core, of the BL 0851+202.}
        \figsetgrpend
        %
        % Number 15
        \figsetgrpstart
        \figsetgrpnum{7.15}
        \figsetgrptitle{0954+658}
        \figsetplot{{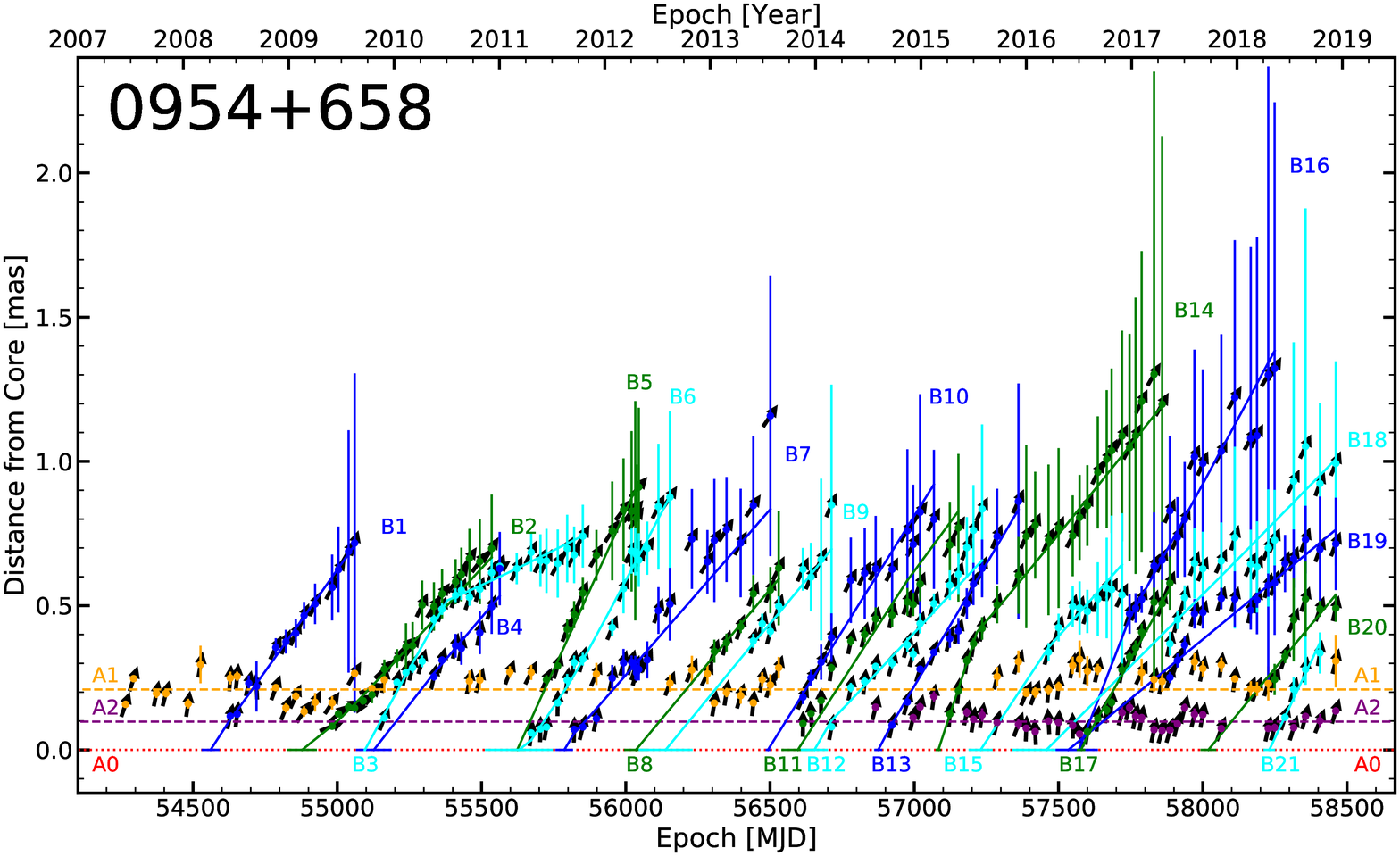}}
        \figsetgrpnote{Separation vs. time of the knots in the jet, relative to the core, of the BL 0954+658.}
        \figsetgrpend
        %
        % Number 16
        \figsetgrpstart
        \figsetgrpnum{7.16}
        \figsetgrptitle{1055+018}
        \figsetplot{{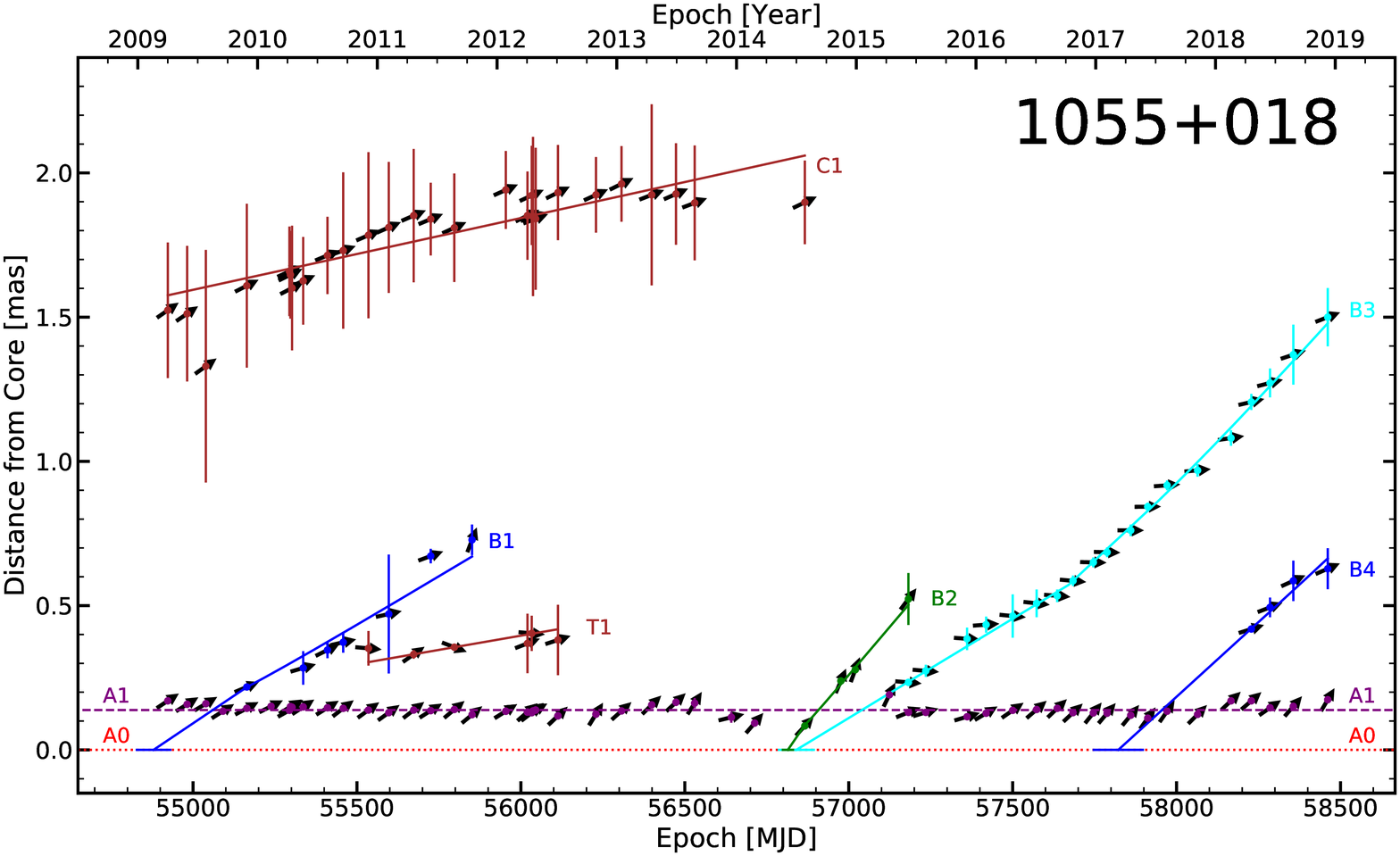}}
        \figsetgrpnote{Separation vs. time of the knots in the jet, relative to the core, of the FSRQ 1055+018.}
        \figsetgrpend
        %
        % Number 17
        \figsetgrpstart
        \figsetgrpnum{7.17}
        \figsetgrptitle{1101+384}
        \figsetplot{{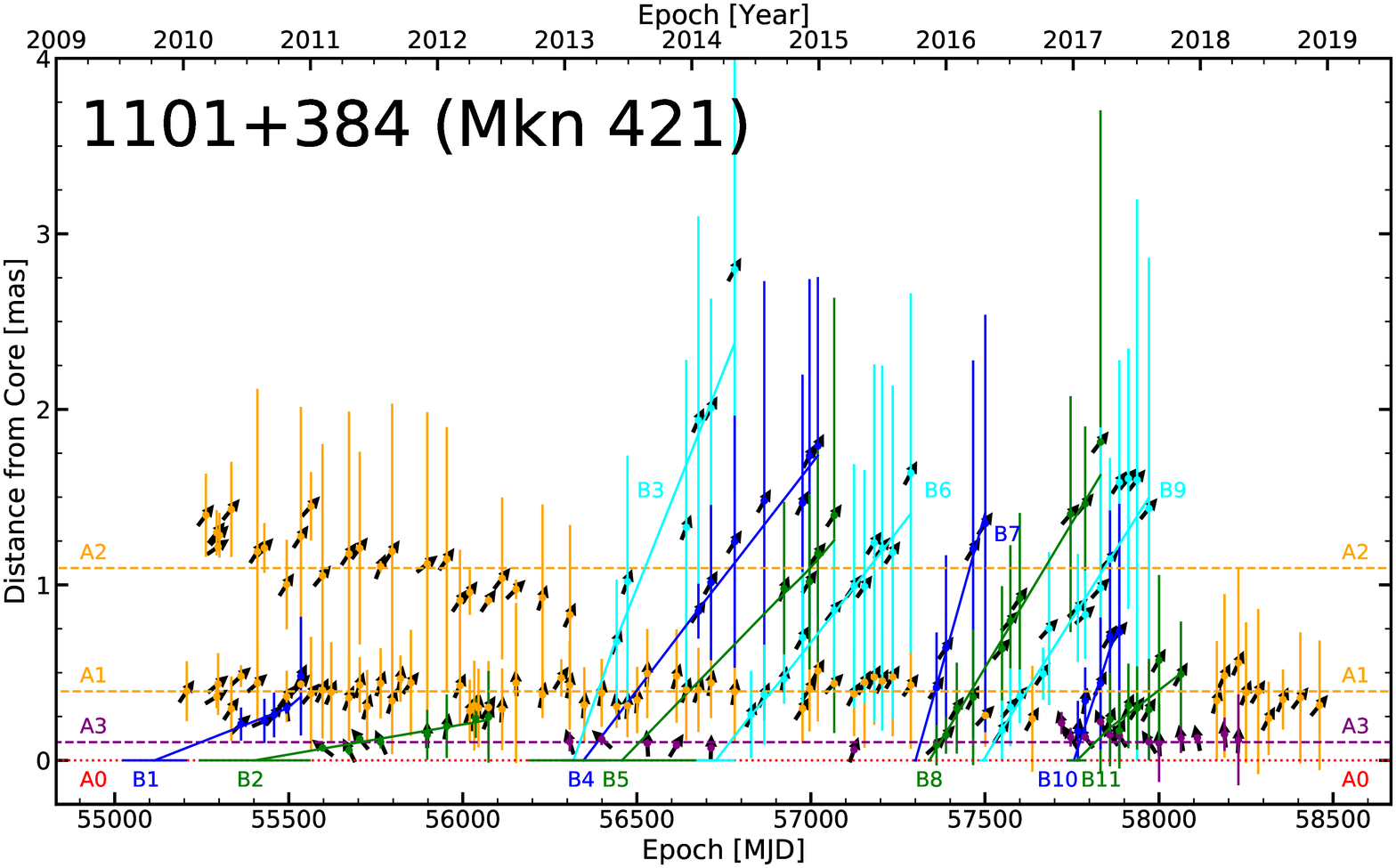}}
        \figsetgrpnote{Separation vs. time of the knots in the jet, relative to the core, of the BL 1101+384.}
        \figsetgrpend
        %
        % Number 18
        \figsetgrpstart
        \figsetgrpnum{7.18}
        \figsetgrptitle{1127-145}
        \figsetplot{{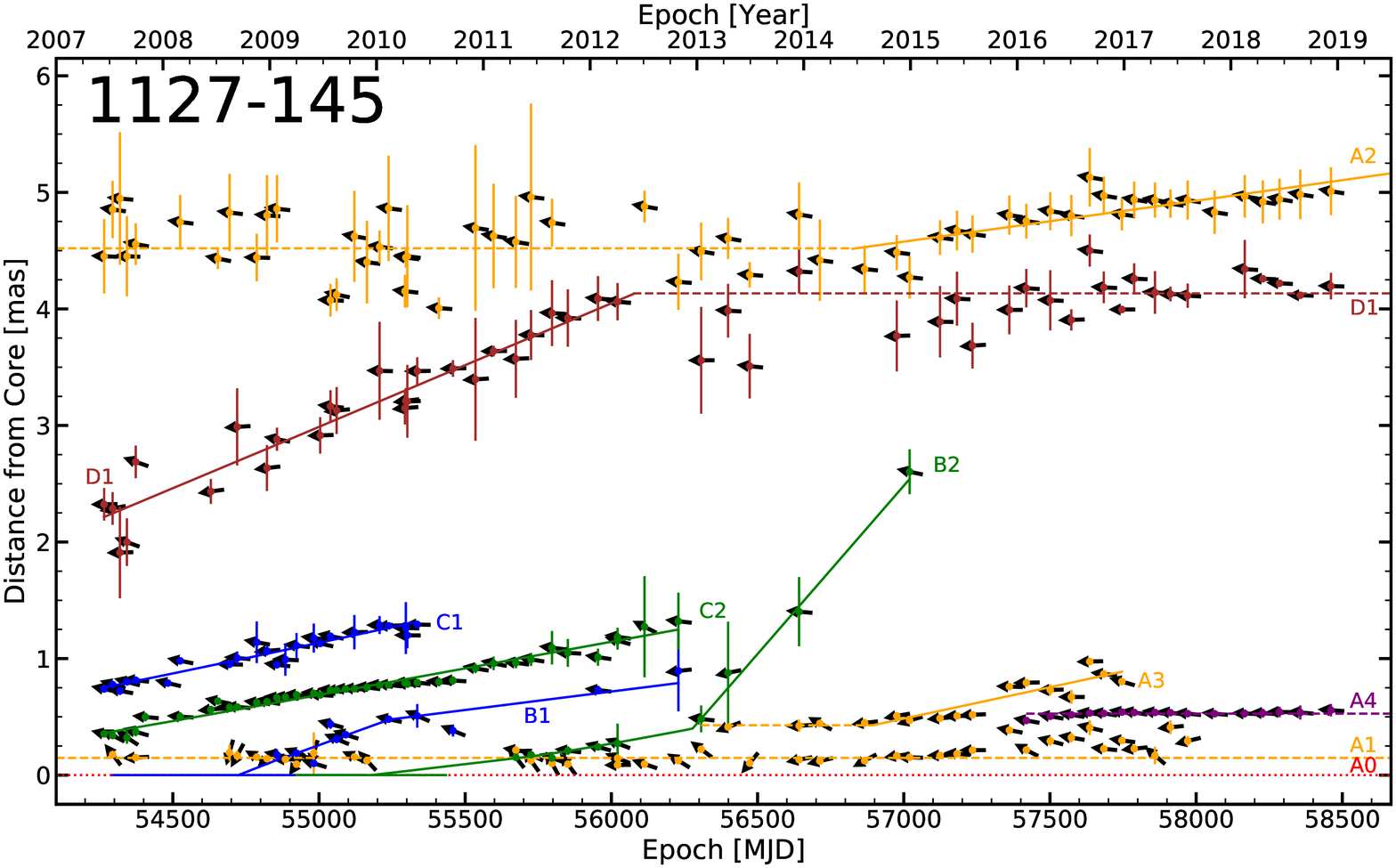}}
        \figsetgrpnote{Separation vs. time of the knots in the jet, relative to the core, of the FSRQ 1127-145.}
        \figsetgrpend
        %
        % Number 19
        \figsetgrpstart
        \figsetgrpnum{7.19}
        \figsetgrptitle{1156+295}
        \figsetplot{{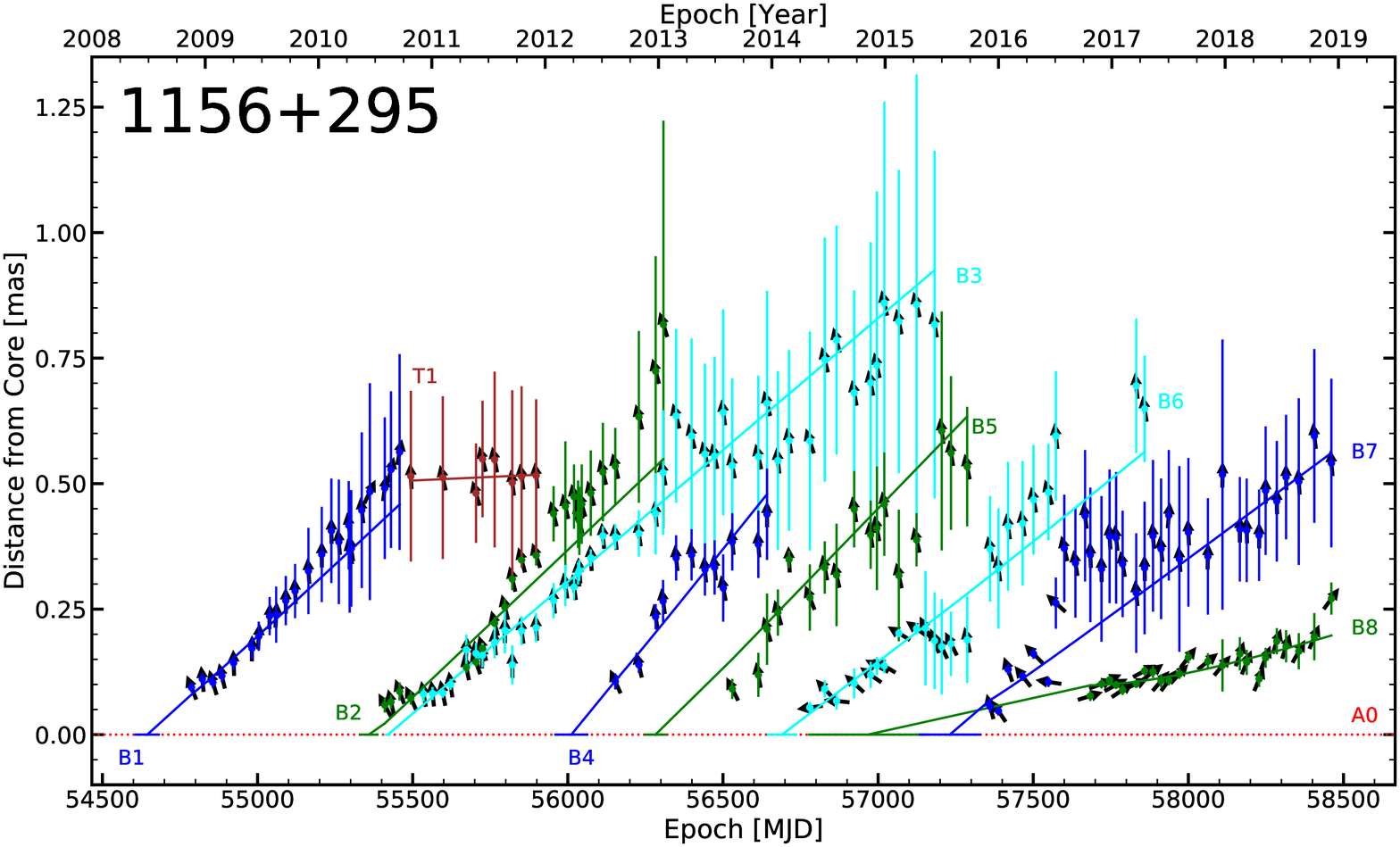}}
        \figsetgrpnote{Separation vs. time of the knots in the jet, relative to the core, of the FSRQ 1156+295.}
        \figsetgrpend
        %
        % Number 20
        \figsetgrpstart
        \figsetgrpnum{7.20}
        \figsetgrptitle{1219+285}
        \figsetplot{{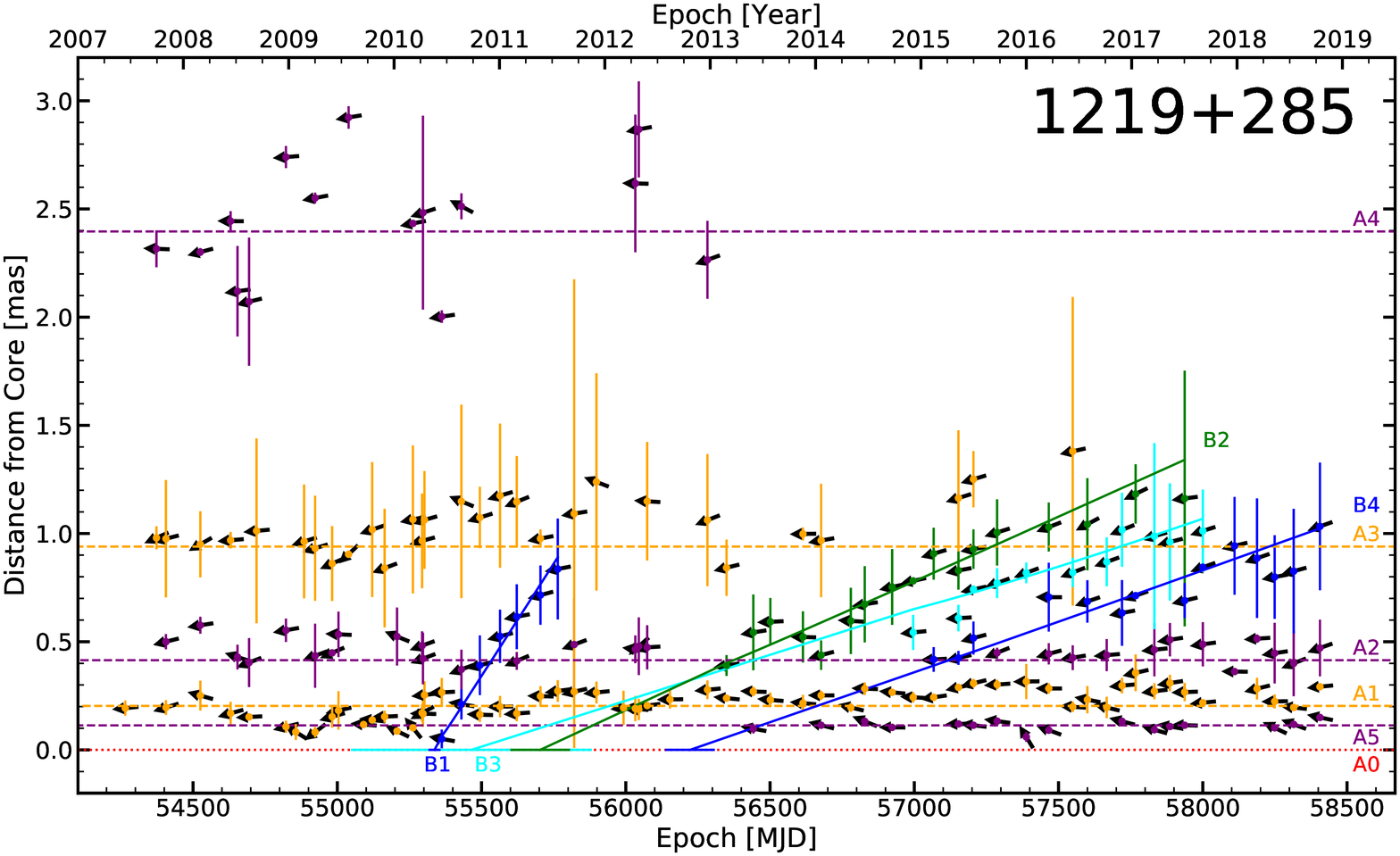}}
        \figsetgrpnote{Separation vs. time of the knots in the jet, relative to the core, of the BL 1219+285.}
        \figsetgrpend
        %
        % Number 21
        \figsetgrpstart
        \figsetgrpnum{7.21}
        \figsetgrptitle{1222+216}
        \figsetplot{{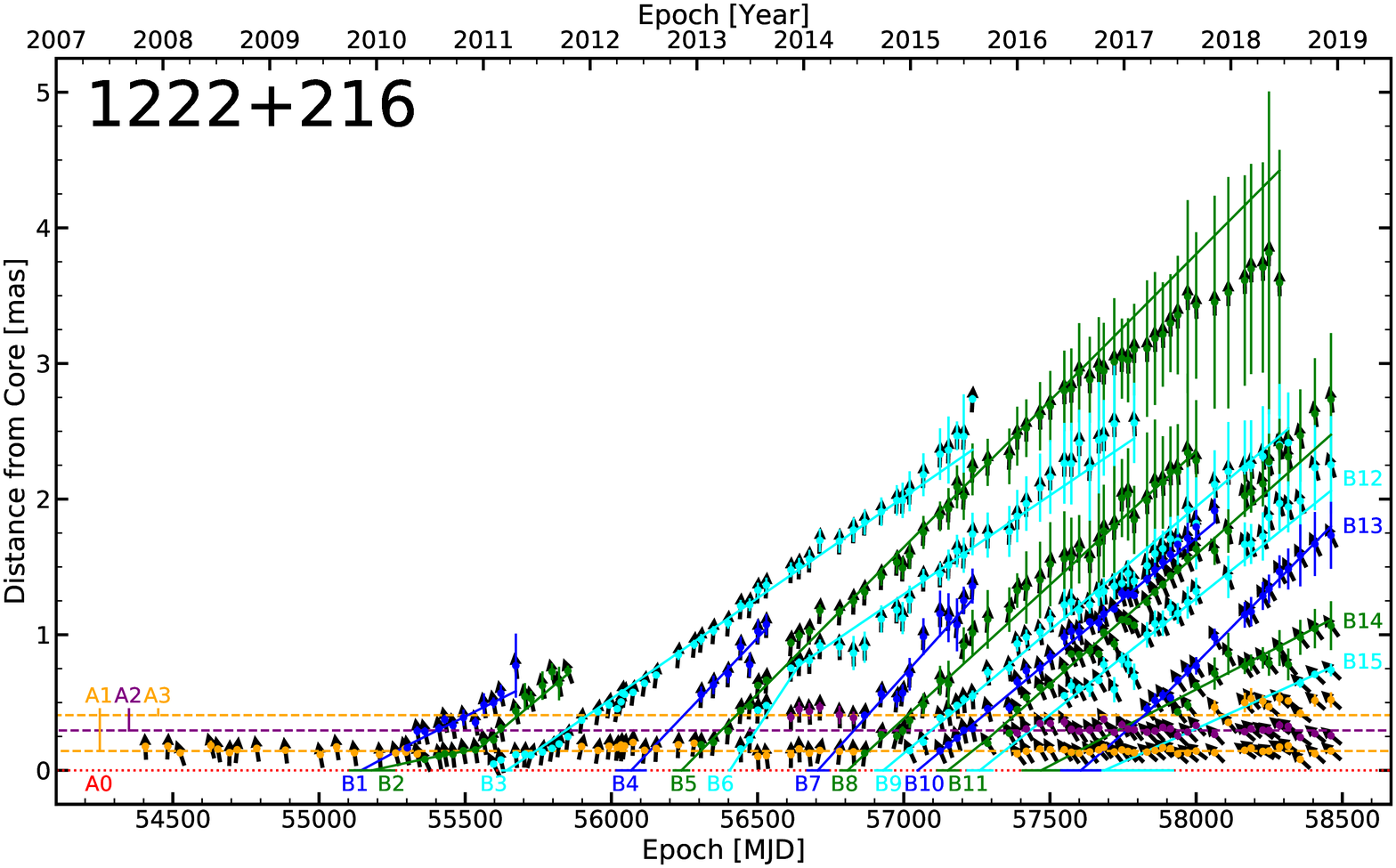}}
        \figsetgrpnote{Separation vs. time of the knots in the jet, relative to the core, of the FSRQ 1222+216.}
        \figsetgrpend
        %
        % Number 22
        \figsetgrpstart
        \figsetgrpnum{7.22}
        \figsetgrptitle{1226+023}
        \figsetplot{{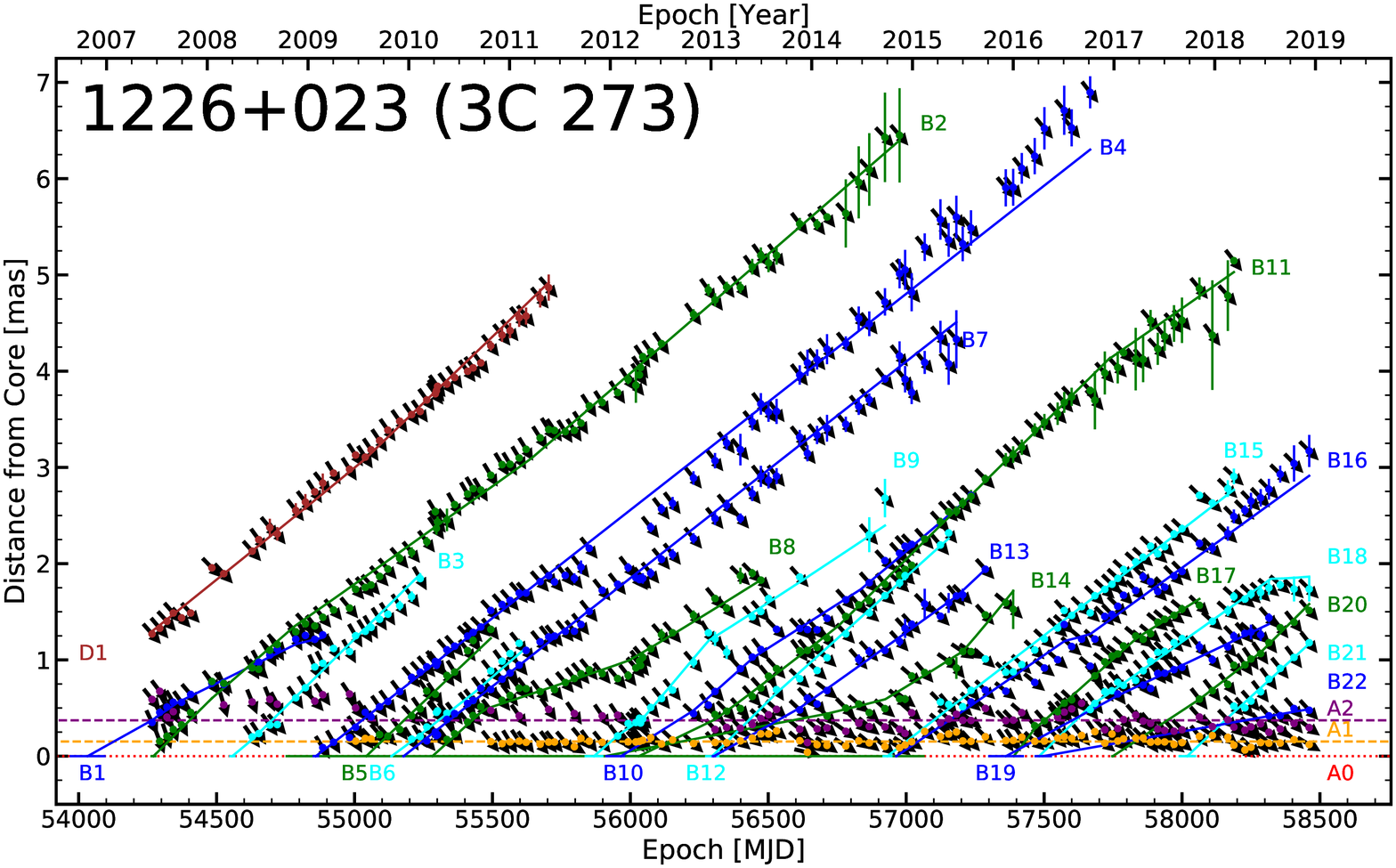}}
        \figsetgrpnote{Separation vs. time of the knots in the jet, relative to the core, of the FSRQ 1226+023.}
        \figsetgrpend
        %
        % Number 23
        \figsetgrpstart
        \figsetgrpnum{7.23}
        \figsetgrptitle{1253-055}
        \figsetplot{{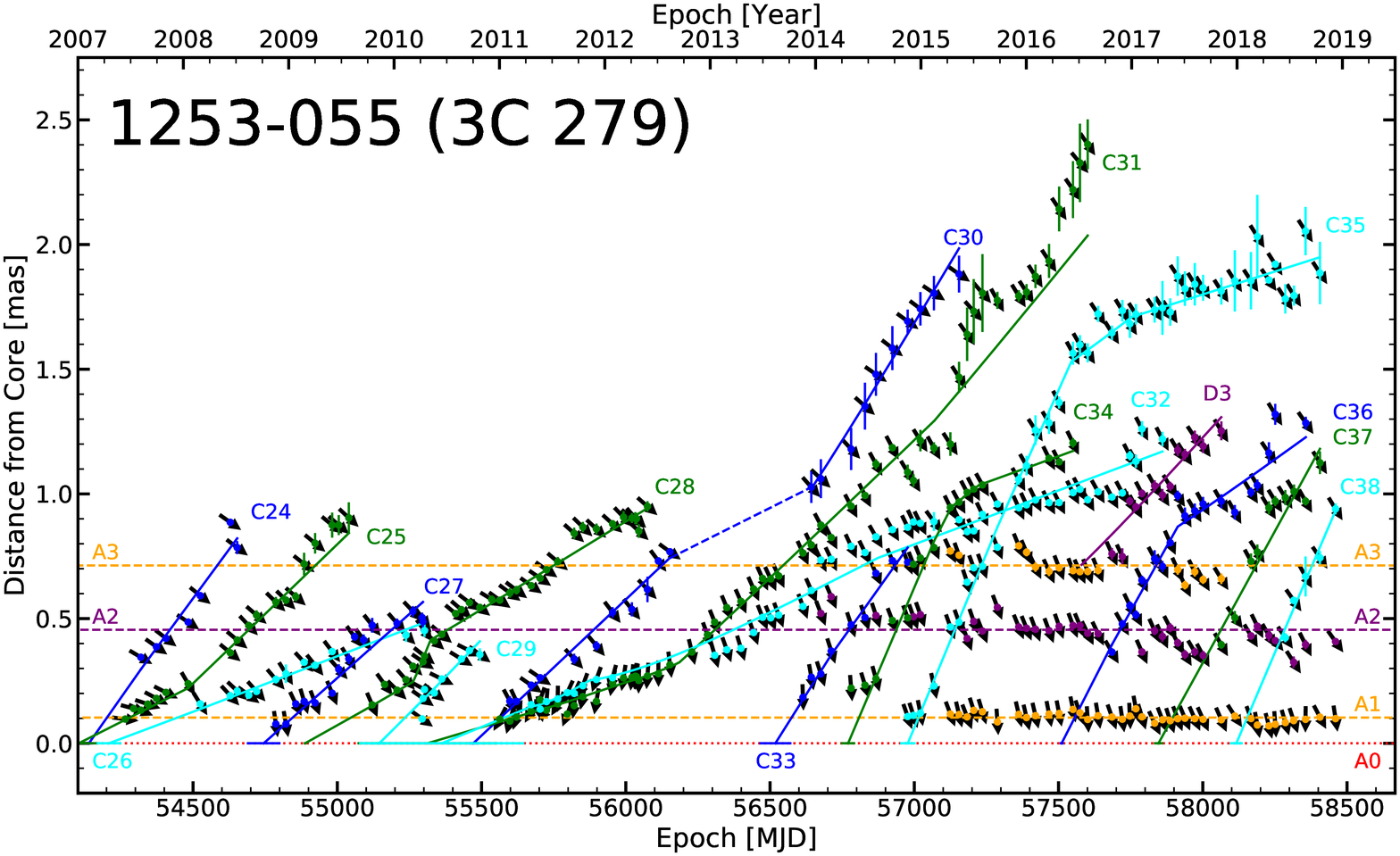}}
        \figsetgrpnote{Separation vs. time of the knots in the jet, relative to the core, of the FSRQ 1253-055.}
        \figsetgrpend
        %
        % Number 24
        \figsetgrpstart
        \figsetgrpnum{7.24}
        \figsetgrptitle{1308+326}
        \figsetplot{{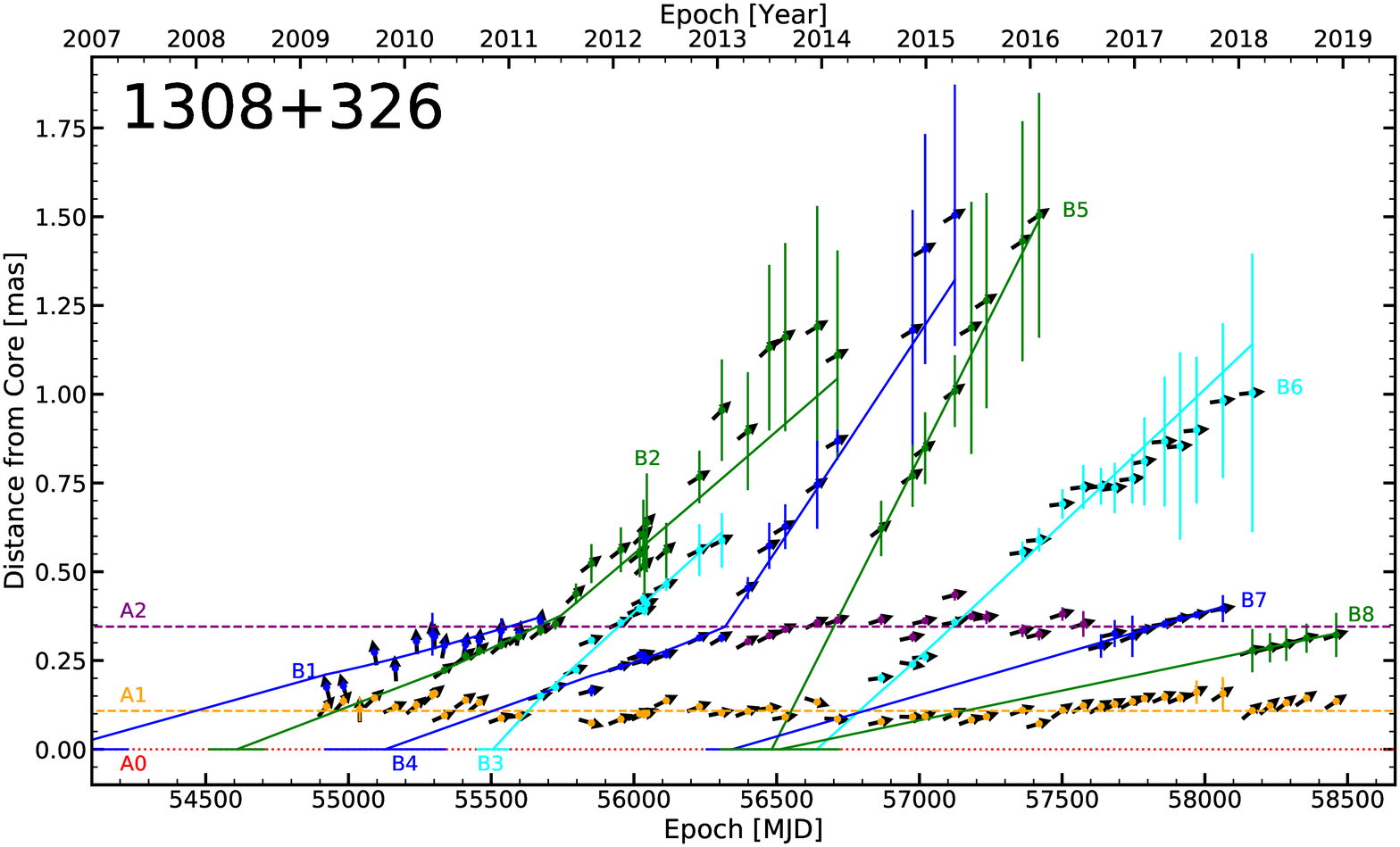}}
        \figsetgrpnote{Separation vs. time of the knots in the jet, relative to the core, of the FSRQ 1308+326.}
        \figsetgrpend
        %
        % Number 25
        \figsetgrpstart
        \figsetgrpnum{7.25}
        \figsetgrptitle{1406-076}
        \figsetplot{{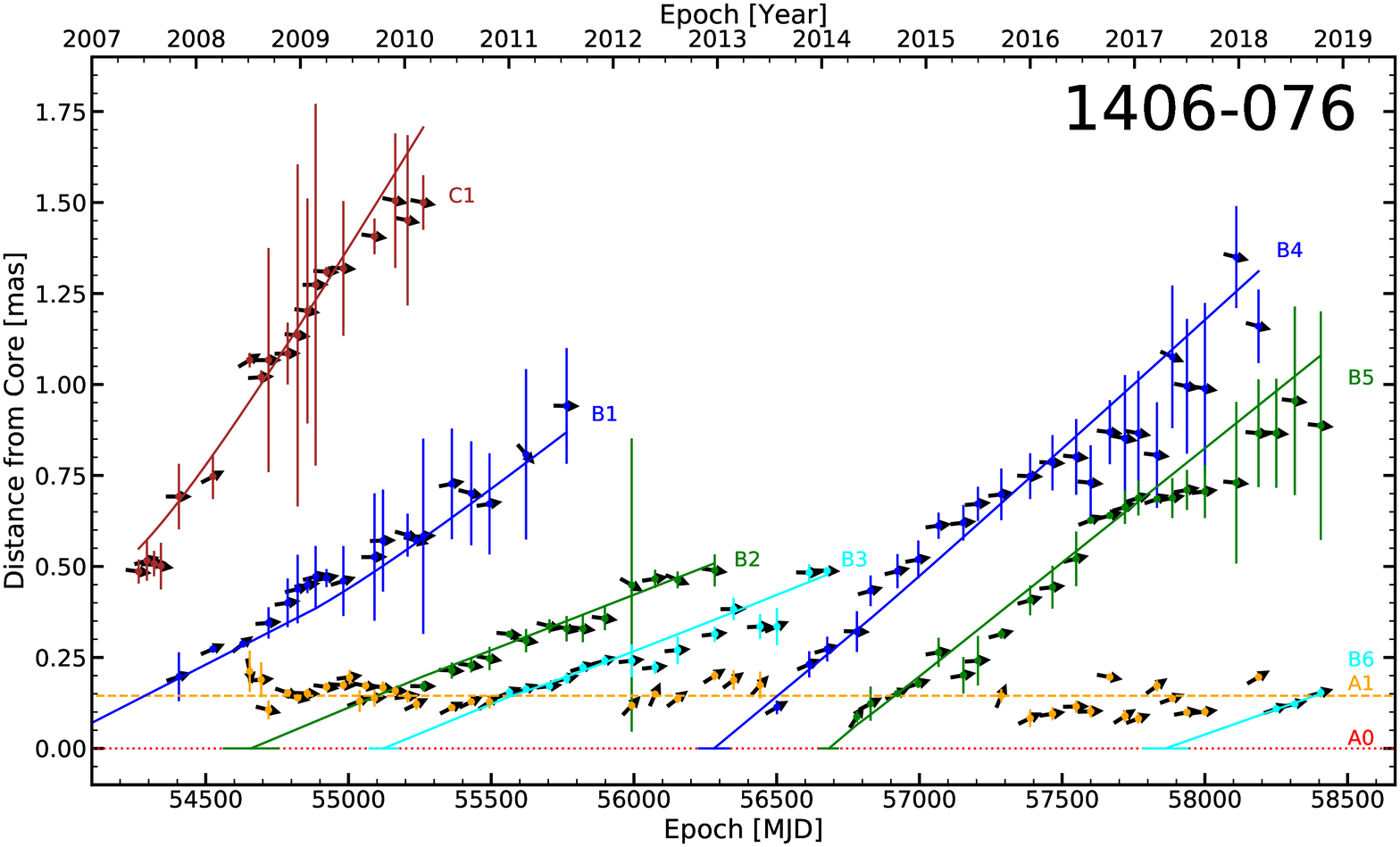}}
        \figsetgrpnote{Separation vs. time of the knots in the jet, relative to the core, of the FSRQ 1406-076.}
        \figsetgrpend
        %
        % Number 26
        \figsetgrpstart
        \figsetgrpnum{7.26}
        \figsetgrptitle{1510-089}
        \figsetplot{{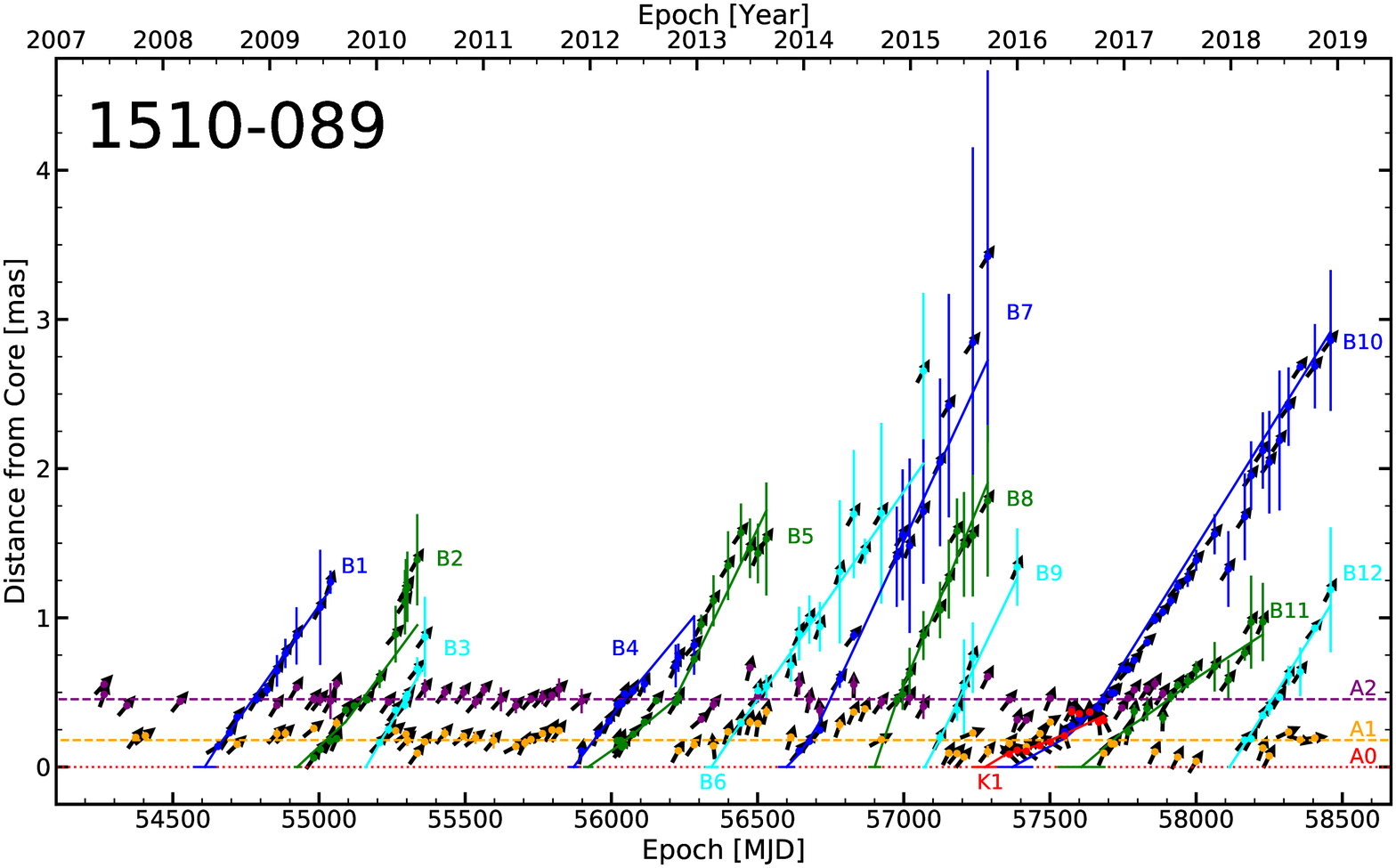}}
        \figsetgrpnote{Separation vs. time of the knots in the jet, relative to the core, of the FSRQ 1510-089.}
        \figsetgrpend
        %
        % Number 27
        \figsetgrpstart
        \figsetgrpnum{7.27}
        \figsetgrptitle{1611+343}
        \figsetplot{{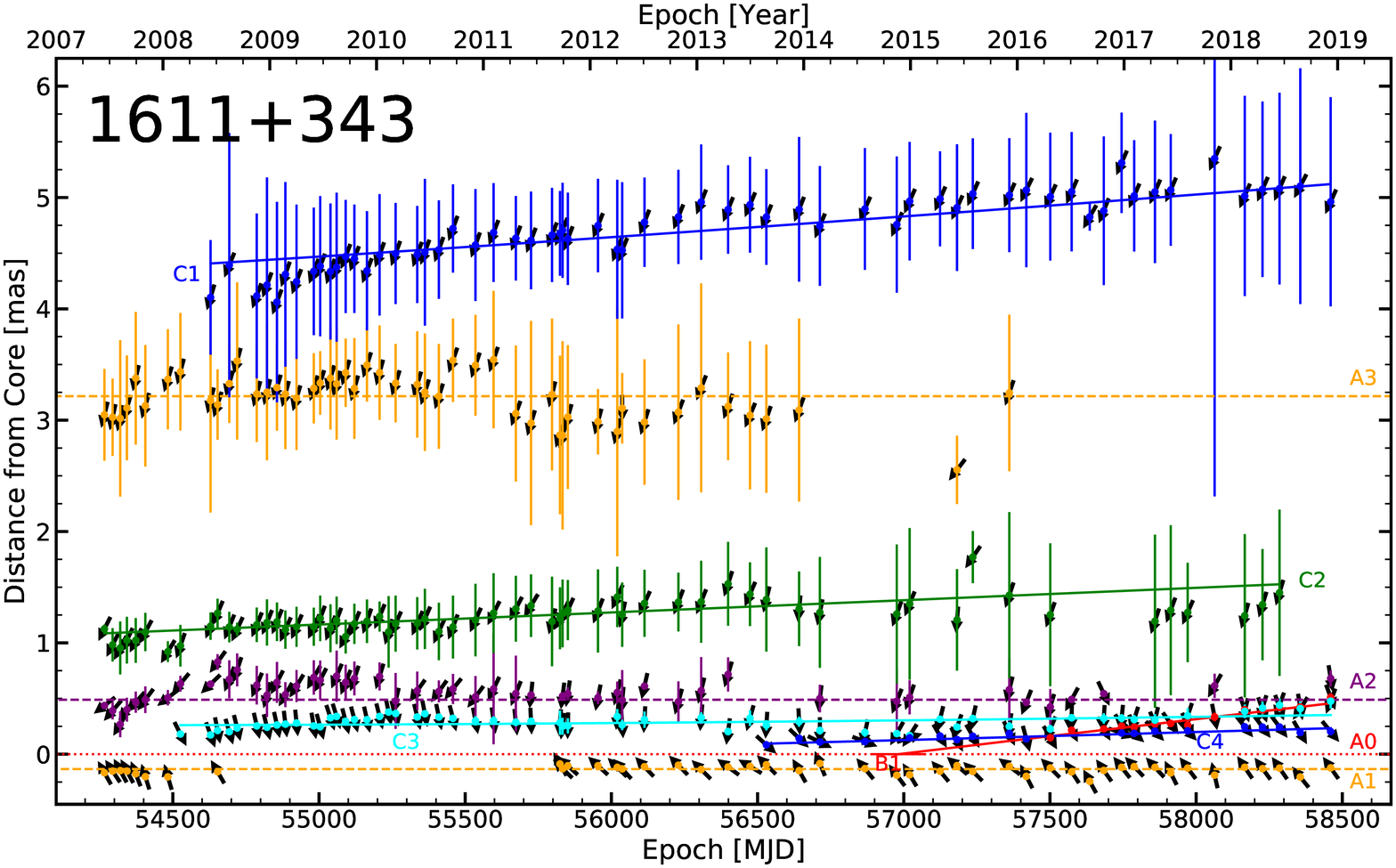}}
        \figsetgrpnote{Separation vs. time of the knots in the jet, relative to the core, of the FSRQ 1611+343.}
        \figsetgrpend
        %
        % Number 28
        \figsetgrpstart
        \figsetgrpnum{7.28}
        \figsetgrptitle{1622-297}
        \figsetplot{{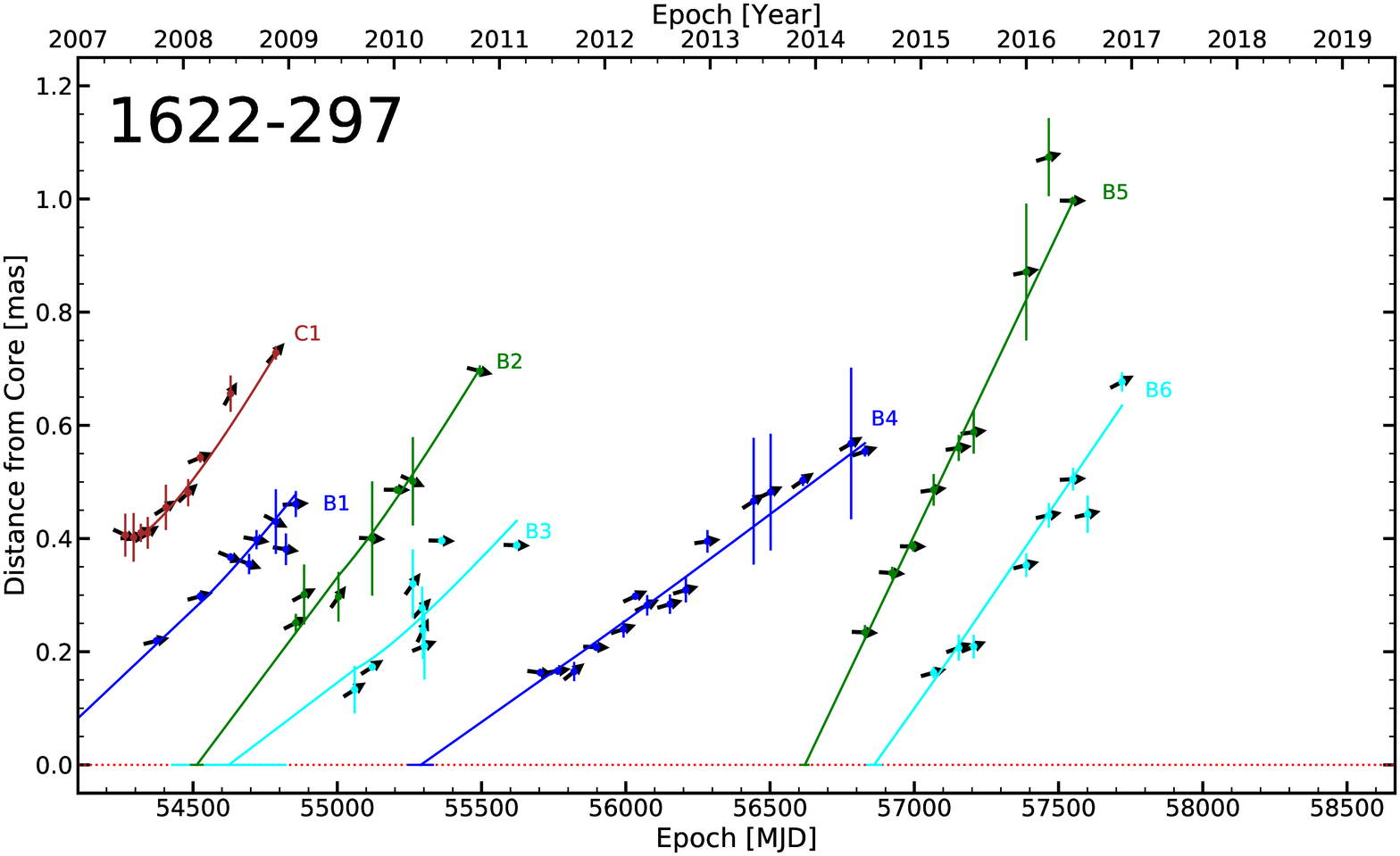}}
        \figsetgrpnote{Separation vs. time of the knots in the jet, relative to the core, of the FSRQ 1622-297.}
        \figsetgrpend
        %
        % Number 29
        \figsetgrpstart
        \figsetgrpnum{7.29}
        \figsetgrptitle{1633+382}
        \figsetplot{{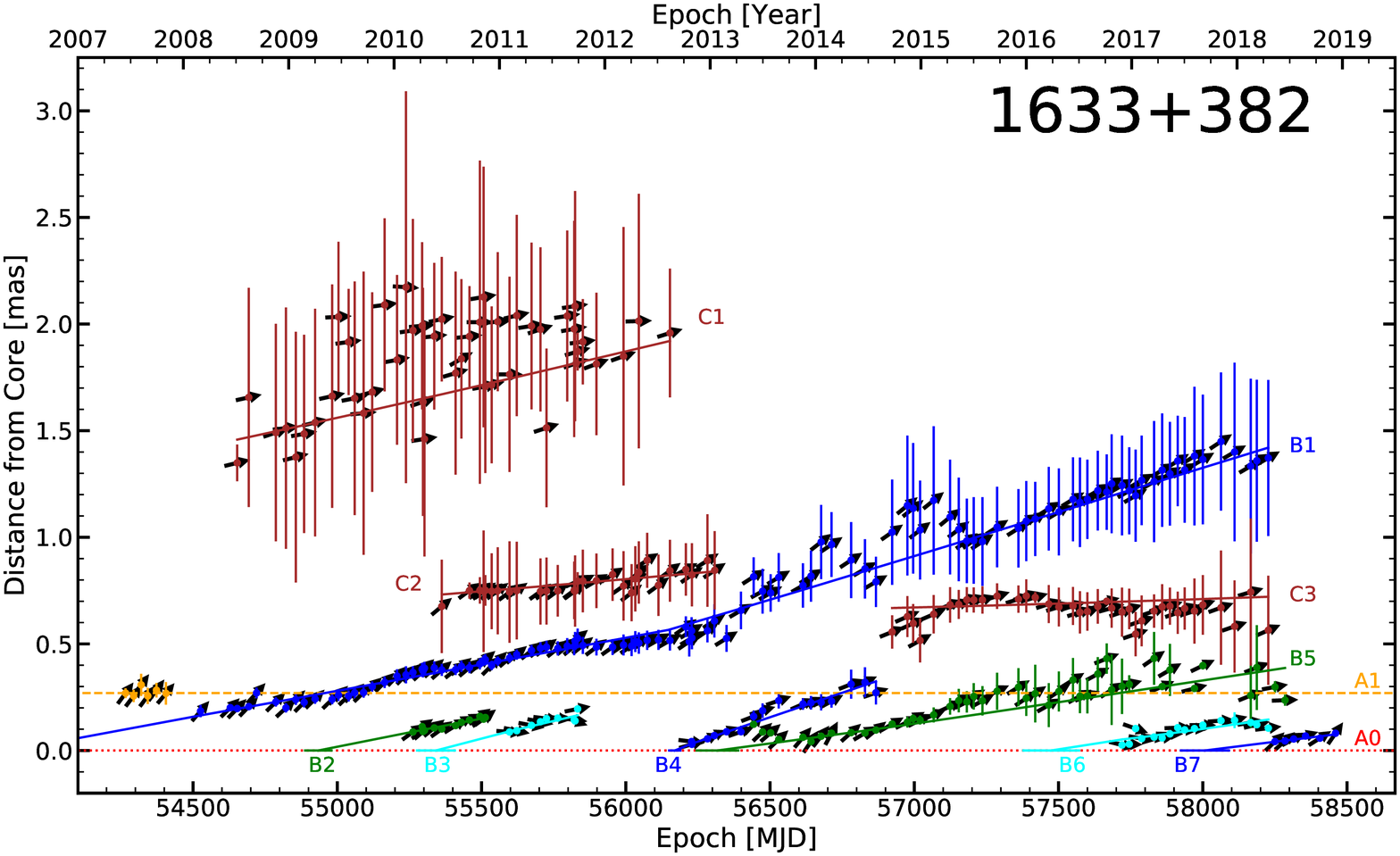}}
        \figsetgrpnote{Separation vs. time of the knots in the jet, relative to the core, of the FSRQ 1633+382.}
        \figsetgrpend
        %
        % Number 30
        \figsetgrpstart
        \figsetgrpnum{7.30}
        \figsetgrptitle{1641+399}
        \figsetplot{{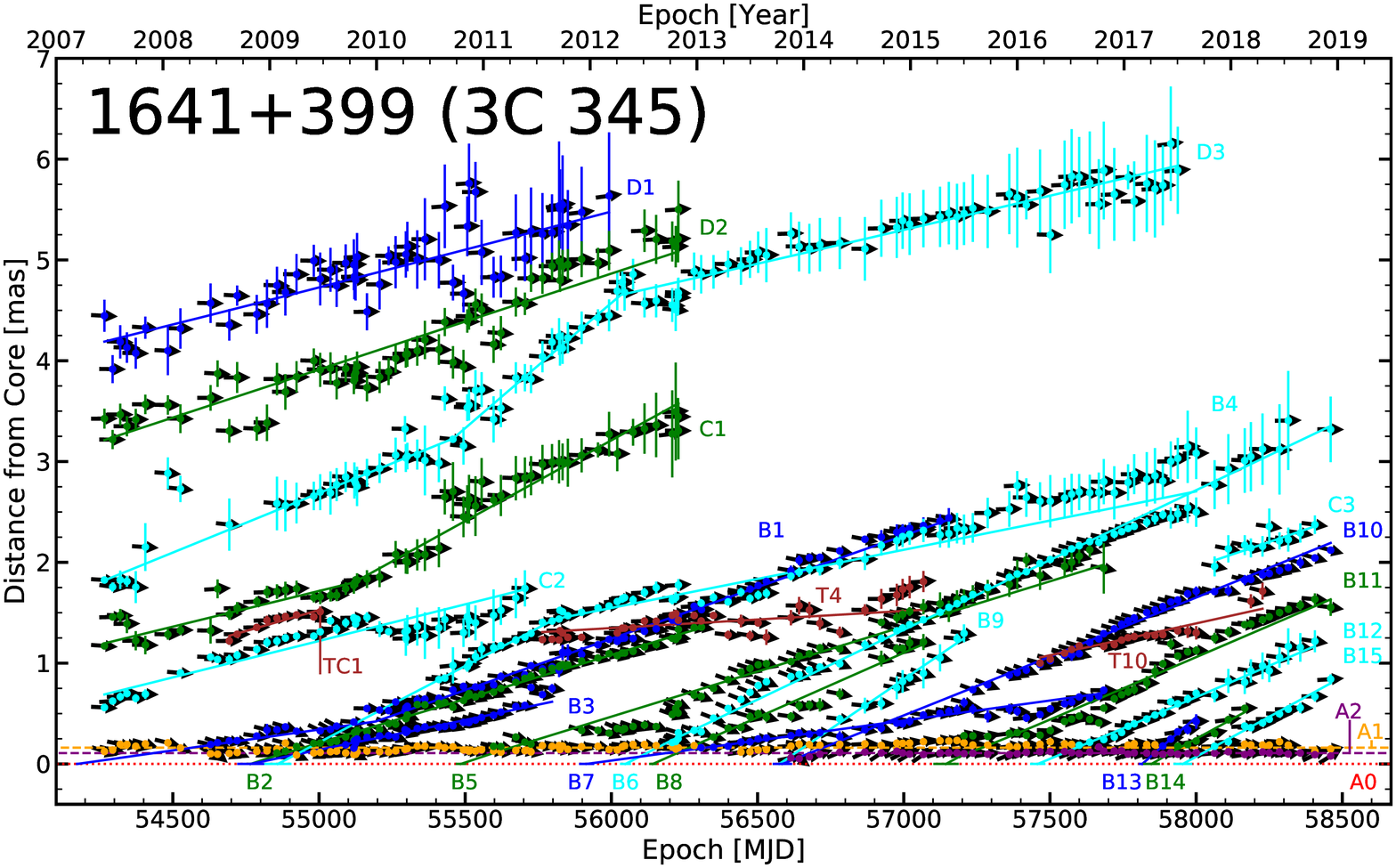}}
        \figsetgrpnote{Separation vs. time of the knots in the jet, relative to the core, of the FSRQ 1641+399.}
        \figsetgrpend
        %
        % Number 31
        \figsetgrpstart
        \figsetgrpnum{7.31}
        \figsetgrptitle{1652+398}
        \figsetplot{{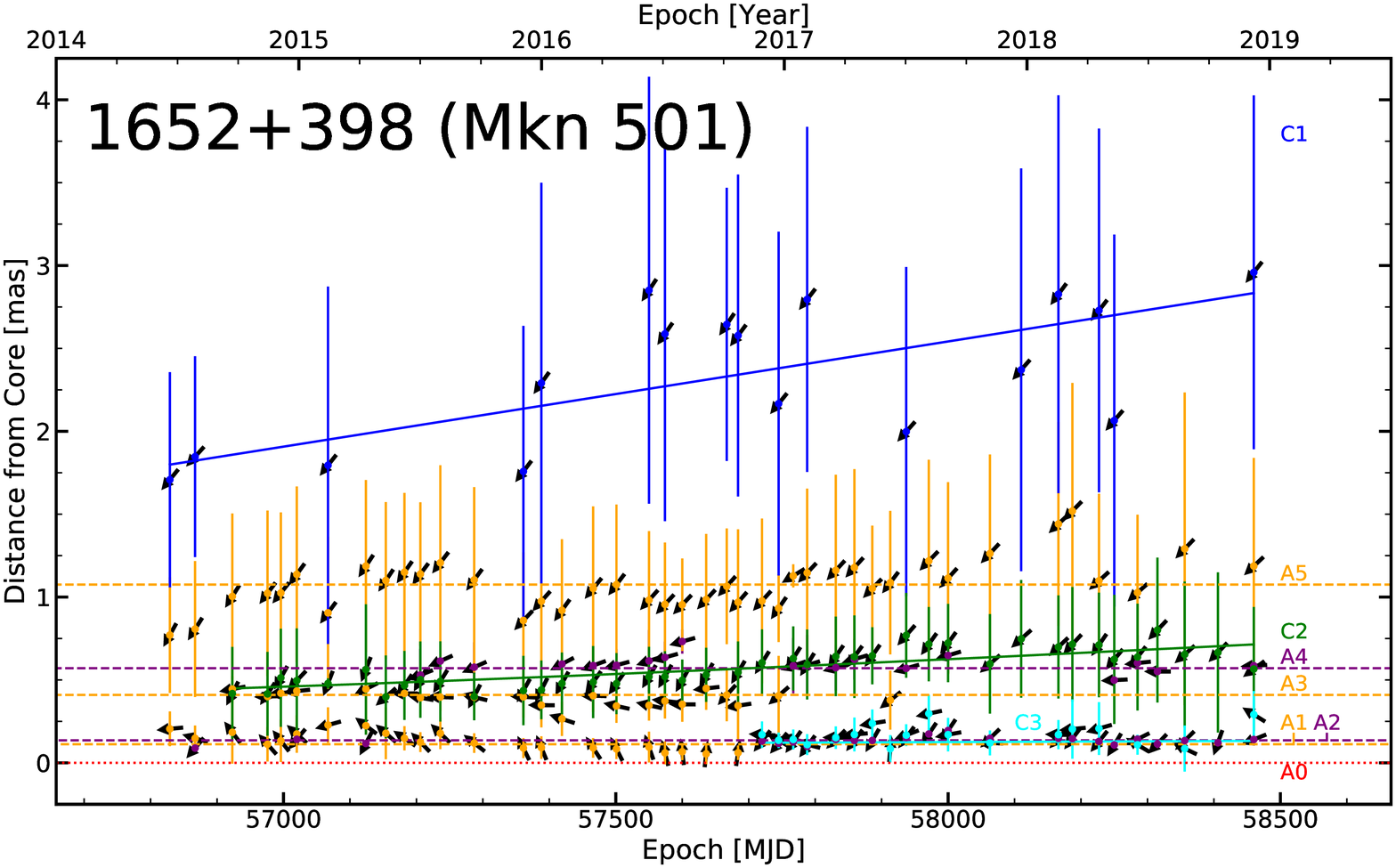}}
        \figsetgrpnote{Separation vs. time of the knots in the jet, relative to the core, of the BL 1652+398.}
        \figsetgrpend
        %
        % Number 32
        \figsetgrpstart
        \figsetgrpnum{7.32}
        \figsetgrptitle{1730-130}
        \figsetplot{{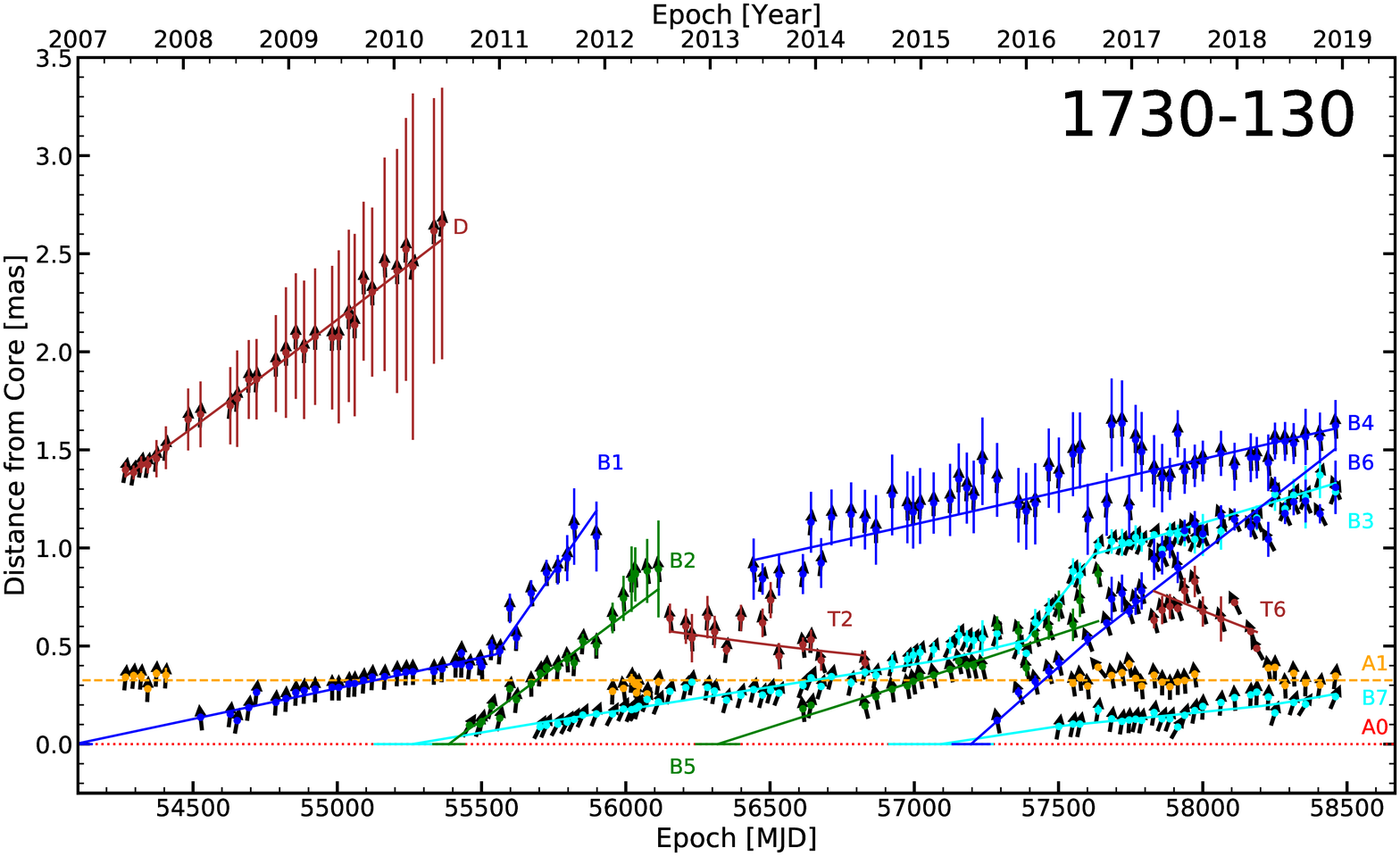}}
        \figsetgrpnote{Separation vs. time of the knots in the jet, relative to the core, of the FSRQ 1730-130.}
        \figsetgrpend
        %
        % Number 33
        \figsetgrpstart
        \figsetgrpnum{7.33}
        \figsetgrptitle{1749+096}
        \figsetplot{{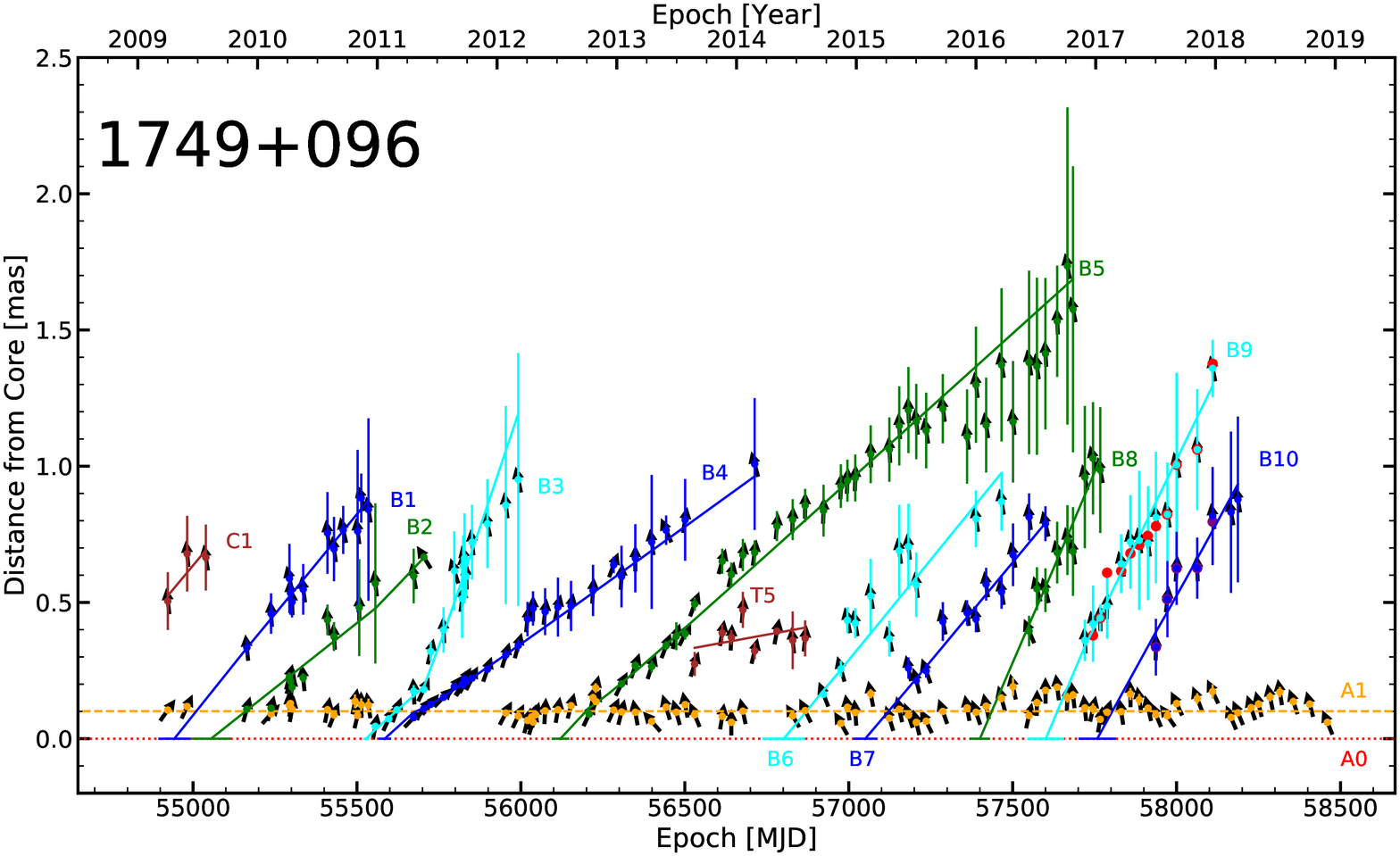}}
        \figsetgrpnote{Separation vs. time of the knots in the jet, relative to the core, of the BL 1749+096.}
        \figsetgrpend
        %
        % Number 34
        \figsetgrpstart
        \figsetgrpnum{7.34}
        \figsetgrptitle{1959+650}
        \figsetplot{{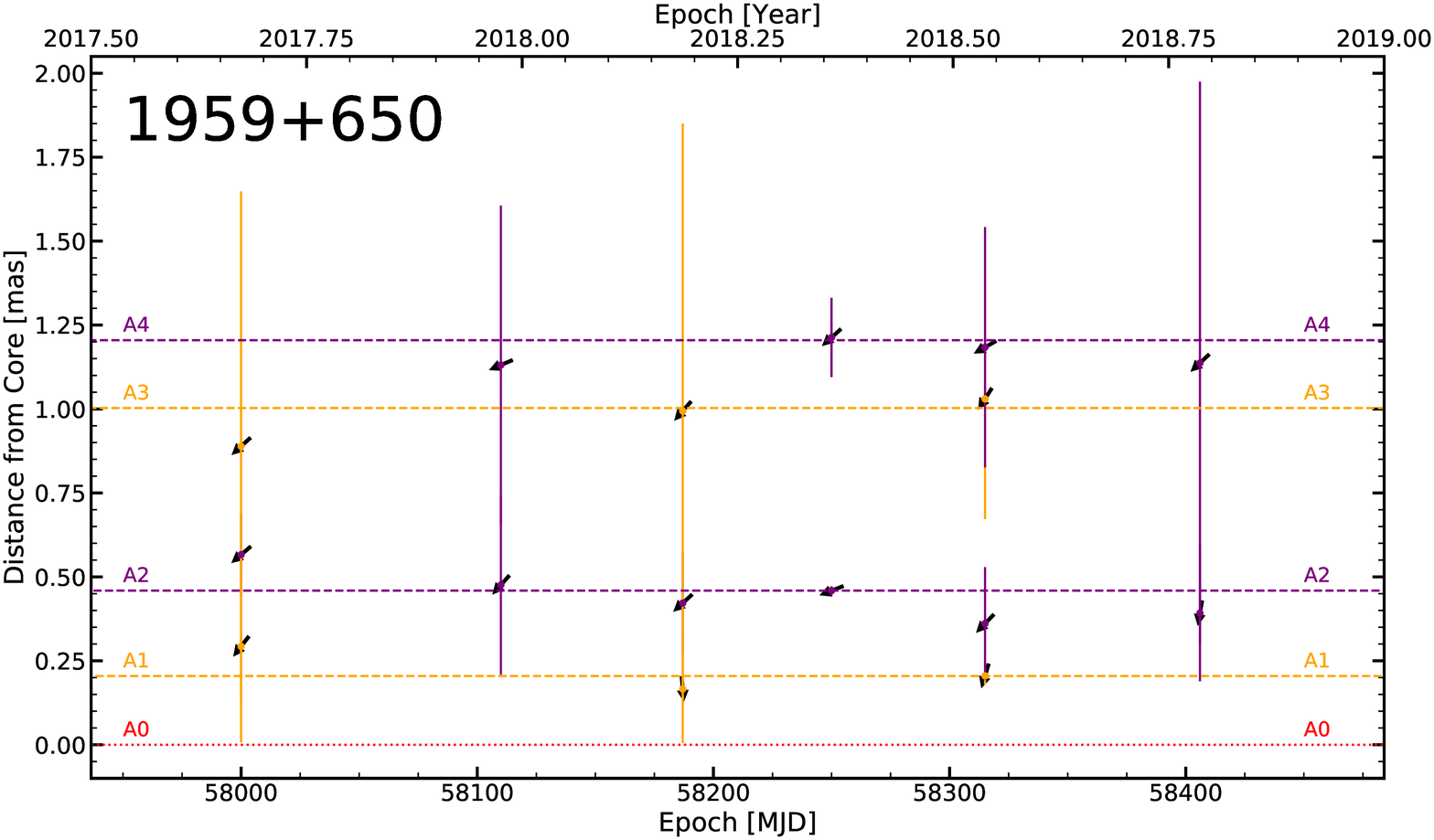}}
        \figsetgrpnote{Separation vs. time of the knots in the jet, relative to the core, of the BL 1959+650.}
        \figsetgrpend
        %
        % Number 35
        \figsetgrpstart
        \figsetgrpnum{7.35}
        \figsetgrptitle{2200+420}
        \figsetplot{{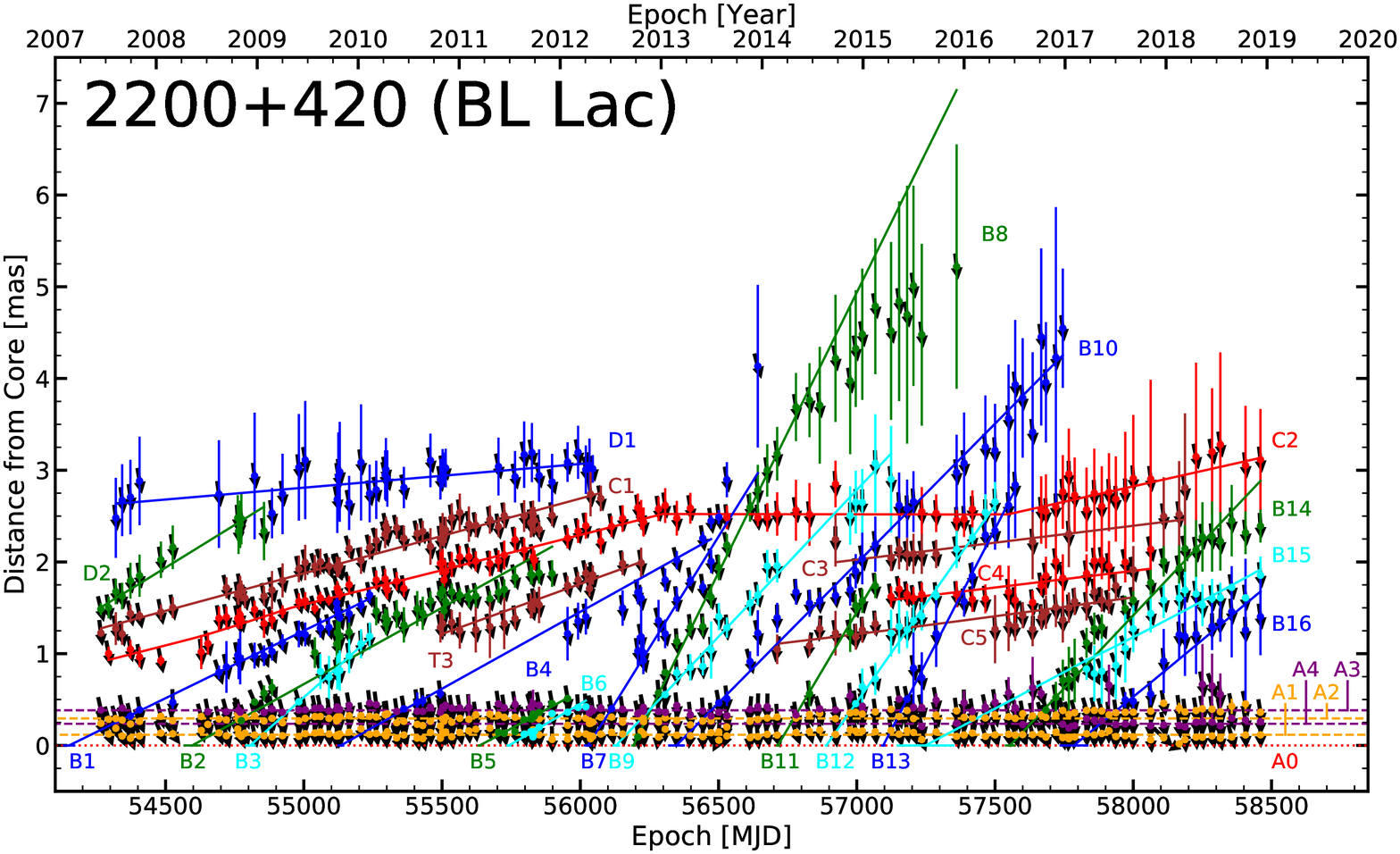}}
        \figsetgrpnote{Separation vs. time of the knots in the jet, relative to the core, of the BL 2200+420.}
        \figsetgrpend
        %
        % Number 36
        \figsetgrpstart
        \figsetgrpnum{7.36}
        \figsetgrptitle{2223-052}
        \figsetplot{{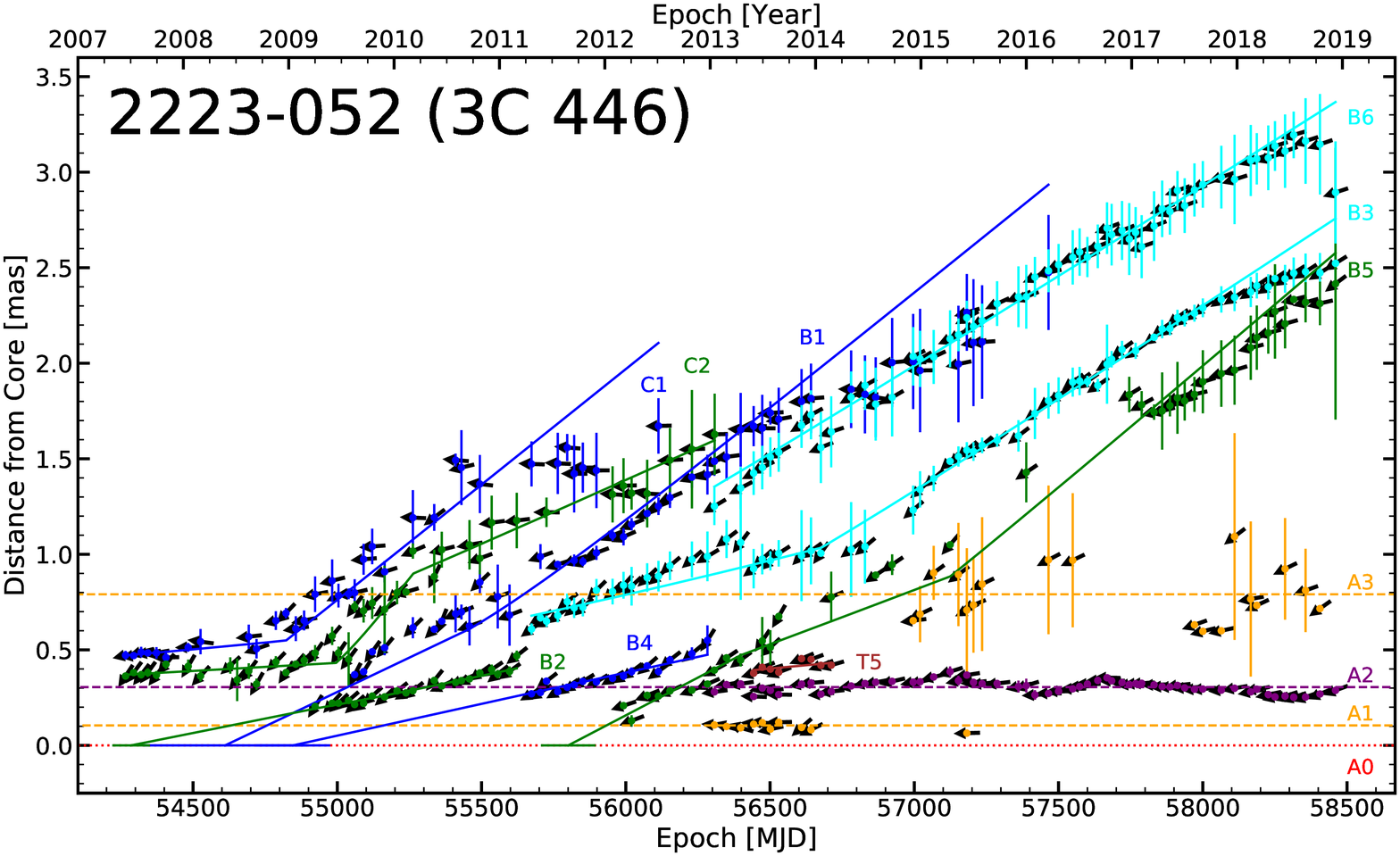}}
        \figsetgrpnote{Separation vs. time of the knots in the jet, relative to the core, of the FSRQ 2223-052.}
        \figsetgrpend
        %
        % Number 37
        \figsetgrpstart
        \figsetgrpnum{7.37}
        \figsetgrptitle{2230+114}
        \figsetplot{{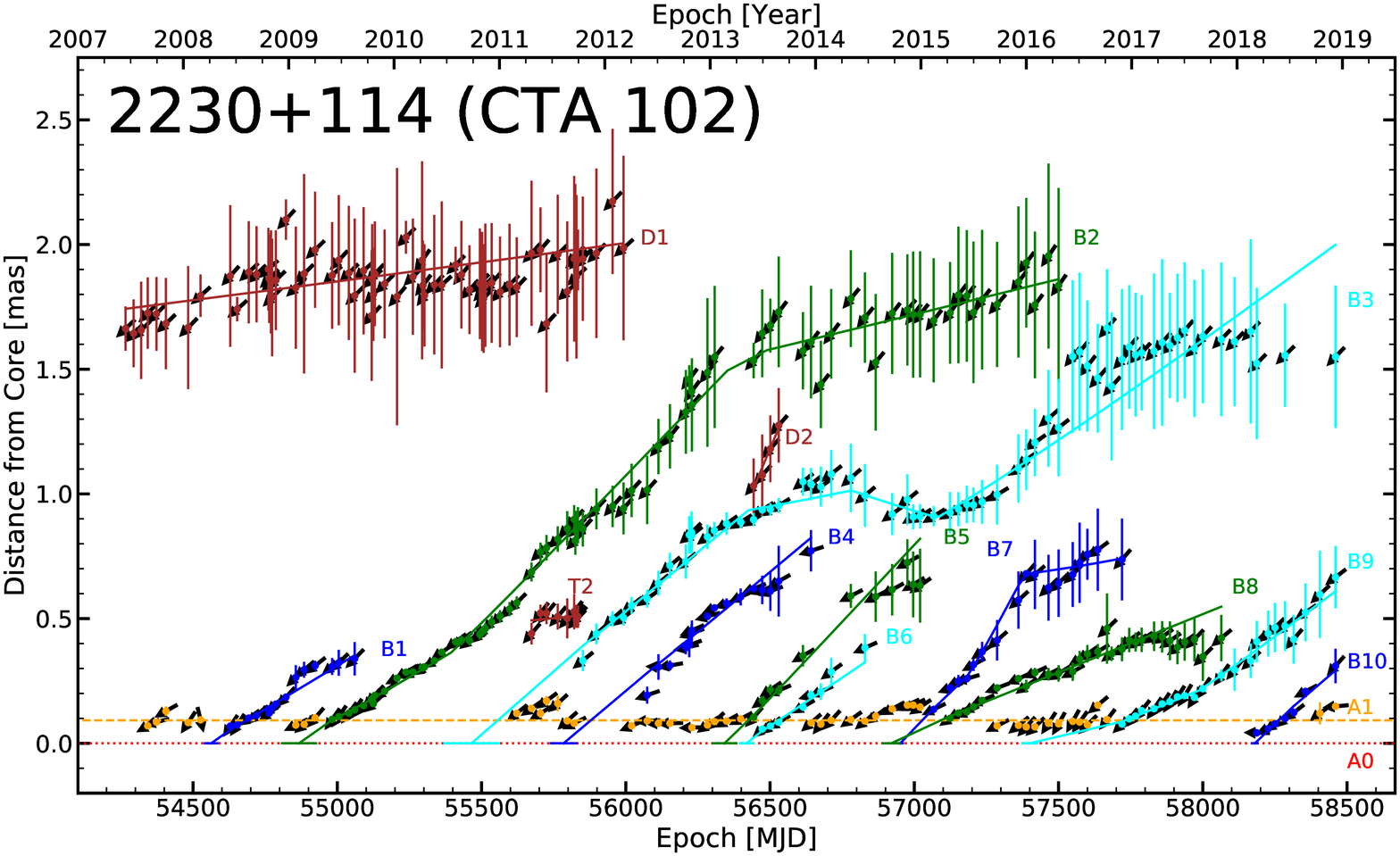}}
        \figsetgrpnote{Separation vs. time of the knots in the jet, relative to the core, of the FSRQ 2230+114.}
        \figsetgrpend
        %
        % Number 38
        \figsetgrpstart
        \figsetgrpnum{7.38}
        \figsetgrptitle{2251+158}
        \figsetplot{{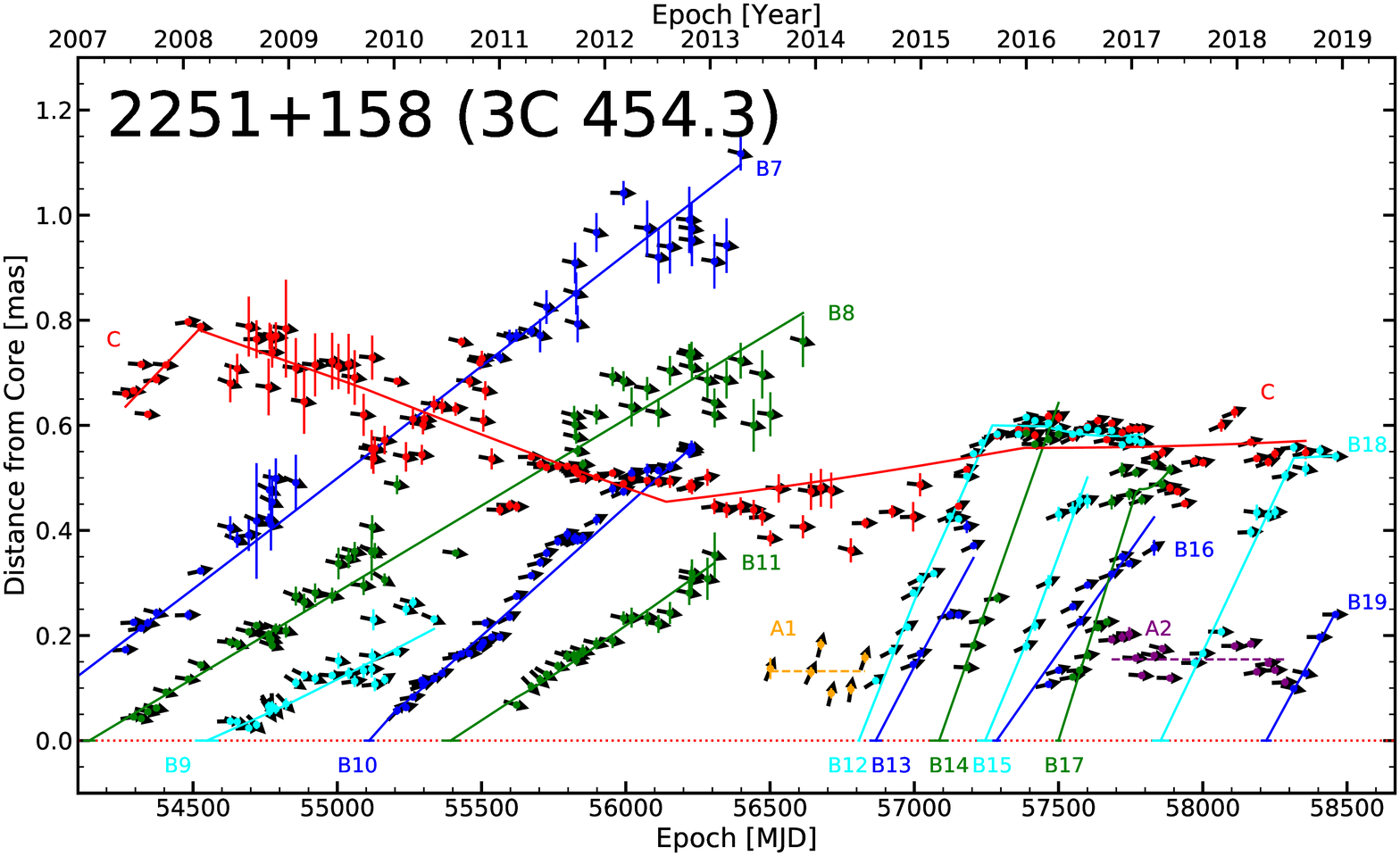}}
        \figsetgrpnote{Separation vs. time of the knots in the jet, relative to the core, of the FSRQ 2251+158.}
        \figsetgrpend
\figsetend
%----

\begin{deluxetable*}{llcccccccc}
    \tablecaption{Velocity in the Jets\label{tab:JetSpeeds}}
    \tablewidth{0pt}
    \tablehead{
    \colhead{Source} & \colhead{Knot} & \colhead{$N_{\text{seg}}$} & \colhead{$t_{\text{s}}$\tablenotemark{a}} & \colhead{$t_{\text{e}}$\tablenotemark{a}} & \colhead{$\mu$} & \colhead{$\Phi$} & \colhead{$\beta_{\text{app}}$} & \colhead{$t_\circ$} & \colhead{Flag} \\
    \colhead{} & \colhead{} & \colhead{} & \colhead{[yr]} & \colhead{[yr]} & \colhead{[mas yr$^{-1}$]} & \colhead{[deg]} & \colhead{[$c$]} & \colhead{[yr]} & \colhead{for $t_\circ$}
    }
    \colnumbers
    \startdata
    0219+428 & B1 & 1 & $2009.56$ & $2010.11$ & $0.915 \pm 0.178$ & $\phantom{-}179.2 \pm \phn 4.0$ & $24.62 \pm 4.80$    & $2008.67 \pm 0.24$ & R \\
             & B2 & 1 & $2010.84$ & $2012.07$ & $0.483 \pm 0.058$ & $-167.0 \pm \phn 4.9$           & $13.01 \pm 1.56$    & $2009.24 \pm 0.27$ & R \\
             & B3 & 1 & $2012.07$ & $2013.29$ & $0.665 \pm 0.070$ & $-168.5 \pm 11.7$               & $17.89 \pm 1.87$    & $2011.93 \pm 0.01$ & XY \\
             & B4 & 1 & $2014.15$ & $2015.93$ & $0.751 \pm 0.056$ & $-170.2 \pm \phn 1.4$           & $20.22 \pm 1.50$    & $2013.25 \pm 0.08$ & XY \\
             & B5 & 1 & $2013.96$ & $2016.31$ & $0.634 \pm 0.050$ & $-168.4 \pm \phn 2.1$           & $17.07 \pm 1.35$    & $2013.80 \pm 0.03$ & XY \\
             & B6 & 1 & $2017.85$ & $2018.94$ & $0.605 \pm 0.067$ & $-165.6 \pm \phn 5.7$           & $16.28 \pm 1.79$    & $2016.12 \pm 0.28$ & R \\
    0235+164 & B1 & 1 & $2007.66$ & $2008.62$ & $0.533 \pm 0.025$ & $-\phn18.3 \pm \phn 3.5$        & $26.58 \pm 1.27$    & $2007.45 \pm 0.01$ & XY \\
             & B2 & 1 & $2008.62$ & $2009.62$ & $0.258 \pm 0.024$ & $\phantom{-}140.3 \pm \phn 4.9$ & $12.88 \pm 1.18$    & $2008.15 \pm 0.11$ & R \\
             & B3 & 1 & $2010.58$ & $2013.96$ & $0.051 \pm 0.004$ & $\phantom{-}147.1 \pm \phn 4.8$ & $\phn2.55 \pm 0.21$ & $2008.10 \pm 0.32$ & R \\
             & B4 & 1 & $2013.49$ & $2016.31$ & $0.107 \pm 0.014$ & $-124.2 \pm \phn 8.9$           & $\phn5.32 \pm 0.68$ & $2012.62 \pm 0.26$ & R \\
             & B5 & 1 & $2016.08$ & $2017.67$ & $0.183 \pm 0.017$ & $\phantom{-}\phn66.7\pm\phn4.8$ & $\phn9.13 \pm 0.87$ & $2015.54 \pm 0.15$ & XY \\
             & B6 & 1 & $2016.91$ & $2018.65$ & $0.120 \pm 0.013$ & $\phantom{-}\phn13.4\pm\phn8.0$ & $\phn6.00 \pm 0.63$ & $2015.42 \pm 0.51$ & R \\
             & B7 & 1 & $2018.30$ & $2018.79$ & $0.191 \pm 0.046$ & $\phantom{-}\phn52.3\pm 16.3$   & $\phn9.55 \pm 2.28$ & $2016.55 \pm 0.95$ & R \\
    0336--019& B1 & 1 & $2007.75$ & $2009.15$ & $0.617 \pm 0.032$ & $\phantom{-}\phn81.4\pm\phn1.4$ & $28.57 \pm 1.50$    & $2007.30 \pm 0.02$ & XY \\
             & B2 & 1 & $2008.81$ & $2009.50$ & $0.227 \pm 0.003$ & $\phantom{-}\phn81.2\pm\phn4.3$ & $10.50 \pm 0.13$    & $2008.37 \pm 0.16$ & R \\
             & B2 & 2 & $2009.50$ & $2010.45$ & $0.650 \pm 0.001$ & $\phantom{-}\phn86.9\pm\phn1.5$ & $30.11 \pm 0.05$    & $\ldots$           & $\ldots$ \\
    \enddata
    \tablenotetext{a}{Knots with a single motion segment have no error in their start and end times as these are the first and last observation of the knot, respectively. The typical error on the segment start and end times for knots fit with multiple segments is $\sigma_{T} \lesssim 0.01$ yr and are determined from the covariance matrices of fit parameters in the least-squares piece-wise linear fitting.}
    \tablecomments{(Table~\ref{tab:JetSpeeds} (529 lines) is published in its entirety in the machine-readable format. A portion is shown here for guidance regarding its form and content.)}
\end{deluxetable*}

\subsection{Moving Feature Properties}
\label{subsec:MovingFeatureProps}

Here we describe the properties of the moving features in the jets of sources in each of the three subclasses of blazars. Because knot \emph{C} of the FSRQ 2251+158 (3C\ 454.3) shows motion despite being a quasi-stationary feature (see $\S$\ref{subsec:QuasiStationaryFeatures}), we include it in the analysis of 426 moving knots. Since any individual knot can have multiple motions over different time periods, we analyze a total of 529 knot speeds.

Table~\ref{tab:JetSpeeds} presents the speeds of the moving knots as follows:
1---name of the source;
2---designation of the component;
3---segment number of the fit to the knot motion, $N_{\text{seg}}$ (1 for purely linear motion in both the $X$ and $Y$ direction, up to 5\footnote{If best fit in a given dimension is with a 3-segment broken linear fit, then there are 2 breakpoints in that direction of motion. If both the \emph{X} and \emph{Y} motions are fit with a 3-segment broken linear fit, there are 4 total breakpoints and 5 line segments.});
4---start date of the segment, $t_{\text{s}}$, in yr;
5---end date of the segment, $t_{\text{e}}$, in yr;
6---proper motion, $\mu$, and its associated $1\sigma$ uncertainty, in mas yr$^{-1}$;
7---direction of motion, $\Phi$, and its associated $1\sigma$ uncertainty, in degrees;
8---apparent speed, $\beta_{\text{app}}$, and its associated $1\sigma$ uncertainty, in units of $c$;
9---epoch of ejection, $t_\circ$, and its associated $1\sigma$ uncertainty (only available for the first segment of a knot)in yr; and
10---flag indicating whether $t_\circ$ was calculated to the combined $X$ and $Y$ dimension fit (XY), or if a fit to the $R$ coordinate of the first segment was necessary (R, see $\S$\ref{subsec:KnotIdentitySpeed}). If the segment was not the first segment for a particular knot, the flag is NA.

\begin{figure*}
    \figurenum{8}
    \begin{center}
        \includegraphics[width=\textwidth]{{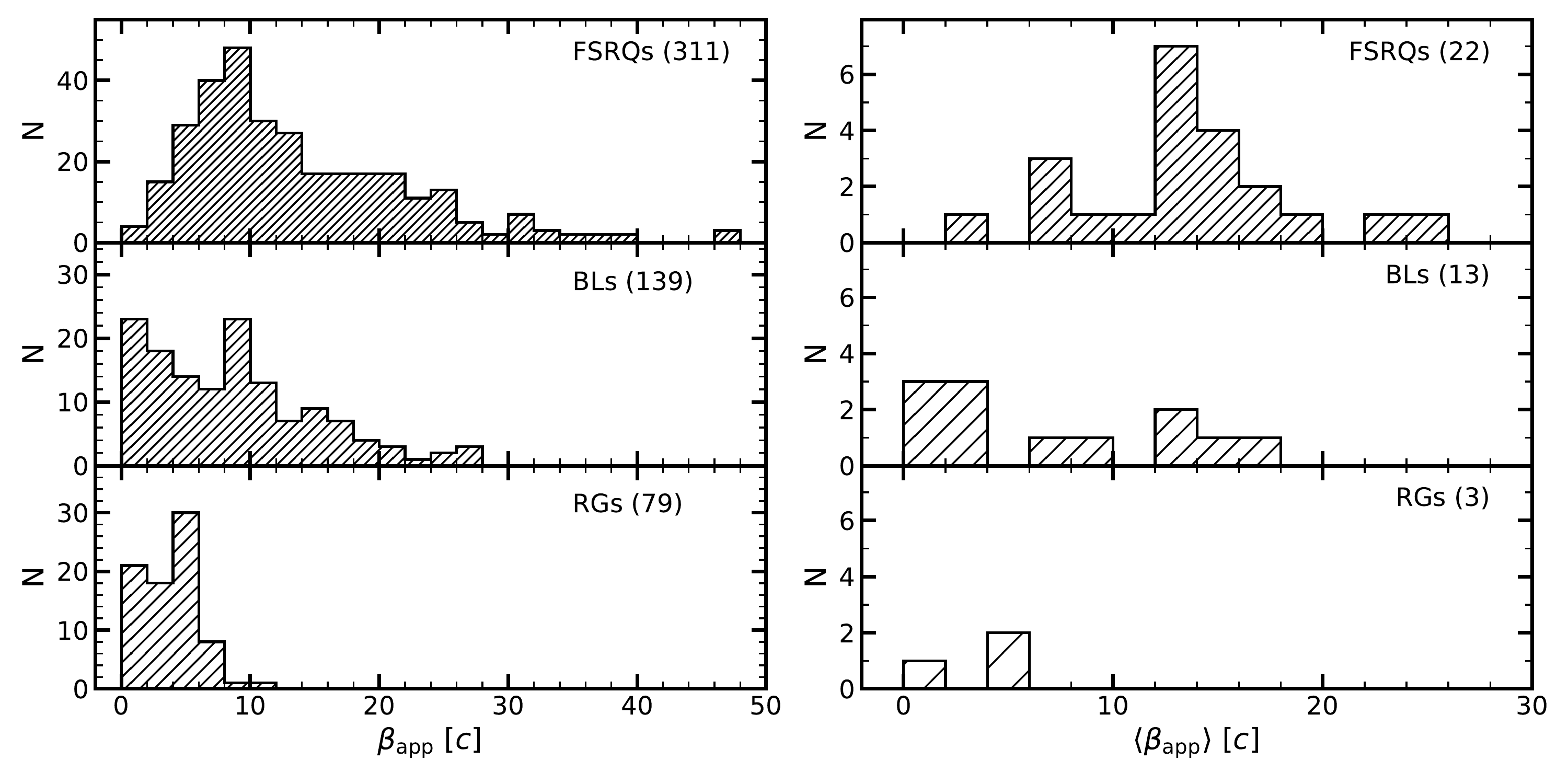}}
        \caption{Distributions of the apparent speeds of all moving knots (left), and weighted average of the apparent speed for each source (right) detected in FSRQs (top), BLs (middle), and RGs (bottom).
        \label{fig:BetaHist}}
    \end{center}
\end{figure*}

Figure~\ref{fig:BetaHist} (left) shows the distributions of the apparent speeds of the features for FSRQs, BLs, and RGs separately for the entire observing period from 2007 June to 2018 December for ease of comparison with previous literature. An increase in the number of moving knots by a factor of $\sim3$ has dramatically filled in the FSRQ and BL distributions over those presented in \citet{Jorstad2017}. The main distribution of apparent speeds of features in the jets of FSRQs covers a wide range of $\beta_{\text{app}}$, from $0.5c$ to $\sim50c$. Two knots (\emph{B1} of 0528+134 and \emph{D2} of 2230+114) have segments where the speed of the knot exceeds $50c$. In order to limit the size of the figure, these have been included in the last bin. They represent a small percentage (0.6\%) of the total number of knots and thus do not affect the conclusions. The FSRQ distribution peaks in the $8$-$10c$ bin but has a very long tail, extending continuously with at least one knot in each bin up to the $38$-$40c$ range. The apparent speeds in the jets of BLs exhibit a very different distribution, with a peak in the $0$-$2c$ bin and a similarly-sized peak in the $8$-$10c$ bin. The distribution does not extend to $\beta_{\text{app}}$ as high as for FSRQs, with the maximum BL speed of $\sim27c$ occurring for a segment of the knot B3 of 1749+096. The distribution of apparent speeds of RG knots is limited to much lower values than that of the other two subclasses. The peak between $4c$ and $6c$ corresponds to 0415+379 and 0430+052, with the peak from $0c$ to $2c$ coming primarily from 0316+413 (3C 84).

In order to more appropriately determine the differences in the distributions of apparent speed for FSRQs, BLs, and RGs, we calculate a single average apparent speed, $\langle \beta_{\mathrm{app}} \rangle$, for each source as the weighted average of the apparent speeds of all knots in that source. The distributions of these average apparent speeds are shown in Figure~\ref{fig:BetaHist} (right). We see that the FSRQs have a well-determined peak in the distribution with $\langle \beta_{\mathrm{app}} \rangle \sim 12$-$14c$. The BLs have a very wide dispersion of average apparent speeds. With just three sources, it is easy to see (with Table~\ref{tab:JetSpeeds}) that the knots of 0316+413 are, on average, subluminal, while those of 0415+379 and 0430+052 are superluminal. Despite the small numbers of sources (especially for RGs), we have performed a KS test between the distributions. The BL and RG distributions are the most similar, with $\mathcal{D} = 0.500$ and $p = 0.525$, likely due to the small number of sources. However, moderately-significant differences are seen between the FSRQ:BL ($\mathcal{D} = 0.492,\ p = 0.029$) and FSRQ:RG ($\mathcal{D} = 0.955,\ p = 0.003$) distributions. Taking all of the results of Figure~\ref{fig:BetaHist} in consideration, we see that the knots of FSRQs have, on average, higher apparent speeds than those of BLs, with RGs having the lowest apparent speeds.

\begin{figure}[t]
    \figurenum{9}
    \begin{center}
    \includegraphics[width=0.475\textwidth]{{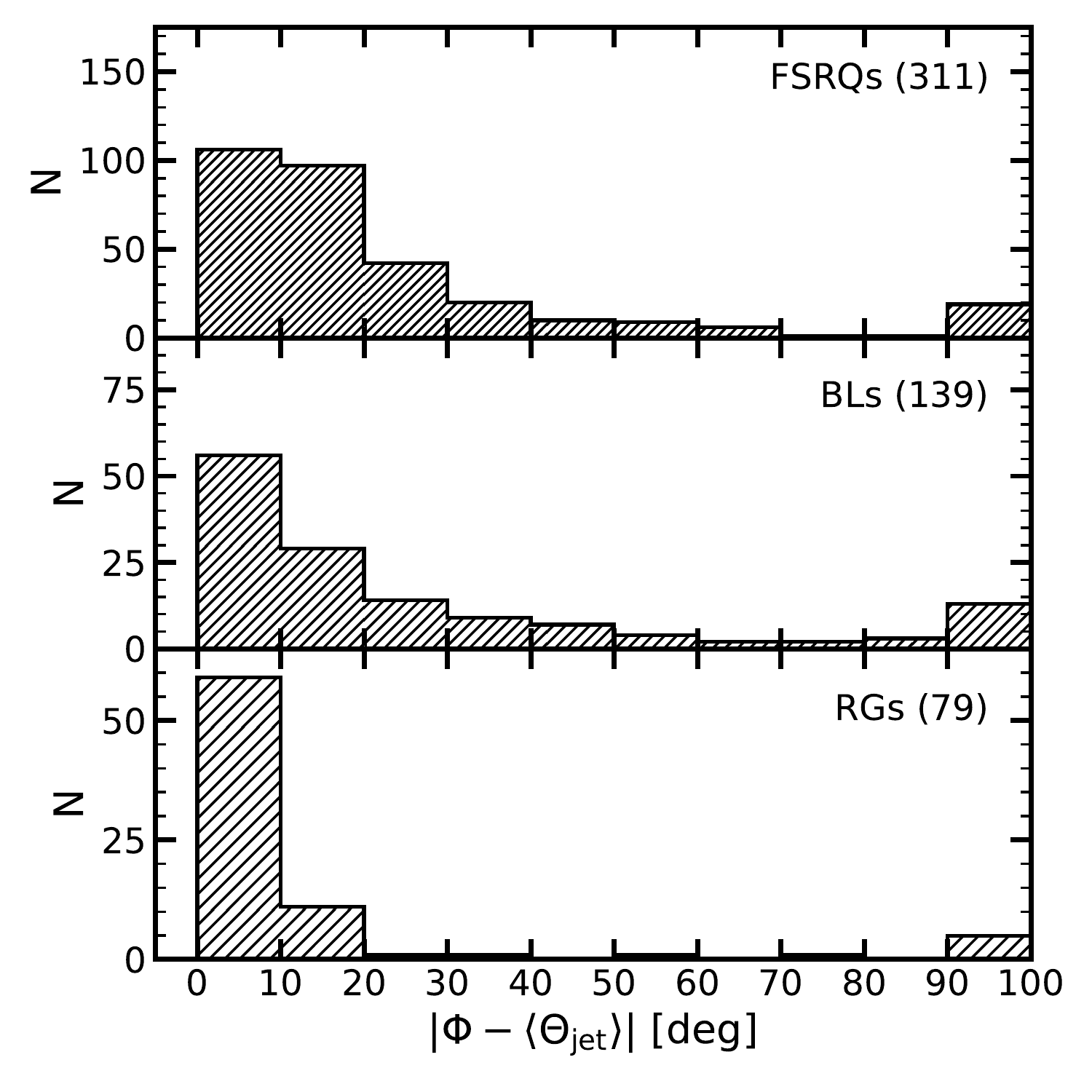}}
    \caption{Distributions of the differences between the velocity vector, $\Phi$, and the jet axis, $\langle \Theta_{\text{jet}} \rangle$, in FSRQs (top), BLs (middle), and RGs (bottom).$|\Phi - \langle \Theta_{\text{jet}} \rangle | > 90\degr$ indicates slow or upstream motion of the knot segment.\label{fig:VelocityDistribution}}
    \end{center}
\end{figure}

We analyze the direction of the velocity vector of each knot by following the same approach as in \citet{Jorstad2017}, calculating the difference between the direction $\Phi$ of the apparent velocity (given in Table~\ref{tab:JetSpeeds}) and the average position angle of the jet, $\langle \Theta_{\text{jet}} \rangle$ (listed in Table~\ref{tab:JetStructure}).
Instead of determining an average value of this parameter for each source, we consider the difference between $\Phi$ and $\langle \Theta_{\text{jet}} \rangle$ for a given knot to be an independent measure of the possible dispersion of the position angles for each subclass of blazar, as all knots are required to determine the width/dispersion of a jet.
Figure~\ref{fig:VelocityDistribution} plots the distributions of $|\Phi - \langle \Theta_{\text{jet}} \rangle |$ for FSRQs, BLs, and RGs. The peak of each distribution lies within the first bin ($<10\degr$) and represents 34.1\%, 40.3\%, and 74.7\% of the total number of knots in FSRQs, BLs, and RGs respectively.

In order to measure the deviation from unidirectional motion, we first calculate the average uncertainty in the motion direction as $\langle \sigma(\Phi)\rangle = \sum_{i=1}^{N_k} \sigma(\Phi_i) / N_k$, where $N_k$ is the total number of knot speeds measured in each subclass. BLs have the largest dispersion of knot motions, with $\langle \sigma(\Phi)\rangle = 7\fdg2$, while FSRQs and RGs have a dispersion $\langle \sigma(\Phi)\rangle = 4\fdg2$ and $4\fdg8$, respectively. The much larger dispersion in this work compared to those presented in \citet{Jorstad2017} is likely caused by a significantly larger number of measured speeds based on the revised method for calculating motion.
In order to determine whether this dispersion is from varying directions of the velocity vectors of knots or from knots smaller than the jet cross-section being displaced from the jet axis by various distances, we calculate the average position-angle dispersion of knots in a subclass as $\langle \sigma(\langle\Theta\rangle)\rangle = \sum_{i=1}^{N_k} \sigma(\langle \Theta \rangle_i) / N_k$. Based on this, FSRQs have the widest range of average knot position angles, with $\langle \sigma(\langle \Theta \rangle) \rangle = 9\fdg9$, while BLs and RGs have smaller dispersions of $8\fdg8$ and $5\fdg7$, respectively.
We conclude that, while the knots of FSRQs are more inclined to move along the same direction in the jet than are the knots in the jets of BLs, FSRQs have wider projected opening angles. The latter can be connected with a smaller viewing angle and/or a wider intrinsic opening angle for FSRQs.
FSRQs and BLs possess a similar percentage of knots in which the velocity vector deviates from the jet axis by a value between $30\degr$ and $60\degr$, 12.5\% and 14.4\% respectively. In contrast, only $2.5\%$ of the knots of RGs have motions that fall within this deviation range.

Among superluminal knots there are 5, 13, and 19 segments in FSRQs, BLs, and RGs, respectively, with very slow ($<1c$) or even upstream motion corresponding to $|\Phi - \langle \Theta_{\text{jet}} \rangle| > 90\degr$. These are shown in the last bin of the distribution, since they represent a small fraction of the total number of knot segments.
A KS test indicates that the FSRQ and BL distributions are \emph{not} significantly different ($\mathcal{D} = 0.138,\ p = 0.046$). However, the FSRQ and BL distributions \emph{are} statistically different from the RG distribution ($\mathcal{D} = 0.439,\ p =1.8\times10^{-11}$ for FSRQs vs RGs, and $\mathcal{D} =0.369,\ p =1.2\times10^{-6}$ for BLs vs RGs).

\subsection{Quasi-Stationary Features}
\label{subsec:QuasiStationaryFeatures}

\begin{figure*}[t]
    \figurenum{10}
    \begin{center}
        \includegraphics[width=0.85\textwidth]{{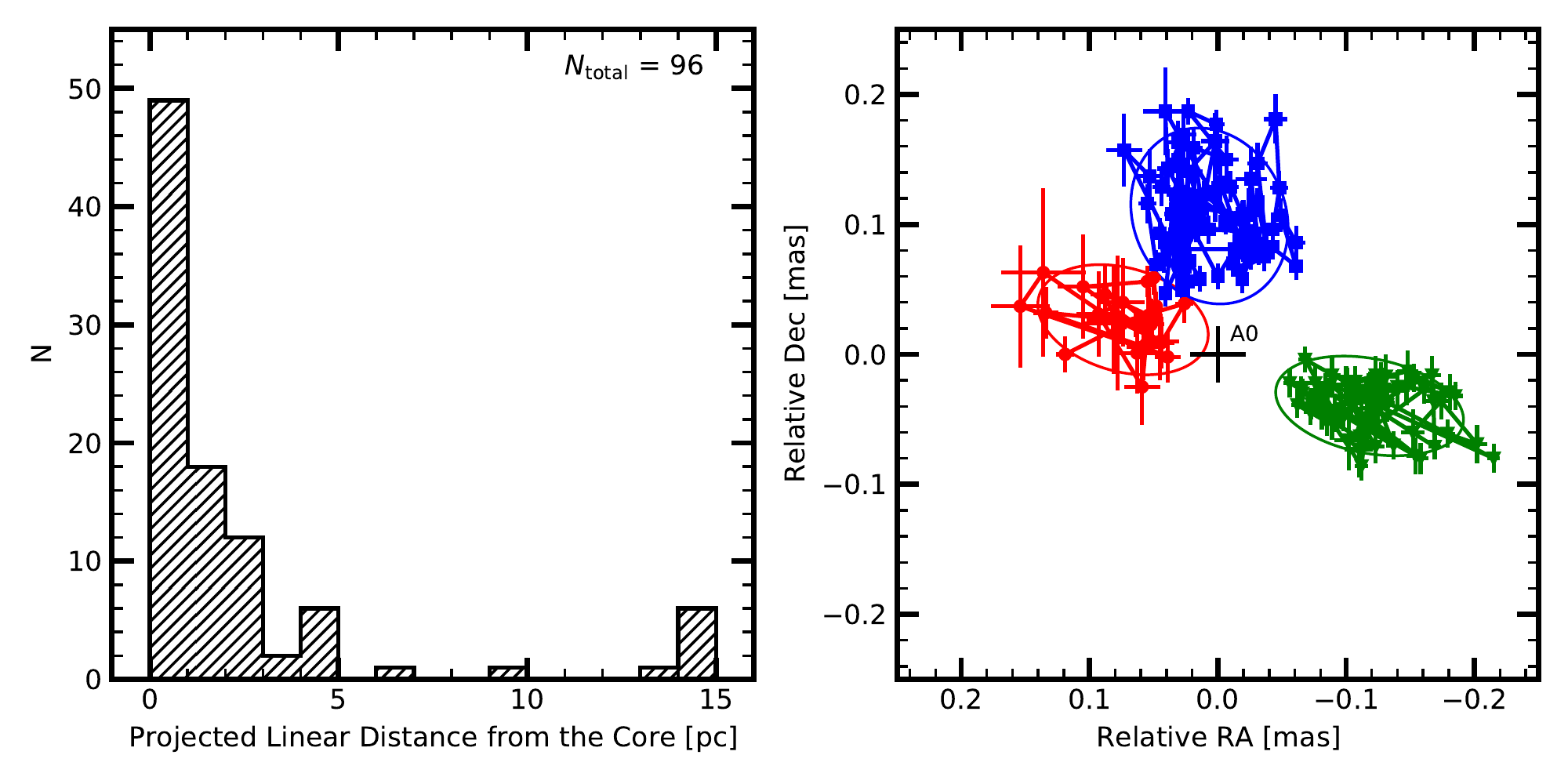}}
        \caption{\emph{Left:} distribution of the projected distances from the core to the stationary features in our sample. \emph{Right:} trajectories of the stationary features \emph{A1} in 0528+134 (FSRQ, red circles), \emph{A2} in 1749+096 (BL, blue squares), and \emph{A1} in 0430+052 (RG, green triangles) with respect to the core, \emph{A0} (the black cross at position (0,0)). The solid lines show $2\sigma$ confidence ellipses used to characterize the dispersion and orientation of the knot motion\label{fig:stationarydist}}
    \end{center}
\end{figure*}

We have identified 96 features with no statistically significant motion \citep[as defined in][]{Jorstad2017}. Of these, 45 are in FSRQs, 38 in BLs, and 13 in RGs. All sources in our sample except 0235+164 (BL), 0316+413 (RG), 1156+295, and 1622$-$297 (FSRQs) continue to contain at least one --- but usually multiple --- stationary feature in addition to the core. On average, FSRQs possess $\sim2$ and BLs $\sim3$ such features. Our sample size of RGs is small, however, we see 7 and 5 stationary features in 0415+379 and 0430+052, respectively. Given the sub-relativistic speeds of knots in 0316+413, it is hard to determine what features are truly stationary.
Almost half of the stationary features in our sample have appeared or disappeared over time, indicating that while these features may be stationary in position, they are often transient features over timescales of $\sim10$ years. Given that there is no way to average the properties of stationary features for a particular source into one overarching ``typical" stationary feature, we treat each stationary feature as an independent feature when making comparisons between the subclasses of blazar.

\begin{deluxetable*}{ccrcccccr}
    \tablecaption{Quasi-Stationary Jet Features\label{tab:Stationary}}
    \tablewidth{0pt}
    \tablehead{
    \colhead{Source} & \colhead{Knot} & \colhead{$N$} & \colhead{$\langle R \rangle$} & \colhead{$X_{\text{cen}}$} & \colhead{$Y_{\text{cen}}$} & \colhead{$A$}   & \colhead{$B$}   & \colhead{$\varpi$} \\
    \colhead{} & \colhead{} & \colhead{} & \colhead{[pc]} & \colhead{[mas]} & \colhead{[mas]} & \colhead{[mas]} & \colhead{[mas]} & \colhead{[deg]}
    }
    \colnumbers
    \startdata
    0219+428  & A1 & 52 & $\phn0.97 \pm 0.23$ & $-0.04$           & $-0.16$           & $0.08$ & $0.06$ & $21.9$ \\
              & A2 & 36 & $\phn2.11 \pm 0.34$ & $-0.08$           & $-0.35$           & $0.15$ & $0.10$ & $-127.7$ \\
              & A3 & 50 & $\phn3.71 \pm 0.51$ & $-0.07$           & $-0.64$           & $0.19$ & $0.11$ & $-21.7$ \\
              & A4 & 56 & $13.20 \pm 0.51$    & $-0.12$           & $-2.31$           & $0.18$ & $0.11$ & $-3.3$ \\
    0336--019 & A1 & 36 & $\phn1.23 \pm 0.31$ & $\phantom{-}0.14$ & $\phantom{-}0.02$ & $0.10$ & $0.08$ & $13.9$ \\
              & A2 &  7 & $14.64 \pm 0.84$    & $\phantom{-}1.77$ & $\phantom{-}0.69$ & $0.28$ & $0.21$ & $16.0$ \\
              & A3 &  8 & $\phn2.15 \pm 0.23$ & $\phantom{-}0.27$ & $\phantom{-}0.07$ & $0.11$ & $0.04$ & $-46.4$ \\
              & A4 & 11 & $\phn3.14 \pm 0.46$ & $\phantom{-}0.39$ & $\phantom{-}0.11$ & $0.17$ & $0.09$ & $-41.8$ \\
    \enddata
    \tablecomments{(Table~\ref{tab:Stationary} (96 lines) is published in its entirety in the machine-readable format. A portion is shown here for guidance regarding its form and content.)}
\end{deluxetable*}

Figure~\ref{fig:stationarydist}, left, shows the distribution of the average projected linear distances from the core of all stationary features in our sample. Five stationary components that appear beyond 15 pc from the core are included in the last bin. The distribution has a prominent peak at distances $<1$ pc from the core, as in the distribution presented in \citet{Jorstad2017}. In addition, there continues to be no statistical difference in the distributions of FSRQs, BLs, and RGs, with a KS test yielding $\mathcal{D} = 0.159$ and $p = 0.601$ for the distributions of FSRQs vs BLs, $\mathcal{D} = 0.260$ and $p = 0.423$ for FSRQs vs RGs, and finally $\mathcal{D} = 0.221$ and $p = 0.642$ for BLs vs RGs.

Despite the ``stationary feature" classification, the brightness centroids of many of these components do exhibit distinguishable motion. The locations of the majority of features tend to fluctuate about their mean reported positions. Figure~\ref{fig:stationarydist}, right, shows three examples of the trajectories of stationary features, one in each subclass. The range of motion in both right ascension and declination is larger than the positional errors can account for, and the loci of $(X,Y)$ positions of a given stationary feature form a pattern that indicates a preferred direction of motion for the fluctuations. In order to characterize the shifts in position, we fit confidence ellipses to the motions of each stationary feature. Table~\ref{tab:Stationary} gives the parameters of the elliptical fits as follows:
1---source name;
2---knot designation;
3---number of observations, $N$, of a knot;
4---projected linear distance from the core to $\langle R \rangle$ of the knot, in parsecs;
5---mean position of the knot along the $X$-axis (relative R.A.), $X_{\text{cen}}$, in mas;
6---mean position of the knot along the $Y$-axis (relative DEC), $Y_{\text{cen}}$, in mas;
7---semimajor axis length, $A$, in mas;
8---semiminor axis length, $B$, in mas; and
9---orientation angle of semimajor axis relative to north, $\varpi$, in degrees. Figure~\ref{fig:stationarydist}, right shows sample $2\sigma$ confidence ellipses for the given knots, where $2\sigma$ indicates that 95\% of the observed data would lie within the ellipse if the data were normally distributed.

\begin{figure}
    \figurenum{11}
    \begin{center}
        \includegraphics[width=0.45\textwidth]{{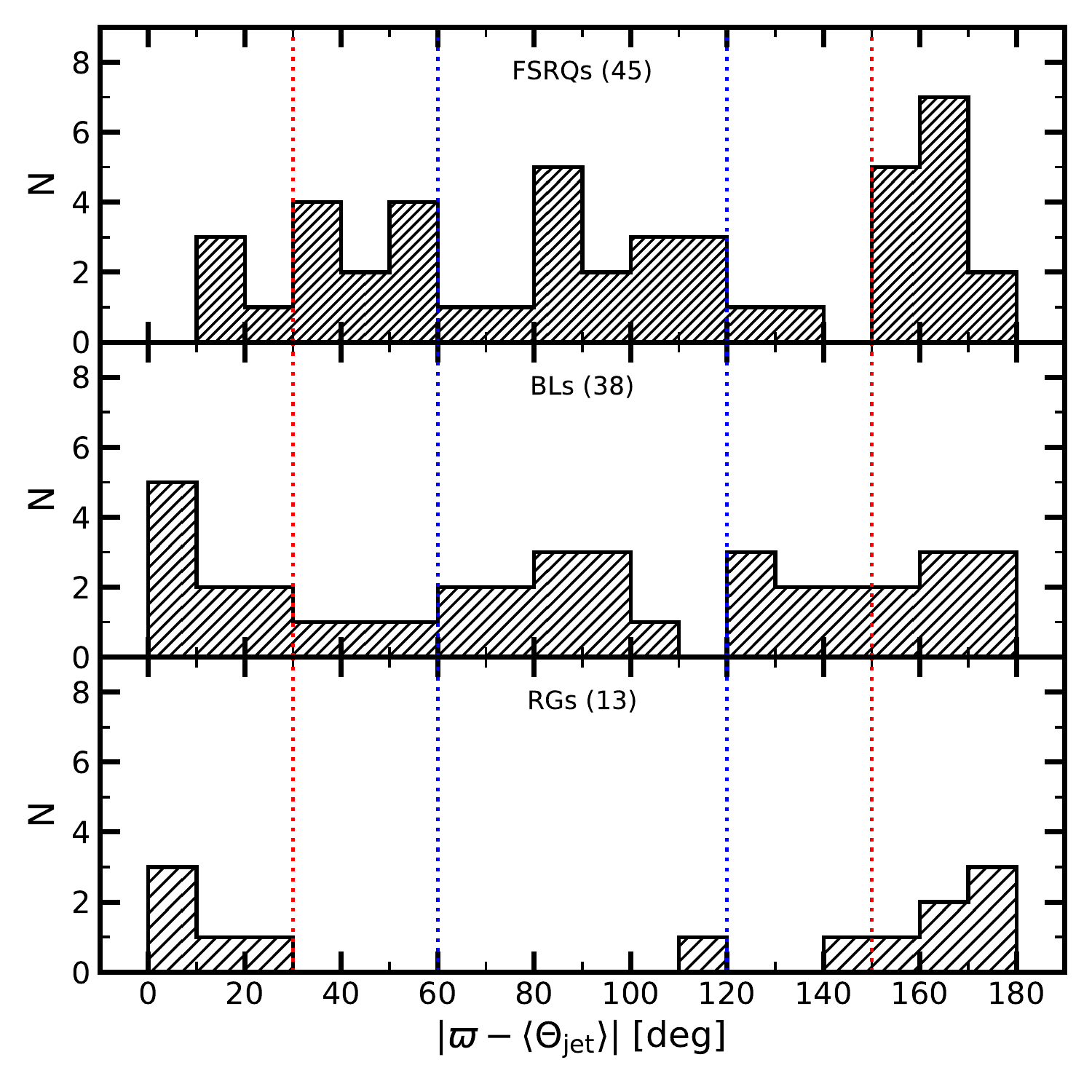}}
        \caption{Distributions of the differences between the major axis orientation angles, $\varpi$, and $\langle \Theta_{\text{jet}} \rangle$, in FSRQs (top), BLs (middle), and RGs (bottom). See the text for details.
        \label{fig:orientationangle}}
    \end{center}
\end{figure}

In this characterization of the knot motion dispersion, $\varpi$ represents the angle of the motion along the major axis. We compare this direction of motion with the direction of the jet through $|\varpi - \langle \Theta_{\text{jet}} \rangle|$, as defined in Table~\ref{tab:JetPAs}. For sources with $\langle \Theta_{\text{jet}} \rangle$ fit by a linear trend or a spline, we average the trend over the dates of observation of the particular stationary feature. Figure~\ref{fig:orientationangle}, left, shows the distributions of this difference angle separately for FSRQs, BLs, and RGs.
A KS test between the distributions indicates that there is no statistical difference between the subclasses, with the FSRQ and BL distributions being the most similar ($\mathcal{D} = 0.162$, $p =0.578$), while  $\mathcal{D} = 0.318,$ $p = 0.200$ for FSRQs versus RGs and $\mathcal{D} = 0.277$, $p = 0.357$ for BLs versus RGs.

\begin{figure}
    \figurenum{12}
    \begin{center}
        \includegraphics[width=0.45\textwidth]{{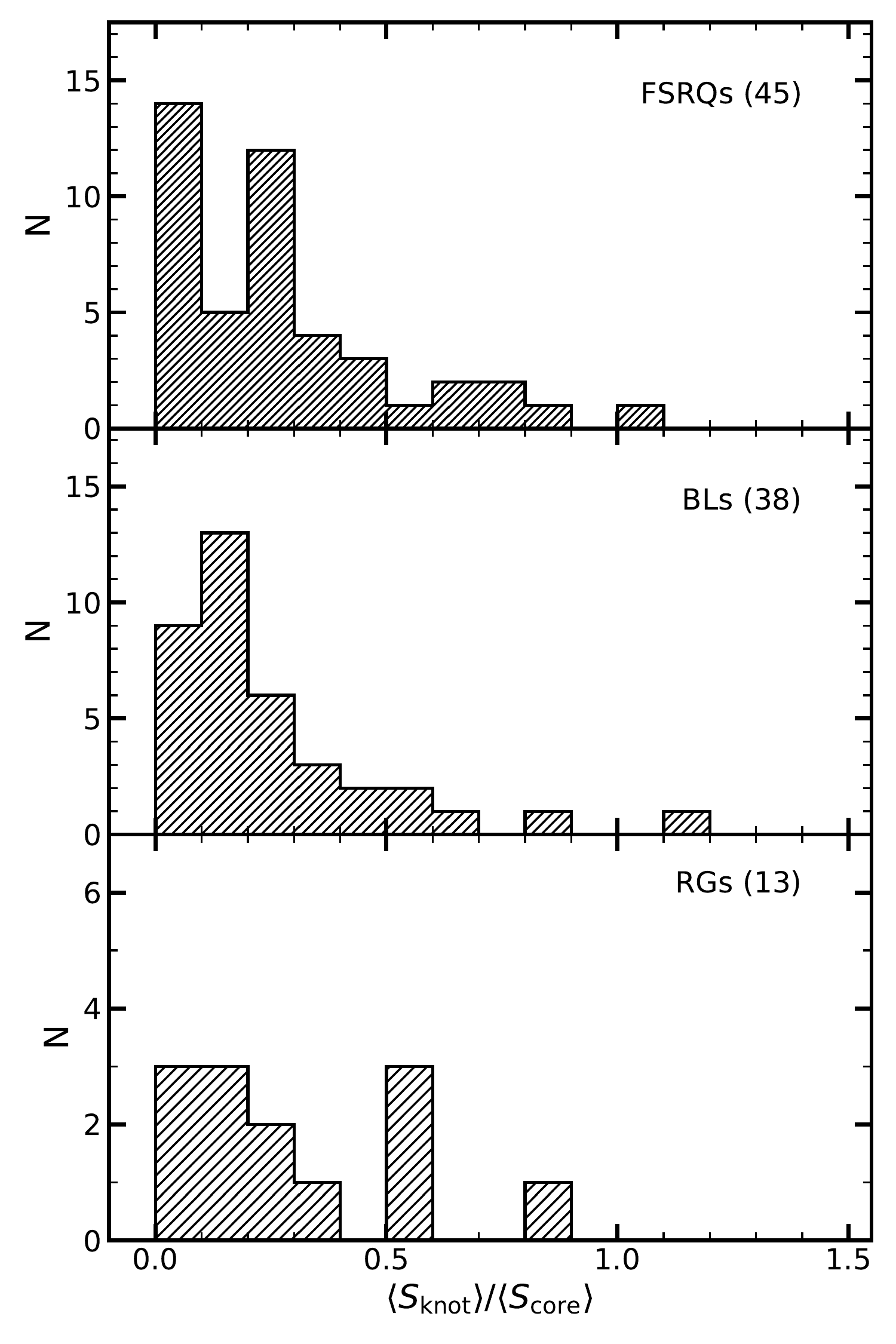}}
        \caption{Distributions of the relative flux densities of quasi-stationary features for the different subclasses---FSRQs (top), BLs (middle), and RGs (bottom).\label{fig:StatFluxDist}}
    \end{center}
\end{figure}

\begin{figure*}
    \figurenum{13}
    \begin{center}
        \includegraphics[width=0.75\textwidth]{{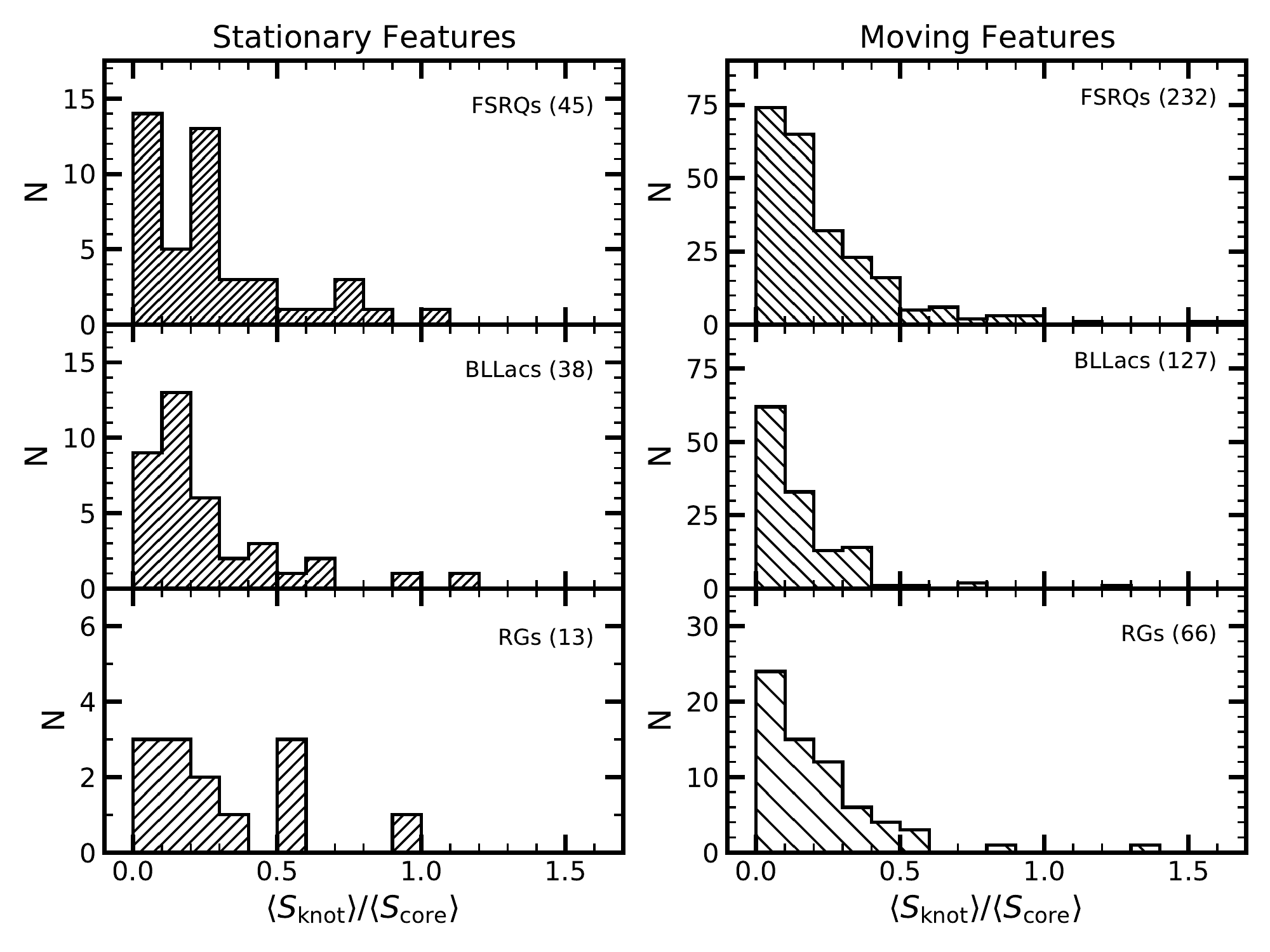}}
        \caption{Distributions of the average flux of a knot relative to the average flux of the core, $\langle S_{\text{core}} \rangle$, for the stationary (left) and moving (right) features in FSRQs (top), BLs (middle), and RGs (bottom).
        \label{fig:StatMoving}}
    \end{center}
\end{figure*}

We are able to separate the entire range of $|\varpi - \langle \Theta_{\text{jet}} \rangle|$ into three categories: (1) fluctuations along the jet, with $|\varpi - \langle \Theta_{\text{jet}} \rangle | \leq 30\degr$ or $|\varpi - \langle \Theta_{\text{jet}} \rangle | > 150\degr$, (2) fluctuations transverse to the jet direction, $60\degr < |\varpi - \langle \Theta_{\text{jet}} \rangle| \leq 120\degr$, and (3) oblique fluctuations with intermediate angle differences, $30\degr < |\varpi - \langle \Theta_{\text{jet}} \rangle| \leq 60\degr$ or $120\degr < |\varpi - \langle \Theta_{\text{jet}} \rangle| \leq 150\degr$. Figure~\ref{fig:orientationangle} shows the divisions between parallel and oblique cases with red dotted lines, while the division between oblique and transverse fluctuations is shown with blue dotted lines.
In FSRQs, 40\% (18/45) of knots fluctuate along the jet direction, 26.7\% (12/45) oblique to the jet, and 33.3\% (15/45) transverse to the jet. The situation is similar for BLs: 44.7\% (17/38), 26.3\% (10/38), and 28.9\% (11/38), respectively. This is supported by the KS test, which indicates that the FSRQ and BL distributions are not significantly different. The RG distribution, however, shows a clear preference for motions along the jet direction, with 11 of the 13 knots (92\%) moving predominately along the jet direction. While the number of RG stationary features has increased over those detected in \citet{Jorstad2017}, there are still only a relatively small number of such knots, so that the KS test is inconclusive.

\begin{deluxetable*}{llcccccccc}
    \tablecaption{Acceleration in the Jets\label{tab:JetAccel}}
    \tablewidth{0pt}
    \tablehead{
    \colhead{Source} & \colhead{Knot} & \colhead{$N_{\text{acc}}$} & \colhead{$t_{\mathrm{start}}$\tablenotemark{a}} & \colhead{$t_{\mathrm{end}}$\tablenotemark{a}} &  \colhead{$\dot{\mu}_\parallel$} & \colhead{$\dot{\mu}_\perp$} & \colhead{$R_{\dot{\mu}}$} & \colhead{$\eta_\parallel$} & \colhead{$\eta_\perp$} \\
    \colhead{} & \colhead{} & \colhead{} & \colhead{[yr]} & \colhead{[yr]} & \colhead{[mas yr$^{-2}$]} & \colhead{[mas yr$^{-2}$]} & \colhead{[mas]}  & \colhead{[yr$^{-1}$]} & \colhead{[yr$^{-1}$]}
    }
    \colnumbers
    \startdata
    0316+413  & C1a & 1 & $2014.69$ & $2017.58$ & $\phantom{-}0.127 \pm 0.003$ & $\phantom{-}0.015 \pm 0.023$ & 2.86 & $\phantom{-}0.538 \pm 0.016$ & $\phantom{-}0.064 \pm 0.098$ \\
              & C10 & 1 & $2016.32$ & $2016.89$ & $-1.778           \pm 0.073$ & $-0.209           \pm 0.307$ & 1.19 & $-1.558           \pm 0.069$ & $-0.183           \pm 0.269$ \\
              & C10 & 2 & $2016.89$ & $2018.04$ & $\phantom{-}0.404 \pm 0.037$ & $\phantom{-}0.047 \pm 0.077$ & 1.24 & $\phantom{-}1.827 \pm 0.289$ & $\phantom{-}0.213 \pm 0.349$ \\
    0336--019 &  B2 & 1 & $2009.16$ & $2009.98$ & $\phantom{-}0.508 \pm 0.013$ & $\phantom{-}0.102 \pm 0.064$ & 0.25 & $\phantom{-}4.145 \pm 0.119$ & $\phantom{-}0.832 \pm 0.522$ \\
              &  B3 & 1 & $2011.63$ & $2013.88$ & $\phantom{-}0.075 \pm 0.002$ & $\phantom{-}0.015 \pm 0.009$ & 0.28 & $\phantom{-}1.094 \pm 0.039$ & $\phantom{-}0.219 \pm 0.131$ \\
              &  B4 & 1 & $2015.33$ & $2016.94$ & $\phantom{-}0.222 \pm 0.005$ & $\phantom{-}0.044 \pm 0.026$ & 0.46 & $\phantom{-}2.056 \pm 0.051$ & $\phantom{-}0.407 \pm 0.241$ \\
    \enddata
    \tablenotetext{a}{We take the typical uncertainties in the estimated start and end dates of the acceleration region to be similar to the uncertainty in the break points for the piece-wise linear fitting, which are $\sigma_{t} \lesssim 0.01$ yr. The uncertainties in the acceleration are dominated by the uncertainty in the speeds rather than the time period of acceleration.}
    \tablecomments{(Table~\ref{tab:JetAccel} (104 lines) is published in its entirety in the machine-readable format. A portion is shown here for guidance regarding its form and content.)}
\end{deluxetable*}

Almost all of the apparent motions of the quasi-stationary features appear to be minor fluctuations about the average positions. A clear trend in the motion is apparent only for knot \emph{C} of the FSRQ 2251+158. Over $\sim10$ years of observations, this feature started $\sim0.7$ mas from the 43 GHz core in 2007, after which the separation decreased to a minimum of $\sim0.45$ mas in 2013. Since then, the separation has been increasing more gradually. It is unclear whether this motion is related to changes in the intensity distribution of this extended feature, or to bright knots that were ejected from the core and later merged with \emph{C}. An alternative possibility is that the change in separation between the core and \emph{C} in 2251+158 is due to the motion of the core itself, contrary to the assumption made in this work that every core is at an essentially stationary position in the jet.

Figure~\ref{fig:StatFluxDist} shows distributions of the average flux density of the stationary features, $\langle S_{\text{knot}} \rangle$, normalized by the average flux density of the core, $\langle S_{\text{core}} \rangle$ for all three subclasses. A KS test confirms the distributions for each subclass are similar, with $\mathcal{D} = 0.201,\ p = 0.319$ for FSRQs vs.\ BLs, $\mathcal{D} = 0.244,\ p = 0.501$ for FSRQs vs.\ RGs, and $\mathcal{D} = 0.223,\ p = 0.628$ for BLs vs.\ RGs.

Finally, we investigate whether the stationary and moving features in the jets have distributions of knot:core flux density ratios that are statistically different within a subclass. We calculate the average $\langle S_{\text{knot}} \rangle$, median, and maximum flux density of each knot and normalize them by the average flux density of the core, $\langle S_{\text{core}} \rangle$. Figure~\ref{fig:StatMoving} displays the distributions of flux ratios for stationary features (left) and moving features (right) separately for each subclass. There is no statistical difference between the flux ratios of the stationary and moving features in RGs according to a K-S test ($\mathcal{D} = 0.333,\ p = 0.133$). The hypothesis of different distributions in FSRQs is similarly rejected ($\mathcal{D} = 0.191,\ p = 0.110$). In BLs, however, there is a moderately-significant result that the stationary features of BLs have higher flux ratios than do the moving features ($\mathcal{D} = 0.275$, $p = 0.019$).
The flux distributions and KS statistics for the average and maximum flux ratios of stationary and moving components show a similar trend, but are not included here for brevity.

\subsection{Knot Acceleration and Deceleration}
\label{subsec:KnotAccels}

Between 2007 and 2018, 75 of the 425 (17.6\%) moving knots detected in our sample exhibit non-ballistic apparent velocities, which we define as motions that are best fit by multiple line segments in a piece-wise fashion. We also include the quasi-stationary feature \emph{C} of the FSRQ 2251+158 in the following analysis, as it has been reliably observed to have motion around the average stationary value. Across the three subclasses, we have detected acceleration in 55 features in 16 FSRQs, 9 features in 6 BLs, and 11 features in the 3 RGs.
Due to the nature of our piece-wise linear fits, the trajectories of many of these knots contain multiple acceleration regions. This totals 104 individual acceleration regions in the jets of 25 blazars in our sample. Given that the multiple acceleration regions may have different properties, we consider each region to be an independent measurement of possible acceleration within a particular subclass of blazar, and do not calculate a ``typical" acceleration for each source.  Table~\ref{tab:JetAccel} gives the accelerations of each region in each knot as follows:
1---name of the source;
2---designation of the component;
3---acceleration region number, $N_{\text{acc}}$; starting with 1 for the first region present (between moving segments 1 and 2) in a knot trajectory and reaching a maximum of 4 for a knot which is best fit by three linear segments in both the $X$ and $Y$ direction;
4---beginning date of the acceleration, $t_{\mathrm{start}}$;
5---end date of the acceleration, $t_{\mathrm{end}}$;
6---the acceleration parallel the jet, $\dot{\mu}_\parallel$, and its $1\sigma$ uncertainty in mas yr$^{-2}$;
7---the acceleration perpendicular to the jet, $\dot{\mu}_\perp$, and its $1\sigma$ uncertainty in mas yr$^{-2}$;
8---$R_{\dot{\mu}}$, the distance from the core to the position of the knot based on the estimated time break point (from the piece-wise fit), in mas;
9---the normalized acceleration parallel to the jet, $\eta_\parallel$, and its $1\sigma$ uncertainty in yr$^{-1}$; and
10---the normalized acceleration perpendicular to the jet, $\eta_\perp$, and its $1\sigma$ uncertainty in yr$^{-1}$ (see below for definitions of $\eta_{\parallel}$ and $\eta_\perp$).
As discussed in $\S$\ref{subsec:JetPAs}, the accelerations parallel and perpendicular to the jet were determined using the averaged position angle for sources with constant or wide jets and using the linear or spline trend to calculate the jet position angle at the time of acceleration for those with time variable jet positions. In this way we separate the time variability of the jet as a whole from the change in directions of knots relative to the bulk flow.

\begin{figure*}[t]
    \figurenum{14}
    \begin{center}
        \includegraphics[width=0.8\textwidth]{{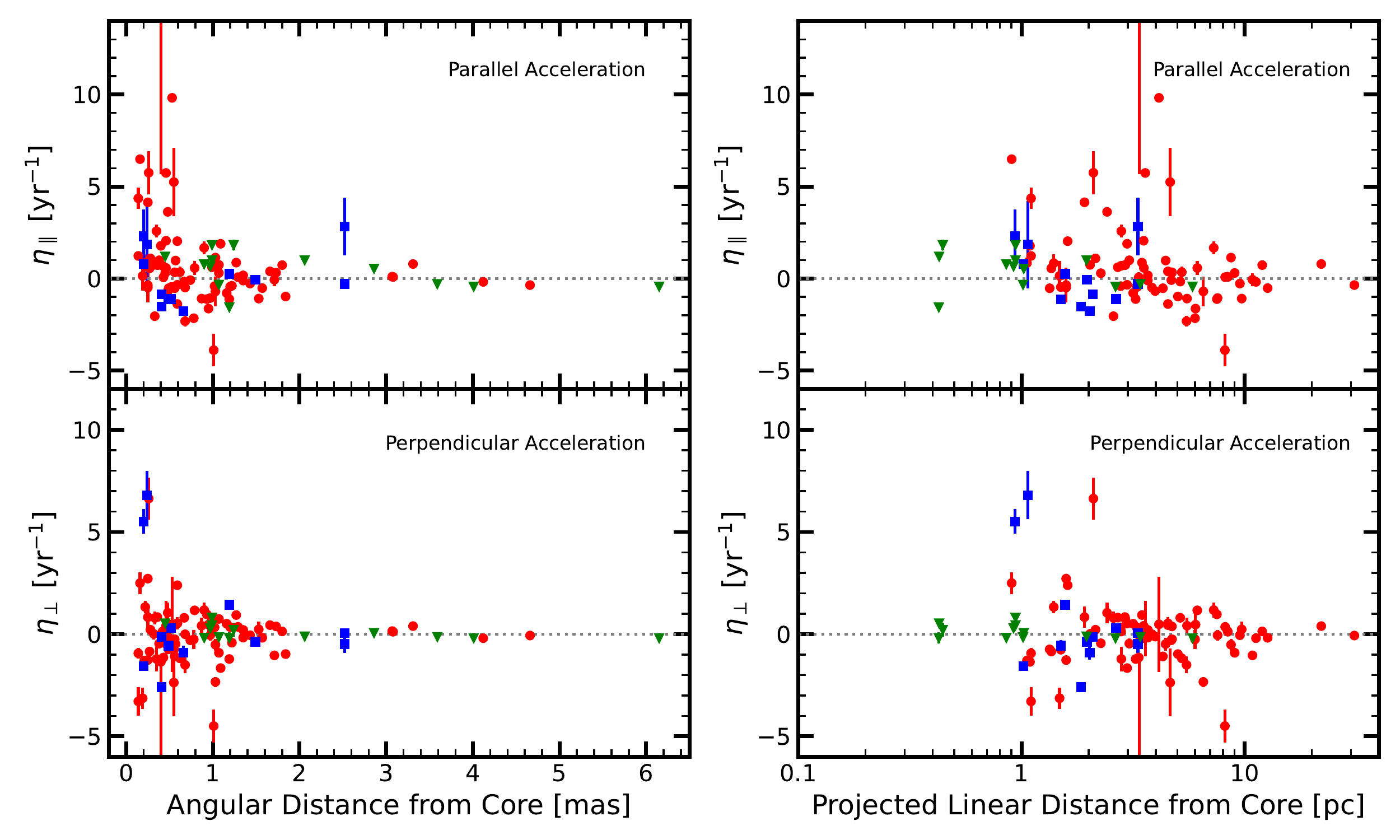}}
        \caption{Left: relative accelerations parallel (top) and perpendicular (bottom) to the jet of moving knots in FSRQs (red circles), BLs (blue squares), and RGs (green triangles) vs. their angular distances from the core of the jet. Right: the same accelerations vs. the average projected linear distance downstream of the core.
        \label{fig:Accelerations}}
    \end{center}
\end{figure*}

The accelerations found in the jets of RGs are primarily in 0415+379 (8/11, 72.7\%). Several knots in this source show very complex motions, requiring more than two acceleration regions. Accelerations occur in only one knot in 0430+052. In contrast, only a few BLs exhibit any acceleration, and only one BL (2200+420) contains a knot (\emph{C2}) that shows complex enough motion to have more than two acceleration zones. A much wider variety of knots are found to have accelerated in FSRQs, but again a few sources dominate in terms of the number of regions (such as 1226+023, with 21 regions for 13 knots).

Due to the fact that the 3D motion is projected onto the two-dimensional plane of the sky,
we cannot distinguish whether the acceleration in a given knot is connected with an intrinsic change in speed (and thus in the Lorentz factor $\Gamma$) or with a change in the intrinsic viewing angle $\Theta_\circ$. To resolve this ambiguity, we follow the statistical approach suggested by \citet{Homan2009} and followed by \citet{Jorstad2017}: If, in a flux-limited sample of beamed jets, the observed relative parallel acceleration exceeds $\sim60\%$ of the observed relative perpendicular acceleration (averaged over the sample), then the observed accelerations are caused more by variations in the Lorentz factors of the knots than by changes in the jet direction.

We compute the relative accelerations in our jets following a formalism similar to that of \citet{Homan2009}, adapted for our fitting procedure: the acceleration parallel to the jet is $\eta_{\parallel} = (1+z) \dot{\mu}_\parallel / \mu_{i}$, while acceleration perpendicular to the jet direction is $\eta_\perp = (1+z) \dot{\mu}_\perp / \mu_{i}$, where $\mu_i$ in both cases refers to the pre-acceleration proper motion of the knot. Averaging over the entire sample, we find $\langle \eta_\parallel \rangle = 0.678$
yr$^{-1}$ and $\langle \eta_\perp \rangle = -0.056$ yr$^{-1}$, thus, $|\langle \eta_\parallel \rangle| > 0.6 |\langle \eta_\perp \rangle|$ and the observed accelerations are likely caused by changes in the Lorentz factors rather than changes in the jet direction.

One advantage of the piece-wise linear fits used in this work \citep[instead of the polynomial fits presented in][]{Jorstad2017} is that, since the accelerations can be seen to take place at specific locations in the jet, it is straightforward to determine the distance from the 43 GHz core to the acceleration zone. Figure~\ref{fig:Accelerations}, left, shows the computed relative accelerations with respect to the angular distance from the core to the acceleration region, while the right panel shows the same relative accelerations as functions of the average projected linear distance from the core to the acceleration region. While the accelerations in RGs tend to appear to occur closer to the core than the accelerations in FSRQs and BLs, this is due to the observational bias that the RG redshifts are smaller than those of the other subclasses. Given that there are no distinct differences in relative acceleration between the subclasses, we consider all accelerations together.

\begin{figure*}[t]
    \figurenum{15}
    \begin{center}
        \includegraphics[width=0.9\textwidth]{{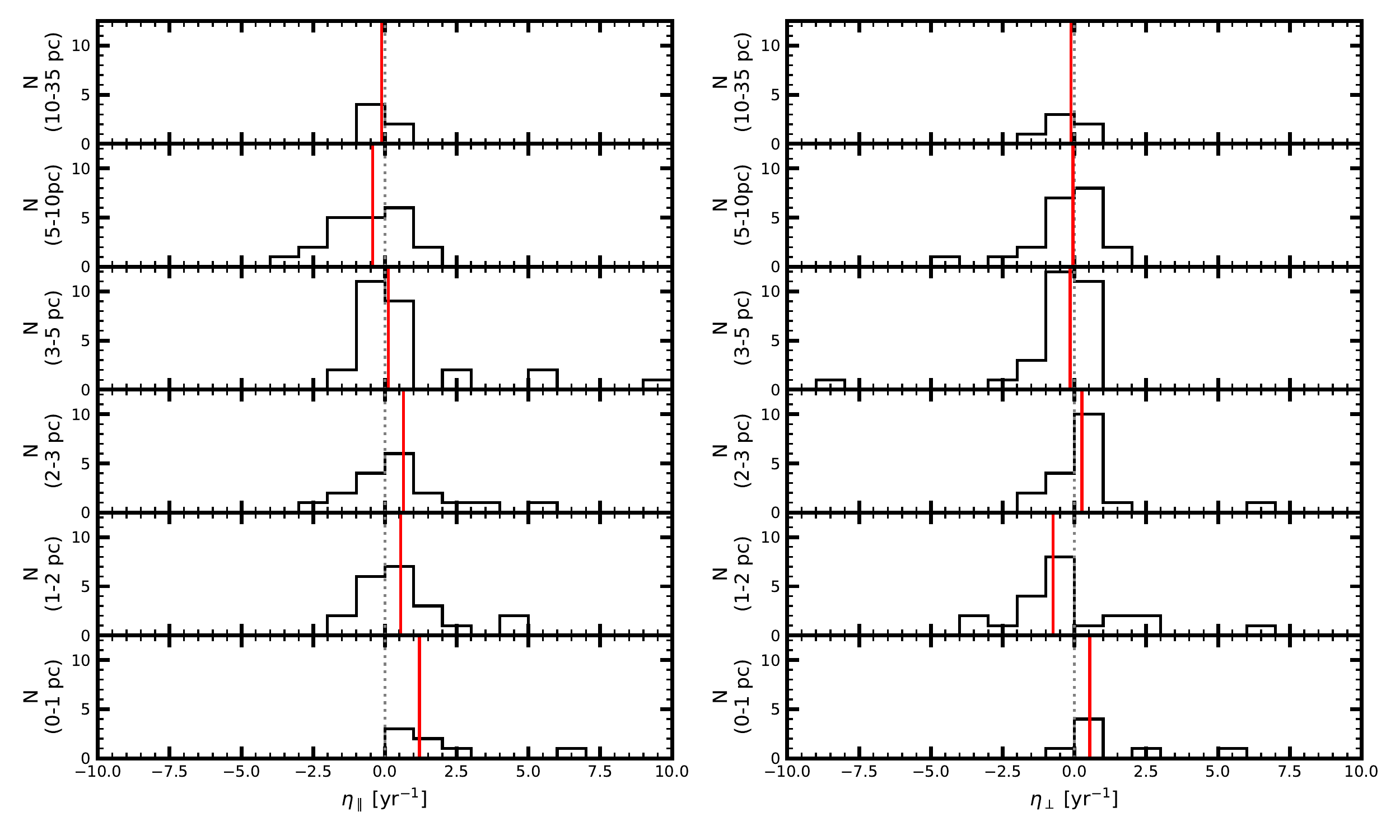}}
        \caption{Histograms of relative knot accelerations (left) parallel and (right) perpendicular to the jet axis for all sources, separated by distance from the 43 GHz core. Red vertical lines in all panels indicate the median acceleration, while the gray dotted vertical lines indicate no acceleration.
        \label{fig:HistRidgeline}}
    \end{center}
\end{figure*}

We bin both $\eta_\parallel$ and $\eta_\perp$ for all sources by distance from the 43 GHz core of the jet, and construct histograms of the accelerations in Figure~\ref{fig:HistRidgeline}. Distance bins are chosen so that a significant number of knot accelerations would be present in each bin ($\gtrsim 10$) while also maintaining adequate distance spacing. The left panels display $\eta_\parallel$, while those on the right show $\eta_\perp$. In the distributions, the red vertical lines indicate the median values, which are listed in Table~\ref{tab:HistRidgelineMedians}.

\begin{deluxetable}{ccc}
    \tablecaption{Median Values of Normalized Accelerations by Distance from the Core.\label{tab:HistRidgelineMedians}}
    \tablewidth{0pt}
    \tablehead{
    \colhead{Distance Bin} & \colhead{$\eta_\parallel$} & \colhead{$\eta_\perp$} \\
    \colhead{[pc]} & \colhead{[yr$^{-1}$]} & \colhead{[yr$^{-1}$]}
    }
    \startdata
    0--1   & $\phantom{-}1.20$ & $\phantom{-}0.53$ \\
    1--2   & $\phantom{-}0.55$ & $-0.74$           \\
    2--3   & $\phantom{-}0.65$ & $\phantom{-}0.26$ \\
    3--5   & $\phantom{-}0.12$ & $-0.15$           \\
    5--10  & $-0.43$           & $-0.05$ \\
    10--35 & $-0.12$           & $-0.12$ \\
    \enddata
\end{deluxetable}

From inspection of the distributions of $\eta_\parallel$, it is apparent that accelerations close to the core are positive, increasing the speed at which knots flow down the jet axis. As the distance from the core increases, the magnitude of this acceleration generally decreases. Once knots reach distances of $\sim5$-10 pc from the core, most accelerations are negative. While this work at 43 GHz probes projected distances of $\sim0$-$35$ pc from the core, data at 15 GHz \citep{Homan2015} probes farther out, up to $\sim100$ pc projected distance from the core. The data at longer wavelengths supports our results here, with a general turnover from positive to negative/no acceleration beyond $\sim10$-20 pc projected distance. This result indicates that, although the overall diversity in apparent speeds likely requires a variety of different shock strengths if the knots are shock waves \citep[e.g.,][]{Marscher1985}, it is common for accelerations to occur close to the core and for constant speeds or decelerations to dominate farther downstream ($\sim10^2$ pc deprojected distance).

The distributions of $\eta_\perp$ indicate a different behavior of the knot motions perpendicular to the jet axis. Beyond $\sim2$ pc from the core, the acceleration distributions are consistent with low, if any, acceleration transverse to the jet axis. However, initially there appears to be positive acceleration, resulting in knot motion away from the jet axis. This could be due to bends in the jet close to the core or a widening of the jet flow. The motion is reversed for distances $\sim1$-2 pc, where there is negative acceleration. This could be an indication of recollimation of the jet flow, such that the jet width narrows or remains constant downstream of this point. Stacking of the 43 GHz images of jets can be used to investigate the jet width \citep[e.g.,][]{Pushkarev2017, Kovalev2020, Casadio2021} and test whether the acceleration trends are actually related to the jet width.

\begin{figure*}
    \figurenum{16}
    \begin{center}
        \includegraphics[width=0.85\textwidth]{{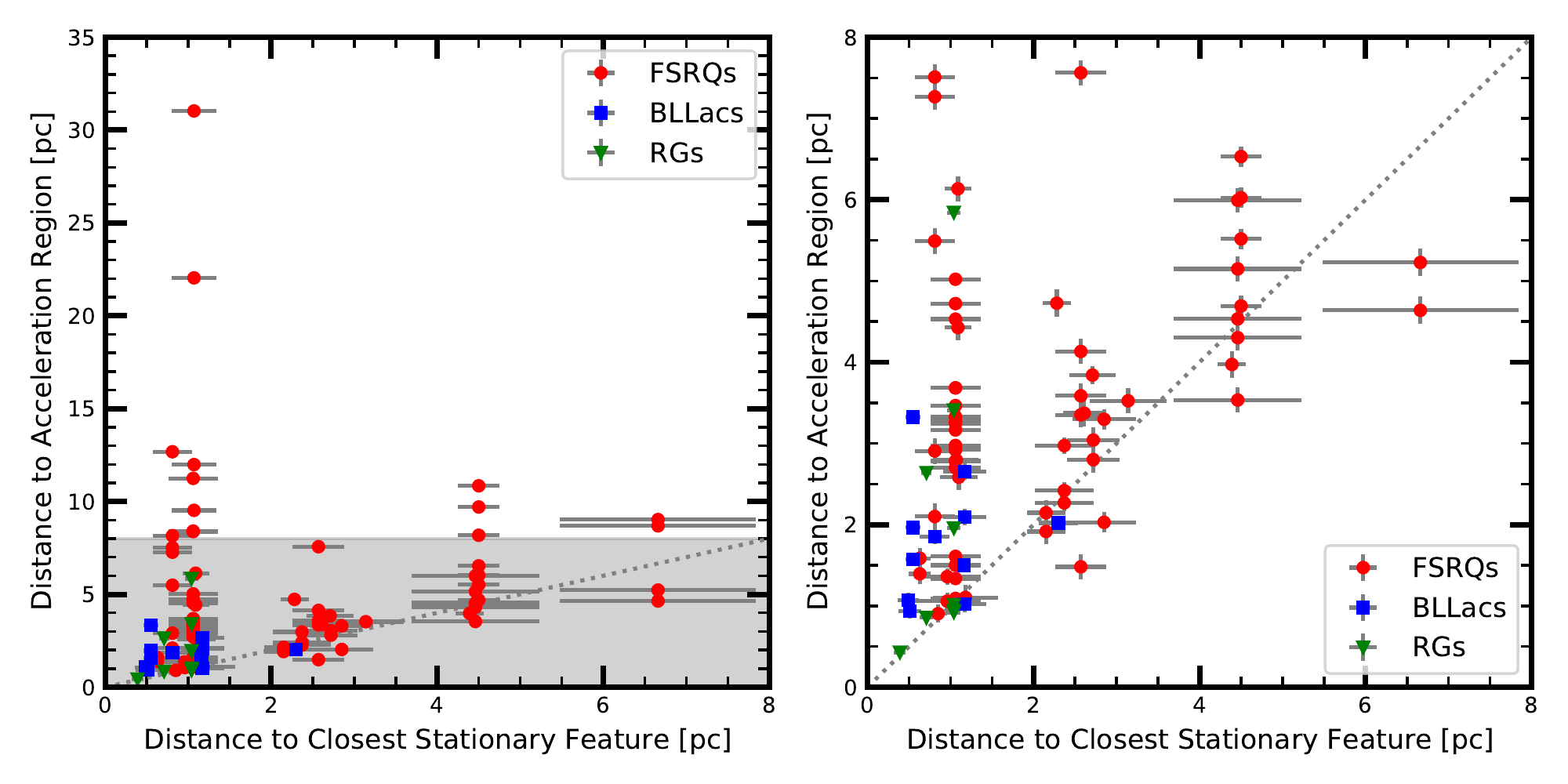}}
        \caption{Left: Distances from the 43 GHz core to the acceleration region and stationary feature closest to that acceleration region. The vertical segments of data correspond to sources that contain multiple knots with acceleration zones beyond the last stationary feature. The gray shaded region delineates the portion of the figure that is expanded for better viewing in the right panel. Error bars are shown in gray in both panels.\label{fig:AccelStat}}
    \end{center}
\end{figure*}

An important question is whether accelerations are related to the stationary features in the jets, as one might expect if the features represent standing shocks. To test this, Figure~\ref{fig:AccelStat} displays the relationship between the locations of each acceleration region and the closest stationary feature. The gray dotted line depicts a 1:1 relationship between the two, i.e., where the positions of acceleration and a stationary feature are the same. We have used the standard deviation of the stationary feature position as the uncertainty in the distance from the core to the closest stationary feature. We have assigned a positional error of 0.02 mas ($\sim0.2$ times our best resolution) to the distance from the core to the acceleration region.

Very few pairs of acceleration region and stationary feature fall below the 1:1 relationship line. This indicates that most accelerations occur at or beyond stationary features. An example of this trend is readily seen in the FSRQ 0336$-$019, where knots \emph{B2, B3}, and \emph{B4} appear to accelerate after passing through stationary features \emph{A1} and \emph{A4}. Such behavior is expected if the stationary features are structural components in the jet that can interact with moving features, such as recollimation shocks. The interpretation of recollimation shocks is further supported by the acceleration distributions for projected distances $\sim1$-4 pc, which are predominately negative for $\eta_\perp$.

%-------------------------------------------------------------------------------
%---   PHYSICAL PARAMETERS OF THE JETS   ---------------------------------------
%-------------------------------------------------------------------------------

\section{Physical Parameters of Jets}
\label{sec:JetPhysicalParams}

The trajectories, apparent velocities, and light curves derived for moving features can be used to compute important physical parameters in the parsec-scale jets of the blazars in our sample. These parameters are: variability Doppler factor $\delta_{\text{var}}$, Lorentz factor $\Gamma$ of the motion, viewing angle $\Theta_\circ$, and opening semi-angle $\theta_\circ$ of the jet. We restrict the following analysis to the most ``reliable'' features, which we define as those with a number of images in which the feature is visible $N \geq 6$, and ejection times within our observed period of VLBA monitoring. The resulting sample includes 191, 101, and 46 knots in FSRQs, BLs, and RGs, respectively. Each knot can have multiple values of speed, depending on the number of line segments used in the piece-wise linear fit. This leads to a total of 262, 109, and 58 potential measurements of the physical parameters in the jets of FSRQs, BLs, and RGs, respectively.

\subsection{Variability Timescale and Doppler Factor}
\label{subsec:tauvar}

We follow the formalism developed by \citet{Jorstad2005} and modified in \citet{Jorstad2017} to calculate the timescale of variability, $\tau_{\text{var}}$, and variability Doppler factor, $\delta_{\text{var}}$, for the moving knots in our sample. In this method, we assume that for the majority of superluminal knots the flux density at 43 GHz declines from radiative energy losses of the relativistic electrons rather than adiabatic cooling. If this decay (calculated in the rest frame of the jet plasma) is also shorter than the light-travel time, as we assume, the Doppler factor can be determined as

\begin{equation}
    \delta_{\text{var}} \approx \frac{16 sD_{\text{Gpc}}}{\tau_{\text{var}}(1+z)},
    \label{eqn:delta}
\end{equation}

\noindent where $D_{\text{Gpc}}$ is the luminosity distance to the source in Gpc, $\tau_{\text{var}}$ is measured in years, $z$ is the redshift of the host galaxy, and $s$ is the angular size of the knot in mas.

In order to estimate $\tau_{\text{var}}$, we make use of the fact that the rise and decay portions of the millimeter-wave light curves of blazars can both be roughly modeled by exponential profiles, so that $\ln S(t) \propto t$ \citep{Terasranta1994, Lister2001}. Such millimeter-wave flares in blazars occur roughly at the same time as the emergence of a new knot from the core, and the decay of the flux density of a knot resembles that of the total flux density \citep{Savolainen2002}. We thus approximate the flux decay of the reliable knots in our sample with an exponential of the form $\ln(S(t) / S_\circ) = k(t - t_{\text{max}})$, where $t_{\text{max}} \leq t$ is the epoch corresponding to the maximum flux density, $S_{\text{max}}$, $S_\circ$ is the flux density of a least-squares fit to the light curve decay at time $t_{\text{max}}$, and $k$ is the slope of the fit. The timescale of variability is then $\tau_{\text{var}} = |1/k|$ yr. While $\tau_{\text{var}}$ for most knots can be determined to within $\sim10\%$, some light curves have no well-defined maximum given the time coverage of the observations, which leads to significant uncertainty in the timescale of variability. Because of this, we remove knots from our reliable sample if $\sigma_{\tau_{\text{var}}}> \tau_{\text{var}}/2$. We also cut from our analysis motion segments beyond the first acceleration zone for knots with complex motion.
This reduced sample consists of estimates for the physical parameters of 162 knots in 22 FSRQs, 80 knots in 11 BLs, and 31 knots in two RGs. The remaining three sources either have no reliable moving knots with defined epochs of ejections (the RG 0316+413 and the BL 1652+398) or no moving features at all identified in this work (the BL 1959+650). While we use the superluminal knots of the BL 1101+384 to determine its jet parameters, the low apparent speeds suggest that we are looking at a sheath of the jet rather than the region closer to the axis (see $\S$\ref{subsec:JetParamAverages}).

\begin{figure}
    \figurenum{17}
    \begin{center}
        \includegraphics[width=0.45\textwidth]{{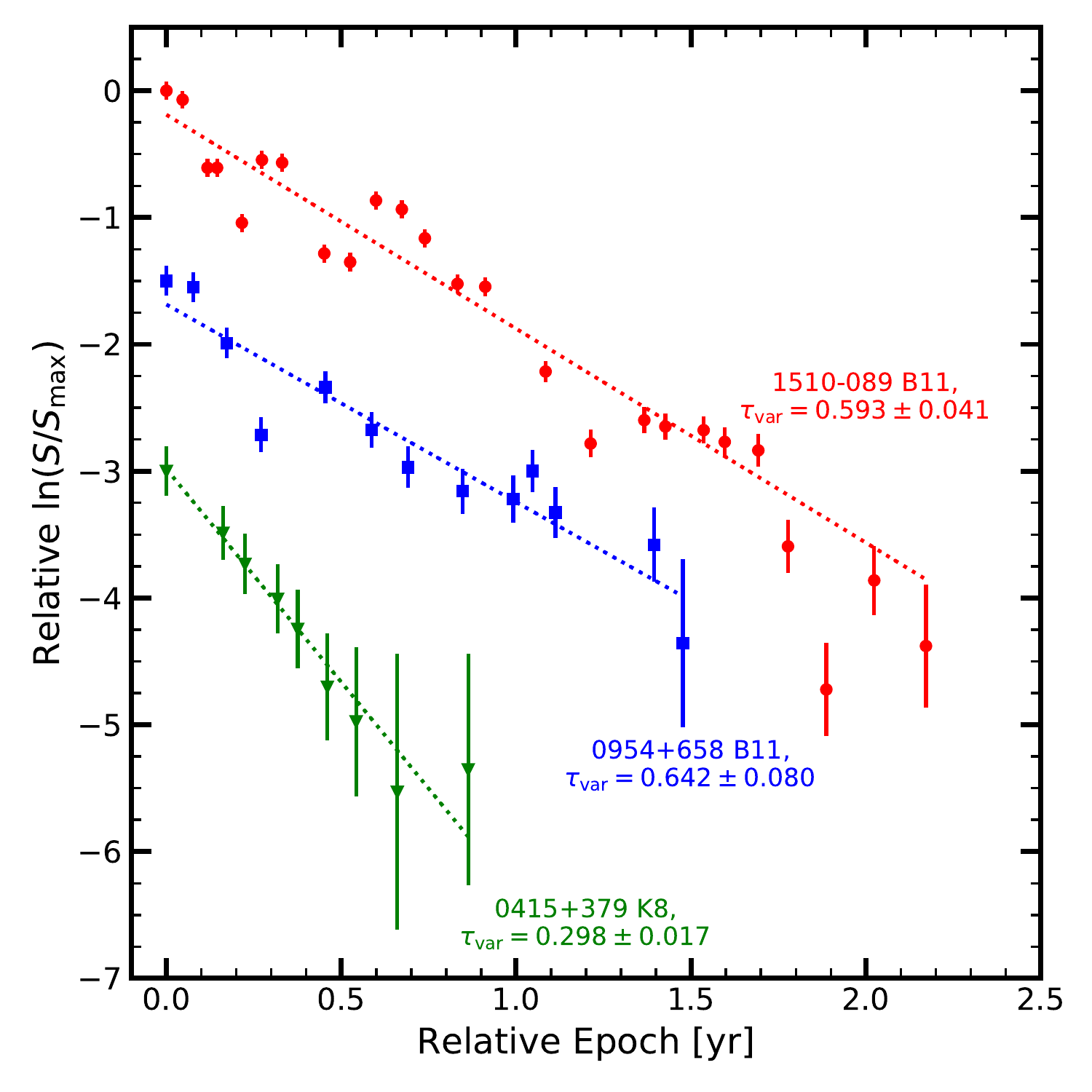}}
        \caption{Light curves of moving knots B11 in the FSRQ 1510$-$089 (red circles), B11 in the BL 0954+658 (blue squares), and K8 in the RG 0415+379 (green triangles). The dotted lines represent the least-square fits to the flux density decays of the knots. The data and fits for B11 and K8 have been shifted by $-1.5$ and $-3.0$ from their normalized values, respectively.\label{fig:ExampleDecay}}
    \end{center}
\end{figure}

Figure~\ref{fig:ExampleDecay} shows the decay light curves and the fits used to calculate $\tau_{\text{var}}$ for a knot in each subclass, shifted for clarity. The distributions of the timescales of variability, corrected for redshift to transform into the host galaxy frame, are similar for all three subclasses, peaking at $\tau_{\text{var}}^{\text{s}} \lesssim 6$ months with the majority ($\gtrsim80\%$) of all knots having a timescale $\tau_{\text{var}}^{\text{s}} < 1$ yr.

\begin{figure}
    \figurenum{18}
    \begin{center}
        \includegraphics[width=0.45\textwidth]{{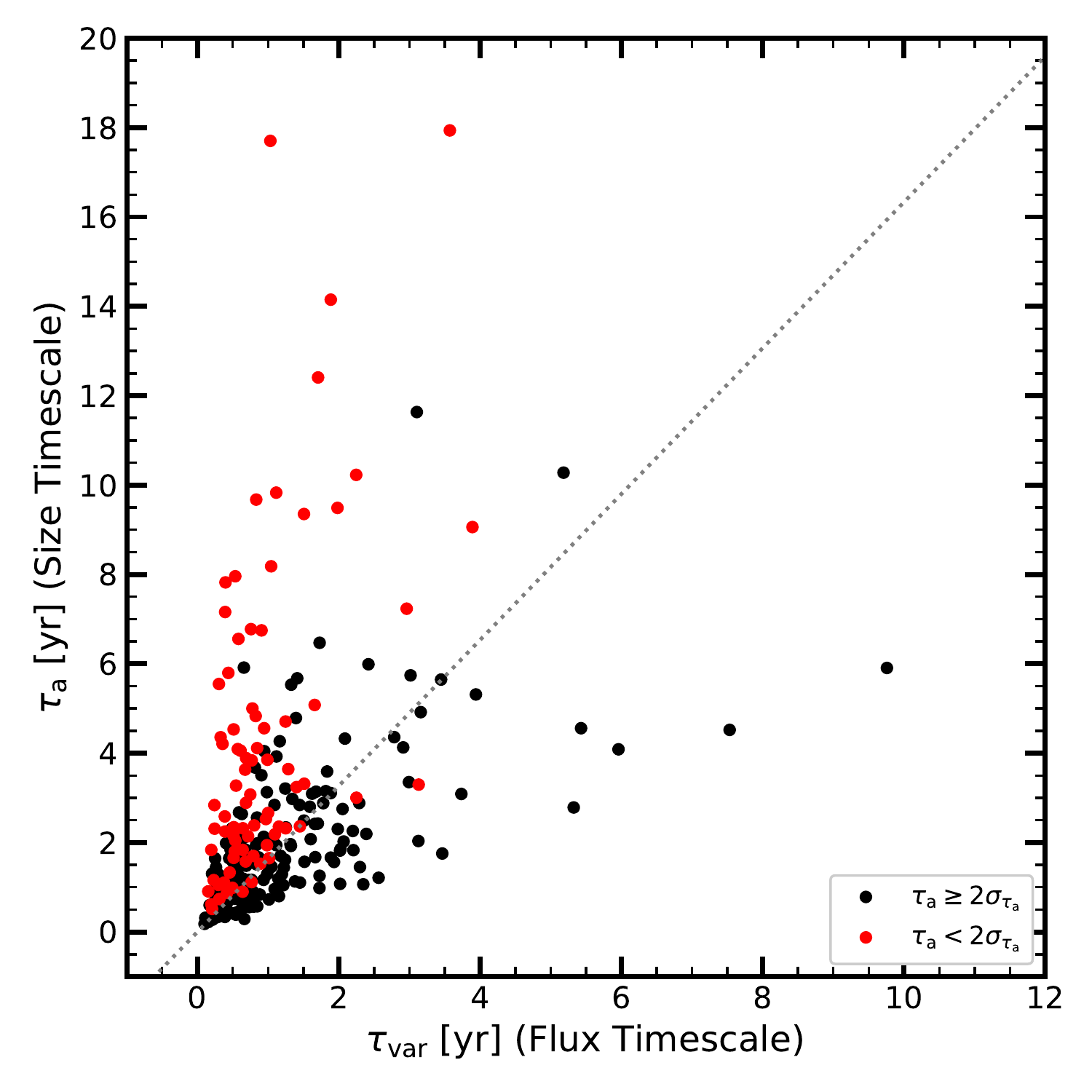}}
        \caption{Timescale of variability of the size vs.\ that of the flux density for superluminal knots in our sample. ``Reliable" estimates of $\tau_{\text{a}}$ (for which $\tau_{\text{a}} \geq 2\sigma_{\tau_{\text{a}}}$) are shown in black, while those do not meet the reliability criterion are shown in red. The solid line marks the approximate relation for an optically thin shock with $\alpha \sim 0.7$ \citep{Marscher1985}. Error bars are omitted for visual clarity.\label{fig:TauA}}
    \end{center}
\end{figure}

In order to test the assumption that the flux density decay of knots at 43 GHz is due primarily to radiative losses rather than adiabatic expansion, we also fit an exponential of the form $\ln(a(t)/a_\circ) = k_{\text{a}}(t-t_{\text{max}})$ to the knot sizes, $a(t)$, over the same epochs as used to calculate the flux timescale of variability. Here, $a_\circ$ is the size of the knot according to a least-squares fit to the data at time $t_{\text{max}}$ (defined above) and $\tau_{\text{a}} = |1/k_{\text{a}}|$ yr. Figure~\ref{fig:TauA} shows the timescale of variability of knot size compared to that of the flux density for individual knots. We have classified knots as having reliable values of the timescale for changes in size if $\tau_{\text{a}} \geq 2\sigma_{\tau_{\text{a}}}$, where $\sigma_{\tau{\text{a}}}$ is the estimated uncertainty. In the figure, the dotted gray line indicates the approximate relation between $\tau_{\text{a}}$ and $\tau_{\text{var}}$ for adiabatic losses for optically thin shocked gas with $\alpha = 0.7$ \citep{Marscher1985}. A substantial majority (188/287, 65.5\%) of knots have $\tau_{\text{var}} < \tau_{\text{a}}$ (above the dotted line), implying that the flux density decay is due predominantly to radiative losses rather than adiabatic losses, and thus that our initial assumption is valid.

The angular size $a$ obtained in the model fits is the FWHM of a presumed Gaussian brightness distribution. We consider a face-on disk to be a more physically appropriate geometry, and therefore convert $a$ to the diameter $s$ of a face-on disk that has a similar interferometric visibility function, $s = 1.6a$, for use in Equation \ref{eqn:delta}. We estimate $a$ as the average over the epochs used to calculate $\tau_{\text{var}}$, so that $a = \sum_{i=1}^n a_i / n$, where $a_i$ is the FWHM size of the knot at each epoch $i$ and $n$ is the number of epochs in the decay phase of the light curve.
Table~\ref{tab:PhysParams} presents the physical parameters of each knot segment as follows:
1---source name;
2---knot designation;
3---timescale of variability of the knot flux density, $\tau_{\text{var}}$, and its estimated $1\sigma$ uncertainty from the least-square fit to the decay of the light curve, in yr;
4---average size of the knot, $\langle a \rangle$, and its propagated $1\sigma$ uncertainty, in mas;
5---number of measurements used to calculate $\tau_{\text{var}}$ and $\langle a \rangle$; $N_t$;
6---Doppler factor of variability for each knot segment, $\delta_{\text{var}}$, and its $1\sigma$ uncertainty;
7---bulk Lorentz factor of each knot segment, $\Gamma$, and its $1\sigma$ uncertainty (see $\S$\ref{subsec:LorentzViewing}); and
8---intrinsic viewing angle of each knot segment, $\Theta_\circ$, and its $1\sigma$ uncertainty, in degrees (see $\S$\ref{subsec:LorentzViewing}).

\begin{figure*}
    \figurenum{19}
    \begin{center}
        \includegraphics[width=\textwidth]{{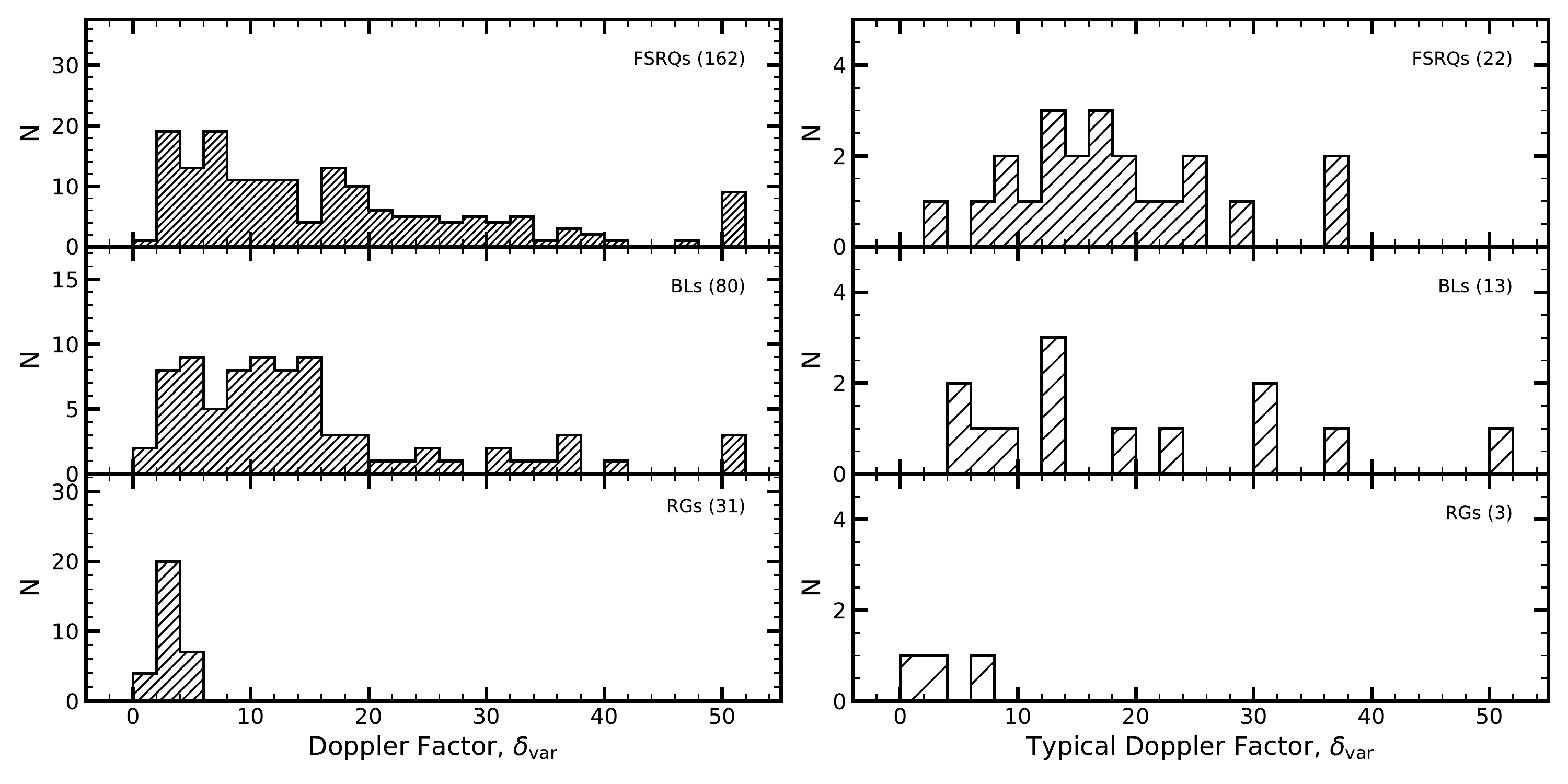}}
        \caption{Distributions of all calculated (left) and typical (defined in $\S$\ref{subsec:JetParamAverages}, right) Doppler factors of variability, $\delta_{\text{var}}$, derived for reliable moving knots in FSRQs (top), BLs (middle), and RGs (bottom). The typical $\delta_{\mathrm{var}}$ for the BL 1101+384 is shown in the last bin to condense the axis.
        \label{fig:DeltaDist}}
    \end{center}
\end{figure*}

For comparison with past literature, Figure~\ref{fig:DeltaDist} (left) displays the distributions of the Doppler factors for all reliable knots derived using the above method.
The Doppler factors of FSRQs and BLs are predominantly $\delta_{\text{var}} < 40$, with only eight and three segments having $\delta_{\text{var}} > 50$, respectively. These few knots are shown in the last bins of each distribution.

To more appropriately compare between the subclasses, we show in Fig.~\ref{fig:DeltaDist} (right) the distributions of $\delta_{\mathrm{var}}$ for ``typical" knots of each source, where the ``typical" knot has been determined through the method outlined in $\S$\ref{subsec:JetParamAverages}. We see that the $\delta_{\mathrm{var}}$ values in RGs are subtantially lower than those of FSRQs and BLs. This result is statistically significant $\sim3\sigma$ when using a KS test to compare the FSRQ and RG distributions ($\mathcal{D} = 0.955,\ p = 0.003$) but not when comparing the BL to RG distribution ($\mathcal{D} = 0.769,\ p = 0.071$) due to the relatively low number of RG sources in our sample. The FSRQ and BL distributions appear very similar (according to a KS test; $\mathcal{D} = 0.311,\ p = 0.333$). However, this simple summary statistic hides an important distinction between the $\delta_{\mathrm{var}}$ values of FSRQs and BLs. More than half (8/13, $61.5\%$) of the BLs have a typical $\delta_{\mathrm{var}} \lesssim 15$, while in FSRQs more than half (12/22, $54.5\%$) of the sources have $\delta_{\mathrm{var}} \gtrsim 15$. This higher probability for FSRQs to have larger $\delta_{\mathrm{var}}$ values is consistent with their observed higher brightness temperatures of jet features as compared to BLs, apparent in Figure~\ref{fig:KnotHists}.

\subsection{Lorentz Factors and Viewing Angles}
\label{subsec:LorentzViewing}

With the apparent speeds and Doppler factors derived for each knot, we can independently calculate the Lorentz factor $\Gamma$ and viewing angle $\Theta_\circ$ with the relations

\begin{equation}
    \Gamma = \frac{\beta_{\text{app}}^2 + \delta_{\text{var}}^2 + 1}{2\delta_{\text{var}}} ; \hspace{0.25cm} \tan{\Theta_\circ} = \frac{2\beta_{\text{app}}}{\beta_{\text{app}}^2 + \delta_{\text{var}}^2 - 1}.
    \label{eqn:GammaTheta}
\end{equation}

\begin{deluxetable*}{clcccccc}
    \tablecaption{Physical Parameters of Jet Features\label{tab:PhysParams}}
    \tablewidth{0pt}
    \tablehead{
    \colhead{Source} & \colhead{Knot}& \colhead{$\tau_{\text{var}}$} & \colhead{$\langle a \rangle$} &\colhead{$N_t$} & \colhead{$\delta_{\text{var}}$} & \colhead{$\Gamma$} & \colhead{$\Theta_\circ$} \\
    \colhead{} &\colhead{} & \colhead{[yr]} & \colhead{[mas]} & \colhead{} & \colhead{} & \colhead{} & \colhead{[$\degr$]}
    }
    \colnumbers
    \startdata
    0219+428 & B1 & $0.535 \pm 0.161$ & $0.27 \pm 0.02$ & 6  & $21.6 \pm 6.7$    & $24.9 \pm 5.6$    & $2.6 \pm 0.7$ \\
             & B2 & $0.944 \pm 0.300$ & $0.40 \pm 0.02$ & 10 & $18.2 \pm 5.9$    & $13.8 \pm 1.8$    & $3.0 \pm 1.3$ \\
             & B3 & $0.639 \pm 0.245$ & $0.15 \pm 0.02$ & 6  & $10.2 \pm 4.0$    & $20.9 \pm 5.4$    & $4.8 \pm 1.0$ \\
             & B4 & $1.511 \pm 0.330$ & $0.43 \pm 0.03$ & 6  & $12.3 \pm 2.8$    & $22.8 \pm 3.5$    & $4.1 \pm 0.5$ \\
             & B5 & $1.240 \pm 0.212$ & $0.28 \pm 0.02$ & 11 & $\phn9.9 \pm 1.8$ & $19.8 \pm 3.0$    & $5.0 \pm 0.5$ \\
    0235+164 & B1 & $0.324 \pm 0.068$ & $0.26 \pm 0.02$ & 6  & $63.3 \pm 13.9$   & $37.2 \pm 5.8$    & $0.7 \pm 0.2$ \\
             & B2 & $0.207 \pm 0.023$ & $0.20 \pm 0.01$ & 9  & $76.7 \pm 9.4$    & $39.4 \pm 4.5$    & $0.2 \pm 0.1$ \\
             & B3 & $1.243 \pm 0.119$ & $0.15 \pm 0.01$ & 28 & $\phn9.3 \pm 0.9$ & $\phn5.1 \pm 0.4$ & $3.2 \pm 0.7$ \\
             & B4 & $0.717 \pm 0.102$ & $0.15 \pm 0.01$ & 13 & $16.4 \pm 2.5$    & $\phn9.1 \pm 1.1$ & $2.1 \pm 0.6$ \\
             & B5 & $0.493 \pm 0.068$ & $0.23 \pm 0.01$ & 13 & $37.9 \pm 5.4$    & $20.0 \pm 2.5$    & $0.7 \pm 0.2$ \\
             & B6 & $0.361 \pm 0.044$ & $0.28 \pm 0.01$ & 13 & $62.5 \pm 8.1$    & $31.6 \pm 4.0$    & $0.2 \pm 0.1$ \\
    \enddata
    \tablecomments{(Table~\ref{tab:PhysParams} (273 lines) is published in its entirety in the machine-readable format. A portion is shown here for guidance regarding its form and content.)}
\end{deluxetable*}

\begin{figure*}
    \figurenum{20}
    \begin{center}
        \includegraphics[width=\textwidth]{{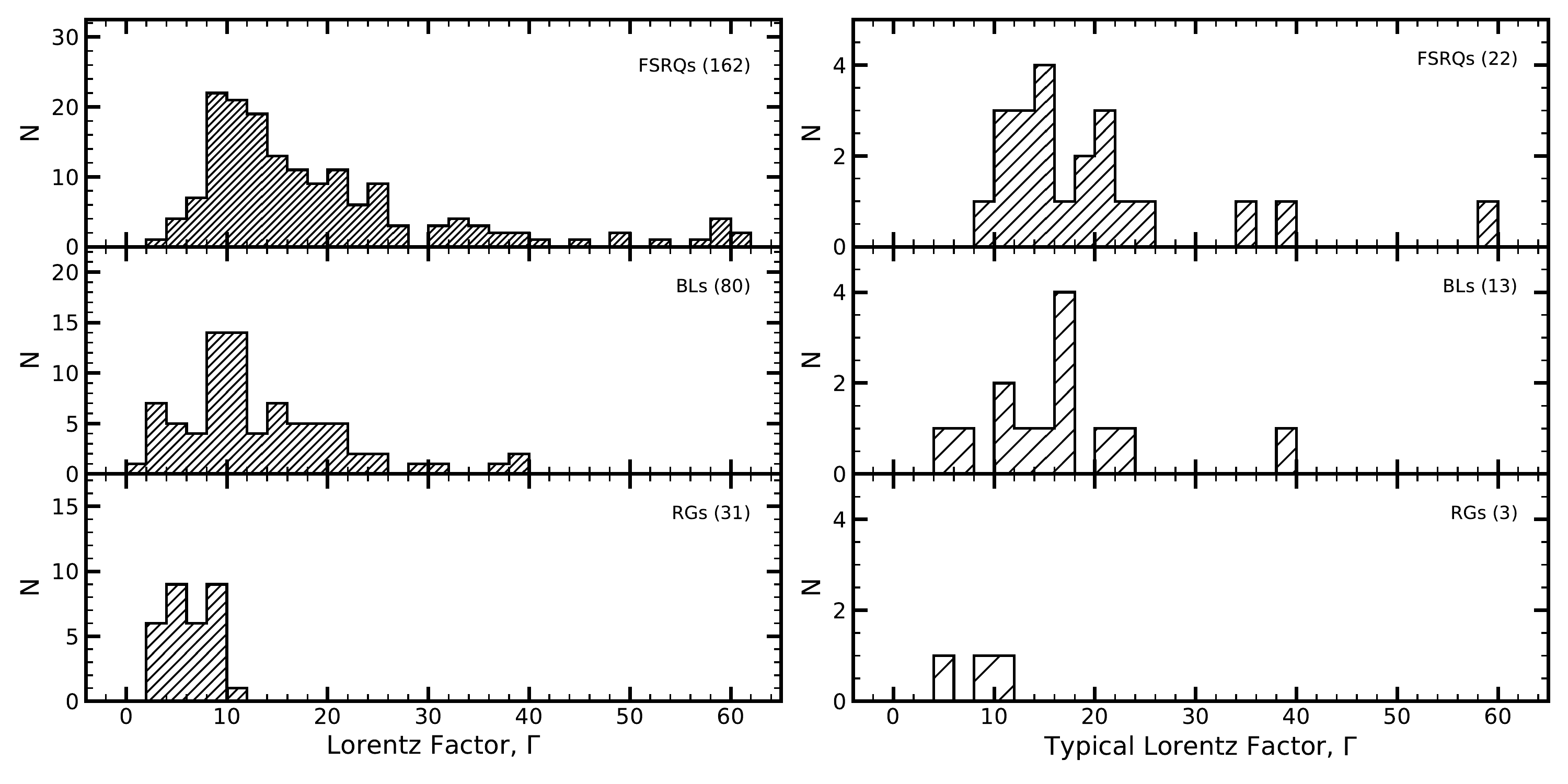}}
        \caption{Distributions of all calculated (left) and typical (defined in $\S$\ref{subsec:JetParamAverages}, right) Lorentz factors, $\Gamma$, in FSRQs (top), BLs (middle), and RGs (bottom), derived from the apparent speeds and variability Doppler factors of reliable knots.\label{fig:Gammadist}}
    \end{center}
\end{figure*}

Figure~\ref{fig:Gammadist}, left, displays the distributions of $\Gamma$ for all reliable knots of each subclass. The FSRQ and BL distributions both peak at $\Gamma \sim 10$, but BLs have a small local peak in the $\Gamma = 2$-$6$ bins, while the FSRQ distribution has a longer high-$\Gamma$ tail. There is a single FSRQ knot for which $\Gamma \geq 60$, \emph{B1} in 0528+134, with $\Gamma = 81 \pm 10$; it is placed in the highest bin in the distribution. The RG distribution is comparatively narrow and low-valued, with all but one of the knots having $\Gamma < 10$. To directly compare the Lorentz factors of FSRQs, BLs, and RGs, we utilize a KS test on the distributions of $\Gamma$ for the ``typical" knots of each source (see $\S$\ref{subsec:JetParamAverages}), shown in Figure~\ref{fig:Gammadist} (right).
Despite the low number of sources, we continue to infer that RGs have the slowest knots ($\Gamma \lesssim 10$), while FSRQs have, on average, faster knots, with a long tail ($\Gamma$ extending up to $\sim60$). Both FSRQ and BL distributions are statistically different from the RG distribution ($\sim2\sigma$), with FSRQ vs.\ RG $\mathcal{D} = 0.955,\ p = 0.003$, while the BL vs.\ RG $\mathcal{D} = 0.846,\ p = 0.036$). The KS test cannot distinguish between the FSRQ and BL distributions, however, with $\mathcal{D} = 0.269$ and $p = 0.495$.

\begin{figure*}
    \figurenum{21}
    \begin{center}
        \includegraphics[width=\textwidth]{{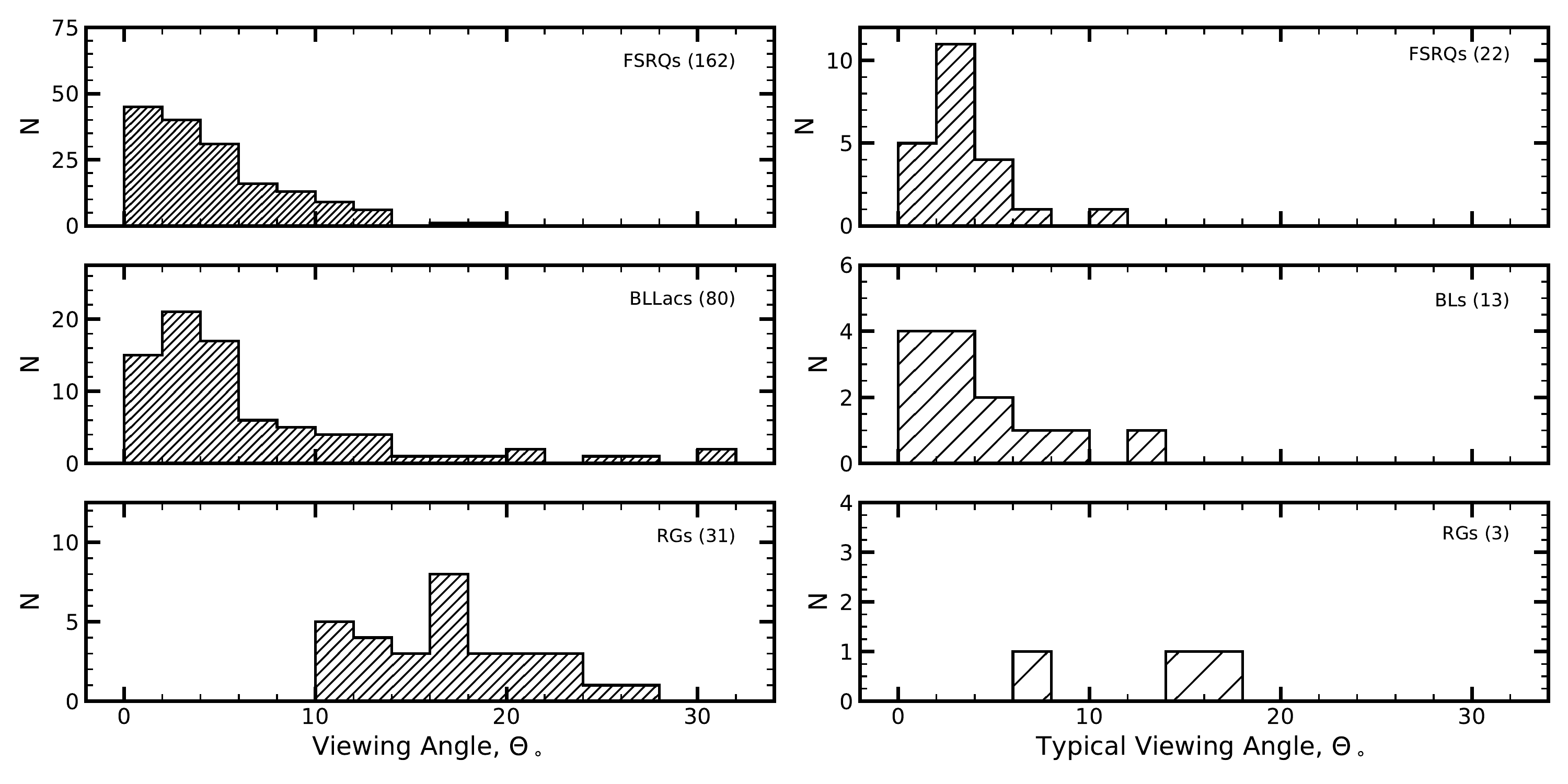}}
        \caption{Distributions of all calculated (left) and typical (defined in $\S$\ref{subsec:JetParamAverages}, right) viewing angles in FSRQs (top), BLs (middle), and RGs (bottom), derived from the apparent speeds and variability Doppler factors of reliable knots. Note the different scales in each panel.\label{fig:ThetaCircDist}}
    \end{center}
\end{figure*}

Figure~\ref{fig:ThetaCircDist} plots the distributions of the viewing angles for all reliable knots in the three subclasses. It is immediately apparent that RGs have a very different distribution than the FSRQs and BLs, peaking in the $16\degr$-$18\degr$ bin, with all knots having $\Theta_\circ > 10$.
Minor differences in can be seen between the FSRQ and BL distributions, such as the longer high-angle tail of the BLs. The peaks are also different, with the FSRQ distribution peaking between $0\degr$ and $2\degr$, while the BL distribution peaks in the $2\degr$-$4\degr$ range. In both subclasses, however, the majority ($>75\%$) of knots have $\Theta_\circ < 10\degr$. Again to compare between the subclasses statistically, we use the viewing angle of a ``typical" knot (see $\S$\ref{subsec:JetParamAverages}) for each source. The distributions for each subclass are shown in Figure~\ref{fig:ThetaCircDist}. The FSRQs are more centrally located around the peak in viewing angle, $\Theta_\circ \sim 2-4\degr$, than the BLs, but with a KS test indicating $\mathcal{D} = 0.231,\ p = 0.678$. The RG distribution is statistically different from the other two subclasses, with $\mathcal{D} = 0.955,\ p = 0.003$ for the FSRQs vs RGs and $\mathcal{D} = 0.769,\ p = 0.071$ for the BLs vs RGs, despite the low number of sources.

\subsection{Average Jet Physical Parameters and Opening Angles}
\label{subsec:JetParamAverages}

\begin{figure}
    \figurenum{22}
    \begin{center}
        \includegraphics[width=0.45\textwidth]{{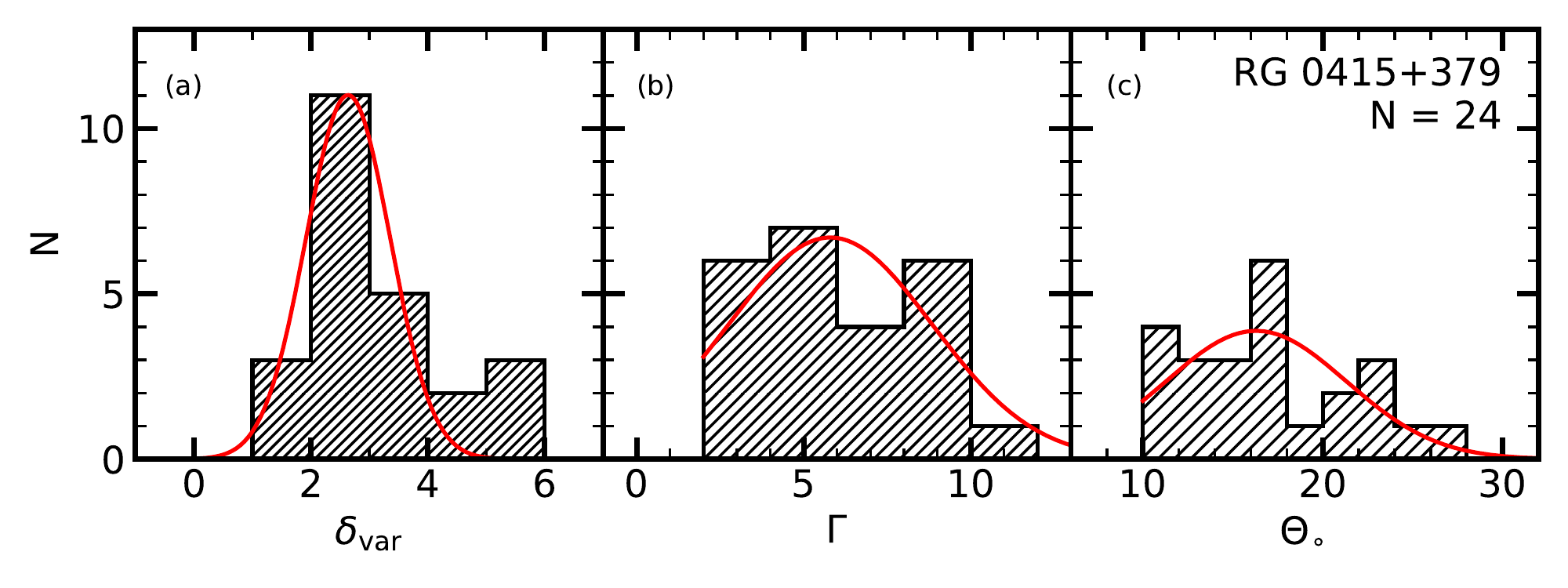}}
        \caption{Distributions of physical parameters $(a)\ \delta_{\text{var}}$, $(b)\ \Gamma$, and $(c)\ \Theta_\circ$ of reliable knots in the RG 0415+379. Red curves represent best-fit Gaussian distributions to the data. See the text for details.}
        \label{fig:3C111Hists}
    \end{center}
\end{figure}

According to Table~\ref{tab:PhysParams}, for any individual source there is a diversity in knot properties, which complicates the determination of the characteristic physical conditions in the jet. The physical parameters of many jets are evidently time-variable, perhaps stochastically determined from a distribution of possible values. As an example, Figure~\ref{fig:3C111Hists} shows the distributions of the physical parameters $\delta_{\text{var}},\ \Gamma$, and $\Theta_\circ$ of the reliable knots of the RG 0415+379. The distributions can be approximated by Gaussian functions, with means $\mu = 2.6,\ 5.8,\ 16.3\degr$ and standard deviations $\sigma = 0.7,\ 3.1,\ 5.0\degr$, respectively, for $\delta_{\text{var}},\ \Gamma$, and $\Theta_\circ$. However, these well-determined distributions are the exception, since the number of knots observed in our study is more limited for most of the sample. The majority of sources have $N \leq 15$, too few to fit Gaussian distributions to the data accurately. Given the knot ejection rates of the sources observed in our sample, another $\sim10$ years of monitoring would be required to observe enough knots to define the distributions of physical parameters for all objects.

Instead of analyzing tentative distributions of physical parameters of jets in our sample, we instead focus on ``representative" values. For each reliable knot, we calculate a metric of overall uncertainty, $\xi\equiv \sqrt{\sigma_{\delta_{\text{var}}}^2+ \sigma_{\beta_{\text{app}}}^2}$, where $\sigma_{\delta_{\text{var}}}$ and $\sigma_{\beta_{\text{app}}}$ are the uncertainties in the variability Doppler factor and proper motion. We focus on these two parameters, since they factor directly into the determination of $\Gamma$ and $\Theta_\circ$ (both given in Eq.~\ref{eqn:GammaTheta}), as well as the opening semi-angle (described below). We then choose as ``representative'' the knot from the reliable sample of each jet that has the optimal combination of relatively high apparent speed and relatively low value of $\xi$. We adopt the criterion $\xi / (\beta + \delta_{\text{var}}) < 0.15$ to select knots with low uncertainty.
Choosing the highest apparent speed with a relatively low uncertainty ensures that the derived physical conditions are consistent with the most extreme, yet robust, behavior of the jet. The statistics of the chosen knots, one per source, would not have differed significantly across the sample had we instead chosen the knots with the highest established Doppler factor and lowest uncertainty. The typical knots are used to determine differences between the distributions of parameters for each subclass in Figures~\ref{fig:BetaHist}, \ref{fig:DeltaDist}, \ref{fig:Gammadist}, and \ref{fig:ThetaCircDist}.

For each jet, we follow \citet{Jorstad2017} in calculating the opening semi-angle in two ways. Method A is based on the relation between the intrinsic and projected opening angles:

\begin{equation}
    \theta_\circ^{\text{A}} = \theta_{\text{p}} \sin{\langle \Theta_\circ\rangle }
    \label{eqn:thetaA}
\end{equation}

\noindent  \citep{Jorstad2005}. Here, $\theta_{\text{p}}$ is the projected opening semi-angle of the jet, and $\langle \Theta_\circ \rangle$ is the representative viewing angle of the jet. We define $\theta_{\text{p}}$ as twice the standard deviation of the knot position angles about a fixed value, regardless of whether the jet position angle is found to vary, as described in \S\ref{subsec:JetPAs}.
We adopt the average uncertainties of $\langle \Theta_{\text{jet}} \rangle$, $\langle \sigma_{\Theta} \rangle$ as the uncertainty of $\theta_{\text{p}}$.

Method B applies to jets with multiple reliable knots. We take the viewing angles of the fastest and slowest knot with low values of $\xi$ (see above) to be the minimum and maximum viewing angles, $\Theta_\circ^{\text{min}}$ and $\Theta_\circ^{\text{max}}$. We then calculate the intrinsic opening semi-angle as

\begin{equation}
    \theta_\circ^{\text{B}} = \frac{1}{2}\left( \Theta_\circ^{\text{max}} - \Theta_\circ^{\text{min}} \right) .
    \label{eqn:thetaB}
\end{equation}

\noindent This definition would be appropriate if the difference in apparent speeds were due mainly to different viewing angles of the knots owing to the brightness centroids of the knots being located at different distances from the jet axis.

Table~\ref{tab:AvePhysParams} gives the typical values of the physical parameters of knots in each source as follows:
1---name of the source;
2---luminosity distance, $D_{\text{Gpc}}$, in Gpc;
3---the knot selected as the representative high-speed knot;
4---Doppler factor of the representative knot, $\delta_{\text{var}}$, and its uncertainty;
5---Lorentz factor of the representative knot, $\Gamma$, and its uncertainty;
6---viewing angle of the representative knot, $\Theta_\circ$, and its uncertainty;
7---projected viewing opening semi-angle, $\theta_{\text{p}}$, of the jet and its uncertainty;
8---opening semi-angle of the jet, $\theta_\circ^{\text{A}}$, derived using method A, and its uncertainty;
9---the knot selected as the representative low-speed knot; and
10---opening semi-angle of the jet, $\theta_\circ^{\text{B}}$, derived using method B, and its uncertainty.

\begin{deluxetable*}{lrcccccccc}
    \tablecaption{Typical Physical Parameters of Jets\label{tab:AvePhysParams}}
    \tablewidth{0pt}
    \tablehead{
    \colhead{Source} & \colhead{$D_{\text{Gpc}}$} & \colhead{Fast Knot} & \colhead{$\delta_{\text{var}}$} & \colhead{$\Gamma$} & \colhead{$\Theta_\circ$} & \colhead{$\theta_{\text{p}}$} & \colhead{$\theta_\circ^{\text{A}}$} & \colhead{Slow Knot} & \colhead{$\theta_\circ^{\text{B}}$} \\
    \colhead{} & \colhead{[Gpc]} & \colhead{} & \colhead{} & \colhead{} & \colhead{[deg]} & \colhead{[deg]} & \colhead{[deg]} & \colhead{} & \colhead{[deg]}
    }
    \colnumbers
    \startdata
    0219+428  & 2.458 & B4 & $12.3 \pm 2.8$     & $22.8 \pm 3.5$     & $\phn 4.1 \pm 0.5$ & $18.0 \pm \phn 5.5$     & $\phn 1.3 \pm 0.4$ & B5 & $\phn 0.4 \pm 0.4$ \\
    0235+164  & 6.121 & B5 & $37.9 \pm 5.4$     & $20.0 \pm 2.5$     & $\phn 0.7 \pm 0.2$ & $13.2 \pm \phn 6.8$     & $\phn 0.2 \pm 0.1$ & B3 & $\phn 1.2 \pm 0.3$ \\
    0316+413\tablenotemark{a}  & 0.077 & $\ldots$ & $\sim 6.9$ & $\sim 4.8$ & $\sim 7.2$ & $19.2 \pm 0.4$           & $\sim 2.4$ & $\ldots$ & $\ldots$ \\
    0336--019 & 5.422 & B5 & $\phn 9.0 \pm 4.1$ & $\phn 9.6 \pm 0.8$ & $\phn 6.3 \pm 2.8$ & $12.8 \pm \phn 4.6$     & $\phn 1.4 \pm 0.8$ & B3 & $\phn 0.8 \pm 1.6$ \\
    0415+379  & 0.215 & K22& $\phn 2.8 \pm 0.2$ & $\phn 8.5 \pm 0.5$ & $15.2 \pm 0.4$     & $\phn 9.8 \pm \phn 2.2$ & $\phn 2.6 \pm 0.6$ & K5 & $11.5 \pm 1.3$     \\
    0420--014 & 5.928 & B3 & $25.2 \pm 3.4$     & $21.2 \pm 0.6$     & $\phn 2.2 \pm 0.4$ & $35.2 \pm \phn 3.7$     & $\phn 1.4 \pm 0.3$ & B5 & $\phn 2.2 \pm 0.6$ \\
    0430+052  & 0.145 & C15& $\phn 1.9 \pm 0.4$ & $10.0 \pm 1.8$     & $17.9 \pm 0.8$     & $\phn 6.4 \pm \phn 2.9$ & $\phn 2.0 \pm 0.9$ & C4 & $\phn 4.0 \pm 0.6$ \\
    0528+134  & 16.109& C2 & $17.3 \pm 2.4$     & $11.7 \pm 0.8$     & $\phn 2.9 \pm 0.6$ & $26.6 \pm \phn 6.0$     & $\phn 1.4 \pm 0.4$ & B3 & $\phn 0.4 \pm 0.4$ \\
    0716+714  & 1.553 & B13& $23.3 \pm 3.6$     & $17.1 \pm 1.1$     & $\phn 2.3 \pm 0.4$ & $38.6 \pm \phn 3.9$     & $\phn 1.5 \pm 0.4$ & B4 & $\phn 0.7 \pm 0.7$ \\
    0735+178  & 2.327 & B3 & $12.2 \pm 3.3$     & $\phn 7.2 \pm 1.4$ & $\phn 3.4 \pm 1.6$ & $27.6 \pm 10.7$         & $\phn 1.6 \pm 1.0$ & B4 & $\phn 1.1 \pm 0.8$ \\
    0827+243  & 6.129 & B6 & $10.8 \pm 1.4$     & $38.0 \pm 3.9$     & $\phn 3.7 \pm 0.2$ & $16.2 \pm \phn 5.8$     & $\phn 1.0 \pm 0.4$ & B7 & $\phn 0.1 \pm 0.3$ \\
    0829+046  & 0.840 & B5 & $19.6 \pm 6.0$     & $13.6 \pm 1.9$     & $\phn 2.6 \pm 1.2$ & $21.2 \pm \phn 5.7$     & $\phn 1.0 \pm 0.5$ & B4 & $\phn 4.4 \pm 3.0$ \\
    0836+710  & 17.185& B3 & $13.5 \pm 0.7$     & $17.8 \pm 0.5$     & $\phn 4.1 \pm 0.2$ & $18.0 \pm \phn 3.3$     & $\phn 1.3 \pm 0.2$ & B6 & $\phn 0.1 \pm 0.3$ \\
    0851+202  & 1.589 & B12& $\phn 6.4 \pm 1.9$ & $11.8 \pm 1.7$     & $\phn 7.9 \pm 1.3$ & $24.6 \pm \phn 1.2$     & $\phn 3.4 \pm 0.6$ & B6 & $\phn 2.5 \pm 1.5$ \\
    0954+658  & 1.969 & B14& $\phn 9.8 \pm 1.6$ & $38.8 \pm 4.7$     & $\phn 3.9 \pm 0.2$ & $19.4 \pm \phn 2.1$     & $\phn 1.3 \pm 0.2$ & B19& $\phn 5.1 \pm 0.8$ \\
    1055+018  & 5.722 & B3 & $16.9 \pm 3.9$     & $12.8 \pm 1.0$     & $\phn 3.2 \pm 1.0$ & $36.8 \pm \phn 6.5$     & $\phn 2.1 \pm 0.7$ & B1 & $\phn 1.4 \pm 1.1$ \\
    1101+384\tablenotemark{a}  & 0.131 & $\ldots$ & $\sim 65.5$ & $\sim 16.4$ & $\sim 1.4$ & $27.8 \pm 14.2$        & $\sim 0.7$         & $\ldots$ & $\ldots$ \\
    1127--145 & 8.142 & C1 & $13.6 \pm 2.5$     & $12.1 \pm 0.5$     & $\phn 4.2 \pm 0.9$ & $20.8 \pm \phn 4.3$     & $\phn 1.5 \pm 0.4$ & B2 & $\phn 3.8 \pm 0.6$ \\
    1156+295  & 4.446 & B2 & $15.0 \pm 2.8$     & $10.1 \pm 1.0$     & $\phn 3.4 \pm 1.0$ & $22.2 \pm \phn 3.8$     & $\phn 1.3 \pm 0.4$ & B3 & $\phn 0.2 \pm 0.5$ \\
    1219+285  & 0.470 & B1 & $\phn 4.5 \pm 1.1$ & $\phn 5.3 \pm 0.3$ & $12.6 \pm 2.6$     & $16.0 \pm 10.6$         & $\phn 3.5 \pm 2.4$ & B3 & $\phn 4.6 \pm 4.1$ \\
    1222+216  & 2.379 & B5 & $\phn 7.0 \pm 0.6$ & $34.7 \pm 2.7$     & $\phn 4.9 \pm 0.1$ & $42.8 \pm \phn 1.9$     & $\phn 3.7 \pm 0.2$ & B3 & $\phn 1.3 \pm 0.1$ \\
    1226+023  & 0.755 & B2 & $\phn 3.2 \pm 0.3$ & $15.7 \pm 1.1$     & $10.9 \pm 0.2$     & $10.8 \pm \phn 0.7$     & $\phn 2.0 \pm 0.1$ & B16& $\phn 1.2 \pm 0.3$ \\
    1253--055 & 3.080 & C35& $\phn 8.2 \pm 0.9$ & $59.6 \pm 5.6$     & $\phn 3.5 \pm 0.1$ & $36.8 \pm \phn 1.0$     & $\phn 2.3 \pm 0.1$ & C31& $\phn 1.0 \pm 0.1$ \\
    1308+326  & 6.591 & B3 & $18.5 \pm 2.9$     & $14.6 \pm 0.7$     & $\phn 3.0 \pm 0.6$ & $25.8 \pm \phn 4.7$     & $\phn 1.3 \pm 0.4$ & B1 & $\phn 7.9 \pm 1.5$ \\
    1406--076 & 10.854& B4 & $18.4 \pm 2.9$     & $18.0 \pm 0.7$     & $\phn 3.1 \pm 0.5$ & $29.0 \pm \phn 7.2$     & $\phn 1.6 \pm 0.5$ & B2 & $\phn 3.1 \pm 0.7$ \\
    1510--089 & 1.919 & B1 & $37.0 \pm 2.7$     & $25.6 \pm 1.1$     & $\phn 1.4 \pm 0.2$ & $16.0 \pm \phn 2.8$     & $\phn 0.4 \pm 0.1$ & B10& $\phn 1.6 \pm 0.6$ \\
    1611+343\tablenotemark{b}  & 10.009& B1 & $25.7 \pm 3.9$     & $14.0 \pm 1.8$     & $\phn 1.2 \pm 0.4$ & $26.2 \pm \phn 3.3$     & $\phn 0.5 \pm 0.2$ & $\ldots$ & $\ldots$           \\
    1622--297 & 5.133 & B5 & $21.9 \pm 2.8$     & $18.0 \pm 0.6$     & $\phn 2.6 \pm 0.4$ & $23.4 \pm \phn 3.3$     & $\phn 1.0 \pm 0.2$ & B4 & $\phn 0.7 \pm 0.6$ \\
    1633+382  & 13.775& B3 & $23.8 \pm 3.9$     & $13.6 \pm 1.8$     & $\phn 1.6 \pm 0.5$ & $18.2 \pm \phn 6.1$     & $\phn 0.5 \pm 0.2$ & B5 & $\phn 0.4 \pm 0.3$ \\
    1641+399  & 3.480 & B4 & $13.2 \pm 1.7$     & $20.8 \pm 1.0$     & $\phn 4.1 \pm 0.3$ & $17.2 \pm \phn 2.6$     & $\phn 1.2 \pm 0.2$ & B1 & $\phn 8.5 \pm 0.3$ \\
    1652+398\tablenotemark{a}  & 0.149 & $\ldots$ & $\sim 31.5$ & $\sim16.4$ & $\sim 0.7$ & $53.8 \pm 11.8$ & $\sim 0.7$ & $\ldots$ & $\ldots$ \\
    1730-130  & 5.817 & B6 & $14.8 \pm 3.7$     & $21.3 \pm 1.9$     & $\phn 3.7 \pm 0.7$ & $27.2 \pm \phn 1.4$     & $\phn 1.8 \pm 0.3$ & B3 & $\phn 2.0 \pm 0.5$ \\
    1749+096  & 1.685 & B7 & $12.2 \pm 1.9$     & $10.8 \pm 0.7$     & $\phn 4.6 \pm 0.8$ & $31.0 \pm \phn 2.3$     & $\phn 2.5 \pm 0.5$ & B5 & $\phn 1.9 \pm 0.5$ \\
    1959+650\tablenotemark{a}  & 0.209 & $\ldots$ & $\sim 30.3$ & $\sim 16.4$ & $\sim 1.0$ & $38.2 \pm 12.9$ & $\sim0.7$ & $\ldots$ & $\ldots$ \\
    2200+420  & 0.311 & B8 & $\phn 4.3 \pm 1.2$ & $14.6 \pm 3.0$     & $\phn 9.5 \pm 0.8$ & $\phn 9.0 \pm \phn 1.8$ & $\phn 1.5 \pm 0.3$ & B6 & $\phn 0.1 \pm 2.0$ \\
    2223-052  & 10.053& B4 & $16.3 \pm 2.3$     & $10.3 \pm 1.0$     & $\phn 2.3 \pm 0.7$ & $29.4 \pm \phn 3.2$     & $\phn 1.5 \pm 0.4$ & B2 & $\phn 0.7 \pm 1.1$ \\
    2230+114  & 6.911 & B4 & $37.5 \pm 6.7$     & $23.6 \pm 2.6$     & $\phn 1.2 \pm 0.4$ & $20.2 \pm \phn 3.6$     & $\phn 0.4 \pm 0.1$ & B2 & $\phn 0.8 \pm 0.2$ \\
    2251+158  & 5.477 & B10& $28.8 \pm 2.2$     & $15.7 \pm 1.0$     & $\phn 1.1 \pm 0.2$ & $24.4 \pm \phn 1.6$     & $\phn 0.5 \pm 0.1$ & B9 & $\phn 0.1 \pm 0.2$ \\
    \enddata
    \tablenotetext{a}{Estimates of the parameters are discussed in $\S$\ref{subsec:SubclassParamAverages}.}
    \tablenotemark{b}{The source contains only one ``reliable'' knot.}
\end{deluxetable*}

\begin{figure}
    \figurenum{23}
    \begin{center}
        \includegraphics[width=0.45\textwidth]{{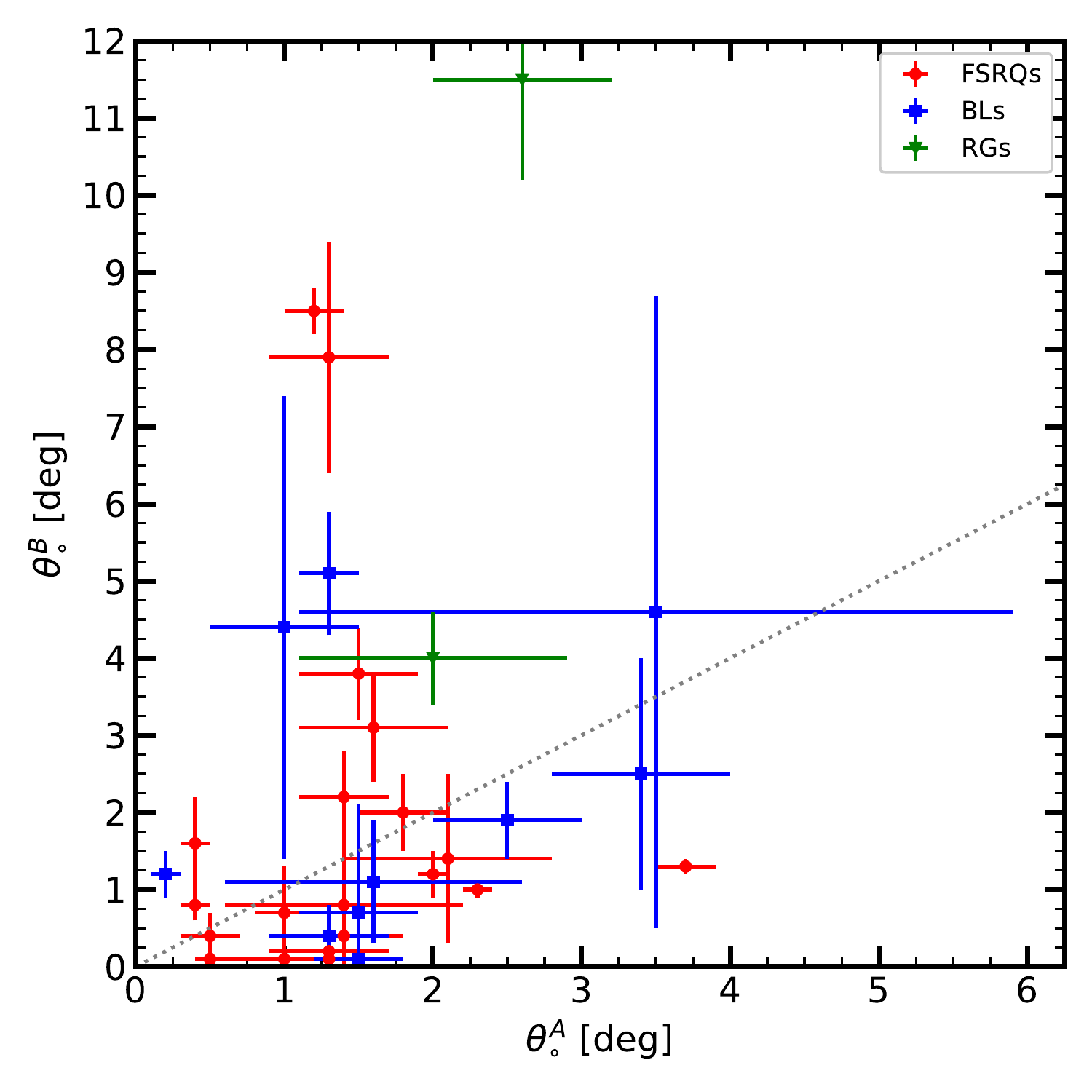}}
        \caption{Comparison between opening semi-angles, $\theta_\circ$ of sample sources, derived using methods A and B (see text for details). The gray dotted line indicates $\theta_\circ^{\text{A}} = \theta_\circ^{\text{B}}$.}
        \label{fig:ThetaAThetaB}
    \end{center}
\end{figure}

Figure~\ref{fig:ThetaAThetaB} presents a comparison between $\theta_\circ^{\text{A}}$ and $\theta_\circ^{\text{B}}$. For the majority of sources, there is good agreement (within the uncertainty) between the two opening semi-angles derived via the different methods. There are a small number of sources for which $\theta_\circ^{\text{B}} > \theta_\circ^{\text{A}}$, in contrast with the large number of sources obeying this uncertainty found in \citet{Jorstad2017}.
This is likely due to the new definition of $\theta_\circ^{\text{B}}$, in which we take the opening semi-angle to be the difference between the fastest and slowest high-significance knot. One can explain this discrepancy for sources that have very small viewing angles, e.g., the BL 0235+164, where the knots have no preferential direction on the plane of the sky. Based on geometrical arguments, this cannot be the case for all sources in our sample.

Larger values of $\theta_\circ^{\text{B}}$ are to be expected if the jets of blazars are structured in a ``spine-sheath" manner \citep[e.g.,][]{Pelletier1989, Sol1989, Celotti2001, Ghisellini2005, Tavecchio2008, DArcangelo2009, Xie2012, Mimica2015}. Such a structure would consist of two flows: a faster, lower-density component streaming along the central axis (spine), and a slower, denser component along the edges of the jet. As the Doppler beaming would be stronger for the faster flow if viewed at a narrow angle, the spine would be expected to dominate the overall jet emission, while the sheath would be more intense at wider viewing angles \cite[e.g., RGs,][]{Bruni2021}.
Evidence for a spine-sheath structure in several blazar jets comes from linear polarization of VLBI images \citep[e.g.,][]{Pushkarev2005, Asada2010, Gabuzda2014, Mertens2016b}, especially when including the space-based \emph{RadioAstron} antenna in the array \citep{Bruni2021}.

It is possible that most knots seen in our images flow down the spine of a jet, \citep[propelled by the rotational energy of the black hole,][]{Blandford1977}, while a relatively small number of knots form along the slower sheath \citep[e.g., from a magnetically driven wind from the accretion disk,][]{Blandford1982}. Several sources in our sample have been inferred to have a spine-sheath structure, such as
0316+413 \citep{Giovannini2018},
1055+018 \citep{Attridge1999, MacDonald2017},
1226+023 \citep{MacDonald2017, Bruni2021},
1510--089 \citep{MacDonald2015},
1652+398 \citep{Giroletti2004, Pushkarev2005},
0836+710, 1253--055, and 2230+114 \citep{MacDonald2017}.
Neither 1652+398 nor 0316+413 have enough reliable superluminal knots for us to determine their jet structure (see $\S$\ref{subsec:SubclassParamAverages} to obtain estimates of their physical parameters). The sources 1055+018 and 1226+023 both have $\theta_\circ^{\text{A}} > \theta_\circ^{\text{B}}$. The spine-sheath structure in 1055+018 was identified from the polarization of the emission rather than the total intensity that we have analyzed here. The inferred layered structure in 1226+023 was based on images with extremely high spatial resolution at lower frequencies (1.6 and 4.8 GHz); it may be the case that we have not observed a slow knot in the sheath of this source. Conclusive evidence for spine-sheath structure in the jets of our sample can be found through an analysis of stacked images and maps of polarized intensity \citep[as has been done for several sources in][]{MacDonald2015, MacDonald2017}, but is beyond the scope of this work.

\subsection{Comparison with Other Physical Parameter Estimations}
\label{subsec:OtherPaperComps}

Several methods exist that use a variety of data at different frequencies to estimate the Doppler factor in blazar jets and subsequently infer the physical parameters of the jets themselves \citep[e.g.,][]{Ghisellini1993, Lahteenmaki1999, Jorstad2005, Fan2013, Fan2014, Chen2018, Lister2019, Zhang2020}.  Reviews of the methods and their results can be found in, e.g., \citet{Liodakis2015, Liodakis2017a, Liodakis2017b}. Here we compare a few of these results with those inferred from our piece-wise kinematic fitting.

\begin{figure*}
    \figurenum{24}
    \begin{center}
        \includegraphics[width=\textwidth]{{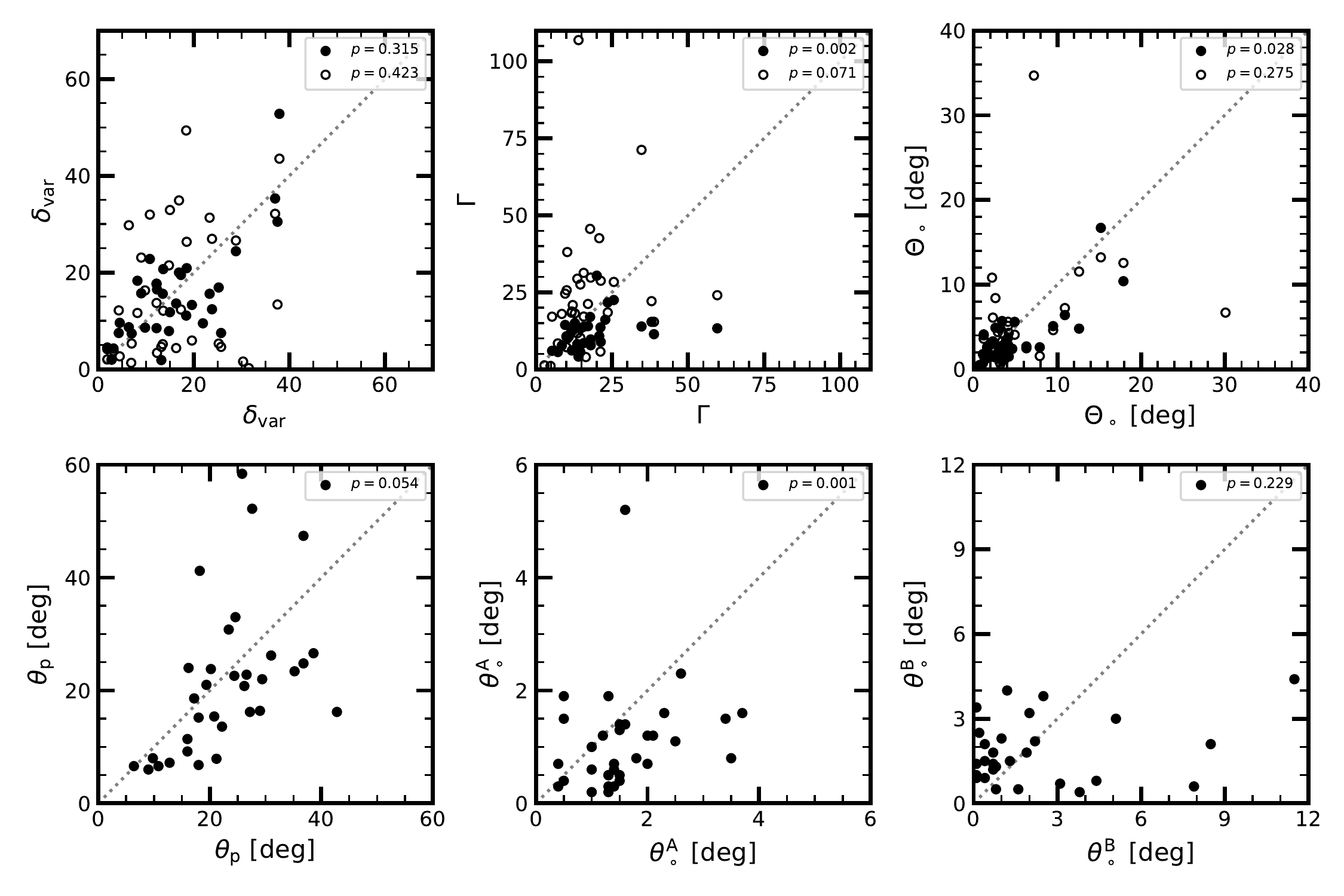}}
        \caption{Comparisons of the physical parameters of blazar jets derived in this work ($x$-axis) with those derived in \citet{Jorstad2017} ($y$-axis, filled circles) and \citet{Liodakis2018} ($y$-axis, open circles). A gray dotted line showing a 1:1 relation is shown in each panel, and the $p$-value of a WRS test between the derived values (from Table~\ref{tab:WRSp}) is given in the legend. \label{fig:ScatterComp}}
    \end{center}
\end{figure*}

In order to quantify whether the distributions of the physical parameters from our work differ from those of others, we employ the Wilcoxon rank-sum test\footnote{The KS test, used elsewhere in this work, is sensitive to differences in the general shape of two distributions (i.e., dispersion, skewness, etc.), whereas the WRS test is sensitive to differences in location (e.g., the median) of the two samples.} \citep[WRS, or, alternatively, the Mann-Whitney $U$-test;][]{Mann1947}. Since we wish to test whether estimation of the physical parameters depends on the method employed, we assume that the parameters are drawn from the same distribution, and thus the WRS test can indicate whether there is a systematic bias in the methodology. The WRS test operates under the null hypothesis that two distributions are drawn from the same sample, and for this test smaller $p$-values indicate a larger likelihood that the distributions are different. We consider two data sets to be different if their WRS $p$-value $<0.005$ ($\sim3\sigma$ confidence level). Figure~\ref{fig:ScatterComp} shows a comparison between the results of this work and those of \citet{Jorstad2017} and \citet{Liodakis2018}, to accompany the indicated $p$-values.

The most direct comparison can be made with \citet{Jorstad2017}, as the current work shares data, methodology, and underlying physical assumptions. In this comparison, we ignore the three sources with no reliable knots for deriving values of physical parameters. The majority of our estimated values of ($\delta_{\text{var}},\ \Gamma,\ \Theta_\circ,\ \theta_{\text{p}},\ \theta_\circ^{\text{A}},$ and $\theta_\circ^{\text{B}}$) agree within the sum of the two estimated $1\sigma$ uncertainties. Table~\ref{tab:WRSp} gives $p$-values between the ``representative" physical parameters given in Table~\ref{tab:AvePhysParams} and the average values from \citet{Jorstad2017}. For most parameters, there is no statistically significant difference in the distributions. However, there is a slight tendency for the representative value of $\Gamma$ defined in this work to be higher than that of \citet{Jorstad2017}. This is likely due to how we define a ``representative" knot as the fastest one with a relatively small $\xi$ value, rather than averaging all knots together as was done in \citet{Jorstad2017}. Similarly, the reported $\theta_\circ^{\text{A}}$ values in this work are slightly larger than those inferred previously, but this is probably due to $\langle \Theta_\circ \rangle$ and $\theta_{\text{p}}$ now being defined from $>10$ years of observation, which samples a wider range of knot behaviors.

\begin{deluxetable}{ccc}
    \tablecaption{Wilcoxon rank-sum test $p$-values between this work and previous results.\label{tab:WRSp}}
    \tablewidth{0pt}
    \tablehead{
    \colhead{Parameter} & \colhead{Comparison Work} & \colhead{$p$-value} \\
    }
    \startdata
    $\delta_{\text{var}}$ & \citet{Jorstad2017} & 0.315 \\
     & \citet{Liodakis2018} & 0.423 \\
    $\Gamma$ & \citet{Jorstad2017} & 0.002 \\
     & \citet{Liodakis2018} & 0.071 \\
    $\Theta_\circ$ & \citet{Jorstad2017} & 0.028 \\
     & \citet{Liodakis2018} & 0.275 \\
    $\theta_{\text{p}}$ & \citet{Jorstad2017} & 0.054 \\
    $\theta_\circ^{\text{A}}$ & \citet{Jorstad2017} & 0.001 \\
    $\theta_\circ^{\text{B}}$ & \citet{Jorstad2017} & 0.229 \\
    \enddata
\end{deluxetable}

Another useful comparison can be made with the work of \citet{Liodakis2018}. These authors modeled the 15 GHz light curves of 1029 blazars and blazar-like AGNs, as observed with the Owen's Valley Radio Observatory \citep[OVRO;][]{Richards2011}, with a series of exponential rise and decay profiles (flares) superposed on a stochastic background. For each flare, a variability brightness temperature was calculated. Since each source can have numerous flares within the light curve, distributions of variability brightness temperatures can be made for each source. Comparison with an expected equilibrium brightness temperature then yields distributions of Doppler factors, $\delta_{\text{var}}$. The authors then used the maximum apparent speeds of radio knots at 15 GHz from the MOJAVE program \citep{Lister2016} to calculate $\Gamma$ and $\Theta_\circ$. Agreement between the two methods would provide important constraints for the Doppler factors and equipartition states in the jets.

All of the sources in the VLBA-BU-BLAZAR program have been observed by OVRO and are included in \citet{Liodakis2018}, except for the FSRQ 1622$-$297, which is below the declination limit for the OVRO monitoring program. Two sources, the BLs 0735+178 and 0954+658, did not have reliable redshift estimates at the time of \citet{Liodakis2018}. These have since been refined in NED to $z = 0.424$ and $0.368$, respectively.
Using the same formalism as \citet{Liodakis2018}, we have updated the Lorentz factors and viewing angles (Eq.~\ref{eqn:GammaTheta}) to be $\Gamma = 8.4 \pm 1.6,\ \Theta_\circ = 3.3 \pm 1.6$ for 0735+178 and $\Gamma = 15.4 \pm 1.2,\ \Theta_\circ = 3.5 \pm 1.1$ for 0954+658.

Comparing the physical parameters $\delta_{\text{var}}$, $\Gamma$, and $\Theta_\circ$ between our current work and \citet{Liodakis2018}, we again see good agreement between the distributions. The majority of estimates of the physical parameters of each jet lie within the sum of the uncertainties between the two samples. WRS $p$-values between the two data sets are given in Table~\ref{tab:WRSp}. No comparison meets our significance threshold, indicating that the distributions are not significantly different. Discrepancies in the values of parameters calculated by \citet{Liodakis2018} and this work are likely due to the difference in observing frequencies, resulting in different parts of the jet being probed.

Finally, we compare our results with modeling of the relation between $\Theta_\circ$ and $\Gamma$. Given that Doppler beaming increases with smaller viewing angles while geometrical likelihood increases with larger angle, the most probable viewing angle is somewhat smaller than that which maximizes the apparent velocity, $\Theta_{\text{max}} = \arcsin{1/\Gamma} \approx 1/\Gamma$ for $\Gamma\gg1$ \citep{Marscher1990}. \citet{Vermeulen1994} found that the most likely angle is $\sim0.5\Theta_{\text{max}}$. Meanwhile, \citet{Lister1997} used Monte Carlo simulations to determine statistical properties of beamed sources in a flux-limited sample, including the viewing angle at which a jet will appear in a flux-limited sample given its Lorentz factor and intrinsic luminosity. They found that the peak of the simulated distribution is $\Theta_\circ \sim 0.4 \Theta_{\text{max}}$, with a considerable number of sources having $\Theta_\circ > \Theta_{\text{max}}$.

\begin{figure}
    \figurenum{25}
    \begin{center}
    \includegraphics[width=0.45\textwidth]{{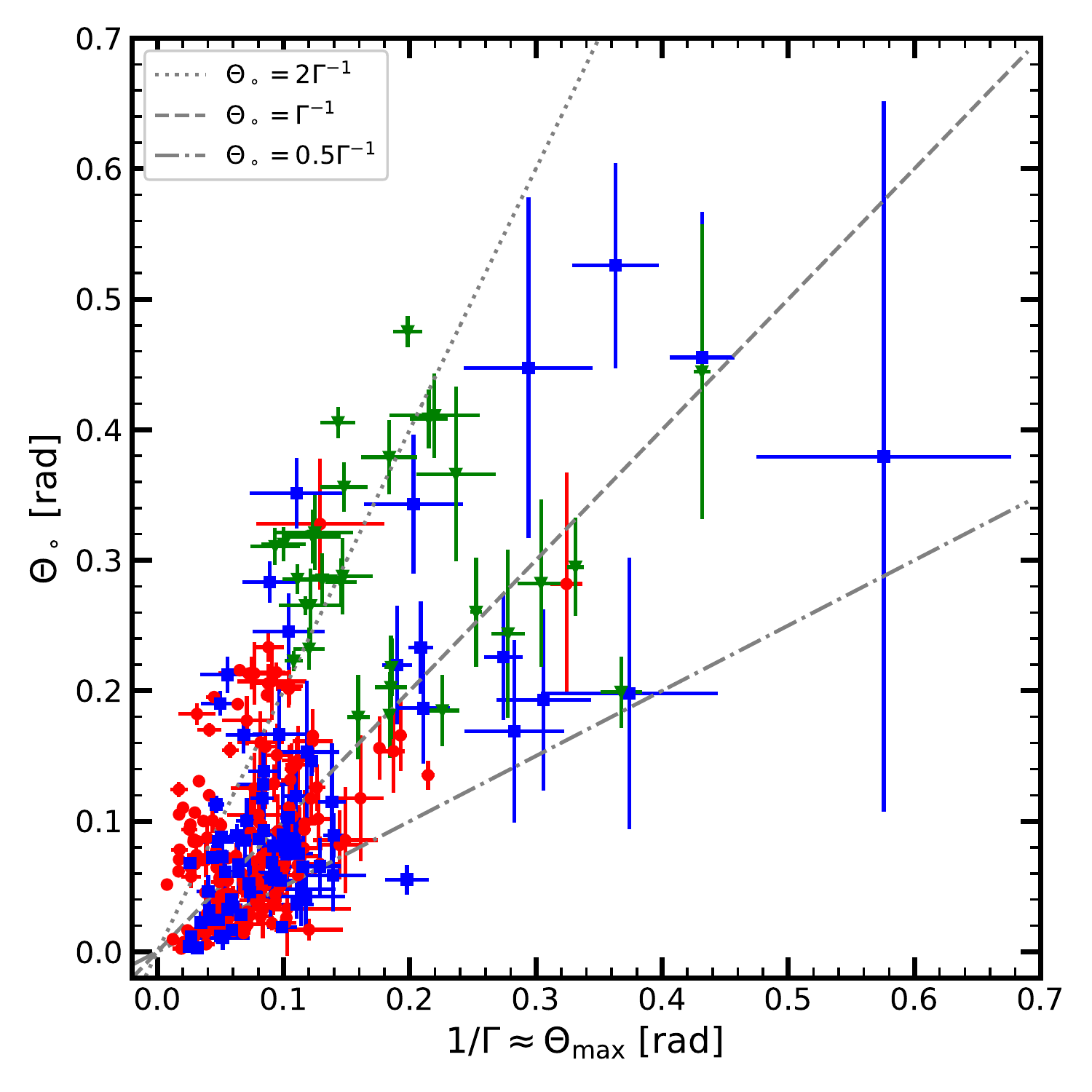}}
    \caption{Observed viewing angle $\Theta_\circ$ vs. theoretical viewing angle $\Theta_{\text{max}} \approx \Gamma^{-1}$ for reliable knots in FSRQs (red circles), BLs (blue squares), and RGs (green triangles). \label{fig:ListerPlot}}
    \end{center}
\end{figure}

To see if these relations hold observationally, we plot the viewing angle of all reliable knots in our sample against $\Gamma^{-1}$ in Figure~\ref{fig:ListerPlot}. The data appear to fall into several distinct groups. To guide the readers eye, we have added three lines indicating values for which $\Theta_\circ = 0.5$, 1, and 2 times $\Theta_{\text{max}}$. The vast majority of knots fall between the 0.5 and 2$\times$ lines and with $\Theta_{\text{max}} \lesssim 0.15$ rad. Two extended populations appear, roughly following the $\Theta_\circ \sim \Theta_{\text{max}}$ and $2\Theta_{\text{max}}$ lines. Almost all of the RG knots appear along these extended populations.

\begin{figure}
    \figurenum{26}
    \begin{center}
    \includegraphics[width=0.45\textwidth]{{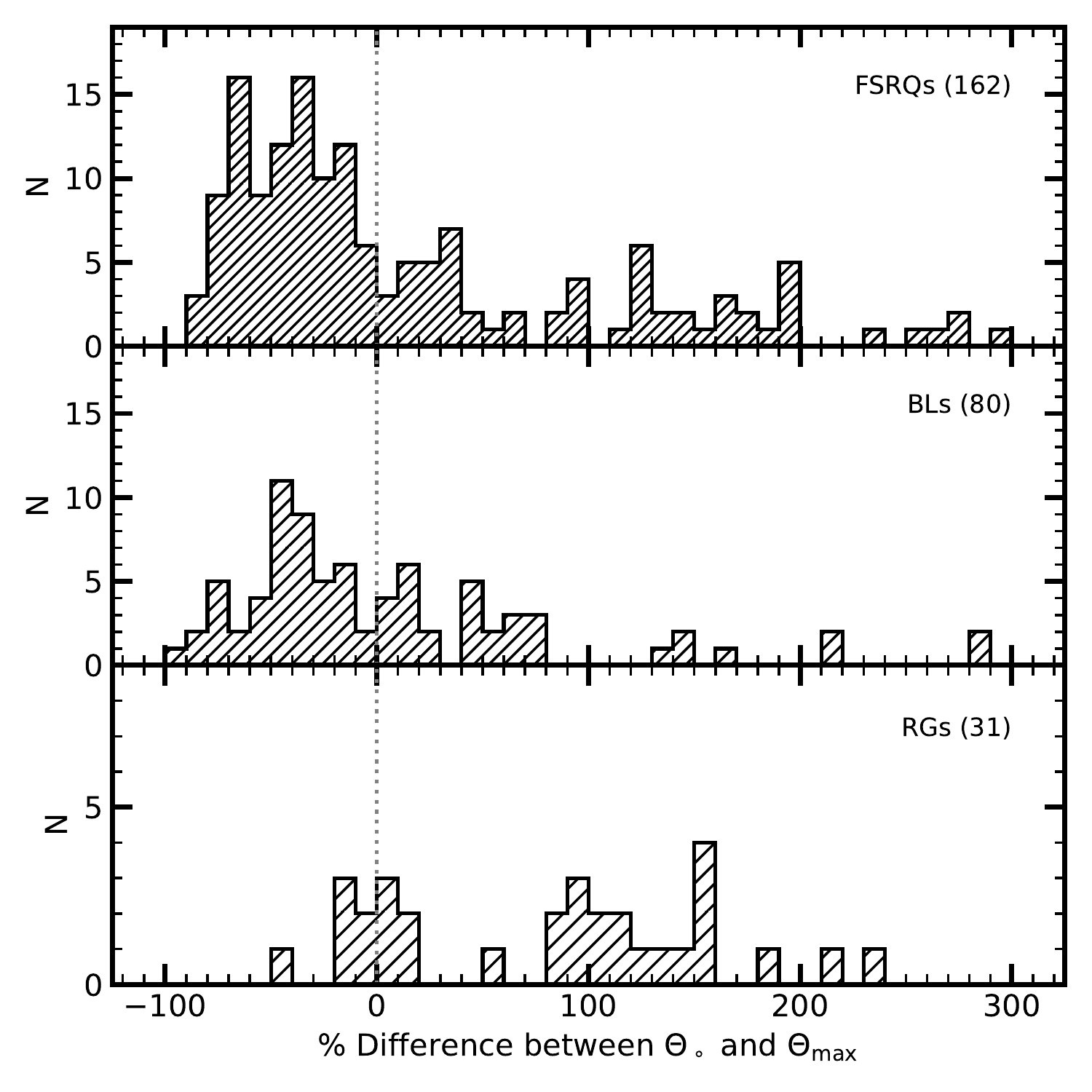}}
    \caption{Distributions for FSRQs, BLs, and RGs of the percent difference between the observed viewing angle $\Theta_\circ$ and expected viewing angle $\Theta_{\text{max}} \approx \Gamma^{-1}$. A grey dotted line has been added at zero difference. \label{fig:ListerHist}}
    \end{center}
\end{figure}

To get a statistical sense of which relation most accurately describes the observations, we have followed \citet{Lister1997} by calculating the percent difference between $\Theta_\circ$ and $\Theta_{\text{max}}$, which we plot as a histogram for each source type in Figure~\ref{fig:ListerHist}. There are 9 FSRQ knots for which the difference is $>300\%$, usually in cases of very high values of $\Gamma$. As these represent a small fraction ($\sim5\%$) of the knots, we have omitted them to reduce the size of the figure. For FSRQs and BLs, we recover the simulated distribution of \citet{Lister1997}. In these two distributions, the peak appears near $-50\%$ difference, but there is a long tail showing that many knots can have $\Theta_\circ > \Theta_{\text{max}}$. The two distributions are similar statistically, with a KS test statistic of $\mathcal{D} = 0.72$ and $p = 0.073$. However, we do not see a similar distribution for the RGs ($\mathcal{D} = 0.460,\ p = 1.5 \times 10^{-5}$ for the FSRQ vs RG and $\mathcal{D} = 0.513,\ p = 6.3 \times 10^{-6}$ for the BL vs RG distributions). There is a wide spread of the percent difference for these sources, and no obvious distribution peak due to the small number of observed knots. The wide spread is an indication that the expected $\Theta_{\text{max}}$ relation with $\Gamma$ is not a good indicator of the true viewing angle for RGs \citep{Lister1997}.

\subsection{Averages of Physical Parameters of AGN Subclasses}
\label{subsec:SubclassParamAverages}

We use the same formalism as \citet{Jorstad2017} to calculate the weighted averages of the physical parameters for each subclass in our sample.
Two values of the average opening semi-angle values are obtained for the two different methods of estimating $\theta_\circ$,  ${\overline{\theta_\circ}}^{\text{A}}$ and ${\overline{\theta_\circ}}^{\text{B}}$. We calculate the uncertainty in the average quantities as the weighted standard deviation of the parameters of the sources in the subclass about these means, where the weights are the inverse of the variance of the parameter for each source. Table~\ref{tab:SubclassParams} gives the result of the averaging as follows:
1---Name of the AGN subclass;
2---number of sources in the subclass, $N_s$, used to calculate the averages;
3---Doppler factor, $\overline{\delta}$, and its weighted standard deviation from the included knots;
4---bulk Lorentz factor, $\overline{\Gamma}$, and its weighted standard deviation;
5---viewing angle, $\overline{\Theta_\circ}$, and its weighted standard deviation;
6---opening semi-angle of the jet according to method A, $\overline{\theta_\circ}^{\text{A}}$, and its weighted standard deviation; and
7---opening semi-angle of the jet according to method B, $\overline{\theta_\circ}^{\text{B}}$, and its weighted standard deviation.

\begin{deluxetable*}{llccccc}
    \tablecaption{Average Physical Parameters of AGN Jets for Different Subclasses\label{tab:SubclassParams}}
    \tablewidth{0pt}
    \tablehead{
    \colhead{Subclass} & \colhead{$N_s$} & \colhead{$\overline{\delta}$} & \colhead{$\overline{\Gamma}$} & \colhead{$\overline{\Theta_\circ}$} & \colhead{$\overline{\theta_\circ}^{\text{A}}$} & \colhead{$\overline{\theta_\circ}^{\text{B}}$} \\
    \colhead{} & \colhead{} & \colhead{} & \colhead{} & \colhead{[deg]} & \colhead{[deg]} & \colhead{[deg]}
    }
    \colnumbers
    \startdata
    FSRQ & 22\tablenotemark{a} & $7.1 \pm 6.1$ & $16.0 \pm 4.7$     & $\phn 4.0 \pm 2.3$ & $1.2 \pm 0.9$ & $1.2 \pm 1.5$ \\
    BL   & 11                  & $3.7 \pm 4.3$ & $\phn 5.5 \pm 4.1$ & $\phn 2.7 \pm 2.2$ & $0.7 \pm 0.8$ & $1.3 \pm 1.1$ \\
    RG   & 2                   & $2.6 \pm 0.5$ & $\phn 8.6 \pm 0.6$ & $15.7 \pm 1.5$     & $2.4 \pm 0.4$ & $5.3 \pm 4.0$ \\
    \enddata
    \tablenotetext{a}{Because the source 1611+343 only has one reliable superluminal knot, there is no estimate for $\theta_\circ^{\text{B}}$ for this source, and 21 sources were used to calculate $\overline{\theta_\circ}^{\text{B}}$.}
\end{deluxetable*}

While the physical parameters of FSRQs and BLs overlap when the standard deviations are included, the expected trend of larger Doppler and Lorentz factors in FSRQs relative to BLs is observed in the current data \citep[][]{Jorstad2005}.
Higher values of $\delta$ in FSRQs are supported by the higher brightness temperatures of moving components. The relative similarity of the viewing and opening semi-angles between the two subclasses \citep[compared to previous results, e.g.,][]{Pushkarev2009, Jorstad2017} can be explained by our choice of the fastest ``typical" knot to represent particular sources. In a spine-sheath jet structure, the fastest knots all come from the spine, resulting in an opening semi-angle that would be smaller than if all knots were considered together (as would be expected in BLs). The physical parameters of RGs are also consistent with the predictions of the unified scheme of AGNs \citep{Urry1995}, with comparable Lorentz factors, smaller Doppler factors, and much larger viewing angles compared to FSRQs and BLs. However, as was pointed out in \citet{Jorstad2017}, these RGs are the brightest \gammaray\ emitters of their subclass, and may not be representative of the subclass as a whole.

Three sources in our sample do not have superluminal knots meeting our reliability criteria described at the beginning of $\S$\ref{sec:JetPhysicalParams}: the RG 0316+413 and the BLs 1652+398 and 1959+650. Thus, we cannot calculate their physical parameters using their knot motions. Instead, we make a rough estimate for the physical parameters of these sources based on two assumptions: (1) the intrinsic opening semi-angle of the jet is equal to ${\overline{\theta_\circ}}^{\text{A}}$, the average of the subclass to which the source belongs, and (2) $\Gamma \theta_\circ \sim 0.2$ is universal for bright \gammaray\ AGNs \citep{ClausenBrown2013,Jorstad2017}. The low apparent speeds of the knots in the BL 1101+384 indicate that we are looking at the sheath of the jet, so we also use the same formalism to estimate the parameters of this source. The estimated parameters for these four sources are given in Table~\ref{tab:AvePhysParams}. The above assumptions result in viewing angles $\sim2 \times$ smaller than the subclass average for these sources. As we have only observed 1959+650 at 6 epochs it is possible that the jet does indeed have physical parameters as estimated above, but more monitoring is necessary. However, the fact that only slow proper motions have been observed in 0316+413, 1101+384, and 1652+398 despite intensive monitoring campaigns \citep[e.g.; this work;][]{Blasi2013, Lister2016} creates a ``Doppler crisis" with their high-energy properties, which require high Doppler factors (and subsequently high proper motions). The decelerations observed in Figure~\ref{fig:HistRidgeline} (5-10 pc from the core for parallel acceleration, and 1-2 pc from the core for perpendicular acceleration) might be able to explain the discrepancy, as high Doppler factors would only be seen closer to the central engine, beyond the opacity limit of 43 GHz observations. It is also possible that a mildly relativistic flow, corresponding to a ``sheath" around the faster central spine, would obscure the faster spine, as seems to be the case in 0316+413 \citep{Nagai2014}.

%-------------------------------------------------------------------------------
%---   DISCUSSION   ------------------------------------------------------------
%-------------------------------------------------------------------------------

\section{Discussion}
\label{sec:Discussion}

\subsection{Brightness Temperatures and Doppler Factors}
\label{subsec:TbandDelta}

The brightness temperature distributions presented in $\S$\ref{subsec:ObsBrightnessTemps} are similar to those expected for unbeamed, incoherent synchrotron radiation produced by relativistic electrons that are in equipartition with the magnetic field $T_{\mathrm{b,eq}} \sim 5 \times 10^{10}$ K \citep{Readhead1994}.
The higher values of $T_{\text{b,obs}}^{\text{s}}$ found in FSRQs and BLs can be partially explained by Doppler boosting: $T_{\text{b,obs}} \propto \delta T_{\text{b,int}}$; the Doppler factor has been inferred to be as high as $\delta\sim 60$ in blazars \citep[e.g.,][]{Liodakis2018}. During high activity states (such as outbursts) of blazars, it is possible that equipartition is violated such that the energy density of radiating particles is greater than that of the magnetic field in the VLBI core \citep[e.g.,][]{Homan2006, Kovalev2016}. Such non-equipartition states could lead to the extremely high ($T_{\text{b,obs}}^{\text{s}} \gtrsim 10^{13}$ K) brightness temperatures found in some cores of some sources.

\begin{figure}
\figurenum{27}
    \begin{center}
        \includegraphics[width=0.45\textwidth]{{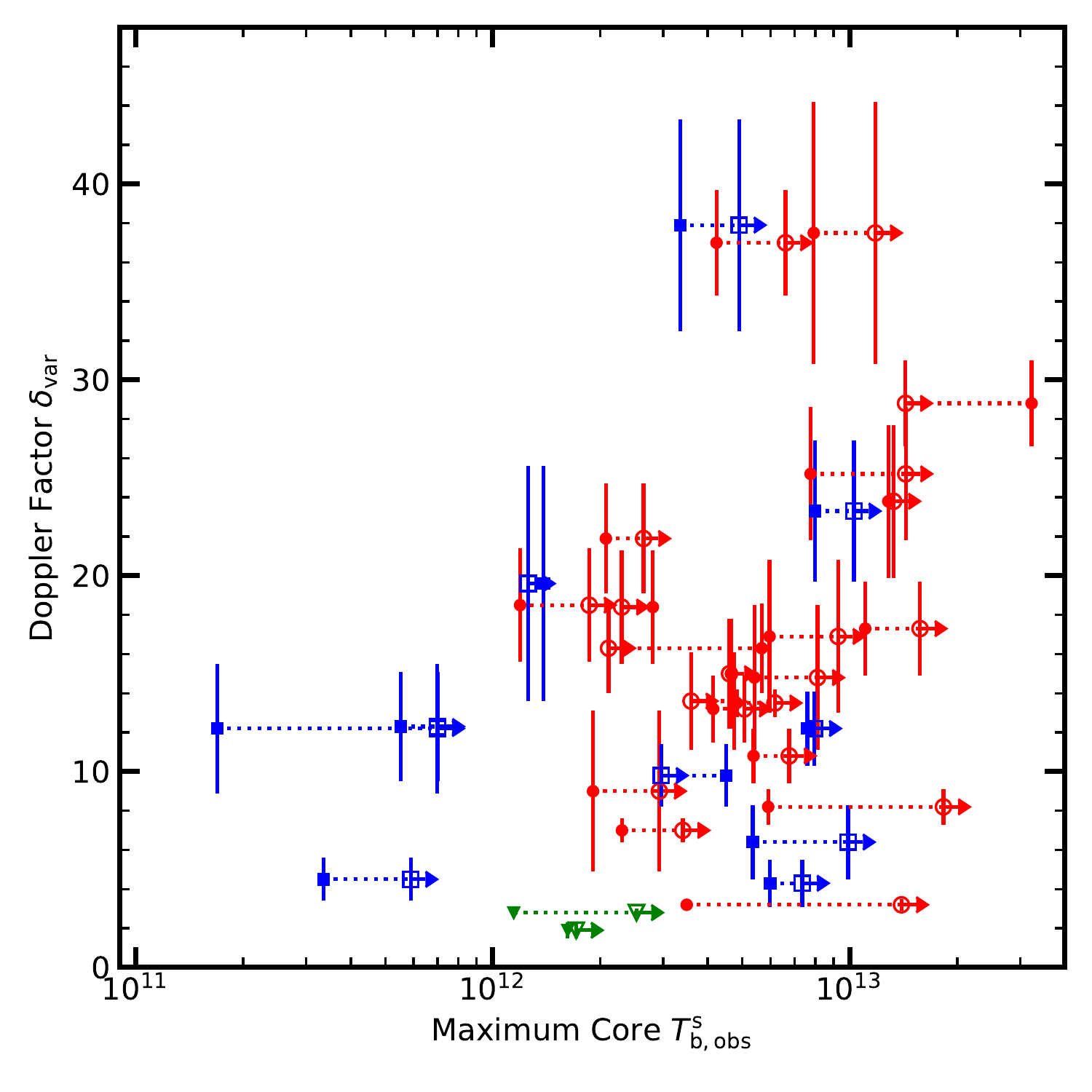}}
        \caption{Typical Doppler factor of variability for each source vs. the maximum brightness temperature of the core, $\emph{A0}$, for resolved (closed symbols) and unresolved (open symbols with right-facing arrows, component size $a \leq 0.02$ mas) epochs. FSRQs are shown in red circles, BLs in blue squares, and RGs in green triangles. \label{fig:DopplerBrightness}}
    \end{center}
\end{figure}

Figure~\ref{fig:DopplerBrightness} compares the typical Doppler factor of variability, $\delta_{\text{var}}$, for each source with the maximum brightness temperature of the core component \emph{A0} in the host galaxy frame, $T_{\text{b,obs}}^s = T_{\text{b,obs}} (1+z)$. In the figure, both the maximum resolved brightness temperature (closed symbols) and lower limit to the maximum brightness temperature (open symbols, based on the observed size of the component; see $\S$\ref{subsec:ModelingTotalIntensityImages}) are plotted. We have connected the resolved brightness temperatures and lower limits in each source with a dotted line to aid the reader. Sources with physical parameters estimated via the method presented in $\S$\ref{subsec:SubclassParamAverages} are omitted. In general, we see that sources with higher maximum brightness temperatures tend to have higher Doppler factors. Also, the maximum brightness temperatures of FSRQs tend to be higher than those of BLs (as expected from the distributions in Figure~\ref{fig:CoreHists}). Taking the three subclasses together, the Doppler factors and maximum brightness temperatures of the cores have a Pearson's $r$ correlation coefficient of $0.367$ for the resolved components. There is more scatter in the data when considering only the lower limits, which have a correlation coefficient of $0.281$ with the maximum brightness temperatures of the cores.
The maximum core brightness temperatures range up to $\sim 10^{13}$ K. These could be indicative of non-equipartition states in the 43 GHz cores of blazars. We see a connection between the \gammaray\ state of blazars and their 43 GHz properties (see $\S$\ref{subsec:GammaRayConnection} below), so these non-equipartition values are expected. However, the moving components all have $T_{\text{b,obs}}^{s} < 5 \times 10^{12}$ K, which can be explained by Doppler boosting with our observed Doppler factor values, $\delta_{\text{var}} \lesssim 60$.

\subsection{Flux Variability}
\label{subsec:Theory}

As discussed in $\S$\ref{subsec:tauvar}, the rise and decay of flux in mm-wave light curves of blazars can be roughly modeled by exponential profiles \citep[e.g.,][]{Terasranta1994, Lister2001, Savolainen2002}. We are thus able to estimate a timescale of variability, $\tau_{\text{var}}$, from the flux decay rate, $k$, as $\tau_{\mathrm{var}} = |1 / k|$ yr. We show in Figure~\ref{fig:TauA} that these decay rates are often primarily due to radiative losses, modulated by light-travel delays, as opposed to adiabatic expansion. As the emission from blazar jets at 43 GHz is dominated by synchrotron radiation, standard synchrotron emission formulae \citep[e.g.,][]{Pacholczyk1970, Longair2011} yield an expected relation between the emitted flux density and distance down the jet $R$, $S_{43} \propto R^{-a}$.
Using the proper motion of a knot, $\mu = dR / dt$, we can estimate the age of a knot as $T_{\text{age}} = \langle R \rangle / \mu$ in the observer's frame. \citet{Homan2002} adopted a decay rate of a knot that is inversely proportional to $T_{\text{age}}$:

\begin{equation}
    k= \frac{(dS_{43} / dt)}{S_{43}} = - a \times \frac{(dR / dt)}{R} = -a / T_{\text{age}}.
    \label{eq:Homan2002}
\end{equation}

\begin{figure}
    \figurenum{28}
    \begin{center}
    \includegraphics[width=0.45\textwidth]{{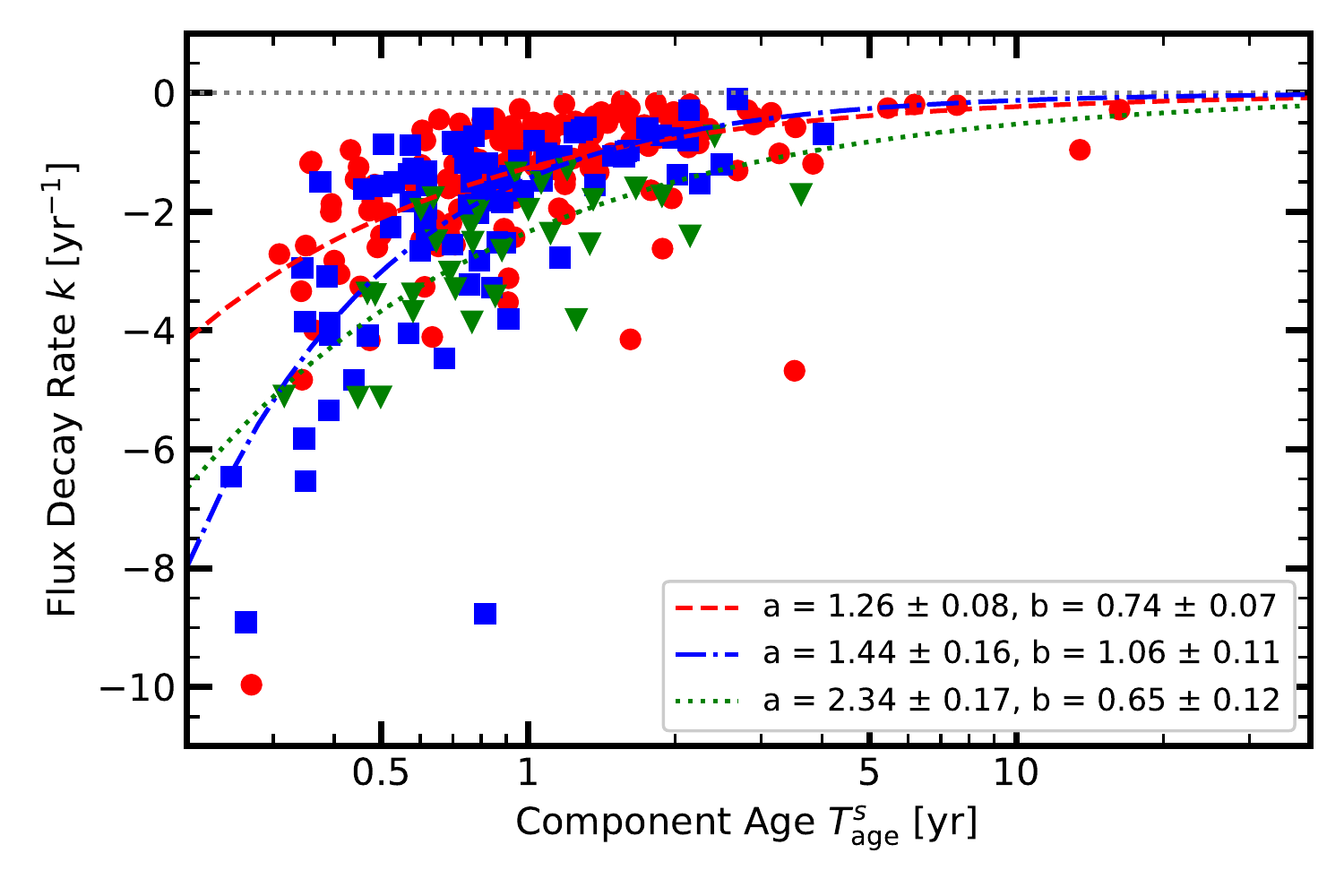}}
    \caption{Flux decay rate $k$ for reliable knots versus the age of the feature in the host galaxy frame, $T_{\text{age}}^s$, for FSRQ (red circles), BL (blue squares), and RG knots (green triangles). Fits to the model $k = -a (T_{\text{age}}^{s})^{-b}$ are plotted in the same color as the subtype of blazar, with values as indicated in the legend.\label{fig:HomanFig}}
    \end{center}
\end{figure}

\noindent We have already calculated the normalized decay rates of the majority of knots in our sample as $(dS_{43}/dt)/S$ (see $\S$\ref{subsec:tauvar}). We revise the relation in Equation \ref{eq:Homan2002} to a more general form: $k= -a T_{\text{age}}^{-b}$, where $a$ and $b$ are constants.
We transform the component ages to the source (i.e., host galaxy) frame ($s$) through $T_{\text{age}}^s = T_{\text{age}} / (1+z)$.
Figure~\ref{fig:HomanFig} shows the decay rate $k^s$ of all knots in our reliable sample versus the age of the feature, $T_{\text{age}}$, separated by blazar subtype. We have fit the generalized equation for the decay rate to each subclass, and show the results by the red dashed, blue dot-dashed, and green dotted curves for FSRQs, BLs, and RGs, respectively, in the figure.

Correcting the age for redshift results in different phenomenological fits for all three subtypes. The BLs have a best-fit exponent $b$ near unity, while FSRQs have $b \sim 0.75$. The exponent for RGs is a factor of 1.6 smaller than for BLs, with $b = 0.65$. The results for BLs and FSRQS are similar to that found in \citet{Homan2002}, despite the authors' assumption that the flux decay of knots is linear rather than exponential.

\begin{figure*}
    \figurenum{29}
    \begin{center}
        \includegraphics[width=0.75\textwidth]{{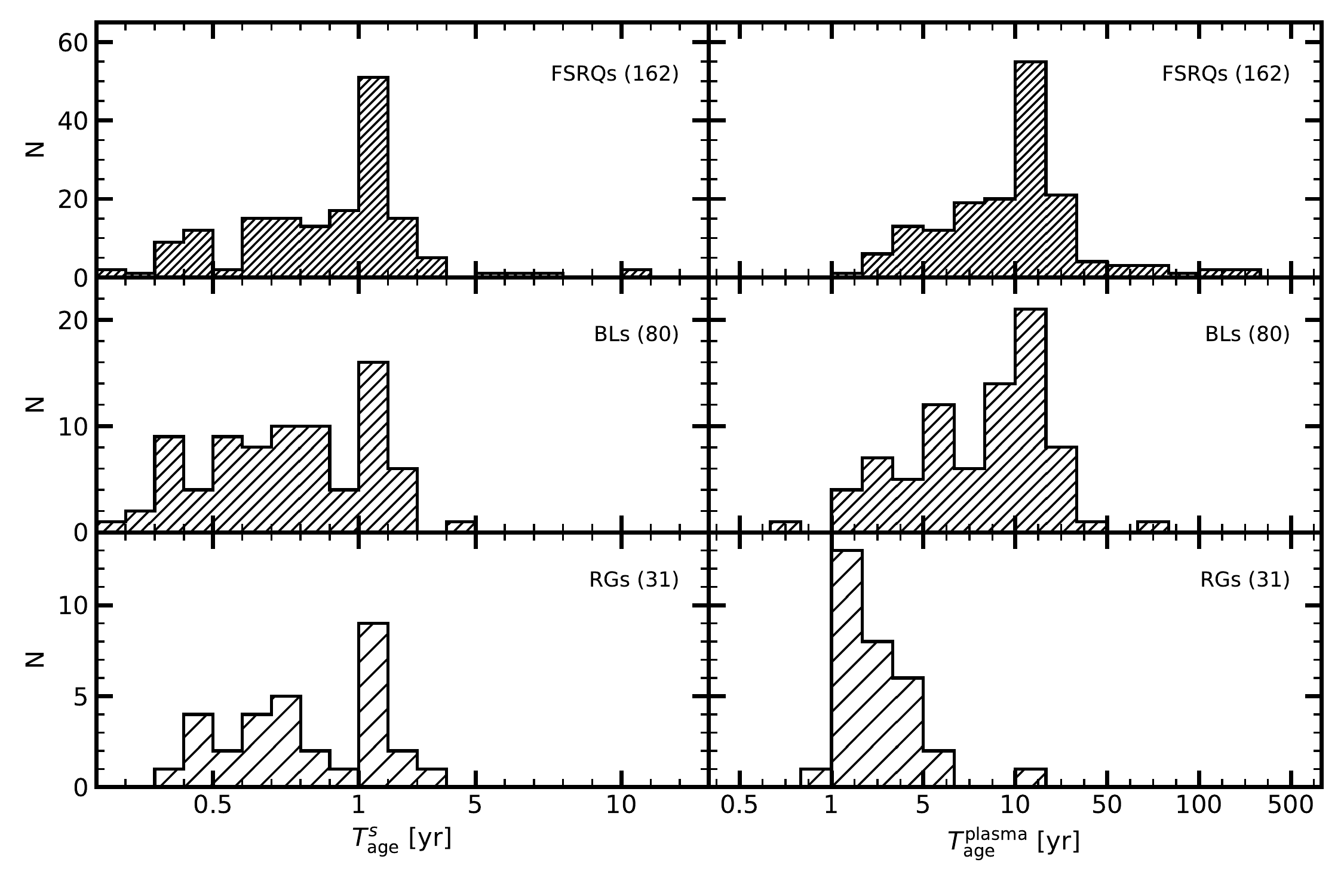}}
        \caption{Distributions of the age of reliable knots in the host galaxy frame, $T_{\text{age}}^s = T_{\text{age}} / (1+z)$ (left), and plasma frame, $T_{\text{age}}^{\text{plasma}} = \delta_{\text{var}} T_{\text{age}} / (1+z)$ (right), in FSRQs (top), BLs (middle), and RGs (bottom). Values in parantheses are the number of knots in each distribution. Note that the age scale is neither linear nor logarithmic.\label{fig:TAgeHists}}
    \end{center}
\end{figure*}

We provide the distributions of knot ages for each subclass in the left panel of Figure~\ref{fig:TAgeHists}. All three distributions are qualitatively similar, with a substantial peak in the distribution for components that live for $\sim1$-$2$ years. There are several knots in FSRQs that are long-lived, with $T_{\text{age}}^s > 5$ yr, while no such knots exist in BLs and RGs. Of the three subtypes, the BL distribution has the largest fraction of short-lived knots, with 70\% of knots having $T_{\text{age}}^s < 1$ yr. A KS test between the distributions confirms the similarities, with $\mathcal{D} = 0.242,\ p = 0.003$ for FSRQs vs BLs, $\mathcal{D} = 0.177,\ p = 0.345$ for FSRQs vs RGs, and $\mathcal{D} = 0.130,\ p = 0.786$ for BLs vs RGs.

If the decay of the flux density of a knot is controlled by the light-crossing time, intrinsically smaller knots should have shorter decay times in the plasma frame. Since the angular sizes of knots are similar for all of the sources (see Table \ref{tab:GaussianComponents}), knots in objects with lower redshifts --- all of the RGs --- should have smaller sizes. Given the lower mean redshift values of the RGs versus the BLs and FSRQs, this effect must be canceled by the higher Doppler factors found for the BLs and FSRQs in order to explain the similar 1-2 year ages of the peaks of the distributions seen in Figure~\ref{fig:TAgeHists}.
To verify this, we calculate the age of components in the plasma frame of each jet by multiplying the redshift-corrected age by the Doppler factor of variability, so that $T_{\text{age}}^{\text{plasma}} = \delta_{\text{var}} T_{\text{age}} / (1+z)$. The distributions for each source type are shown in Figure~\ref{fig:TAgeHists} (right). The FSRQ and BL distributions differ at a $\sim3\sigma$ level, with $\mathcal{D} = 0.244$ and $p = 0.003$. However, both distributions peak at a knot age $\sim10$ yr. A large number of BL knots have shorter ages, while there are more knots in FSRQs with greater ages. Scaling by $\delta_{\text{var}}$ separates the RG distribution from the FSRQ and BL ($\mathcal{D} = 0.813,\ p = 3.2 \times10^{-16}$ and $\mathcal{D} = 0.718,\ p = 4.9 \times10^{-11}$ respectively). The typical intrinsic age of a RG knot is $\sim1$ yr, a factor of ten lower than for FSRQs and BLs, roughly in accord with the ratios of physical sizes of their knots.

It is important to point out the effect that our roughly monthly observation cadence might have on the observed lifetimes of knots. Very bright components in the jets of blazars are easy to identify in the radio maps and fit with Gaussian profiles (as in $\S$2), thus would need to have extremely fast decays ($<1$ month) to not be seen at other epochs. Given the paucity of short-decay knots seen in the distributions in Fig.~\ref{fig:TAgeHists}, this scenario is unlikely. However, faint components are difficult to distinguish from noise, especially if such knots travel down the jet quickly with short decay times (e.g., if their decay were $<0.25$ yr, as we require at least 4 observations of a knot for it to be properly identified). Such knots would be missed in the above analysis. Fast, faint components have been seen in some sources \citep[e.g., the BL 0219+428 (3C 66A);][]{Jorstad2005}, but to detect them requires denser temporal sampling.

\subsection{Gamma-Ray Connection}
\label{subsec:GammaRayConnection}

It is well established that there exists a connection between the \gammaray\ state of blazars and their parsec-scale jet behaviors. \citet{Nieppola2011} have shown that there is a statistically significant correlation between the \gammaray\ and 37 GHz radio light curves. The brightest \gammaray\ flares tend to coincide with with the early stages of a millimeter-wave outburst \citep[][]{LeonTavares2011}. Such trends are not universal: for example, at 15 GHz a significant correlation was only found in 4 of 41 analyzed blazars \citep{MaxMoerbeck2014}. However, the F-GAMMA program has found the radio variations of 54 \gammaray\ bright blazars that lag the \gammaray s, with delays systematically decreasing as the observed frequency changes from centimeter to millimeter/submillimeter \citep{Fuhrmann2014}. With regard to knots traveling down the parsec-scale jet, studies of individual sources have revealed a wide diversity in the behavior of blazars. The majority of \gammaray\ outbursts appear to be connected with the propagation of a knot down the jet \citep[e.g.,][]{Jorstad2001b,Marscher2010, Agudo2011, Jorstad2013b, Morozova2014, Jorstad2017}. In several of these cases, the flux profile of the \gammaray\ outburst can be connected with the passage of superluminal knots through the core or quasi-stationary features \citep[e.g,][]{Agudo2011, Weaver2019, Liodakis2020}. However, these features are usually located parsecs from the central black hole, at odds with the small sizes required by causality arguments from the short timescales of high-energy variability \citep[][]{Aharonian2007, Tavecchio2010, Nalewajko2014}. Another problem is the origin of high-energy outbursts known as ``orphan" flares \citep[e.g.,][]{Krawczynski2004, Marscher2010, MacDonald2015, Sobacchi2021, Yang2021}. During such flares, the \gammaray\ flux may increase by a factior $\sim100$ and yet little variability is seen at longer wavelengths, with no interaction between moving and stationary features observed in VLBA images.

While a full analysis of the relationship between the parsec-scale jet kinematics of the blazars in our sample and their \gammaray\ behavior is beyond the scope of this work, here we present some simple relationships and correlations. It has been shown by the MOJAVE survey that the jets of \gammaray\ bright quasars have faster average apparent motions and brighter parsec-scale cores than those with weaker \gammaray\ fluxes \citep[e.g.,][]{Kovalev2009, Lister2009, Lister2016}. This relation implies that the jets of brighter \gammaray\ quasars have higher Doppler boosting factors. In order to determine if there is a relationship between the observed physical parameters of the sources in our sample and their \gammaray\ flux, we calculate the maximum \gammaray\ flux, $S_{\gamma,\text{max}}^{\text{obs}}$, (ignoring upper-limits) of each source according to their 10-yr \fermi-LAT light curves.

In the analyses that follow, we have corrected the fluxes for redshift effects through $S = S^{\text{obs}} (1+z)^{1+\alpha}$, where $S^{\text{obs}}$ is the observed flux, $z$ is the redshift, and $\alpha$ is the spectral index of the source at the observed frequency. The $(1+z)^\alpha$ factor is the k-correction, while the $(1+z)^1$ factor accounts for the change in frequency interval. In the radio band, $\alpha_{43} \sim 0$. For the \gammaray\ spectral index, we have taken the observed power-law index $p$ from the \fermi\ 4$^{\text{th}}$ Catalog of AGN \citep{Ajello2020}, regardless of which spectral model best-fits each source. Then, $\alpha_\gamma = |p + 1|$.

\begin{figure*}
    \figurenum{30}
    \begin{center}
        \includegraphics[width=0.75\textwidth]{{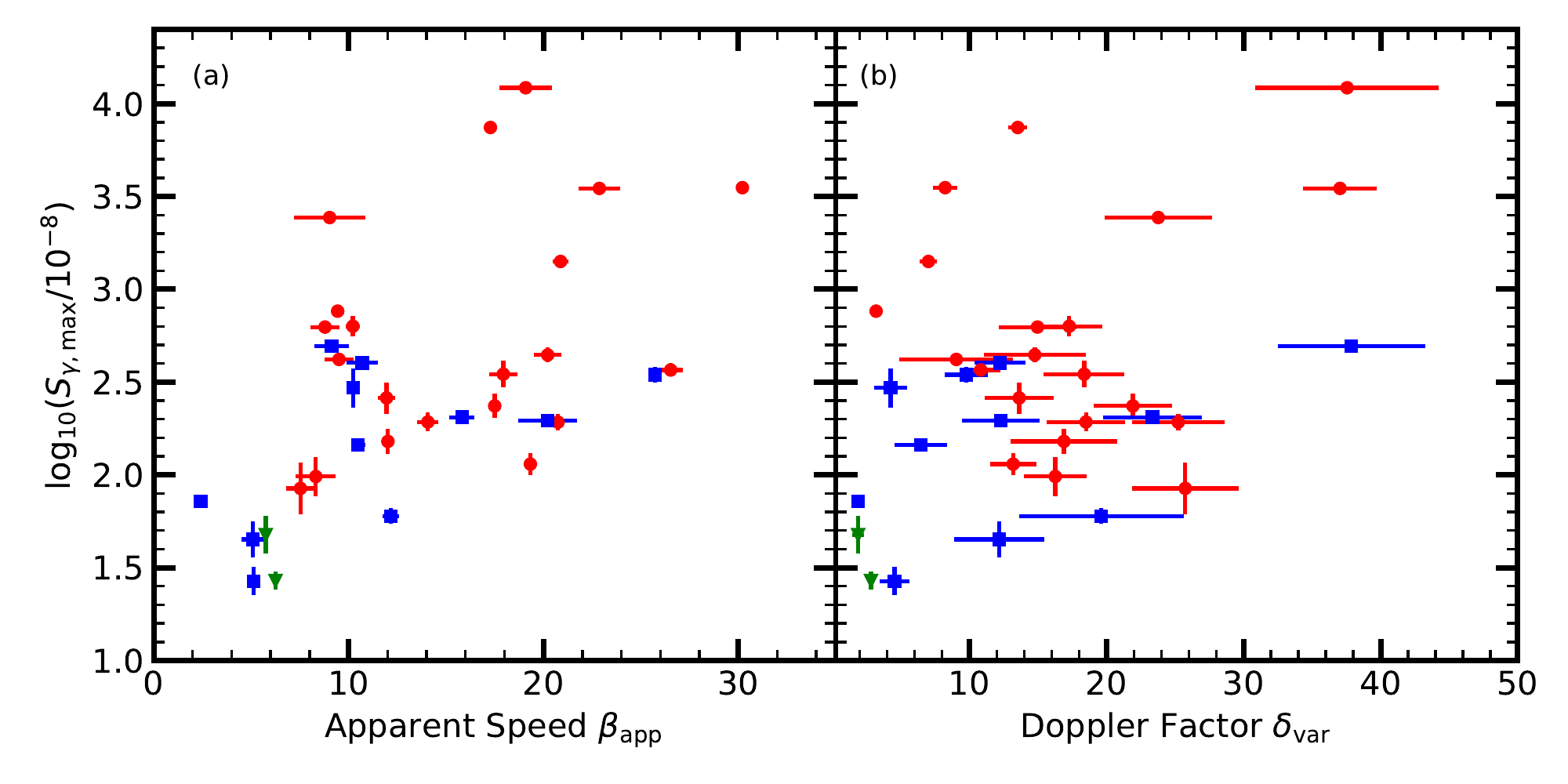}}
        \caption{Maximum \gammaray\ flux $S_{\gamma,\text{max}}$ (scaled by $10^{-8}$ ph cm$^{-2}$ s$^{-1}$) versus \emph{(a)} apparent speed $\beta_{\text{app}}$ and \emph{(b)} Doppler factor $\delta_{\text{var}}$ of the ``typical" knot in FSRQs (red circles), BLs (blue squares), and RGs (green triangles). \label{fig:GammaMaxParams}}
    \end{center}
\end{figure*}

Figure~\ref{fig:GammaMaxParams} shows the relationship between $S_{\gamma,\text{max}}$ and the apparent speed $\beta_{\text{app}}$ and Doppler factor $\delta_{\text{var}}$ of the ``typical" knot of each source. The sample as a whole has a weak positive correlation between the physical parameters in the jet and $S_{\gamma, \text{max}}$, with Pearson's $r$ correlation coefficients of $0.361$ and $0.274$ for $\beta_{\text{app}}$ and $\delta_{\text{var}}$, respectively. The significant scatter is indicative of the diversity in knot behaviors seen in each jet. Our sample consists of too few sources of each type to determine if there is a difference between the subclasses. However, even at the maxima in their \gammaray\ light curves BLs and RGs are in general fainter than FSRQs, with lower speeds and Doppler factors \citep[as expected in the standard AGN unification scheme; e.g.,][]{Urry1995}.

\begin{figure}
    \figurenum{31}
    \begin{center}
    \includegraphics[width=0.45\textwidth]{{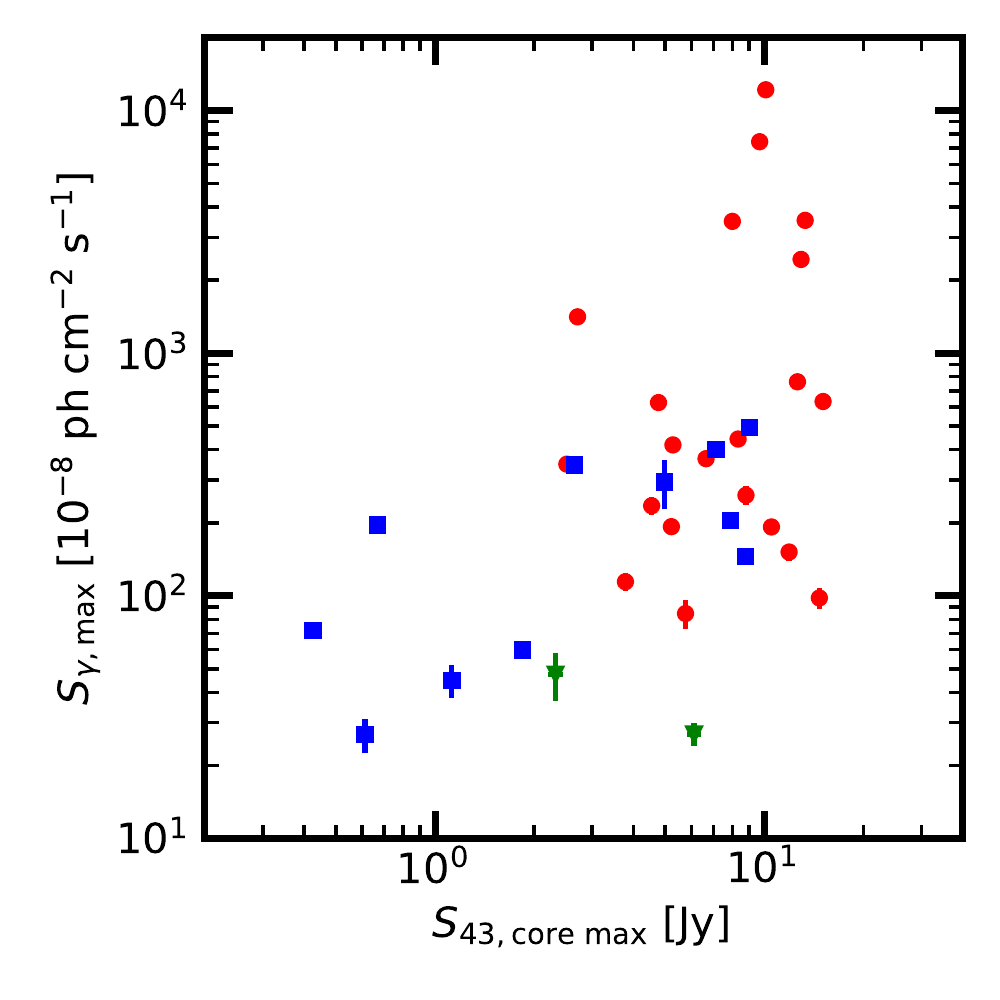}}
    \caption{Maximum \gammaray\ flux, $S_{\gamma,\text{max}}$ versus maximum core flux density at 43 GHz, $S_{43,\text{core\ max}}$, for FSRQs (red circles), BLs (blue squares), and RGs (green triangles). \label{fig:GammavCore}}
    \end{center}
\end{figure}

To further investigate a connection between the \gammaray\ and 43 GHz radio behavior, in Figure~\ref{fig:GammavCore} we show $S_{\gamma,\text{max}}$ compared to the maximum flux density of the core component \emph{A0} in each source in Table~\ref{tab:AvePhysParams} for which the physical parameters were calculated rather than estimated from the $\Gamma \theta_\circ \sim 0.2$ relation. We find a $\sim3\sigma$ positive correlation between $S_{\gamma,\text{max}}$ and the maximum core flux density of each source, with a Pearson's $r$ coefficient of $0.465$. We also see that the results of the core brightness temperature distributions from $\S$\ref{subsec:ObsBrightnessTemps} extend to the maximum core flux densities, with the FSRQ cores generally brighter than the BL and RG cores.

The correlations observed here, while weak, support the conclusion of a connection between the activity states of a blazar at 43 GHz and at \gammaray\ energies. In general, the brighter the 43 GHz core of a source, and the faster the knots in a source appear to move down the jet, the brighter the jet can be at \gammaray s. We plan to perform correlation analyses between the \gammaray\ and 43 GHz core light curves and comparisons between knot time of ejections and \gammaray\ outbursts for the whole sample in a future work.

%-------------------------------------------------------------------------------
%---   Summary   ------------------------------------------------------------
%-------------------------------------------------------------------------------

\section{Summary}
\label{sec:Summary}

We have presented a new analysis of the jet kinematics in 35 blazars and 3 radio galaxies observed with the VLBA at 43 GHz over ten years under the VLBA-BU-BLAZAR program. The total observed time period is 2007 June through 2018 December, resulting in 3705 total intensity images of the jets. The first five years of data were presented in \citet{Jorstad2017}, and are included in this study as well after applying a new analysis procedure. The primary findings and conclusions of the ten years of data are summarized as follows:

\begin{enumerate}
    % 1
    \item We have implemented a new piece-wise linear fitting method to derive the kinematics of moving components in the jets. This algorithm allows us to estimate the distance down the jet at which acceleration (if any) occurs. We have identified 559 distinct emission features, of which 38 are the 43 GHz cores, which are assumed to be stationary. We have further classified 96 components as quasi-stationary, leaving 425 knots that move with respect to the VLBI core. Among these moving knots, 75 exhibit non-linear motion, for which we calculate the acceleration parallel and perpendicular to the jet direction.
    % 2a
    \item To incorporate the presence of lower limits to the brightness temperature of 10.1\% of all components into our analysis, we have used the Kaplan-Meier estimator for the survival function of the cores and knots to derive a median value of the brightness temperature for each source. We plan to provide a deeper analysis of the survival functions of the core and knot brightness temperatures, covering the temporal properties and whether all sources can be described with a similar parameterized survival function, in a future work (Weaver, in prep.).
    % 2b
    \item We find, despite a low significance in a KS test due to a small number of sources, that FSRQs and BLs have more intense cores than those of RGs, but the knots in the extended jets of BLs are significantly less intense than the knots of RGs and FSRQs. The brightness temperature distributions of the non-core components can be explained by Doppler boosting of the expected intrinsic brightness temperature.
    % 3
    \item With 10.5 years of data, we have analyzed parsec-scale jet position angle variations of each source. We have quantitatively placed the jets into four categories. The majority of jets (23) have jet position angles that are constant with time, while a set of six sources has a very wide jet that is constant in position angle. Two sources, the FSRQs 0420$-$014 and 1226+023, have jet position angles that change linearly with time. The remaining seven sources have complex variations of position angle, which we have fit with 3rd-order splines.
    % 4
    \item Due to our piece-wise linear fitting method, any individual moving knot can have multiple motions over different time periods. We therefore have analyzed a total of 529 knot speeds. These speeds range from $0.01c$ to $78c$. The FSRQs generally have the fastest knots, with a distribution that peaks between $8c$-$10c$ and extends to $\sim78c$. The distribution of apparent speeds for BLs peak between $0c$-$2c$, as well as $8c$-$10c$, but only extends up to $\sim30c$. RGs have a very narrow distribution of speeds, only extending to $\sim10c$. We find superluminal motion in at least one knot of the RG 0316+413 of $\beta_{\text{app}} = 1.38c \pm 0.03c$, while the remaining superluminal knots in RGs come from 0415+379 and 0430+052. The moving knots in FSRQs tend to travel in the same direction, while knots of BLs show larger variations in the direction of motion.
    % 5
    \item On average, there exist $\sim2$ quasi-stationary features in the jet of each FSRQ, $\sim3$ in each BL, and $\sim6$ in each RG. Almost half of these features have appeared or disappeared during the 10.5-year observation period, indicating that they may be transient.
    The locations of the majority of stationary features are concentrated within 1 mas of the core (projected), independent of the subclass.
    We have characterized the fluctuations in the positions of these features using confidence ellipses, which show that the stationary features in RGs tend to ``slosh" around the average position in a direction parallel to the jet axis. However, there is no clear preference of directional shifts for the stationary features of BLs and FSRQs.
    % 6
    \item We have detected 104 different acceleration regions in the sample.  Only a few BLs have knots with discernible acceleration. We find that, statistically, these accelerations are consistent with changes in the Lorentz factor of the knots rather than changes in the jet direction. Parallel to the jet axis, we see an initial positive acceleration close to the core of jet ($<5$ pc), while beyond $10$ pc the knots tend to decelerate. Perpendicular to the jet axis, we find an initial positive acceleration for projected distances $0$-$1$ pc from the core and negative acceleration $1$-$2$ pc from the core. Beyond $\sim5$ pc, we see very small magnitudes of acceleration. Similar trends are found farther from the core at longer wavelengths \citep[e.g.,][]{Homan2015}.
    % 7
    \item To test whether interactions with standing shocks have an effect on the moving knots, we have paired the acceleration regions with the closest (on average) stationary features. We have found that few pairs fail to follow an expected 1:1 relation, indicating that accelerations take place at or beyond the stationary features. Such behavior is expected if the stationary features are structural components such as recollimation shocks.
    % 8
    \item We have derived Doppler factors of variability, $\delta_{\text{var}}$, for a subset of reliable knots in the sample, which, along with the measured apparent speeds, have allowed us to calculate the Lorentz factors $\Gamma$ and viewing angle $\Theta_\circ$. The flux decays of the knots in our sample are statistically likely to be caused by radiative losses (modulated by light-travel delays) rather than adiabatic expansion.
    % 9
    \item The Lorentz factors of knots in the jets of our sample range from $2$-$60$, with faster knots observed in FSRQs than in BLs and RGs.
    Our results agree with the expectation that the jets of RGs lie at larger viewing angles with respect to the line of sight, with an average $\Theta_\circ \sim 15\degr$-$30\degr$, compared to the FSRQ and BL jets with peak viewing angles $\sim0\degr$-$4\degr$.
    % 10
    \item For most of the jets in our sample, we see a wide diversity in the apparent speeds and Doppler factors of knots moving down the parsec-scale jet. For sources with many observed knots (such as the RG 0415+379), we see that this diversity can be explained by a distribution of physical parameters from which a knot's parameters can be drawn statistically. Whether the distributions of knot parameters are similar for different sources remains unclear, as many of our sources will likely require a further 5 to 10 years of monitoring for enough knots to be observed to determine their distributions robustly.
    % 11
    \item We have defined the ``typical" knot in each jet by choosing the fastest knot with the smallest uncertainty in $\beta_{\text{app}}$ and $\delta_{\text{var}}$ to represent the jet as a whole. From these typical values, we calculate the Lorentz factor and viewing angle of each jet, and find that, on average, FSRQ jets have higher values of $\delta_{\text{var}}$ and $\Gamma$ compared to BLs, while knots in RGs are slower and have much wider viewing angles and intrinsic opening semi-angles. The physical parameters of each jet as calculated using the ``typical" knot are statistically similar to other estimates of the physical parameters determined through lower-frequency monitoring \citep[e.g.,][]{Liodakis2018}, which probes farther distances from the core.
    % 12
    \item In general, the knots of FSRQs are longer lived than the knots of BLs and RGs, reaching farther projected distances from the core. The knots of FSRQs and BLs have flux decay rates that are related to their age roughly as $k \sim -2 T_{\text{age}}^{-1}$ despite the differences in their age distributions, while RG knots follow $k \sim 2.4 T_{\text{age}}^{-0.65}$. The distribution of ages of all subclasses peak in the 1-2 yr range despite large differenes in  the physical sizes of knots of RGs versus BLs and FSRQs. This can be explained by the lower Doppler factors of the RGs.
    % 13
    \item We find weak correlations between the maximum \gammaray\ flux and the typical apparent speeds and Doppler factors of jets in our sample.
    We also find that the blazars with brighter 43 GHz cores typically reach higher maximum \gammaray\ fluxes. We plan to perform a temporal analysis comparing the ejection times of knots in our sample with \gammaray\ outbursts in a future study.
\end{enumerate}

%-------------------------------------------------------------------------------
%---   ACKNOWLEDGMENTS   -------------------------------------------------------
%-------------------------------------------------------------------------------

\acknowledgments

We thank the anonymous referee for comprehensive and constructive feedback that greatly improved this paper.
The research at Boston University was supported in part by NASA grants
80NSSC17K0649, 80NSSC20K1567, and 80NSSC20K1566 (Fermi Guest Investigator Program),
the NRAO Student Observing Support Program, and
Massachusetts Space Grant 316080.
I.A. acknowledges financial support from the Spanish ``Ministerio de Ciencia e Innovaci\'on'' (MCINN) through grants AYA2016-80889-P and PID2019-107847RB-C44, and through the ``Center of Excellence Severo Ochoa'' award for the Instituto de Astrof\'isica de Andaluc\'ia-CSIC (SEV-2017-0709).
The VLBA is an instrument of the National Radio Astronomy Observatory. The National Radio Astronomy Observatory is a facility of the National Science Foundation operated by Associated Universities, Inc. This research has made use of the NASA/IPAC Extragalactic Database (NED), which is operated by the Jet Propulsion Laboratory, California Institute of Technology, under contract with the National Aeronautics and Space Administration.
The optical data used in this work were obtained using the 1.8 m Perkins Telescope Observatory in Flagstaff, Arizona, USA, which is owned and operated by Boston University.
This publication makes use of data obtained at the Mets\"ahovi Radio Observatory, operated by Aalto University in Finland.
This research made use of Astropy,\footnote{\url{https://www.astropy.org}}, a community-developed core Python package for Astronomy \citep{Astropy2013, Astropy2018}.

%-------------------------------------------------------------------------------
%---   APPENDIX   --------------------------------------------------------------
%-------------------------------------------------------------------------------

\appendix

\section{Survival Analysis of Jet Component Brightness Temperatures}
\label{app:survival}

Treatment of data with lower and upper limits (or, more generally, right- and left-censored data, respectively) in astronomy has been performed in a number of fashions, of which many are biased or incorrect \citep{Feigelson2012}. For example, if the number of limits is small, then they are unlikely to change the result of statistical tests, however, they will bias the results as the number of limits in a sample grows. The most statistically rigorous procedures consider all the data points, both detected values and censored data, to model the properties of a parent distribution of the variable in question under specific mathematical constraints. Such procedures fall under the umbrella of ``survival analysis" \citep{Avni1980,Avni1986,Feigelson1985,Schmitt1985,Isobe1986,LaValley1992}. A few methods for censored data have been developed specifically for astronomical contexts \citep{Akritas1995, Akritas1996}.

Most survival analysis techniques are based on estimating the ``survival function" of an entire data set (observations and censored data), where the survival function of a parameter, $x$, is

\begin{equation}
    \mathcal{S}(x) = \mathrm{Pr}(X > x).
\end{equation}

\noindent That is, $\mathcal{S}(x)$ is the probability that the value of $x$, $X$, is above a specified value. Alternatively, one may estimate the ``hazard function", $\mathcal{H}(x)$:

\begin{equation}
    \mathcal{H}(x) = \frac{\mathrm{d\ ln}\mathcal{S}(x)}{\mathrm{d}x} = \lim_{\Delta x \rightarrow 0} \frac{\mathrm{Pr}(x < X < x + \Delta x | X > x)}{\Delta x}.
\end{equation}

The focus on the survival and hazard function of data in survival analysis is due to their relation to the common statistical concepts of probability density functions (PDFs), $f(x)$, and cumulative distribution functions (CDFs), $F(x)$, of data through $f(x) = \mathcal{S}(x) \mathcal{H}(x)$ and $F(x) = 1 - \mathcal{S}(x)$. Thus, estimates of $\mathcal{S}(x)$ and $\mathcal{H}(x)$ can describe the underlying parent distributions of the data, even in the presence of censored or truncated\footnote{Truncated data occur when no information, not even an upper or lower limit, is known for a quantity of interest, such as in the case of surveys of astronomical objects being truncated at some sensitivity limit.} data, depending on the estimator.

In order to determine a ``typical" brightness temperature for each source in our sample, we estimate the survival function of the log$_{10}$ of the brightness temperature of cores and knots, $\mathcal{S}(\mathrm{log}_{10}\ T_{\mathrm{b,obs}}^{\mathrm{s}})$, using the Kaplan-Meier estimator implemented as part of the \texttt{lifelines}\footnote{\url{https://lifelines.readthedocs.io/en/latest/index.html}} package \citep{DavidsonPilon2019}. The Kaplan-Meier estimator \citep{KaplanMeier1958} is a non-parametric method to estimate $\mathcal{S}(x)$ and was developed to directly take right-censored data into account:

\begin{equation}
    \hat{\mathcal{S}}(x) = \prod_{i : x_i < x} \left(1 - \frac{d_i}{n_i} \right)
\end{equation}

\noindent where $x_i$ is a discrete value of the parameter of interest $x$ with at least one observed or censored data point, $d_i$ is the number of ``events" (observed data) up to that value, and $n_i$ the total number of data points ``known to survive" past $x_i$ (have a value $x > x_i$ or are censored).

\begin{figure*}
    \figurenum{A1}
    \begin{center}
        \includegraphics[width=\textwidth]{{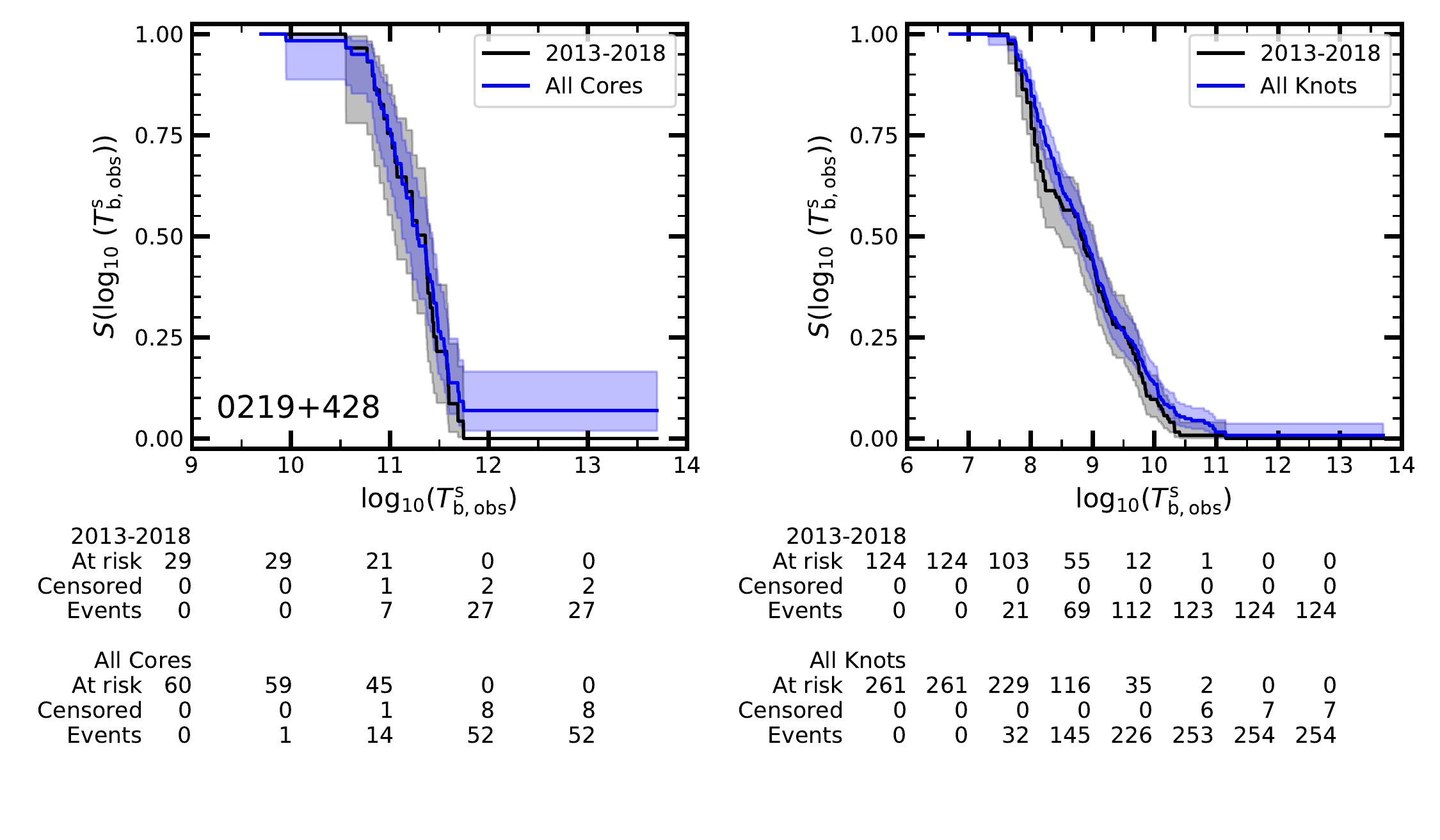}}
        \caption{Kaplan-Meier estimated survival functions of the brightness temperature of cores (left) and knots (right) in the BL 0219+428. The survival function for components observed between 2013 January and 2018 December are shown in black, while those observed between 2007 June and 2018 December are shown in blue. Shaded regions represent 95\% confidence intervals to the survival function. The tables below each panel show the summary statistics of each survival function in the form of the number of unused (``At risk", both observed and censored), censored, and observed (``event") data points at each major interval \citep[included per the recommendation by][]{Morris2019}. \label{fig:KMEstimateSrc}}
    \end{center}
\end{figure*}

Figure~\ref{fig:KMEstimateSrc} shows, as an example, the Kaplan-Meier estimates for the survival function of the cores (left) and knots (right) of the BL 0219+428 for both the time period presented here (2013 January - 2018 December) as well as the total observed time period (2007 June - 2018 December). Figure SET A1 contains similar plots for all sources in our sample. Doubling the number of brightness temperatures used by combining those presented in this work with those presented in \citet{Jorstad2017} yields better estimates for the survival function, increasing the resolution along $T_{\mathrm{b,obs}}^{\mathrm{s}}$ and resulting in narrower confidence intervals along the curve. The two sources added after 2013, the BLs 1652+398 and 1959+650, have the same survival functions presented in the two time periods.

%---- Figure Set A1
\figsetstart
    \label{figset:A1}
    \figsetnum{A1}
    \figsettitle{Core and Knot Brightness Temperature Survival Functions}
        % Number 1
        \figsetgrpstart
        \figsetgrpnum{A1.1}
        \figsetgrptitle{0219+428}
        \figsetplot{{figA1_1.pdf}}
        \figsetgrpnote{Survival functions of the core (left) and knot (right) brightness temperatures for the marked time periods of the BL 0219+428.}
        \figsetgrpend
        %
        % Number 2
        \figsetgrpstart
        \figsetgrpnum{A1.2}
        \figsetgrptitle{0235+164}
        \figsetplot{{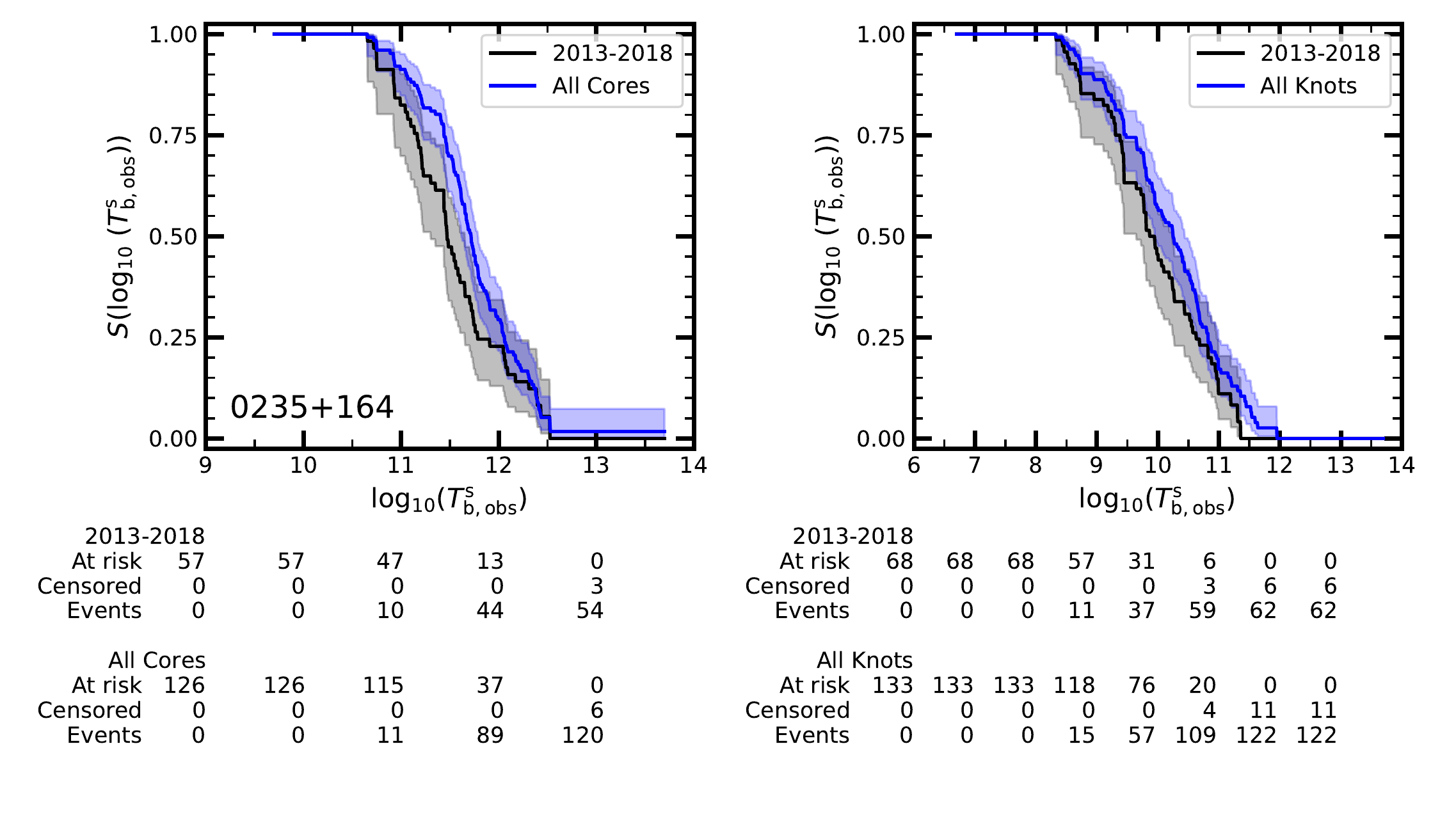}}
        \figsetgrpnote{Survival functions of the core (left) and knot (right) brightness temperatures for the marked time periods of the BL 0235+164.}
        \figsetgrpend
        %
        % Number 3
        \figsetgrpstart
        \figsetgrpnum{A1.3}
        \figsetgrptitle{0316+413}
        \figsetplot{{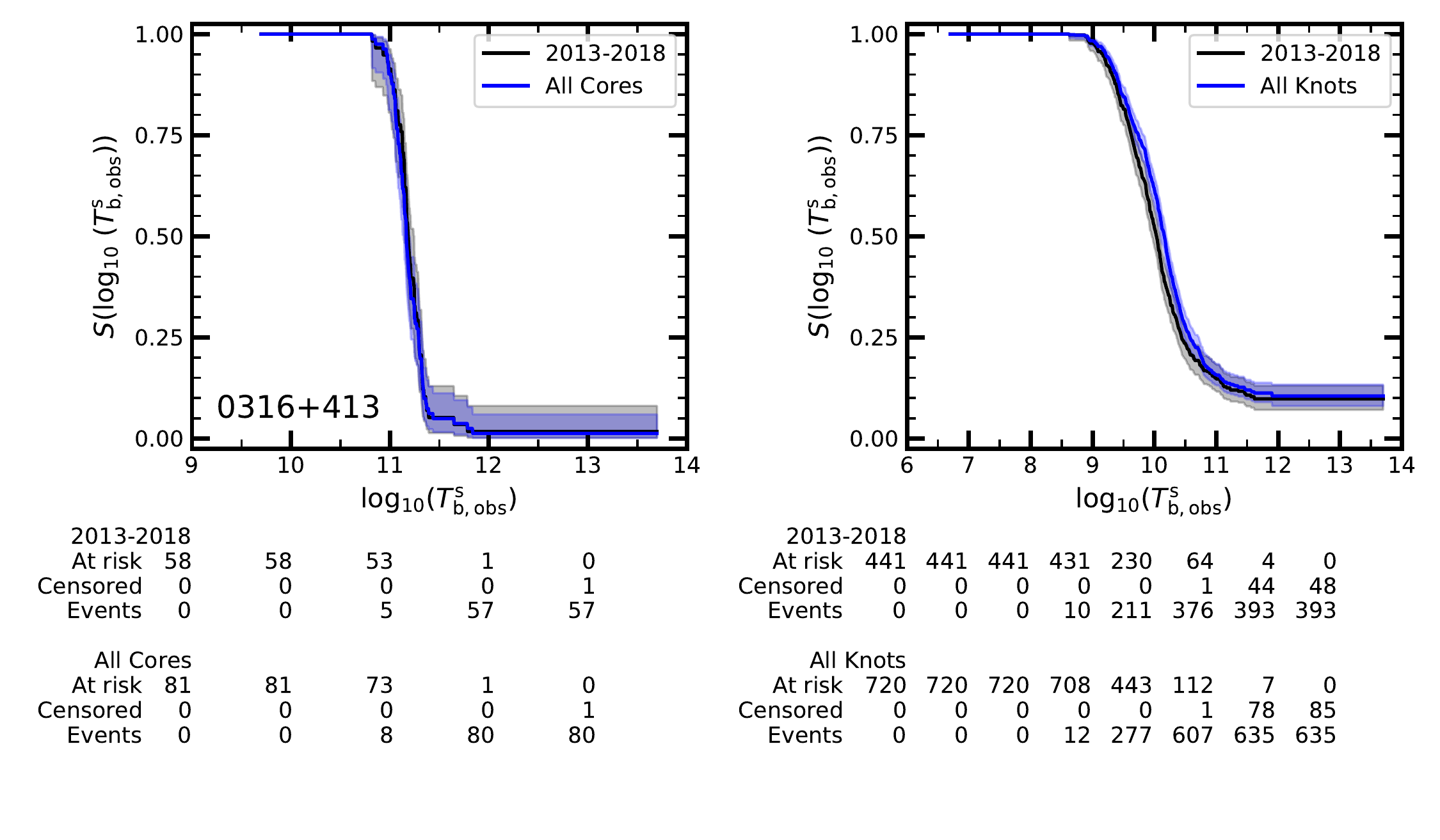}}
        \figsetgrpnote{Survival functions of the core (left) and knot (right) brightness temperatures for the marked time periods of the RG 0316+413.}
        \figsetgrpend
        %
        % Number 4
        \figsetgrpstart
        \figsetgrpnum{A1.4}
        \figsetgrptitle{0336-019}
        \figsetplot{{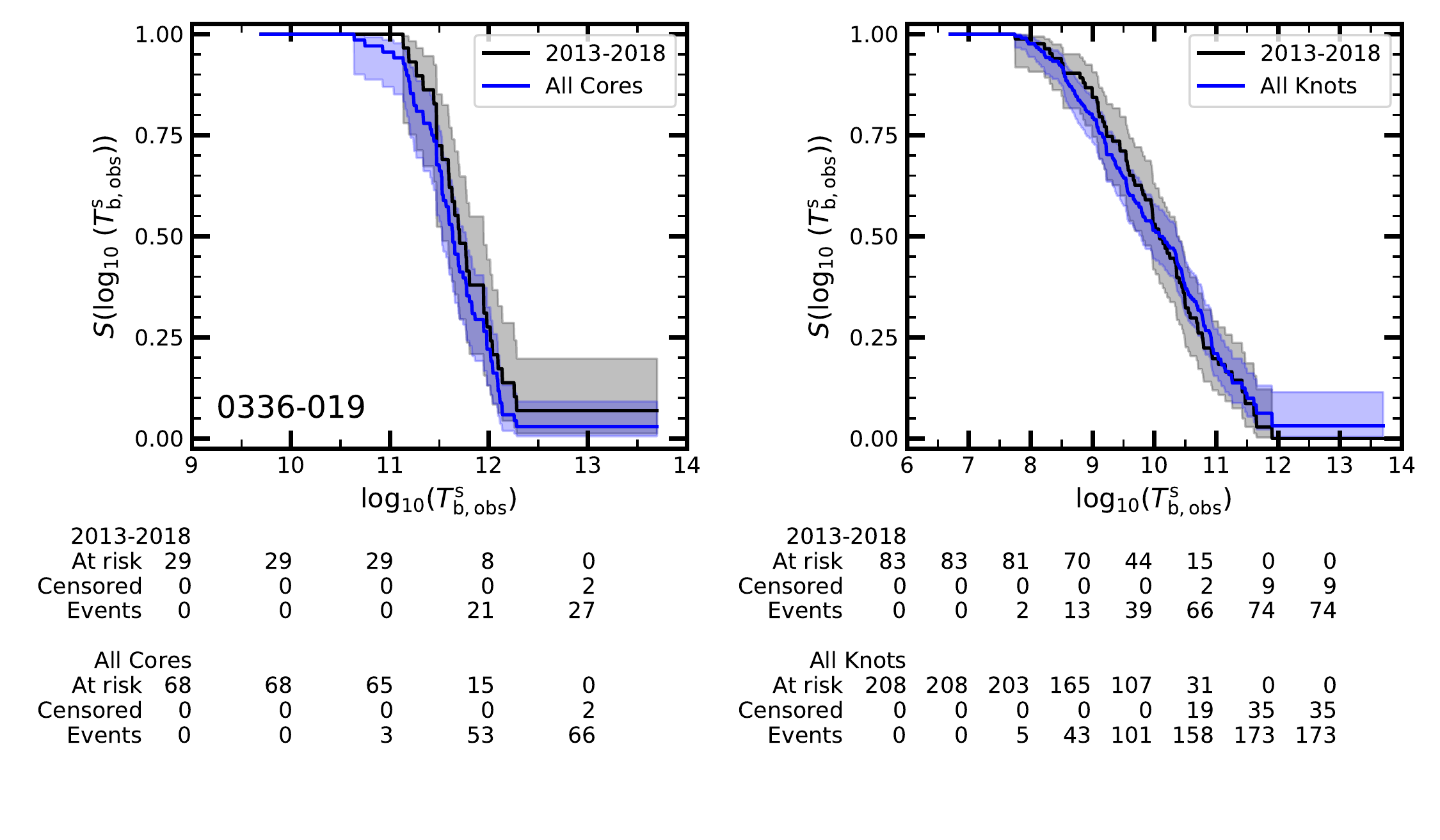}}
        \figsetgrpnote{Survival functions of the core (left) and knot (right) brightness temperatures for the marked time periods of the FSRQ 0336-019.}
        \figsetgrpend
        %
        % Number 5
        \figsetgrpstart
        \figsetgrpnum{A1.5}
        \figsetgrptitle{0415+379}
        \figsetplot{{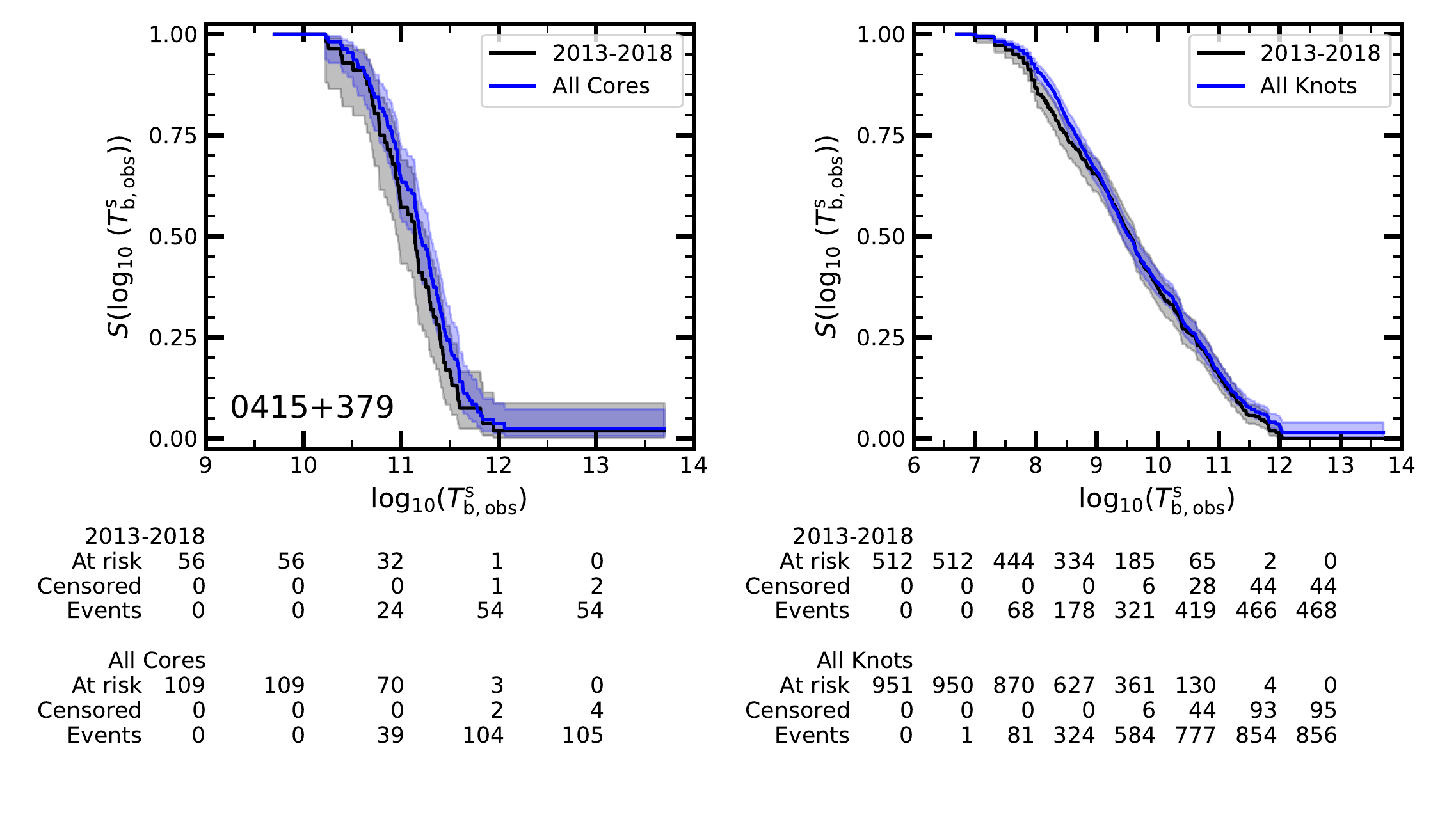}}
        \figsetgrpnote{Survival functions of the core (left) and knot (right) brightness temperatures for the marked time periods of the RG 0415+379.}
        \figsetgrpend
        %
        % Number 6
        \figsetgrpstart
        \figsetgrpnum{A1.6}
        \figsetgrptitle{0420-014}
        \figsetplot{{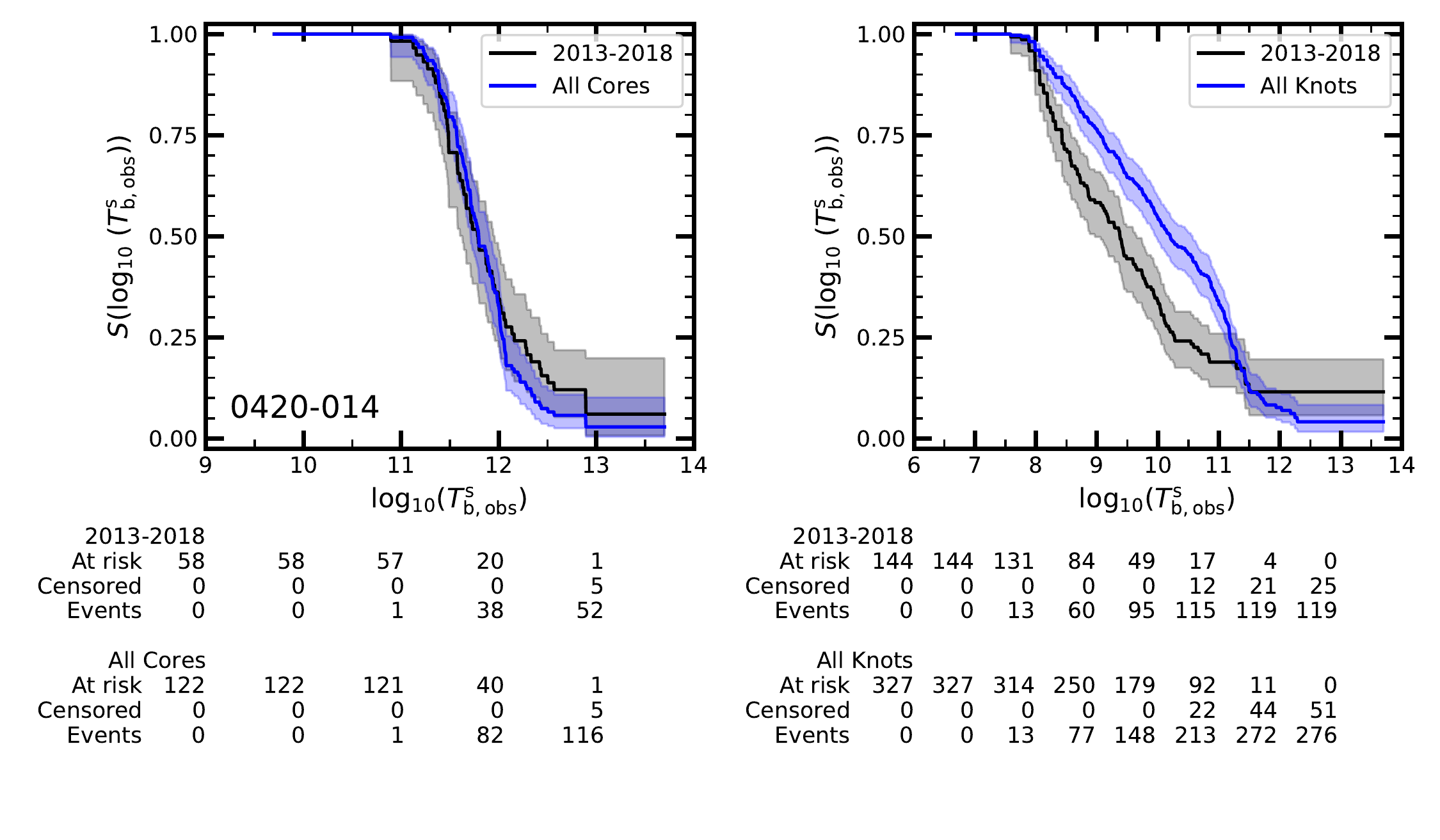}}
        \figsetgrpnote{Survival functions of the core (left) and knot (right) brightness temperatures for the marked time periods of the FSRQ 0420-014.}
        \figsetgrpend
        %
        % Number 7
        \figsetgrpstart
        \figsetgrpnum{A1.7}
        \figsetgrptitle{0420+052}
        \figsetplot{{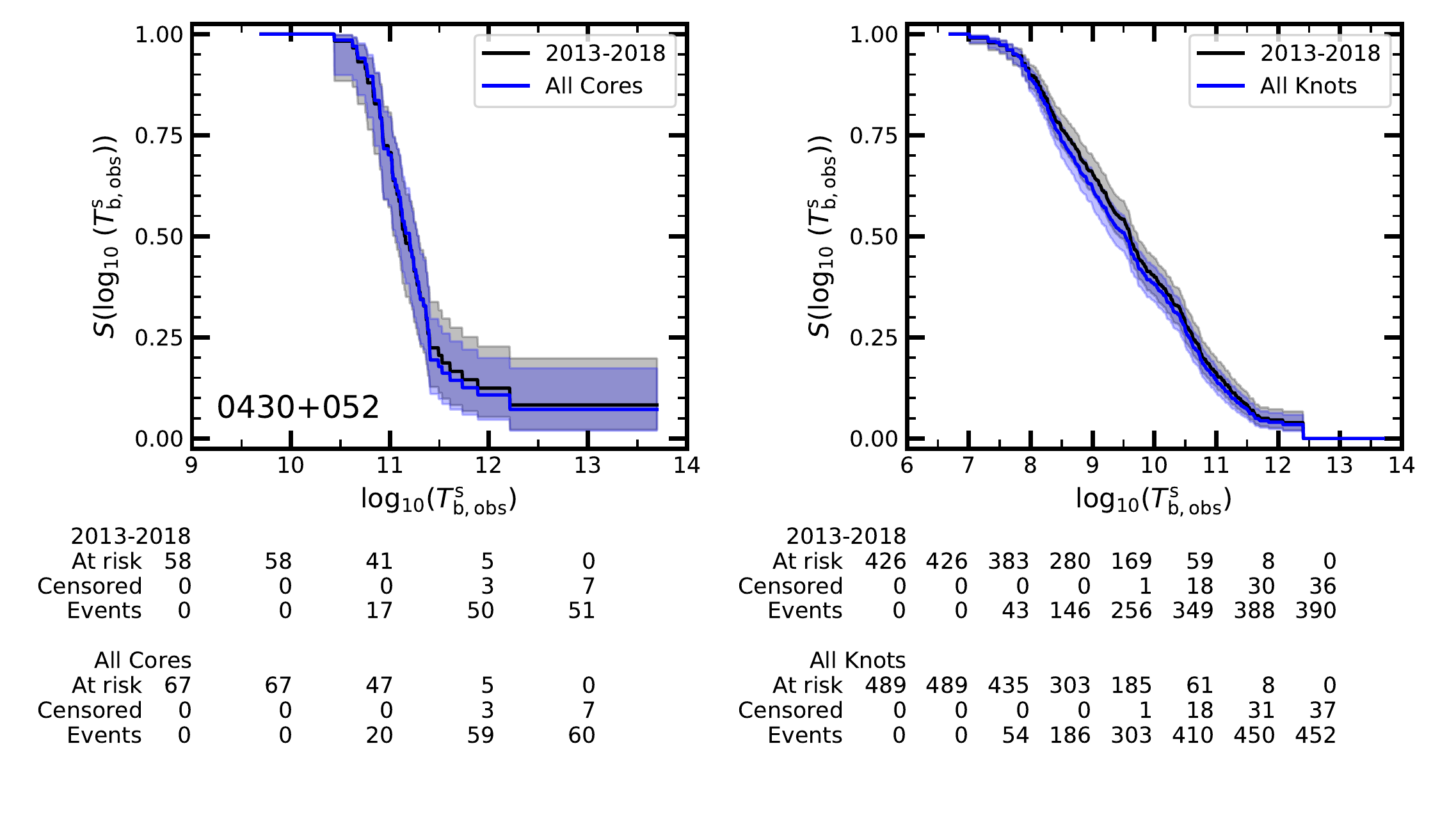}}
        \figsetgrpnote{Survival functions of the core (left) and knot (right) brightness temperatures for the marked time periods of the RG 0430+052.}
        \figsetgrpend
        %
        % Number 8
        \figsetgrpstart
        \figsetgrpnum{A1.8}
        \figsetgrptitle{0528+134}
        \figsetplot{{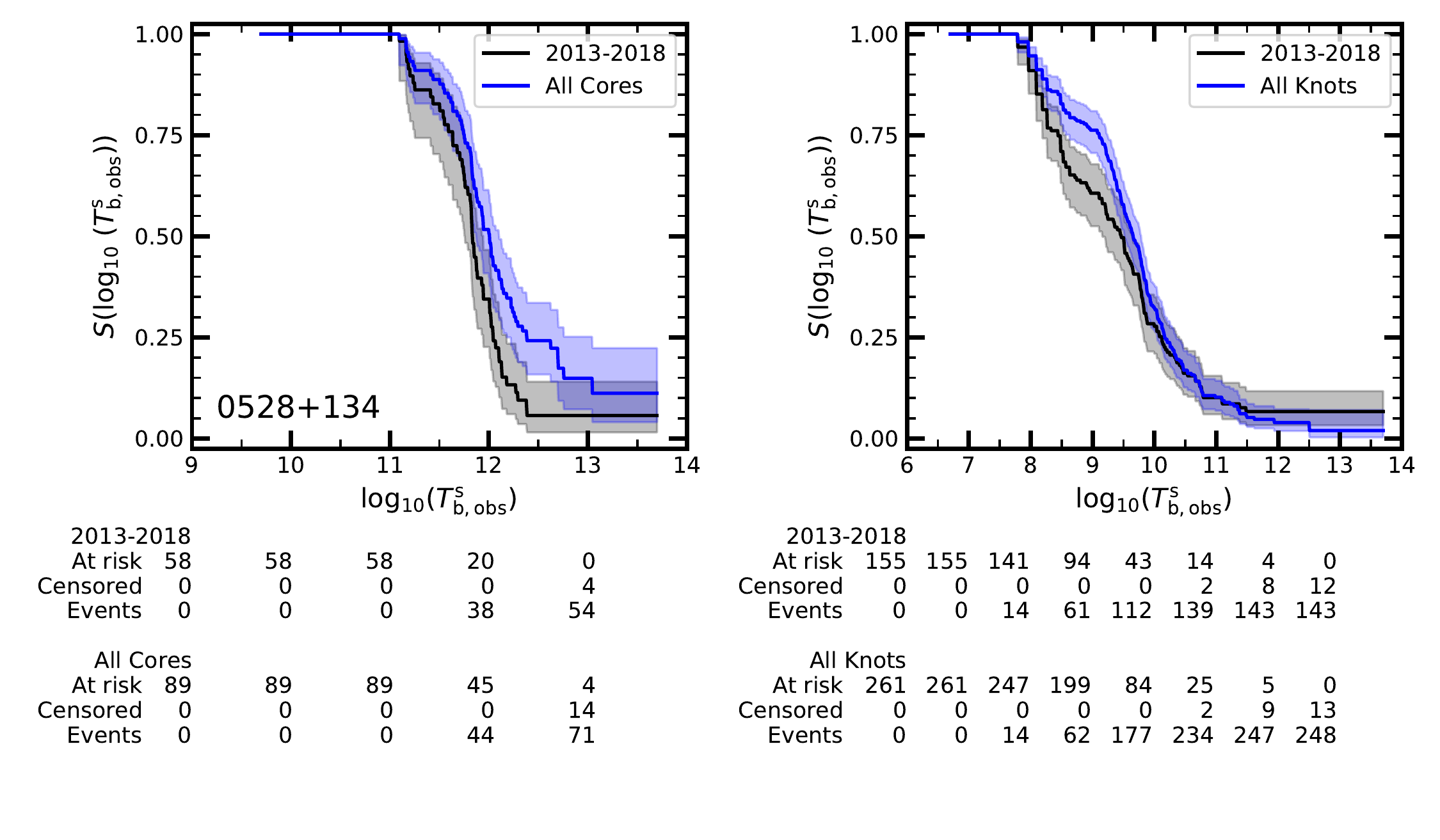}}
        \figsetgrpnote{Survival functions of the core (left) and knot (right) brightness temperatures for the marked time periods of the FSRQ 0528+134.}
        \figsetgrpend
        %
        % Number 9
        \figsetgrpstart
        \figsetgrpnum{A1.9}
        \figsetgrptitle{0716+714}
        \figsetplot{{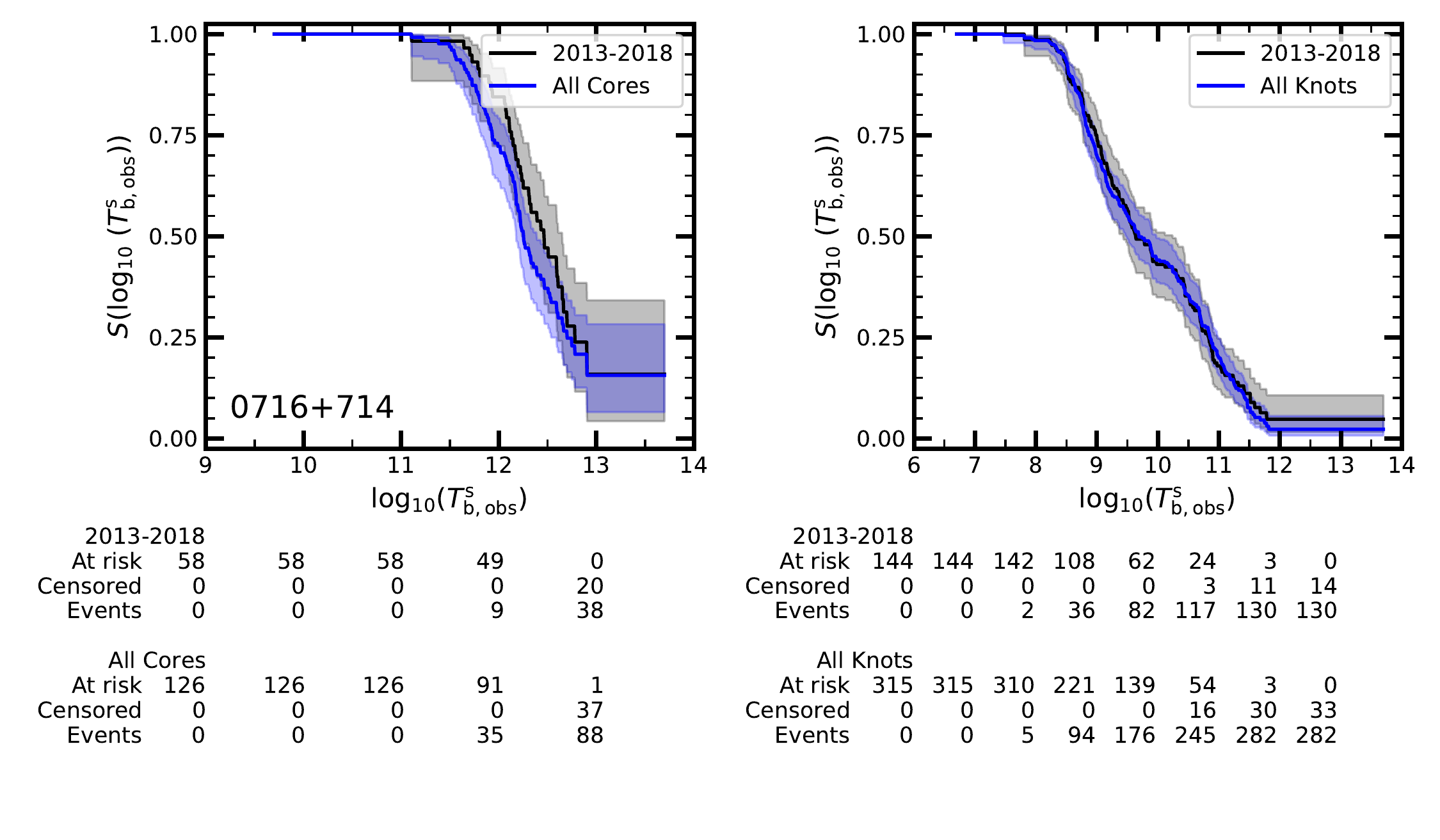}}
        \figsetgrpnote{Survival functions of the core (left) and knot (right) brightness temperatures for the marked time periods of the BL 0716+714.}
        \figsetgrpend
        %
        % Number 10
        \figsetgrpstart
        \figsetgrpnum{A1.10}
        \figsetgrptitle{0735+178}
        \figsetplot{{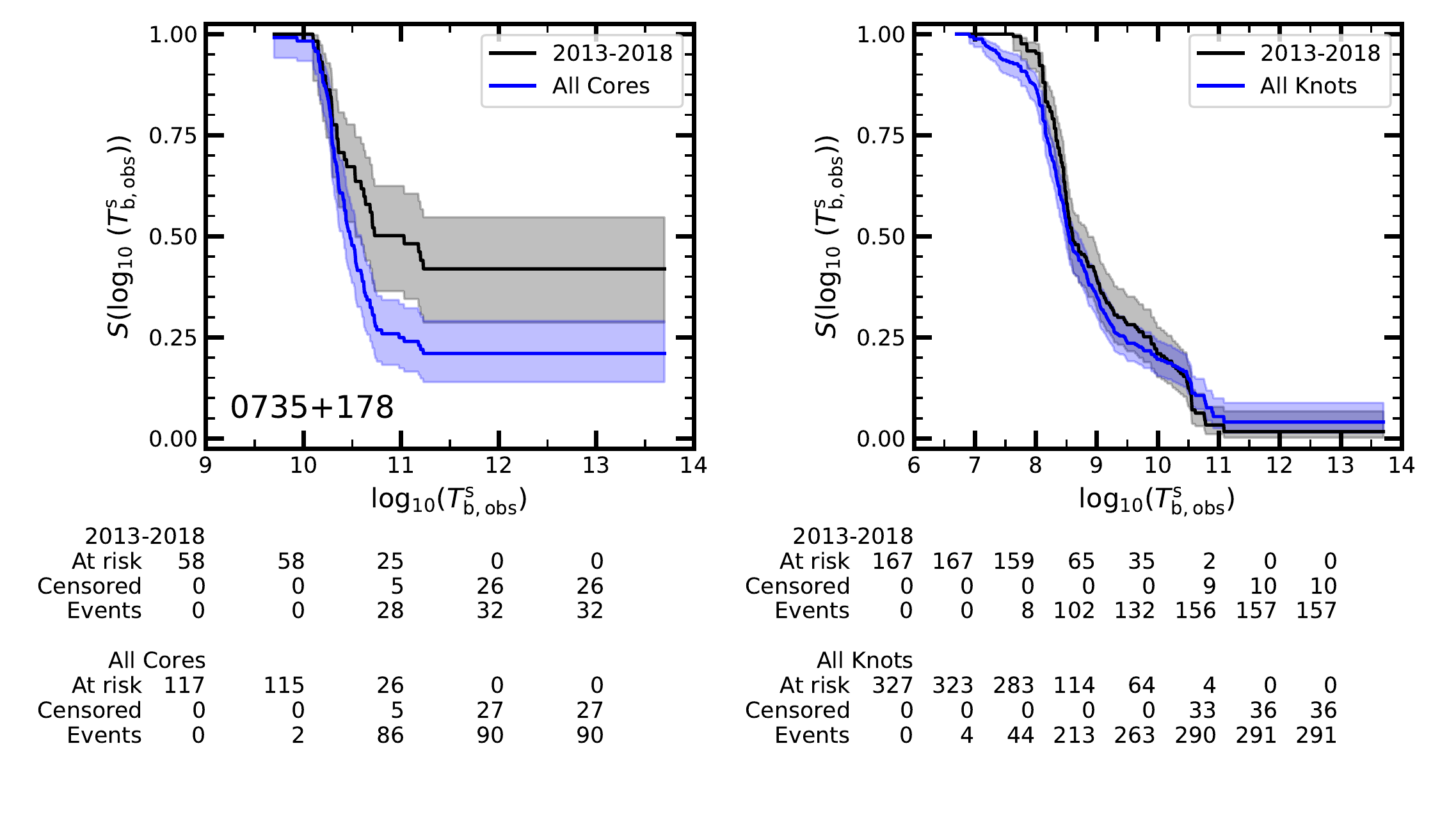}}
        \figsetgrpnote{Survival functions of the core (left) and knot (right) brightness temperatures for the marked time periods of the BL 0735+178.}
        \figsetgrpend
        %
        % Number 11
        \figsetgrpstart
        \figsetgrpnum{A1.11}
        \figsetgrptitle{0827+243}
        \figsetplot{{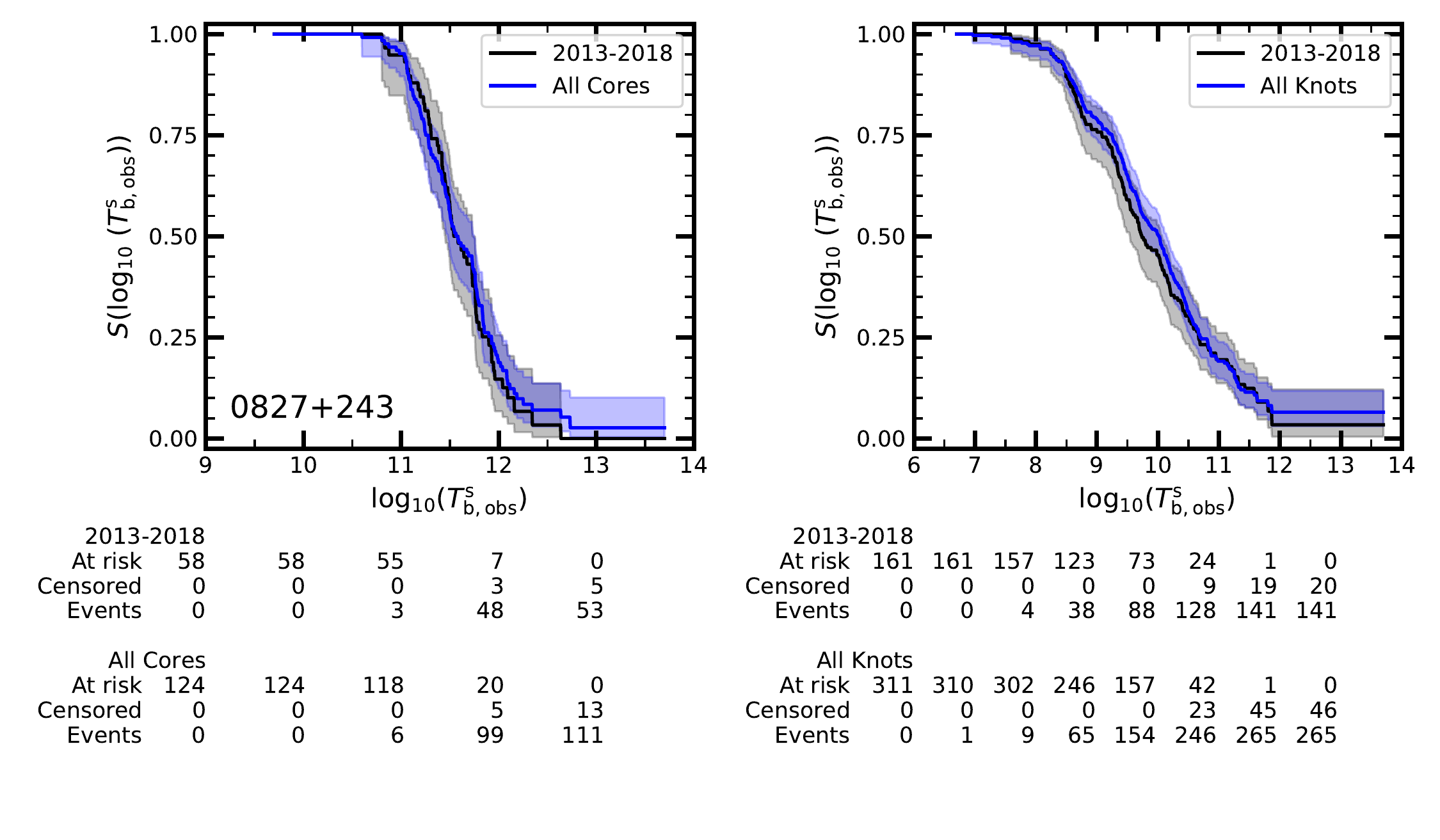}}
        \figsetgrpnote{Survival functions of the core (left) and knot (right) brightness temperatures for the marked time periods of the FSRQ 0827+243.}
        \figsetgrpend
        %
        % Number 12
        \figsetgrpstart
        \figsetgrpnum{A1.12}
        \figsetgrptitle{0829+046}
        \figsetplot{{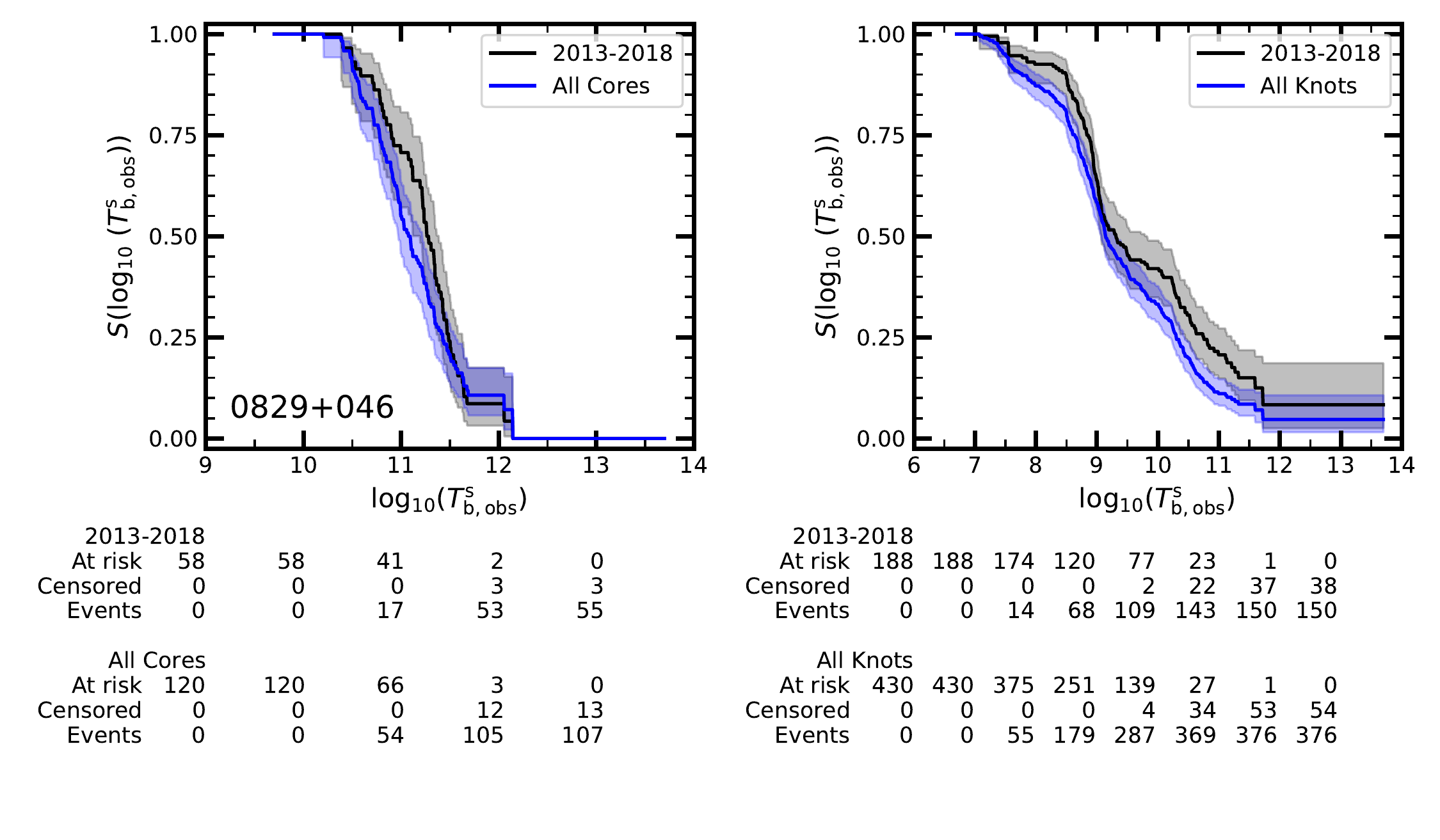}}
        \figsetgrpnote{Survival functions of the core (left) and knot (right) brightness temperatures for the marked time periods of the BL 0829+046.}
        \figsetgrpend
        %
        % Number 13
        \figsetgrpstart
        \figsetgrpnum{A1.13}
        \figsetgrptitle{0836+710}
        \figsetplot{{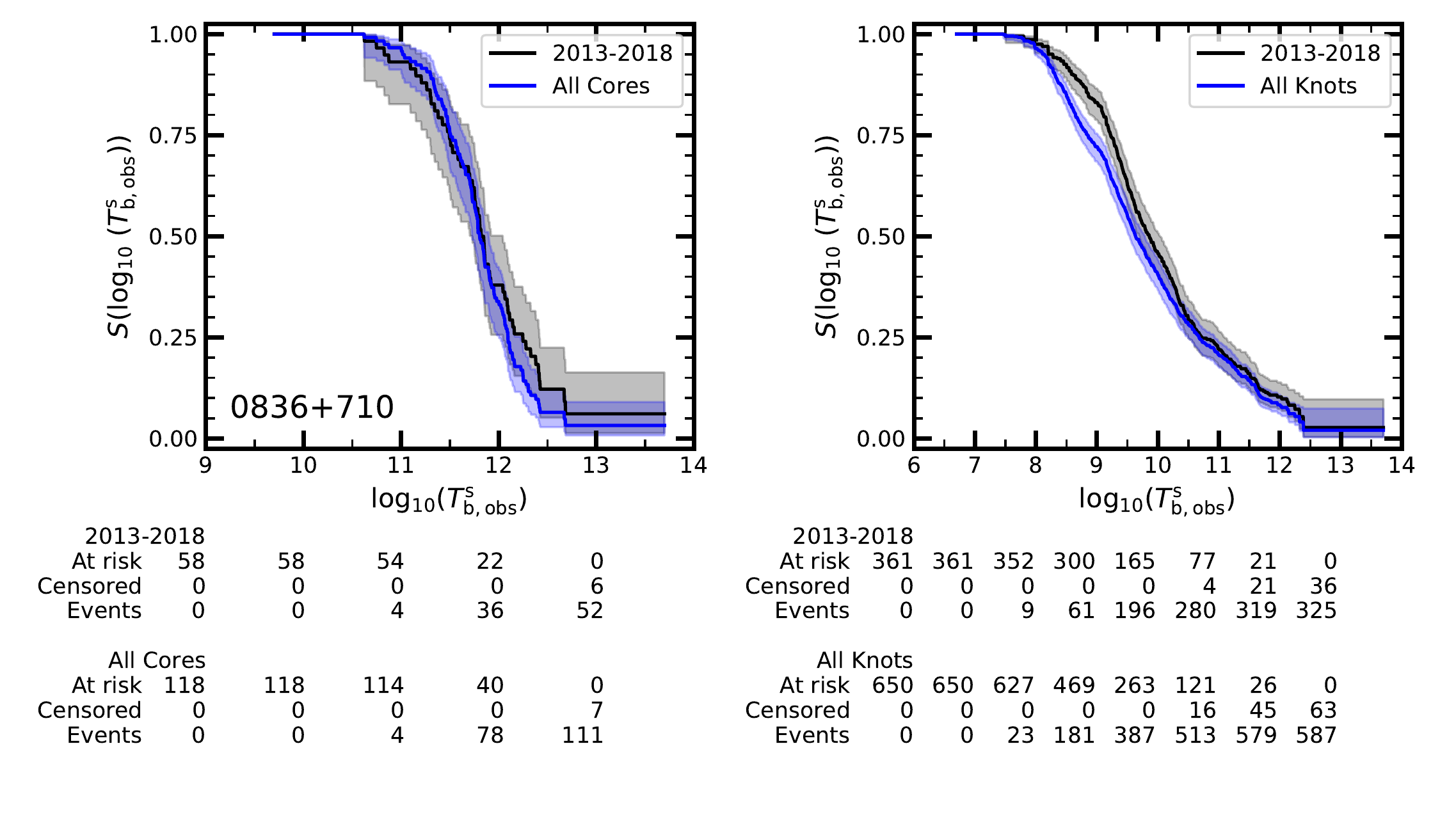}}
        \figsetgrpnote{Survival functions of the core (left) and knot (right) brightness temperatures for the marked time periods of the FSRQ 0836+710.}
        \figsetgrpend
        %
        % Number 14
        \figsetgrpstart
        \figsetgrpnum{A1.14}
        \figsetgrptitle{0851+202}
        \figsetplot{{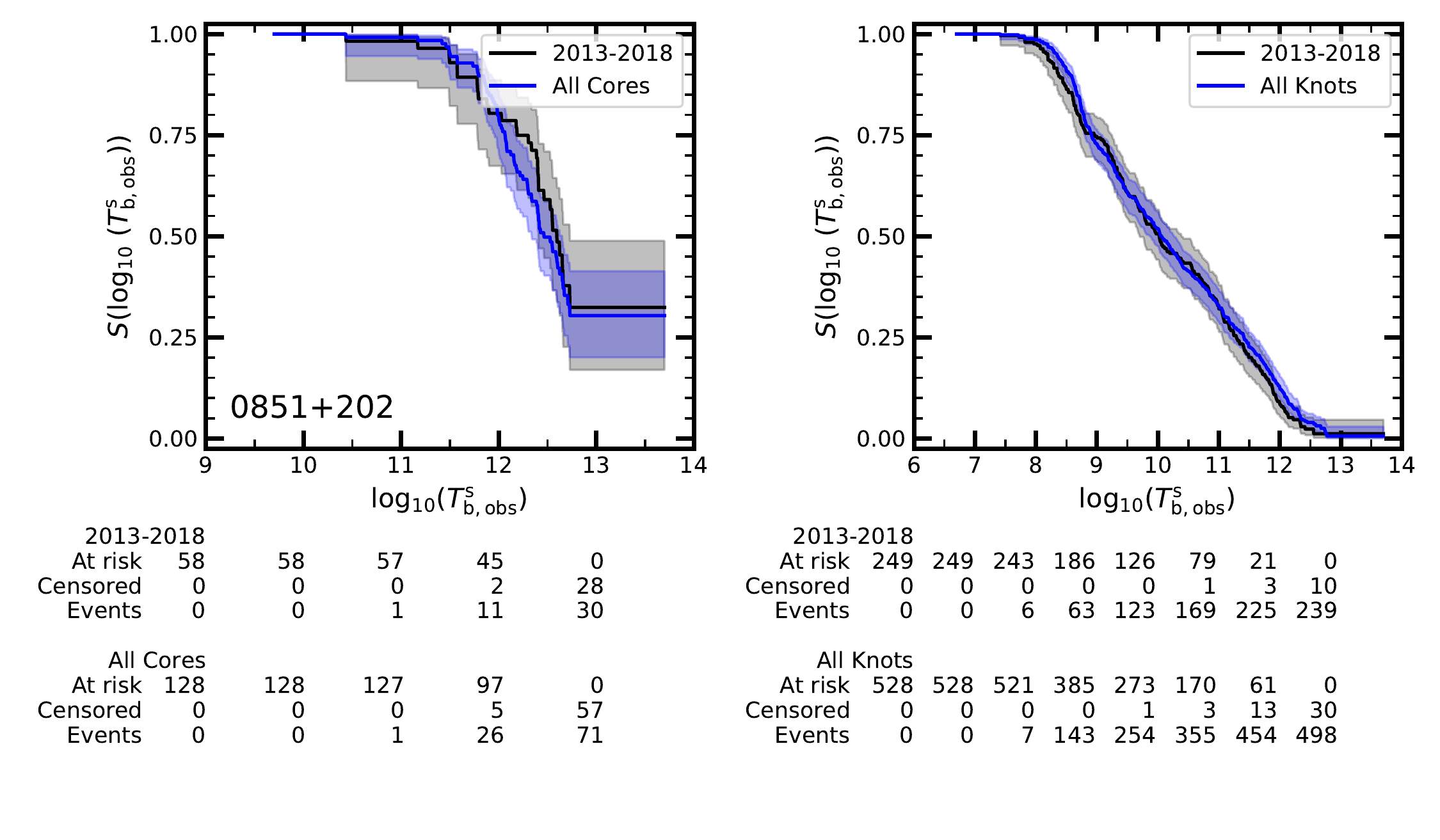}}
        \figsetgrpnote{Survival functions of the core (left) and knot (right) brightness temperatures for the marked time periods of the BL 0851+202.}
        \figsetgrpend
        %
        % Number 15
        \figsetgrpstart
        \figsetgrpnum{A1.15}
        \figsetgrptitle{0954+658}
        \figsetplot{{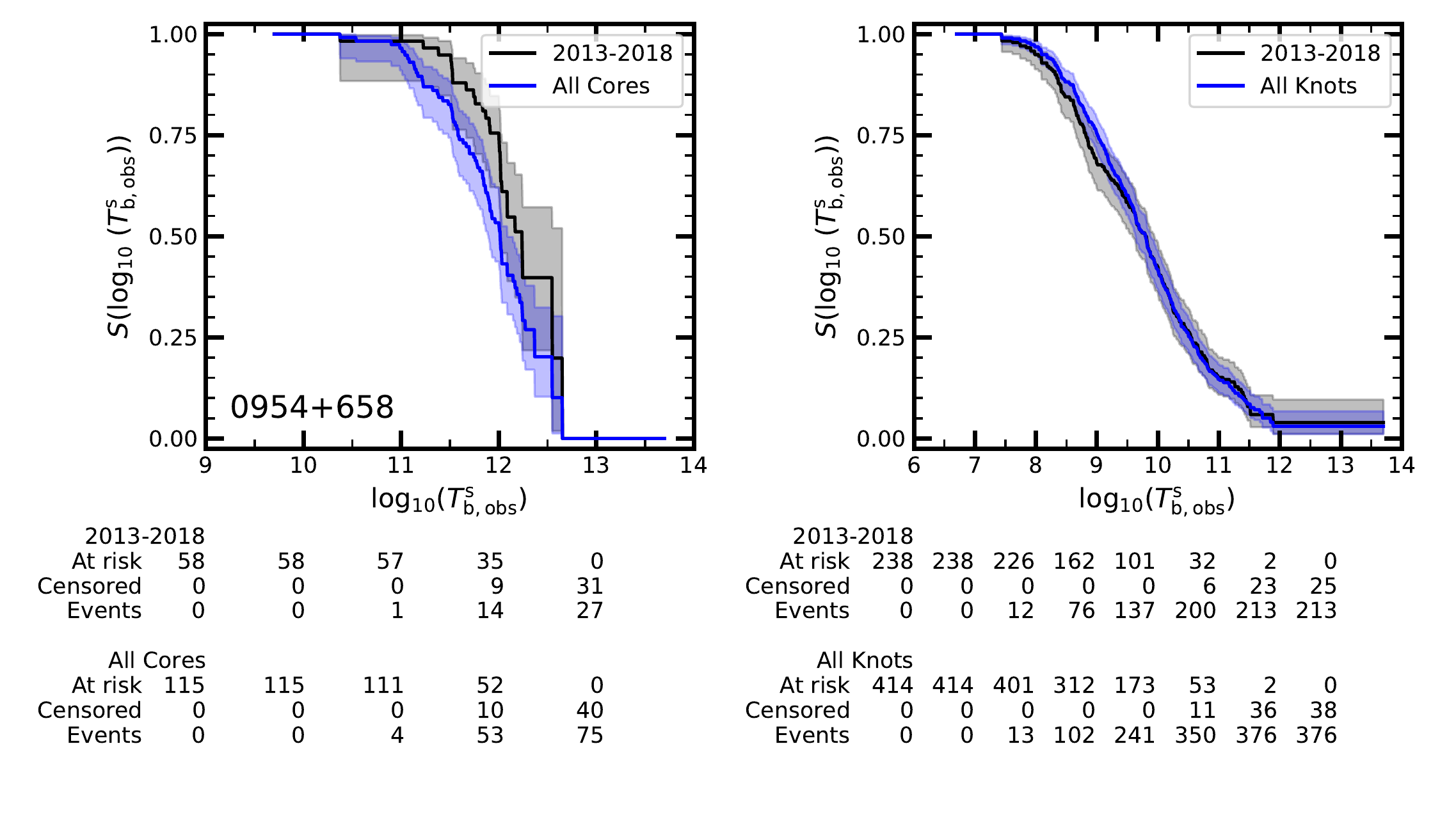}}
        \figsetgrpnote{Survival functions of the core (left) and knot (right) brightness temperatures for the marked time periods of the BL 0954+658.}
        \figsetgrpend
        %
        % Number 16
        \figsetgrpstart
        \figsetgrpnum{A1.16}
        \figsetgrptitle{1055+018}
        \figsetplot{{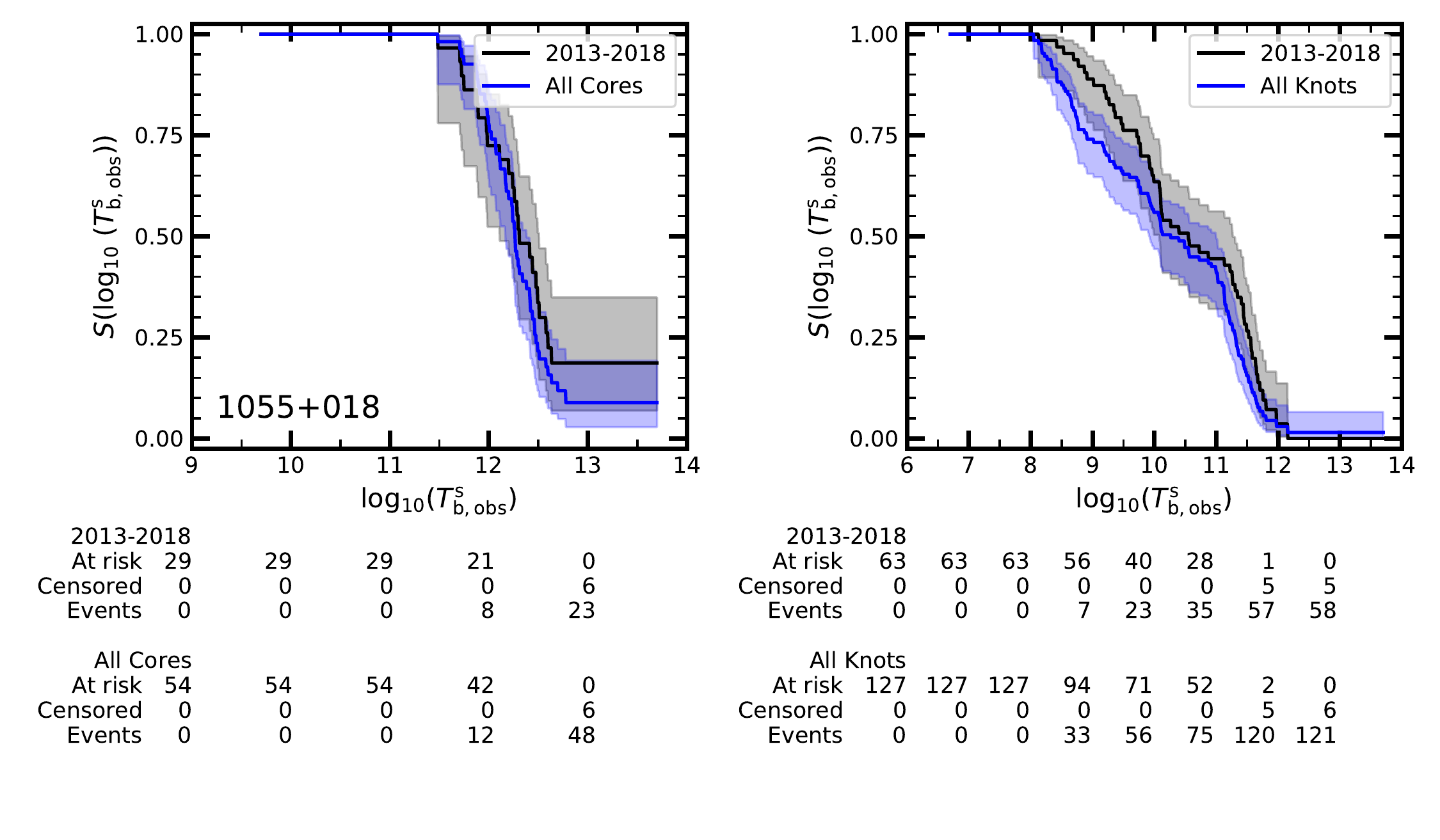}}
        \figsetgrpnote{Survival functions of the core (left) and knot (right) brightness temperatures for the marked time periods of the FSRQ 1055+018.}
        \figsetgrpend
        %
        % Number 17
        \figsetgrpstart
        \figsetgrpnum{A1.17}
        \figsetgrptitle{1101+384}
        \figsetplot{{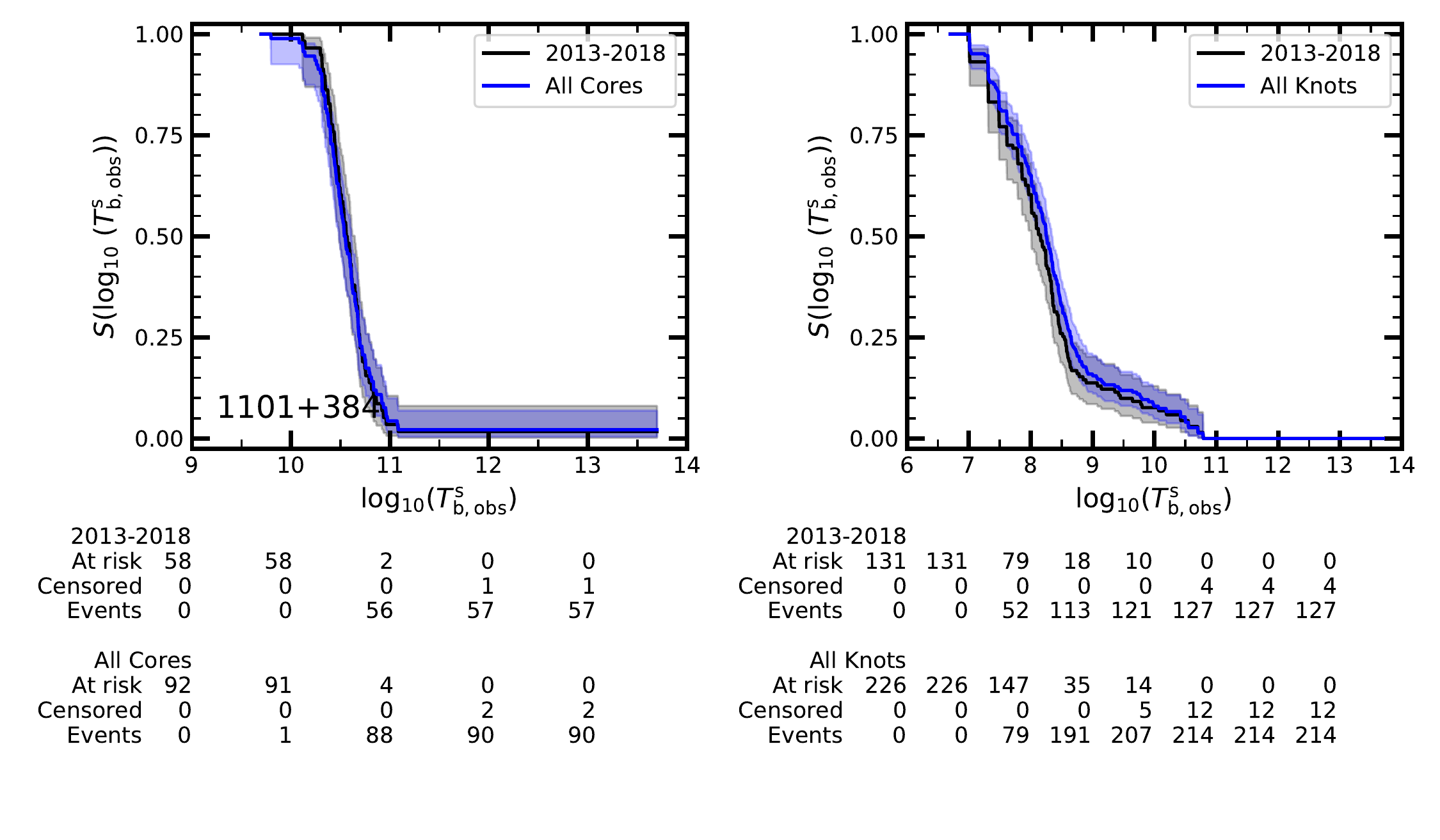}}
        \figsetgrpnote{Survival functions of the core (left) and knot (right) brightness temperatures for the marked time periods of the BL 1101+384.}
        \figsetgrpend
        %
        % Number 18
        \figsetgrpstart
        \figsetgrpnum{A1.18}
        \figsetgrptitle{1127-145}
        \figsetplot{{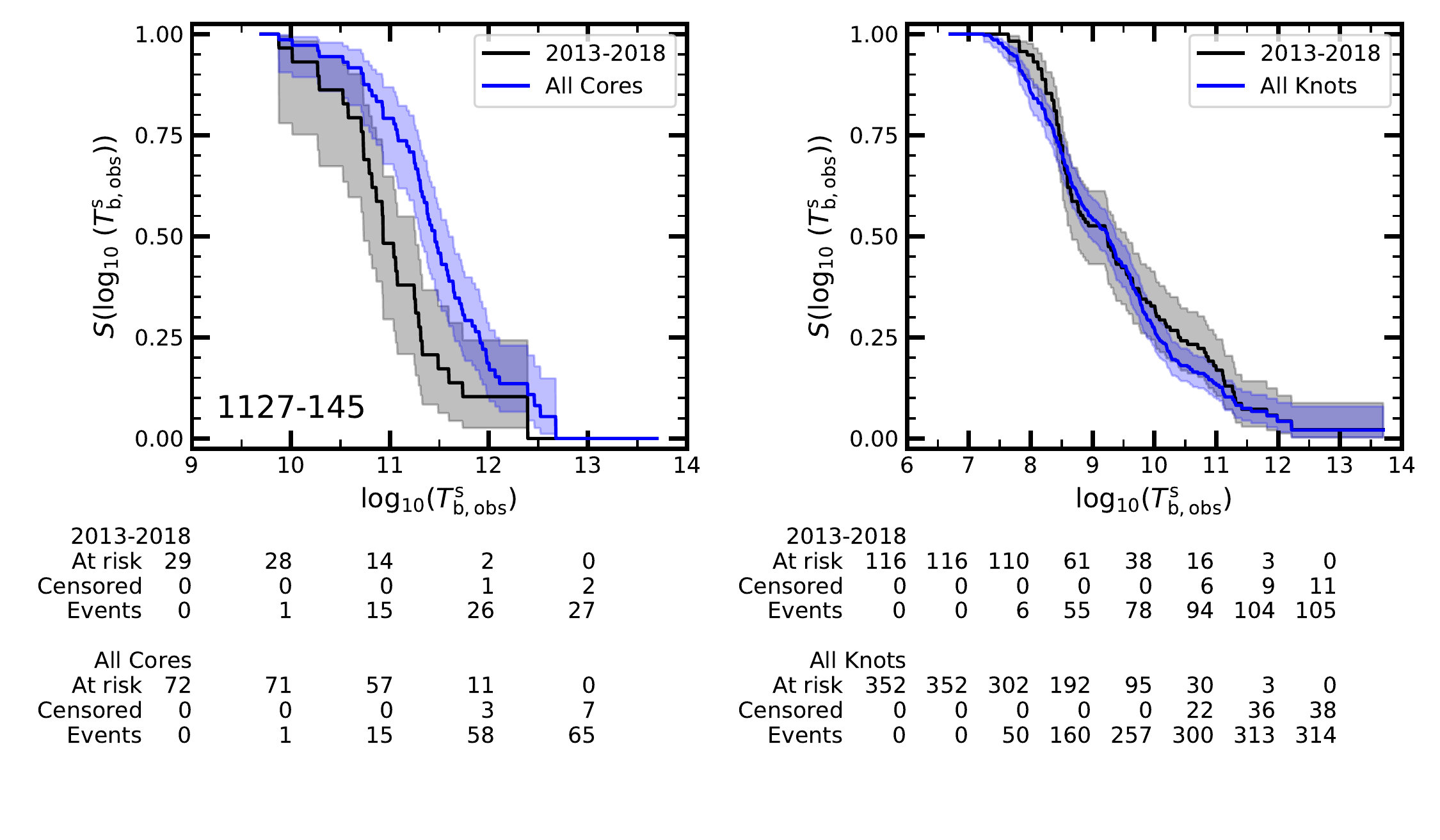}}
        \figsetgrpnote{Survival functions of the core (left) and knot (right) brightness temperatures for the marked time periods of the FSRQ 1127-145.}
        \figsetgrpend
        %
        % Number 19
        \figsetgrpstart
        \figsetgrpnum{A1.19}
        \figsetgrptitle{1156+295}
        \figsetplot{{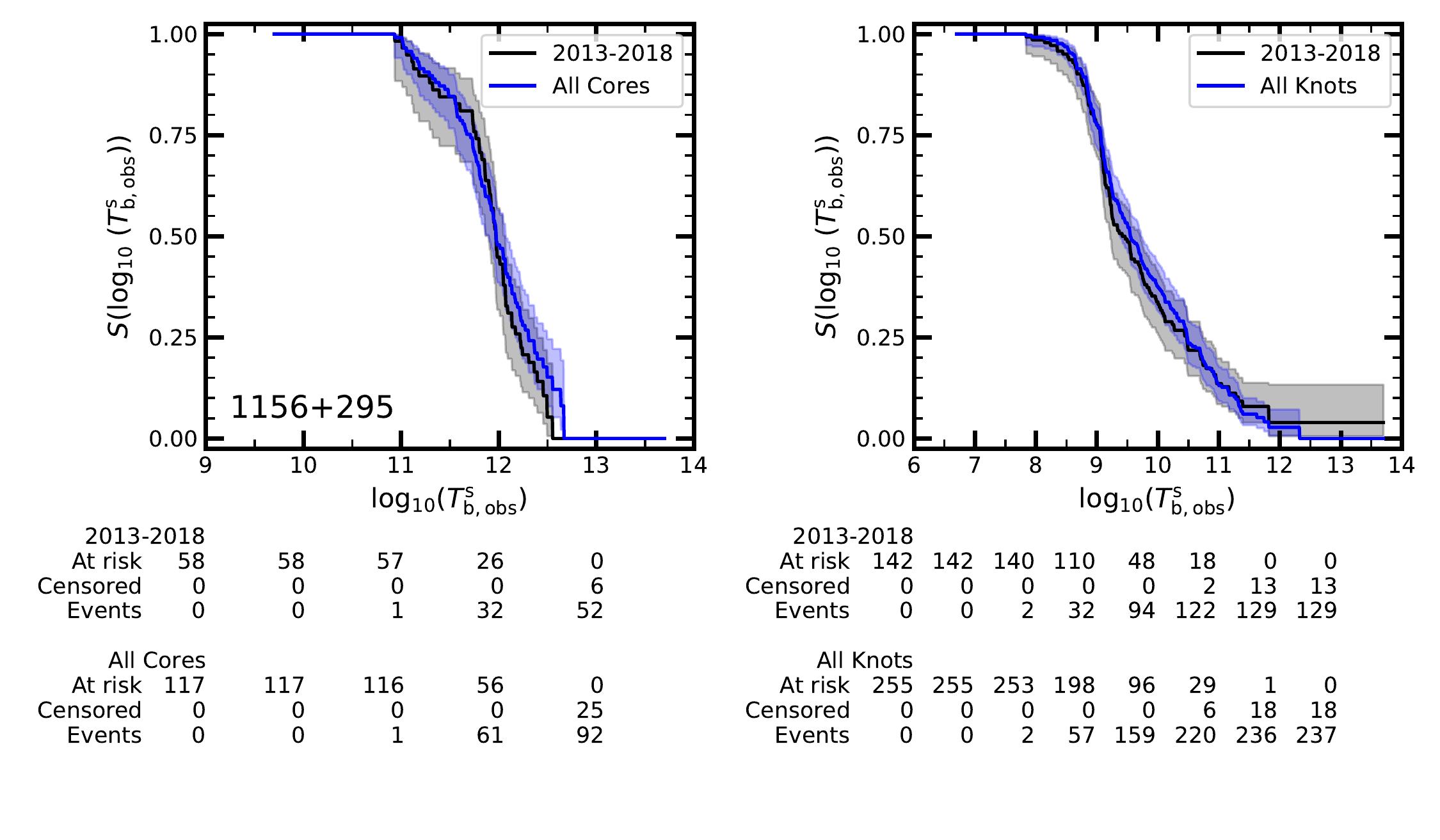}}
        \figsetgrpnote{Survival functions of the core (left) and knot (right) brightness temperatures for the marked time periods of the FSRQ 1156+295.}
        \figsetgrpend
        %
        % Number 20
        \figsetgrpstart
        \figsetgrpnum{A1.20}
        \figsetgrptitle{1219+285}
        \figsetplot{{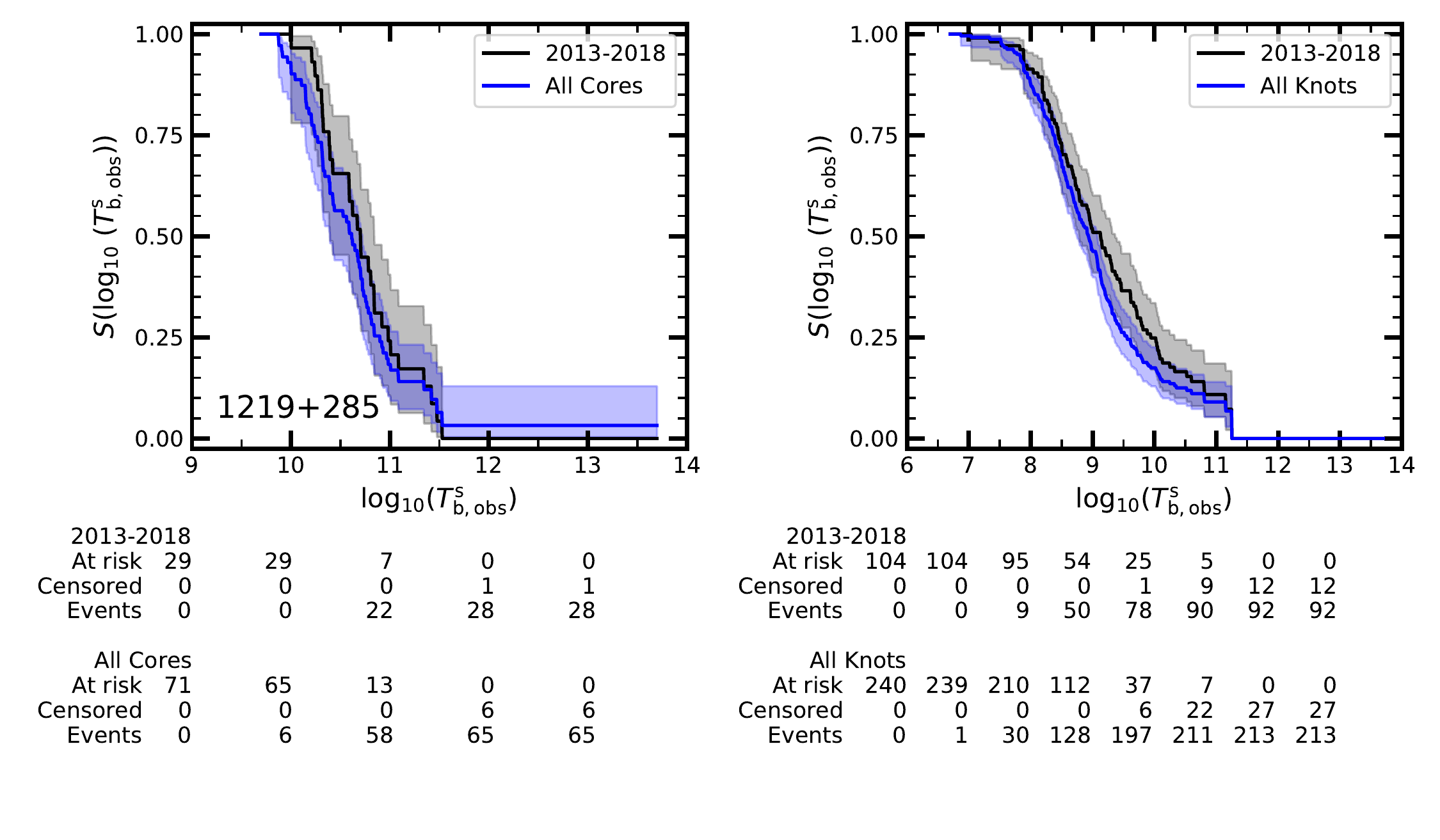}}
        \figsetgrpnote{Survival functions of the core (left) and knot (right) brightness temperatures for the marked time periods of the BL 1219+285.}
        \figsetgrpend
        %
        % Number 21
        \figsetgrpstart
        \figsetgrpnum{A1.21}
        \figsetgrptitle{1222+216}
        \figsetplot{{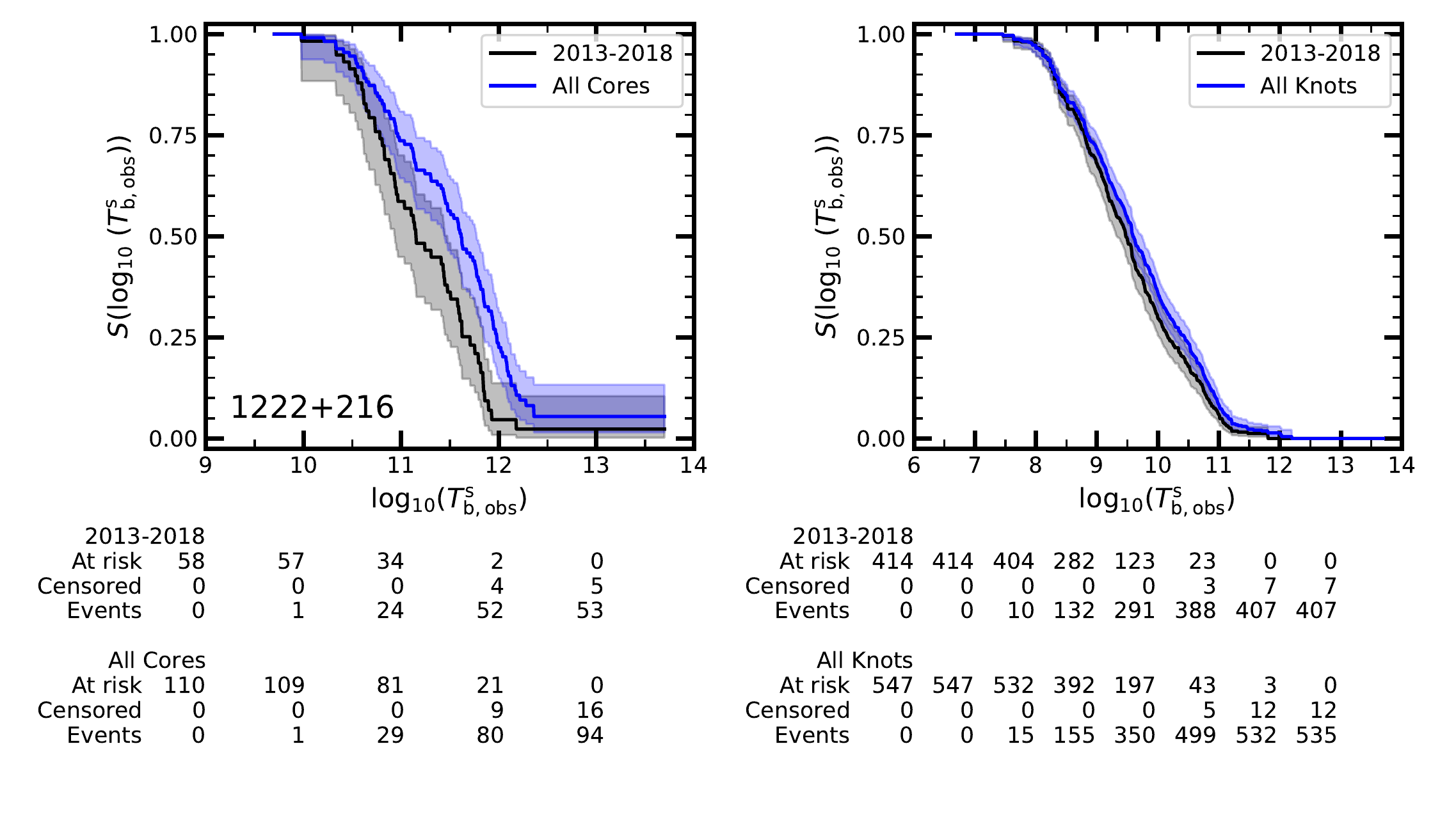}}
        \figsetgrpnote{Survival functions of the core (left) and knot (right) brightness temperatures for the marked time periods of the FSRQ 1222+216.}
        \figsetgrpend
        %
        % Number 22
        \figsetgrpstart
        \figsetgrpnum{A1.22}
        \figsetgrptitle{1226+023}
        \figsetplot{{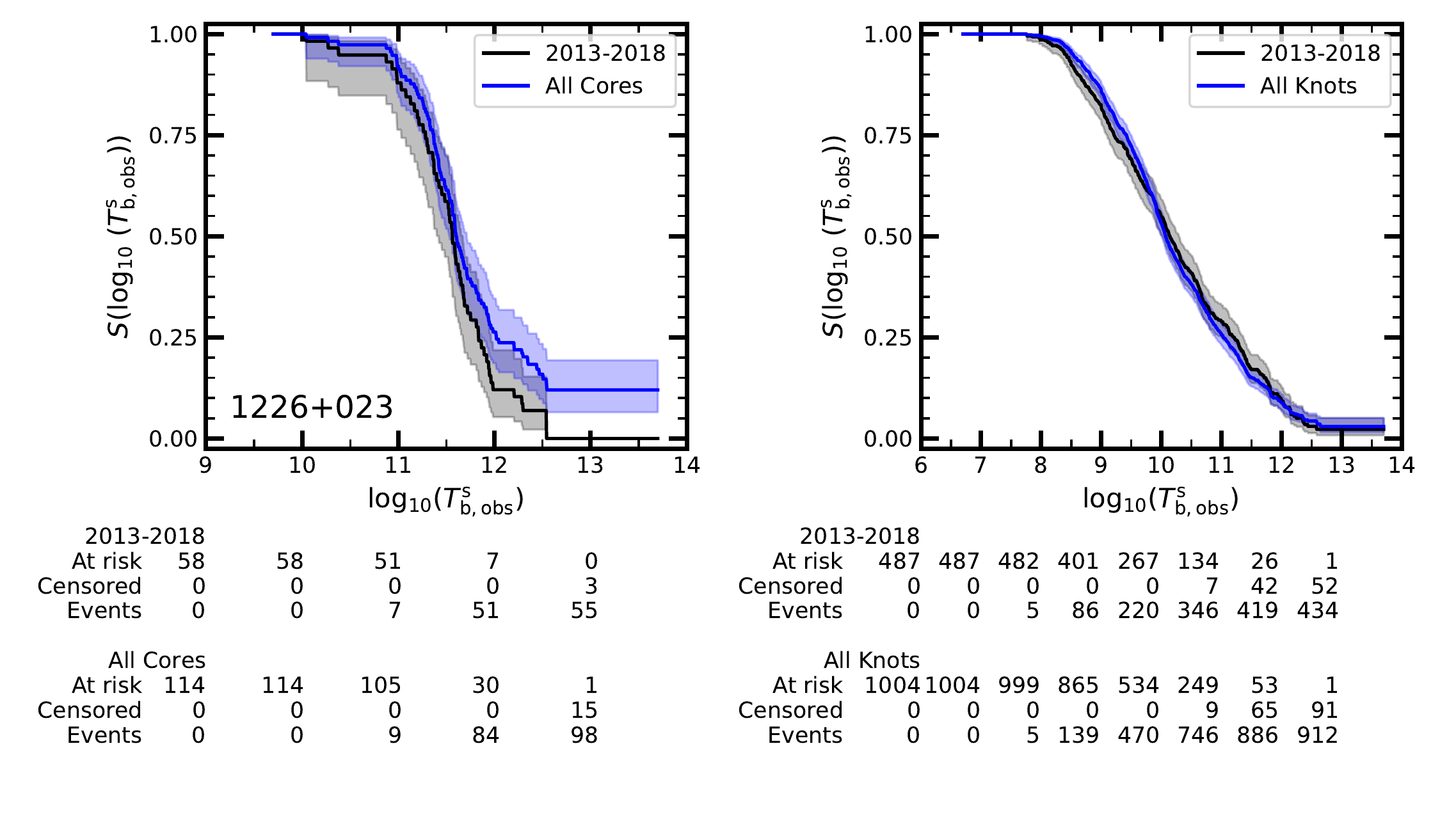}}
        \figsetgrpnote{Survival functions of the core (left) and knot (right) brightness temperatures for the marked time periods of the FSRQ 1226+023.}
        \figsetgrpend
        %
        % Number 23
        \figsetgrpstart
        \figsetgrpnum{A1.23}
        \figsetgrptitle{1253-055}
        \figsetplot{{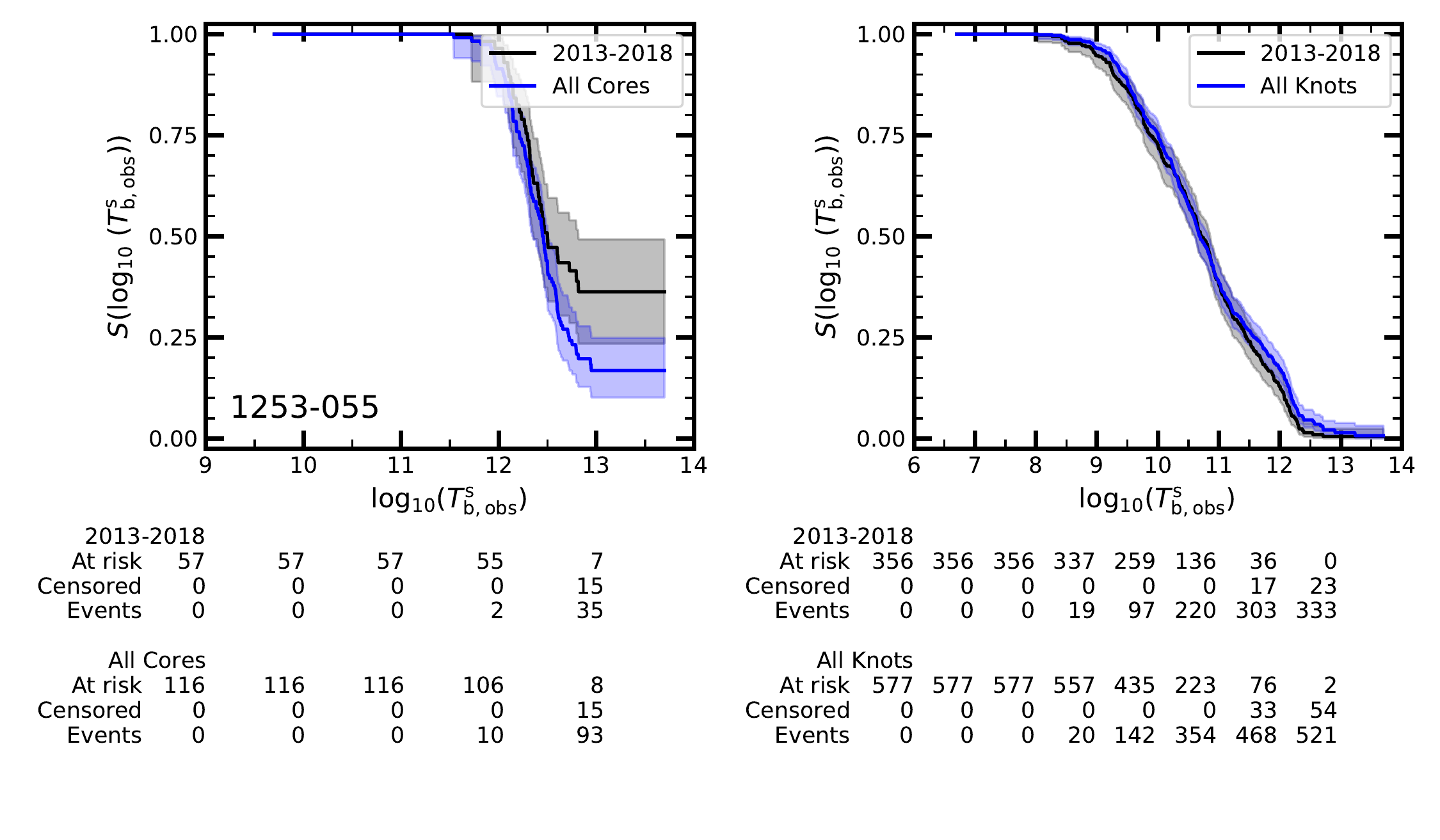}}
        \figsetgrpnote{Survival functions of the core (left) and knot (right) brightness temperatures for the marked time periods of the FSRQ 1253-055.}
        \figsetgrpend
        %
        % Number 24
        \figsetgrpstart
        \figsetgrpnum{A1.24}
        \figsetgrptitle{1308+326}
        \figsetplot{{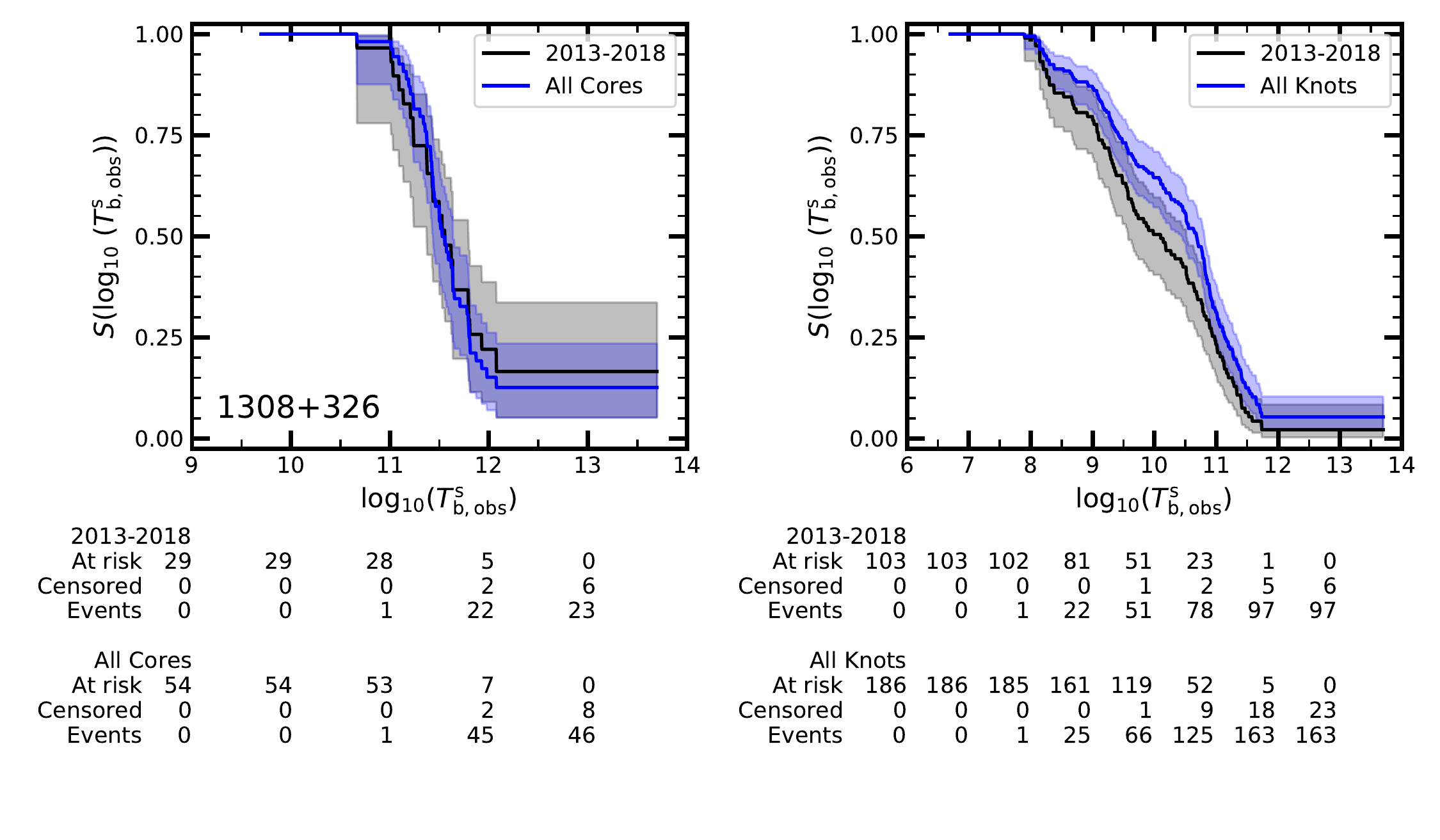}}
        \figsetgrpnote{Survival functions of the core (left) and knot (right) brightness temperatures for the marked time periods of the FSRQ 1308+326.}
        \figsetgrpend
        %
        % Number 25
        \figsetgrpstart
        \figsetgrpnum{A1.25}
        \figsetgrptitle{1406-076}
        \figsetplot{{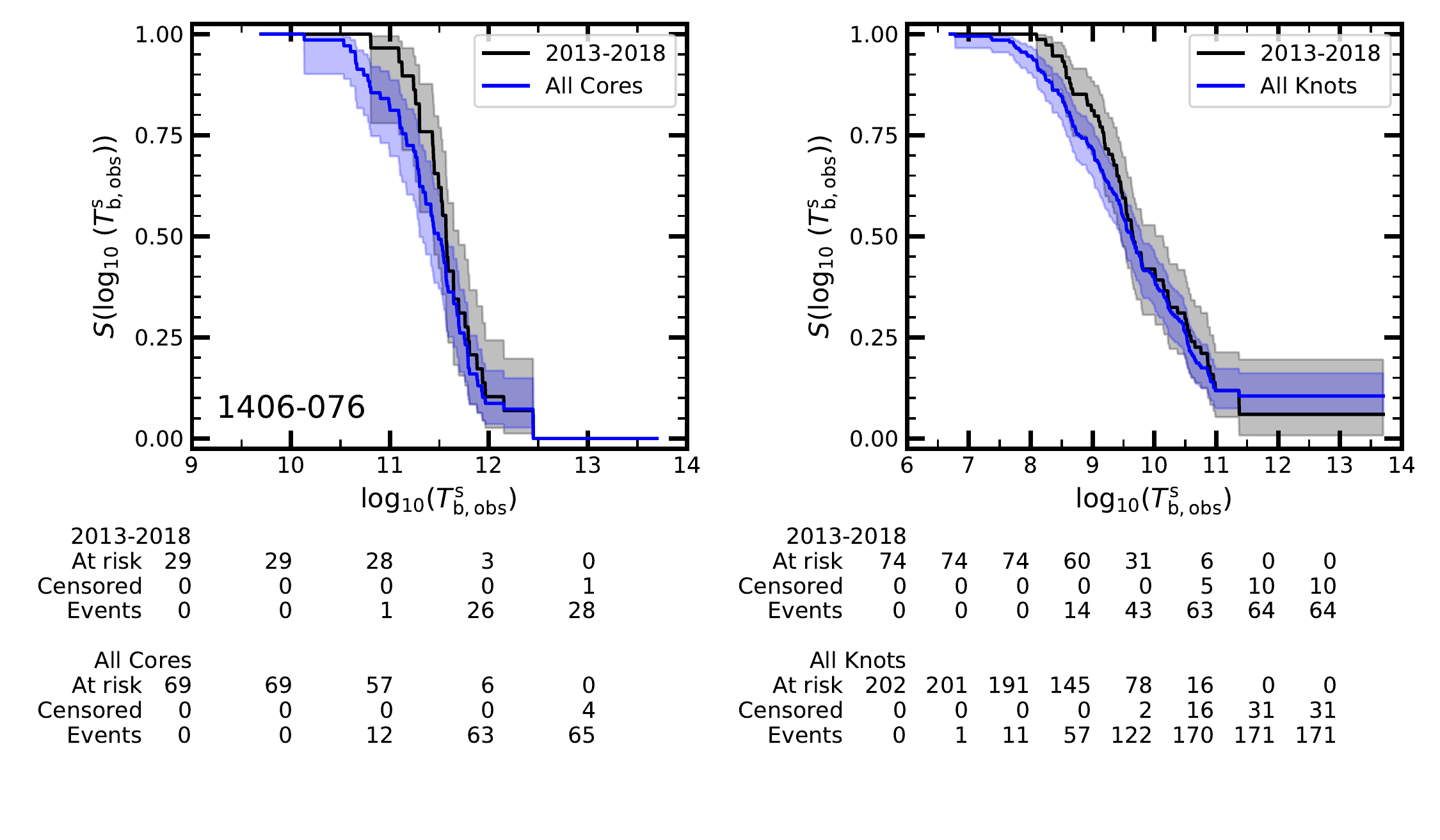}}
        \figsetgrpnote{Survival functions of the core (left) and knot (right) brightness temperatures for the marked time periods of the FSRQ 1406-076.}
        \figsetgrpend
        %
        % Number 26
        \figsetgrpstart
        \figsetgrpnum{A1.26}
        \figsetgrptitle{1510-089}
        \figsetplot{{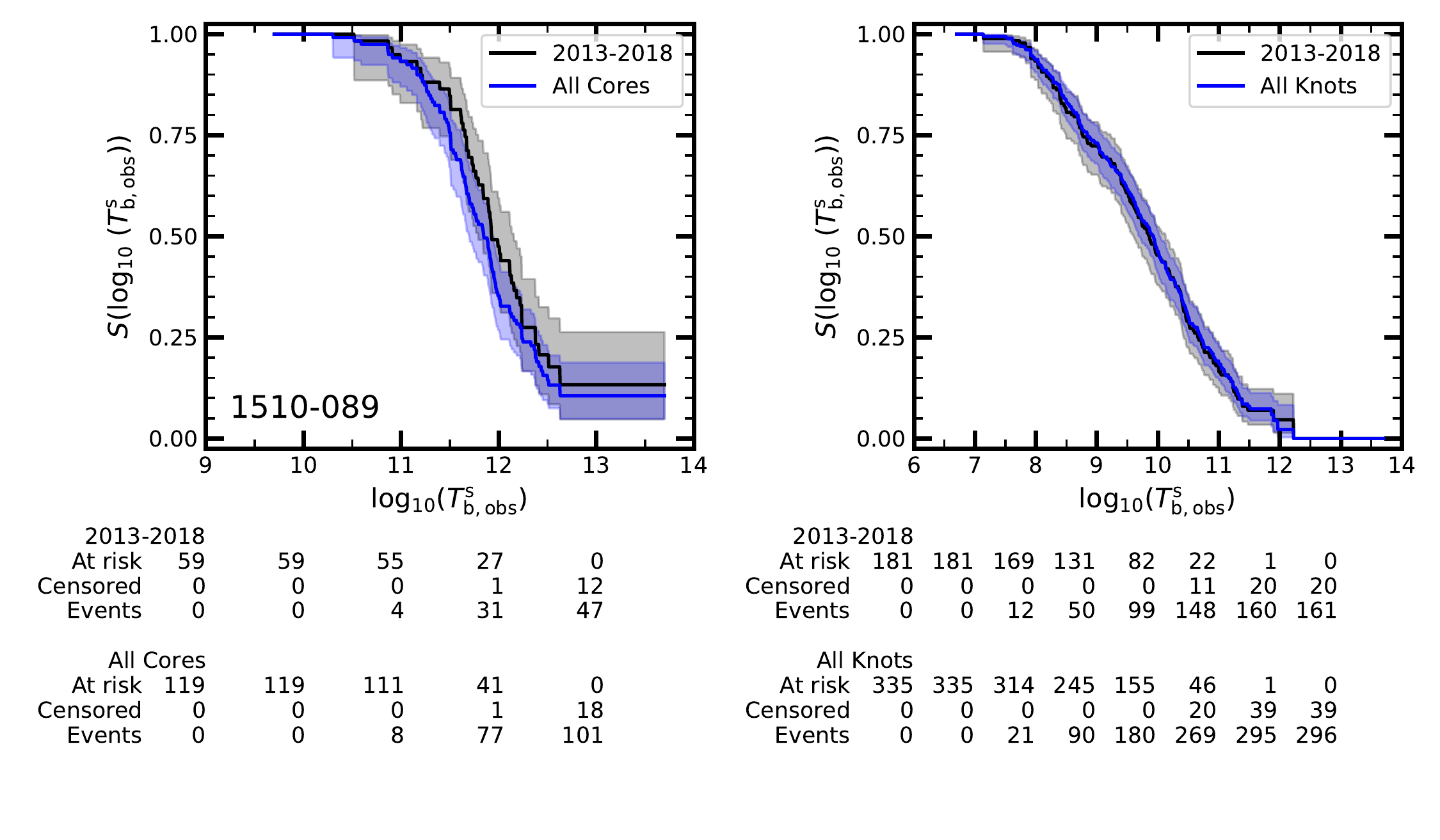}}
        \figsetgrpnote{Survival functions of the core (left) and knot (right) brightness temperatures for the marked time periods of the FSRQ 1510-089.}
        \figsetgrpend
        %
        % Number 27
        \figsetgrpstart
        \figsetgrpnum{A1.27}
        \figsetgrptitle{1611+343}
        \figsetplot{{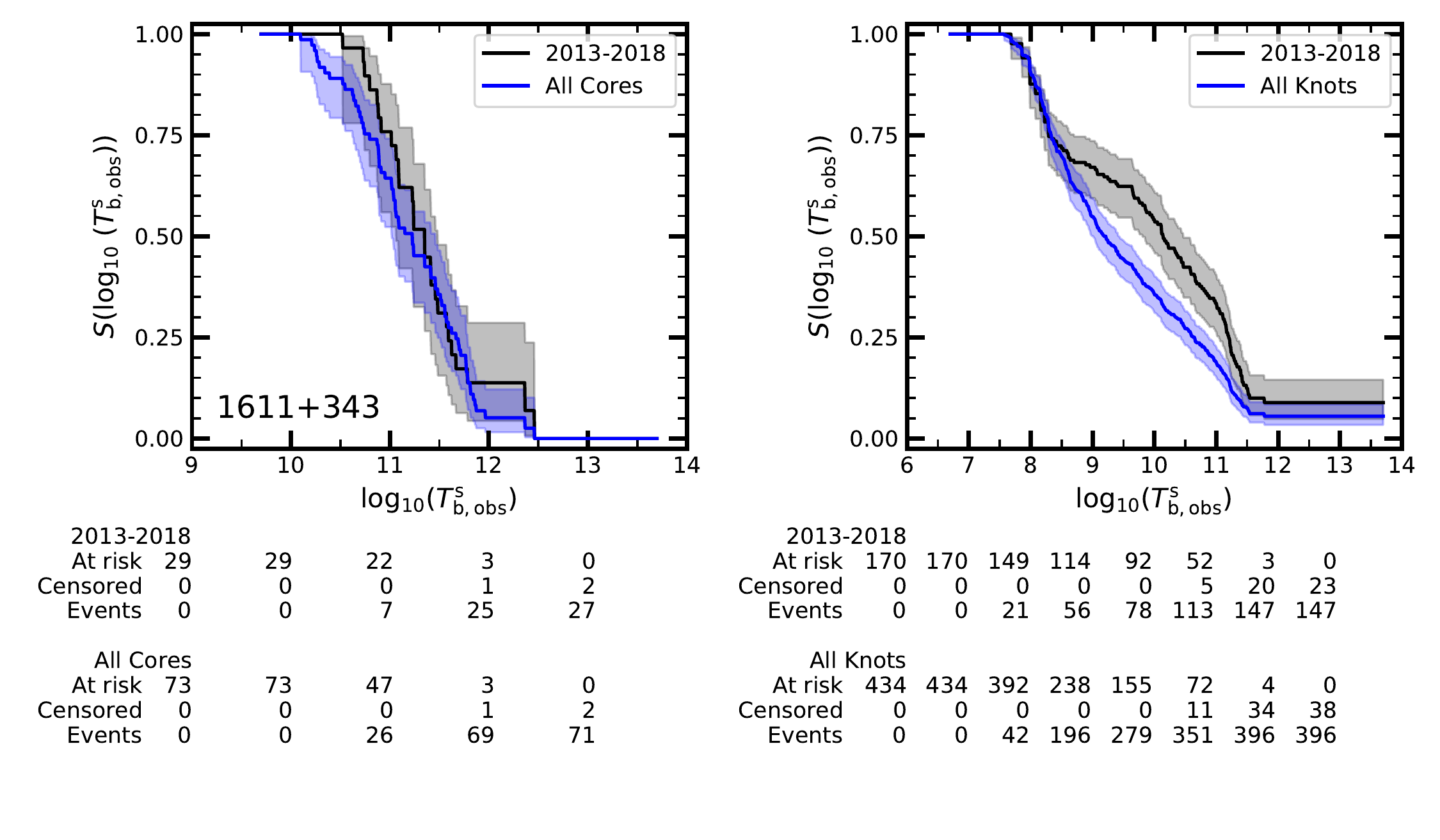}}
        \figsetgrpnote{Survival functions of the core (left) and knot (right) brightness temperatures for the marked time periods of the FSRQ 1611+343.}
        \figsetgrpend
        %
        % Number 28
        \figsetgrpstart
        \figsetgrpnum{A1.28}
        \figsetgrptitle{1622-297}
        \figsetplot{{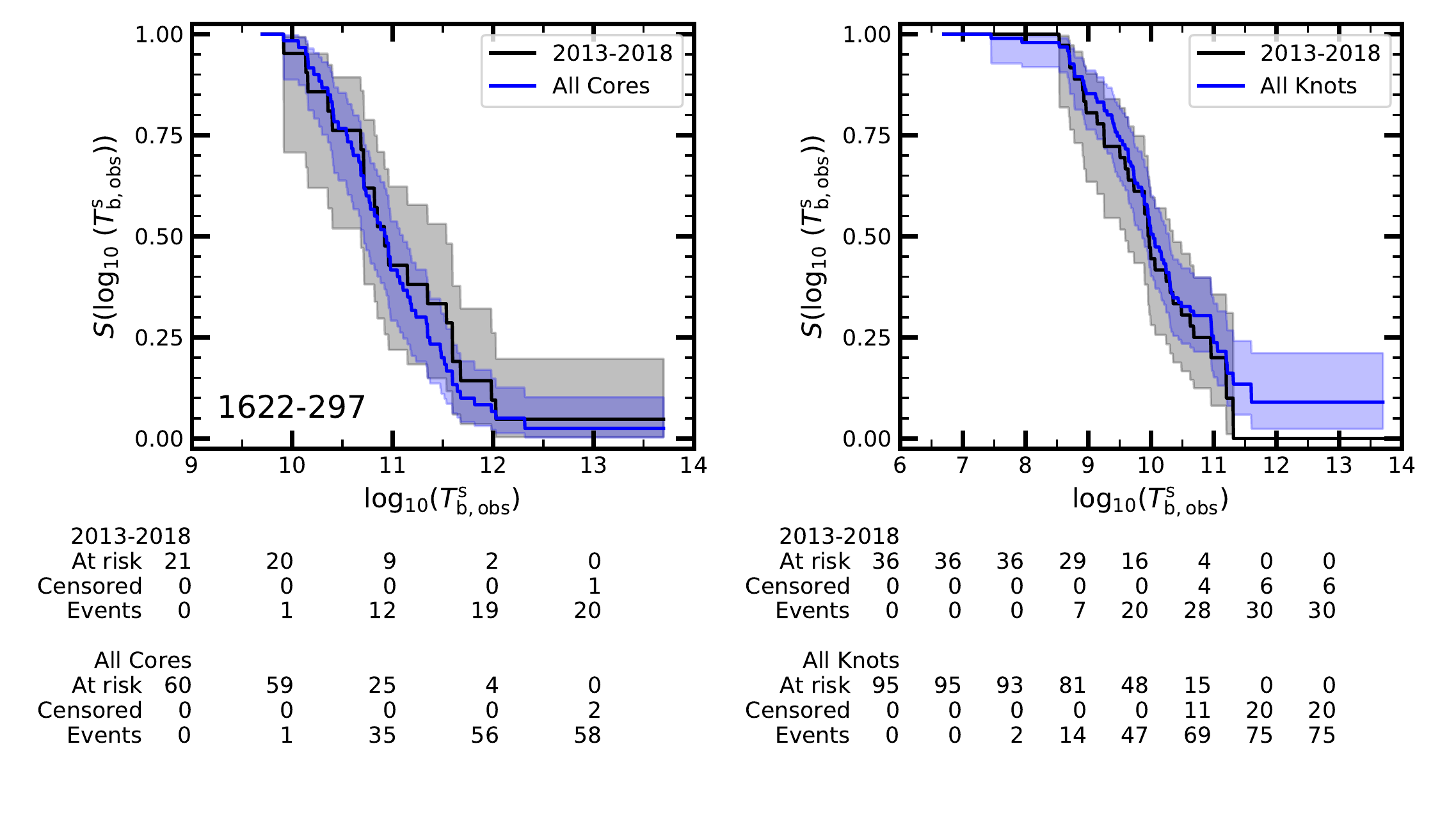}}
        \figsetgrpnote{Survival functions of the core (left) and knot (right) brightness temperatures for the marked time periods of the FSRQ 1622-297.}
        \figsetgrpend
        %
        % Number 29
        \figsetgrpstart
        \figsetgrpnum{A1.29}
        \figsetgrptitle{1633+382}
        \figsetplot{{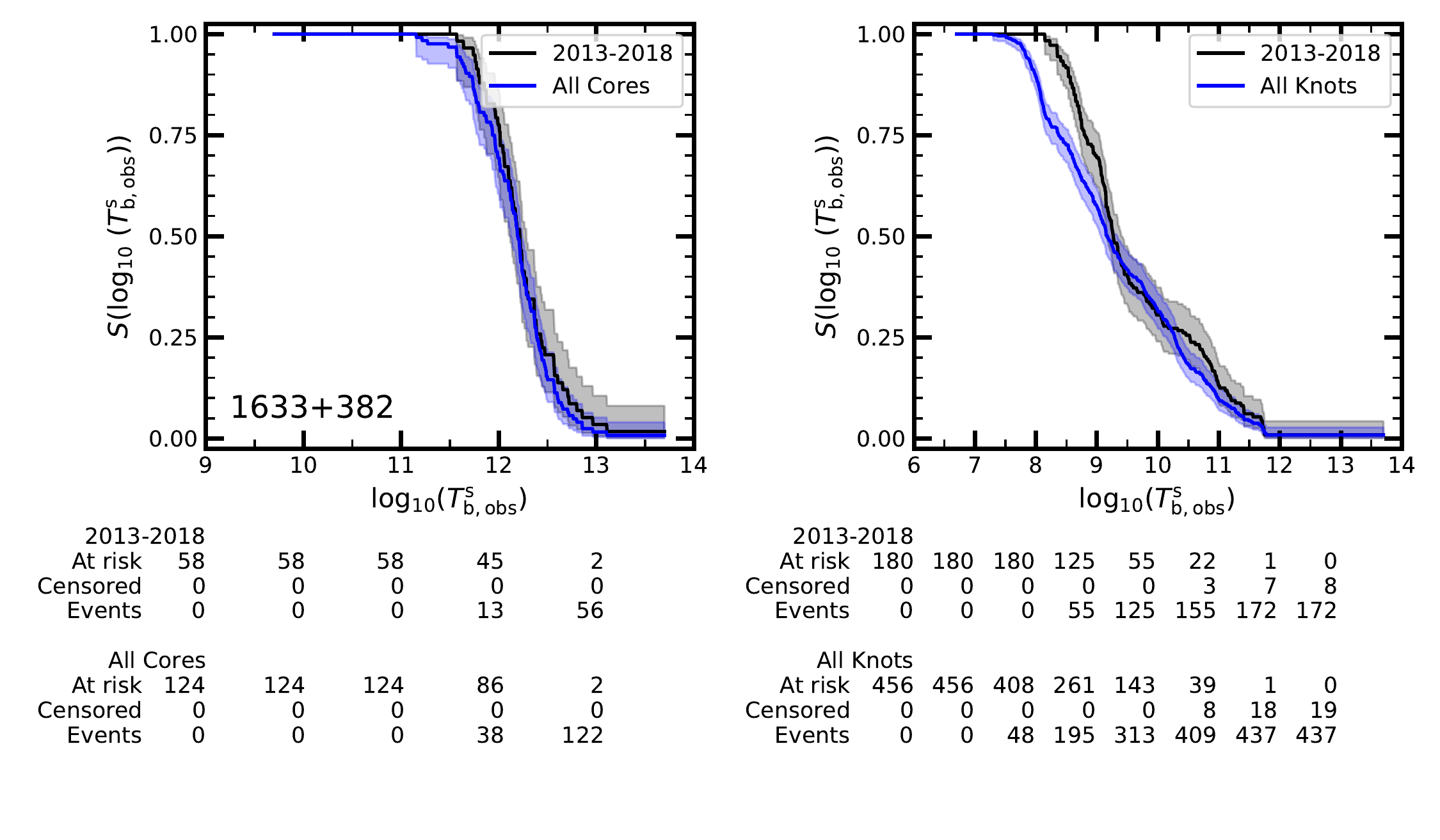}}
        \figsetgrpnote{Survival functions of the core (left) and knot (right) brightness temperatures for the marked time periods of the FSRQ 1633+382.}
        \figsetgrpend
        %
        % Number 30
        \figsetgrpstart
        \figsetgrpnum{A1.30}
        \figsetgrptitle{1641+399}
        \figsetplot{{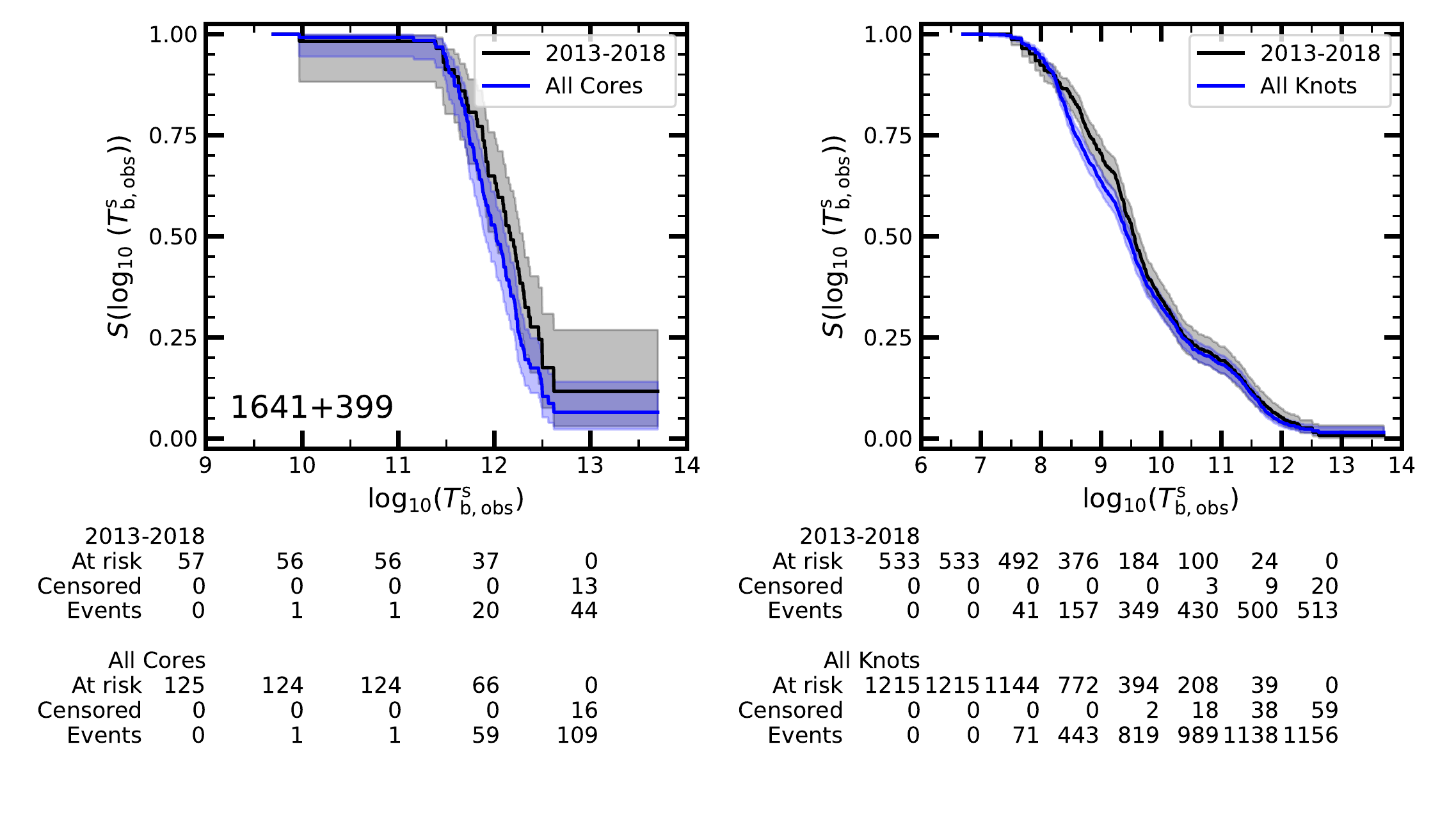}}
        \figsetgrpnote{Survival functions of the core (left) and knot (right) brightness temperatures for the marked time periods of the FSRQ 1641+399.}
        \figsetgrpend
        %
        % Number 31
        \figsetgrpstart
        \figsetgrpnum{A1.31}
        \figsetgrptitle{1652+398}
        \figsetplot{{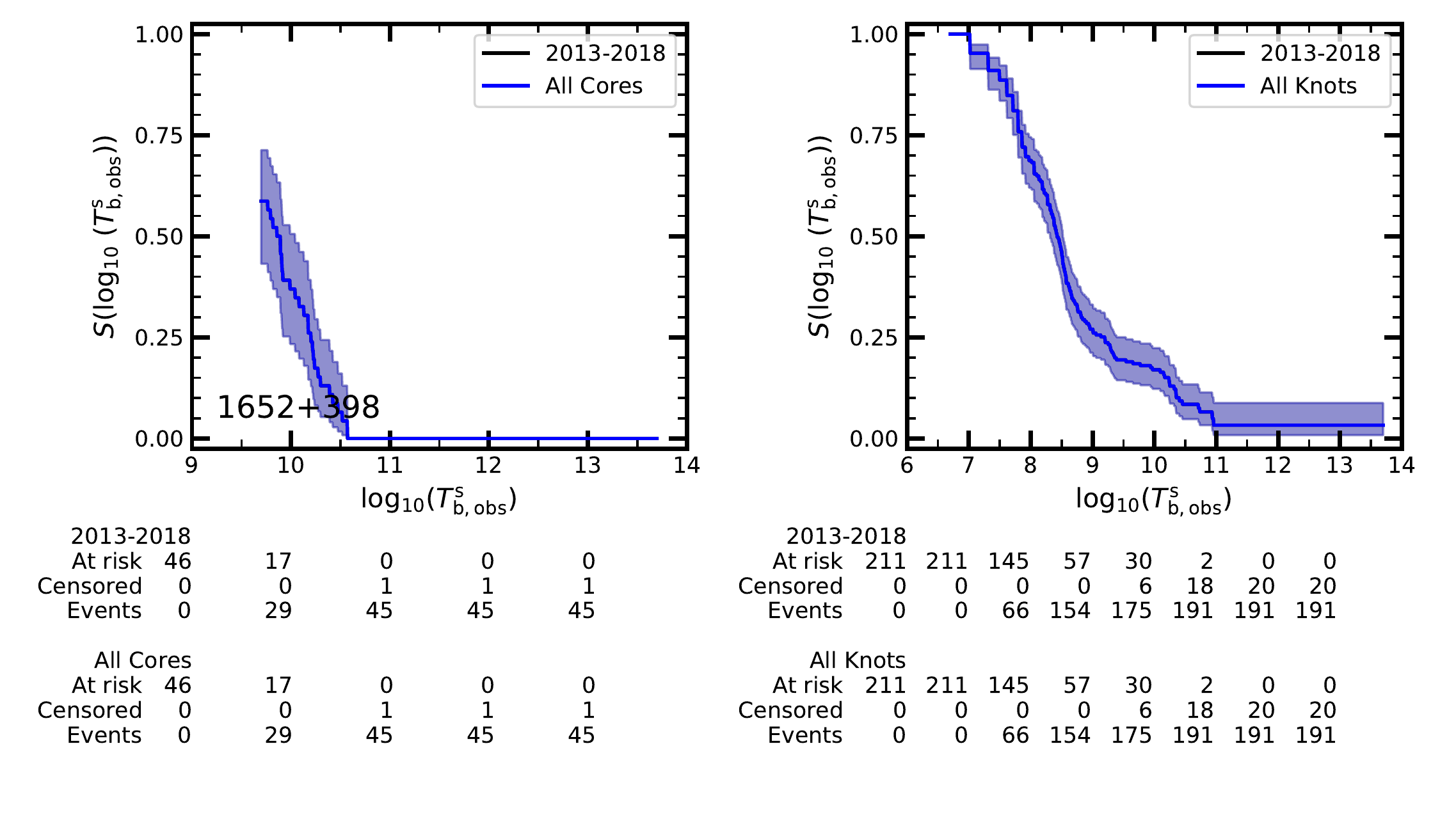}}
        \figsetgrpnote{Survival functions of the core (left) and knot (right) brightness temperatures for the marked time periods of the BL 1652+398.}
        \figsetgrpend
        %
        % Number 32
        \figsetgrpstart
        \figsetgrpnum{A1.32}
        \figsetgrptitle{1730-130}
        \figsetplot{{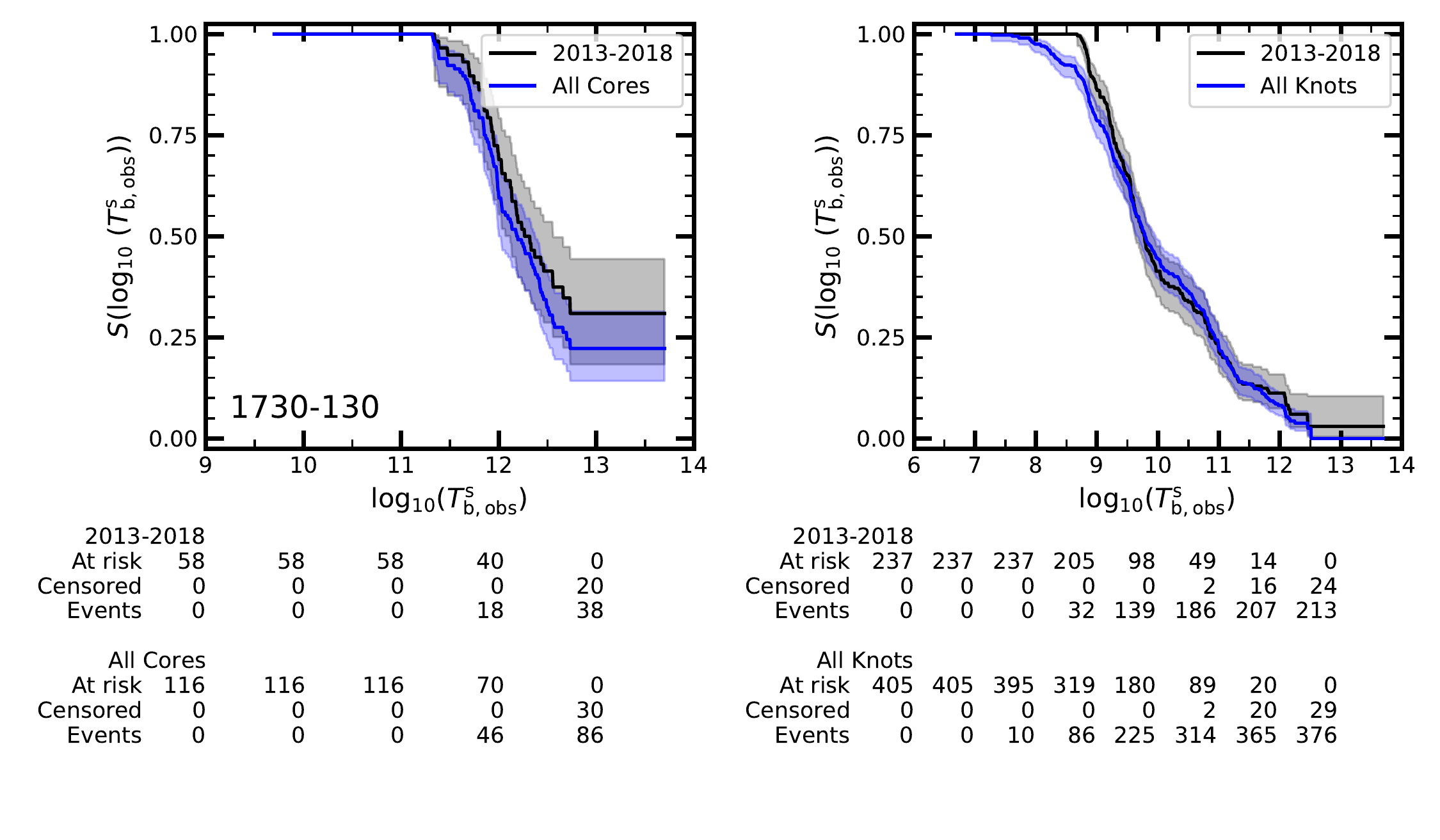}}
        \figsetgrpnote{Survival functions of the core (left) and knot (right) brightness temperatures for the marked time periods of the FSRQ 1730-130.}
        \figsetgrpend
        %
        % Number 33
        \figsetgrpstart
        \figsetgrpnum{A1.33}
        \figsetgrptitle{1749+096}
        \figsetplot{{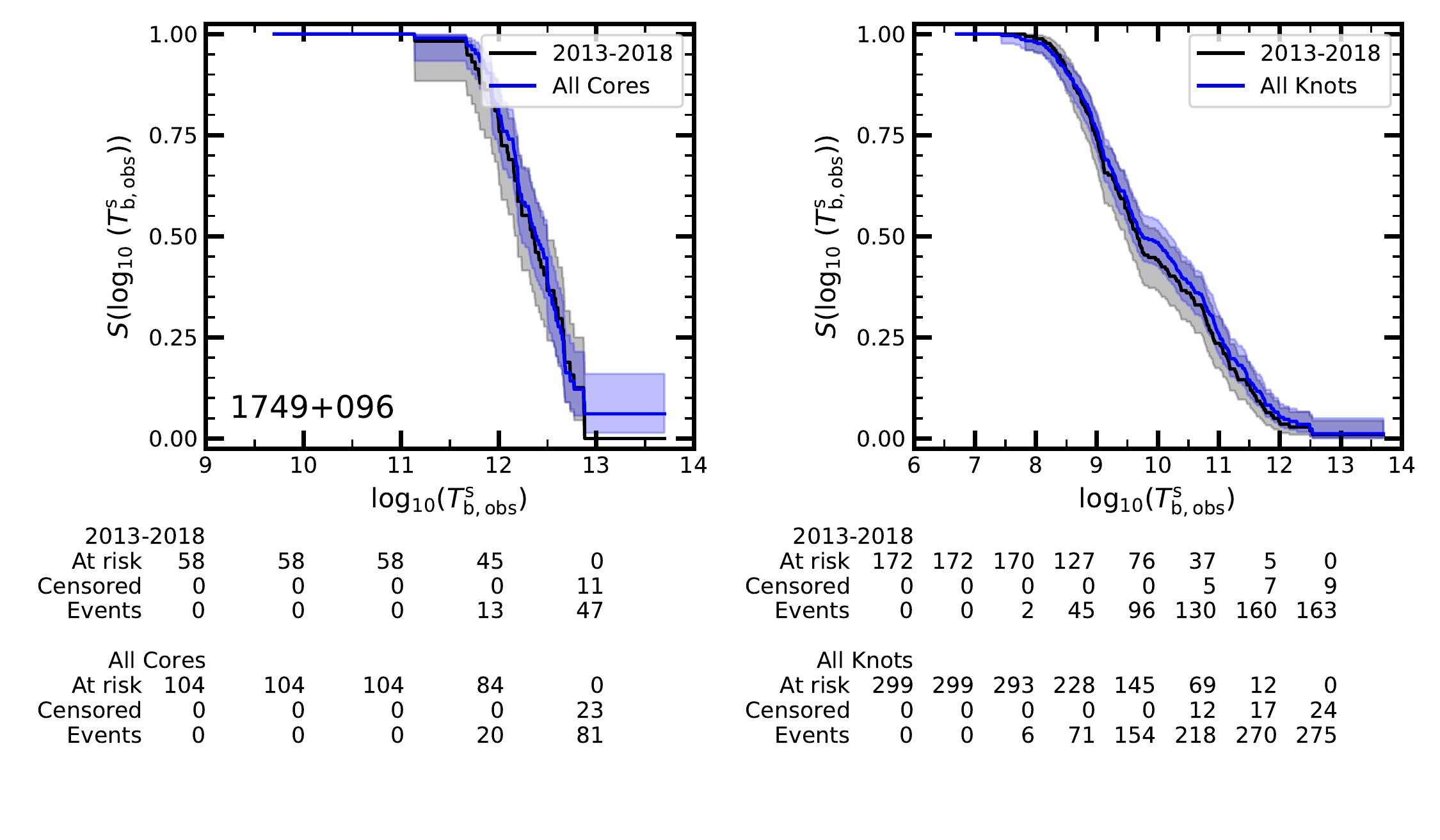}}
        \figsetgrpnote{Survival functions of the core (left) and knot (right) brightness temperatures for the marked time periods of the BL 1749+096.}
        \figsetgrpend
        %
        % Number 34
        \figsetgrpstart
        \figsetgrpnum{A1.34}
        \figsetgrptitle{1959+650}
        \figsetplot{{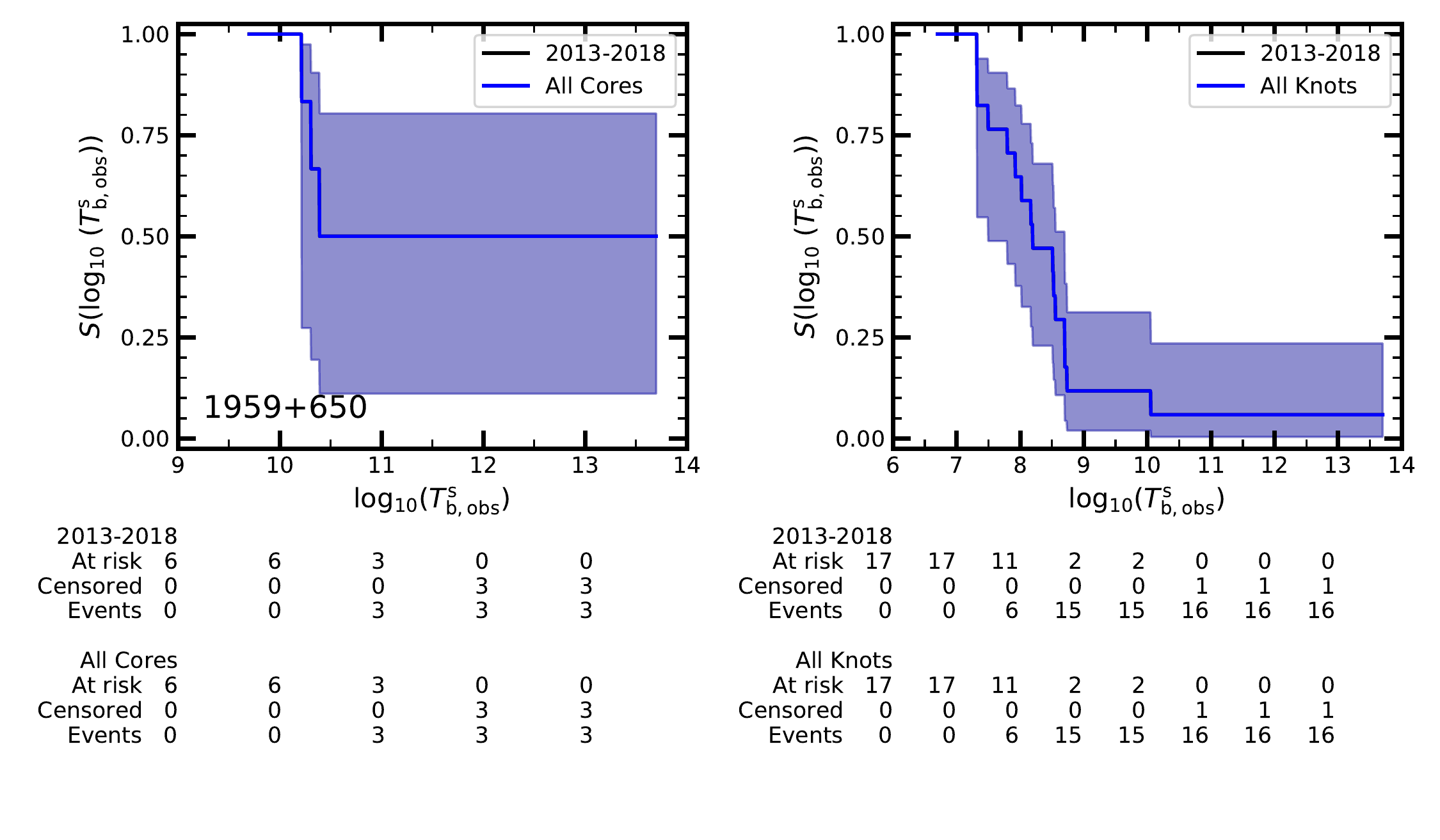}}
        \figsetgrpnote{Survival functions of the core (left) and knot (right) brightness temperatures for the marked time periods of the BL 1959+650.}
        \figsetgrpend
        %
        % Number 35
        \figsetgrpstart
        \figsetgrpnum{A1.35}
        \figsetgrptitle{2200+420}
        \figsetplot{{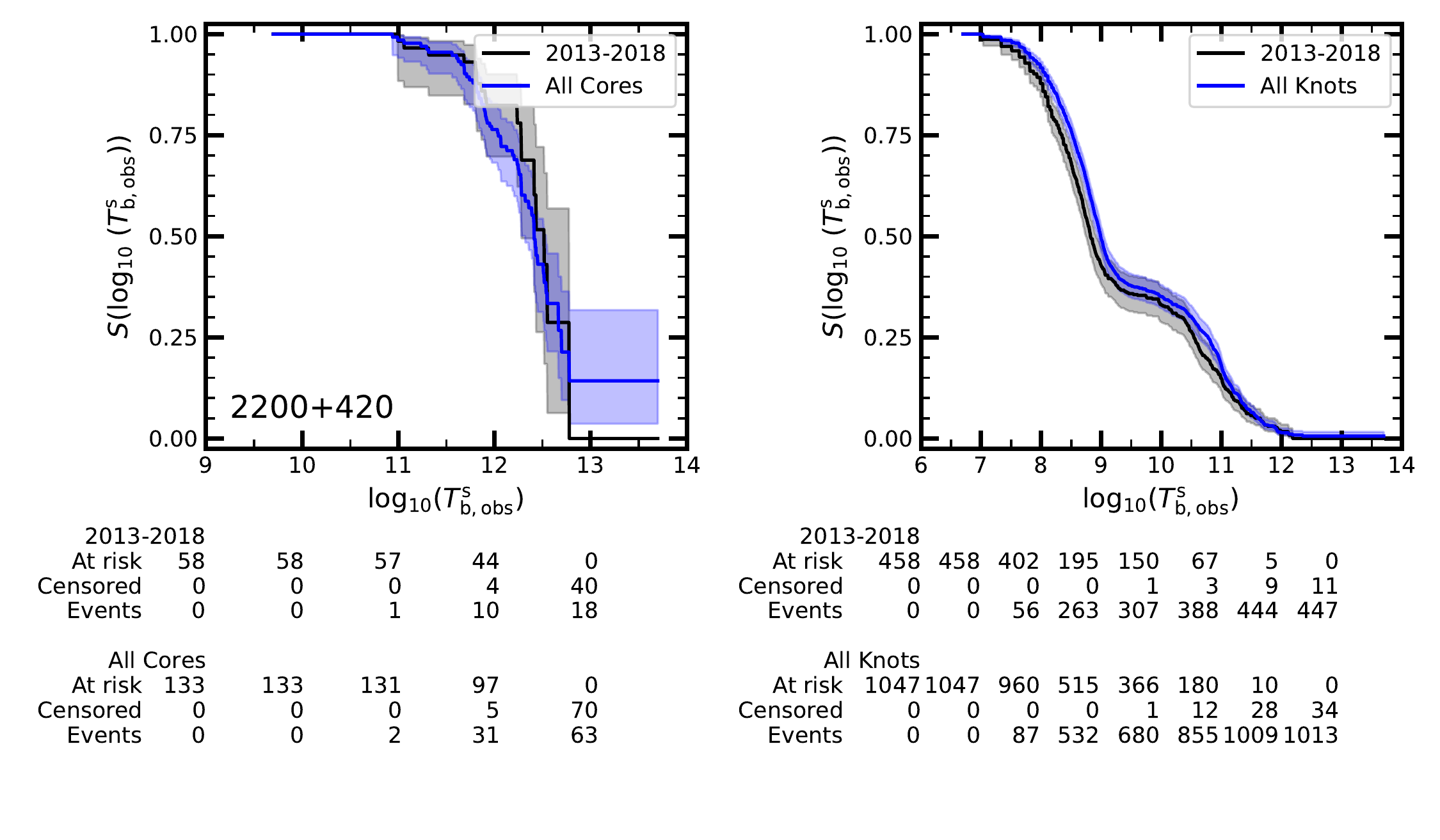}}
        \figsetgrpnote{Survival functions of the core (left) and knot (right) brightness temperatures for the marked time periods of the BL 2200+420.}
        \figsetgrpend
        %
        % Number 36
        \figsetgrpstart
        \figsetgrpnum{A1.36}
        \figsetgrptitle{2223-052}
        \figsetplot{{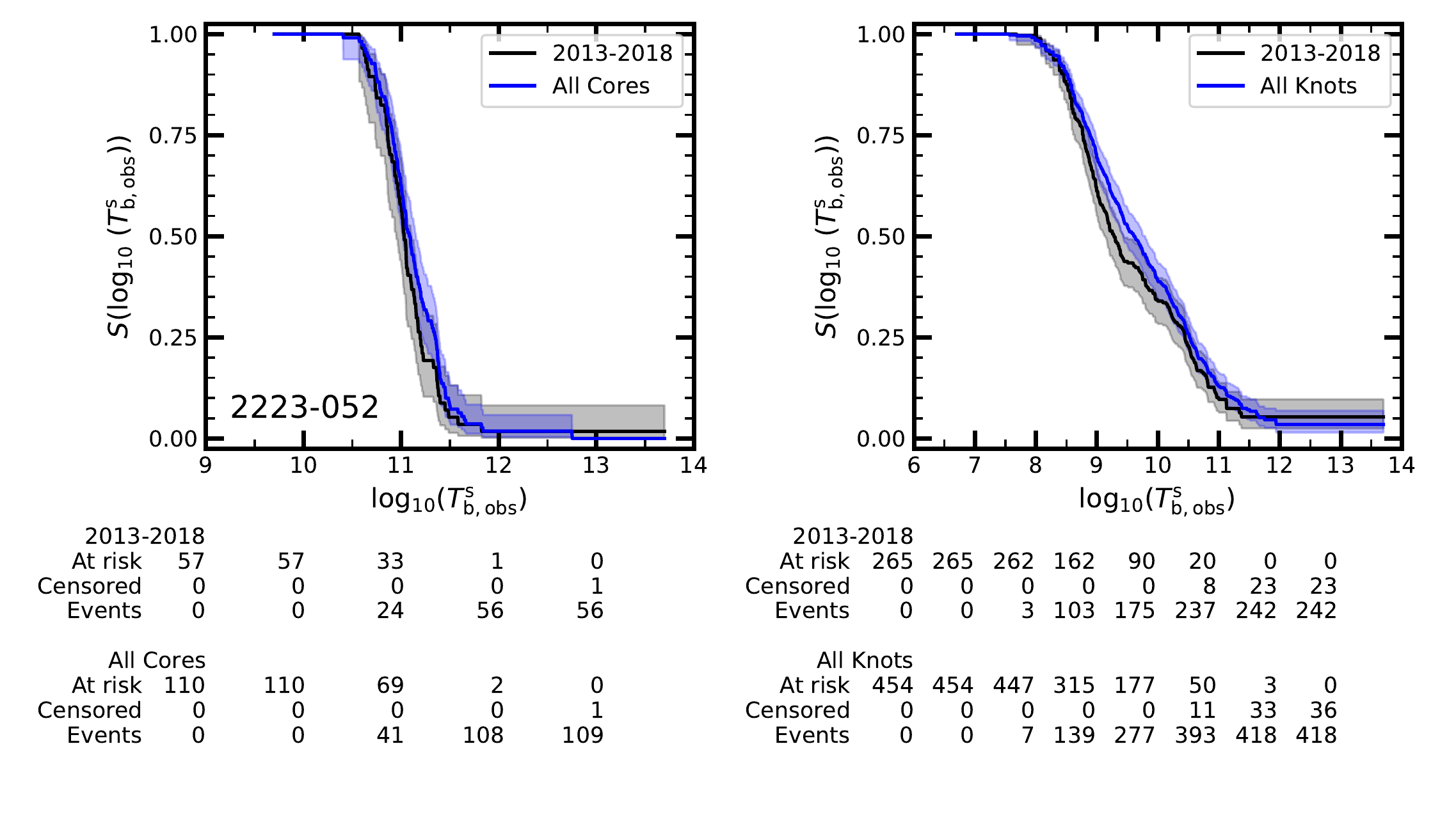}}
        \figsetgrpnote{Survival functions of the core (left) and knot (right) brightness temperatures for the marked time periods of the FSRQ 2223-052.}
        \figsetgrpend
        %
        % Number 37
        \figsetgrpstart
        \figsetgrpnum{A1.37}
        \figsetgrptitle{2230+114}
        \figsetplot{{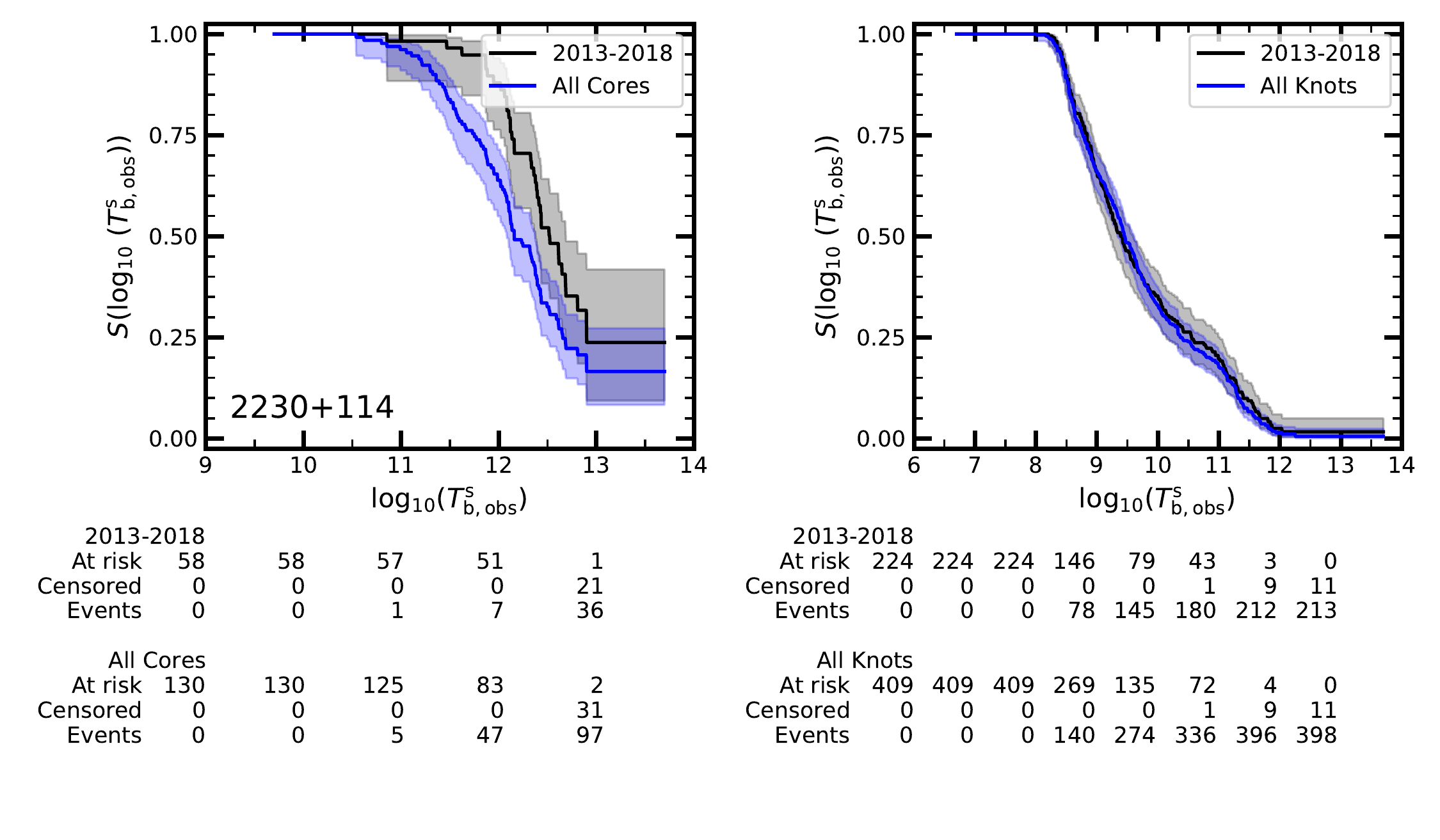}}
        \figsetgrpnote{Survival functions of the core (left) and knot (right) brightness temperatures for the marked time periods of the FSRQ 2230+114.}
        \figsetgrpend
        %
        % Number 38
        \figsetgrpstart
        \figsetgrpnum{A1.38}
        \figsetgrptitle{2251+158}
        \figsetplot{{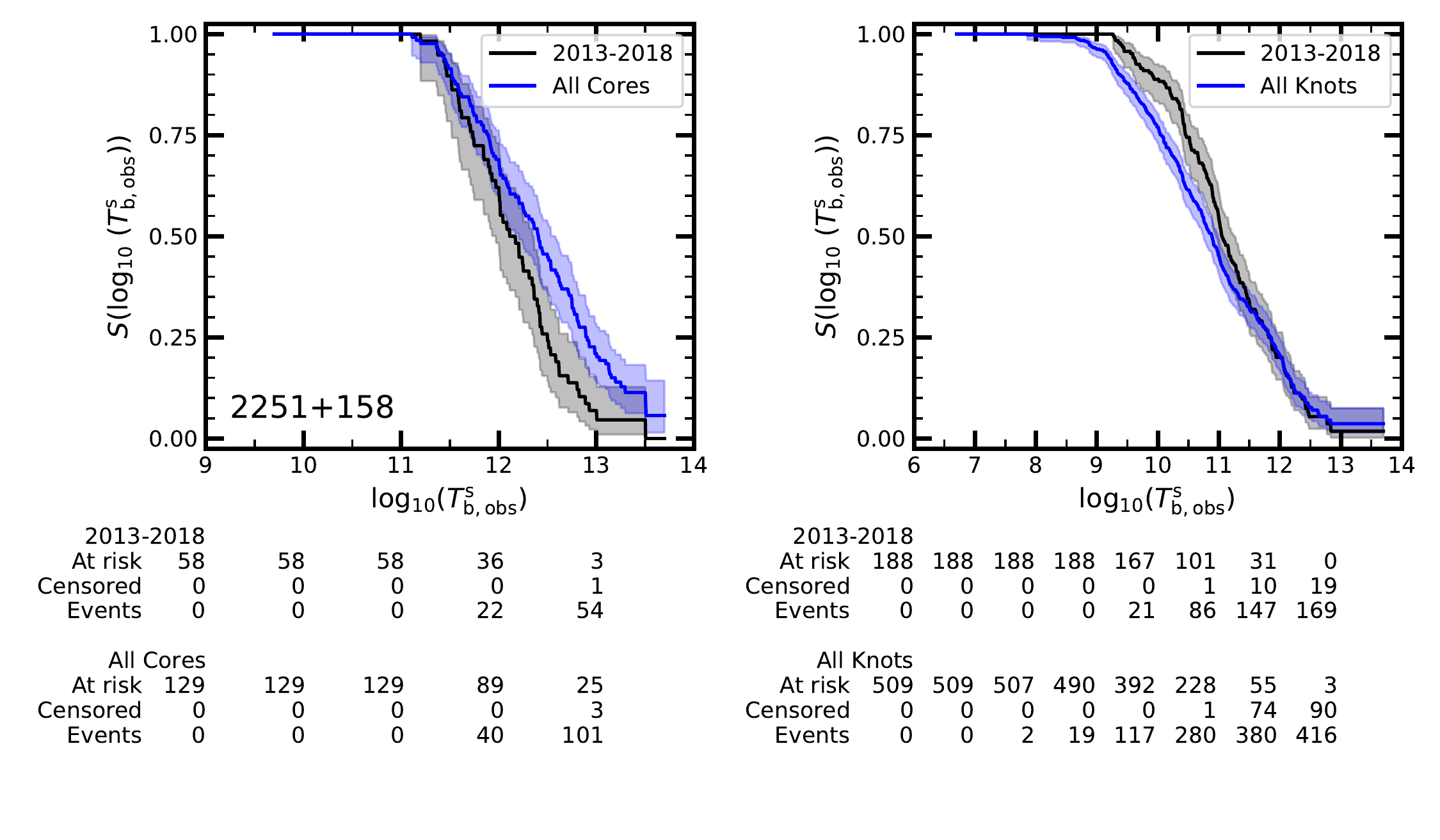}}
        \figsetgrpnote{Survival functions of the core (left) and knot (right) brightness temperatures for the marked time periods of the FSRQ 2251+158.}
        \figsetgrpend
\figsetend
%----

To determine a ``typical" brightness temperature for the core and knots of a source, we calculate the median of the survival function, i.e., where $\mathcal{S}(\mathrm{log}_{10}\ T_{\mathrm{b,obs}}^{\mathrm{s}}) = 0.5$, for each time period. These median brightness temperatures, with 95\% confidence intervals, are given in Table~\ref{tab:BrightnessMedians}. Due to the low number of core components observed for the BL 1959+650 $(N = 6)$, we use the lower confidence interval as a lower limit to the brightness temperature of the core for the source (among all 13 BL objects, a single lower limit will not significantly affect the results).

\begin{deluxetable*}{lll|ll}
  \tablenum{A1}
  \tablecaption{Median Brightness Temperatures of Source Components as Determined Through Survival Analysis\label{tab:BrightnessMedians}}
  \tablewidth{0pt}
  \tablehead{
             \colhead{Source} & \multicolumn{2}{c}{Core Median $T_{\mathrm{b,obs}}^{\mathrm{s}}$} & \multicolumn{2}{c}{Knot Median $T_{\mathrm{b,obs}}^{\mathrm{s}}$} \\
             \colhead{} & \multicolumn{2}{c}{$\times 10^{10}$ [K]} & \multicolumn{2}{c}{$\times 10^{8}$ [K]} \\
             \colhead{} & \colhead{2013-2018} & \colhead{2007-2018} & \colhead{2013-2018} & \colhead{2007-2018}
  }
  \colnumbers
  \startdata
    0219+428 & $\phn 22.94^{+\phn\phn2.68}_{-\phn11.66}$          & $\phn19.08^{+\phn\phn8.00}_{-\phn\phn5.76}$     & $\phn \phn \phn 6.43^{+\phn\phn\phn4.70}_{-\phn\phn\phn3.51}$ & $\phn\phn7.68^{+\phn\phn2.58}_{-\phn\phn2.47}$ \\
    0235+164 & $\phn 29.70^{+\phn15.27}_{-\phn\phn9.54}$          & $\phn52.12^{+\phn\phn9.41}_{-\phn10.35}$        & $\phn \phn 73.47^{+\phn\phn99.32}_{-\phn\phn28.91}$           & $178.45^{+\phn97.42}_{-\phn89.28}$ \\
    0316+413 & $\phn 15.43^{+\phn\phn2.29}_{-\phn\phn1.36}$       & $\phn14.74^{+\phn\phn0.98}_{-\phn\phn0.92}$     & $\phn 108.22^{+\phn\phn14.91}_{-\phn\phn16.13}$               & $144.69^{+\phn12.15}_{-\phn15.45}$ \\
    0336-019 & $\phn 50.70^{+\phn38.27}_{-\phn16.91}$             & $\phn43.34^{+\phn15.41}_{-\phn\phn9.55}$        & $\phn 121.16^{+\phn133.33}_{-\phn\phn55.53}$                  & $121.16^{+113.61}_{-\phn59.63}$ \\
    0415+379 & $\phn 13.82^{+\phn\phn4.07}_{-\phn\phn4.52}$       & $\phn15.87^{+\phn\phn4.30}_{-\phn\phn2.05}$     & $\phn \phn 35.55^{+\phn\phn\phn9.01}_{-\phn\phn\phn9.81}$     & $\phn34.42^{+\phn\phn8.72}_{-\phn\phn8.68}$ \\
    0420-014 & $\phn 62.10^{+\phn29.36}_{-\phn19.15}$             & $\phn62.67^{+\phn18.46}_{-\phn11.03}$           & $\phn \phn 23.75^{+\phn\phn21.53}_{-\phn\phn14.43}$           & $159.39^{+215.45}_{-\phn65.80}$ \\
    0430+052 & $\phn 14.47^{+\phn\phn5.33}_{-\phn\phn3.08}$       & $\phn15.87^{+\phn\phn3.75}_{-\phn\phn4.48}$     & $\phn \phn 40.45^{+\phn\phn13.63}_{-\phn\phn13.43}$           & $\phn34.42^{+\phn\phn8.03}_{-\phn15.16}$ \\
    0528+134 & $\phn 68.72^{+\phn19.42}_{-\phn11.04}$             & $102.16^{+\phn25.30}_{-\phn27.49}$              & $\phn \phn 29.29^{+\phn\phn15.27}_{-\phn\phn14.89}$           & $\phn45.28^{+\phn16.25}_{-\phn11.95}$\\
    0716+714 & $292.24^{+158.50}_{-119.45}$                       & $180.94^{+\phn66.61}_{-\phn29.07}$              & $\phn \phn 43.14^{+\phn132.46}_{-\phn\phn16.12}$              & $\phn49.09^{+\phn46.03}_{-\phn18.84}$ \\
    0735+178 & $\phn 10.77^{+\phn\phn\infty}_{-\phn\phn7.37}$     & $\phn\phn3.02^{+\phn\phn0.89}_{-\phn\phn0.46}$  & $\phn \phn \phn 4.03^{+\phn\phn\phn3.17}_{-\phn\phn\phn0.71}$ & $\phn\phn3.48^{+\phn\phn1.57}_{-\phn\phn0.47}$ \\
    0827+243 & $\phn 38.45^{+\phn15.63}_{-\phn\phn9.82}$          & $\phn36.71^{+\phn19.39}_{-\phn\phn6.46}$        & $\phn \phn 54.96^{+\phn\phn62.35}_{-\phn\phn19.98}$           & $104.78^{+\phn35.31}_{-\phn49.82}$ \\
    0829+046 & $\phn 19.26^{+\phn\phn5.22}_{-\phn\phn3.54}$       & $\phn11.92^{+\phn\phn4.69}_{-\phn\phn2.54}$     & $\phn \phn 20.54^{+\phn\phn47.24}_{-\phn\phn\phn8.67}$        & $\phn13.94^{+\phn\phn6.60}_{-\phn\phn2.07}$ \\
    0836+710 & $\phn 70.65^{+\phn40.35}_{-\phn16.58}$             & $\phn63.84^{+\phn15.81}_{-\phn10.26}$           & $\phn \phn 72.30^{+\phn\phn35.92}_{-\phn\phn21.60}$           & $\phn43.14^{+\phn18.39}_{-\phn\phn8.72}$ \\
    0851+202 & $396.16^{+140.87}_{-139.31}$                       & $292.24^{+158.50}_{-\phn72.65}$                 & $\phn 106.49^{+\phn139.92}_{-\phn\phn55.79}$                  & $115.43^{+\phn68.87}_{-\phn38.31}$ \\
    0954+658 & $174.39^{+276.35}_{-\phn69.36}$                    & $104.06^{+\phn18.78}_{-\phn25.14}$              & $\phn \phn 62.53^{+\phn\phn25.21}_{-\phn\phn23.99}$           & $\phn63.54^{+\phn12.34}_{-\phn20.40}$ \\
    1055+018 & $203.98^{+119.45}_{-\phn75.34}$                    & $184.30^{+\phn56.49}_{-\phn36.58}$              & $\phn 357.13^{+1494.43}_{-\phn235.97}$                        & $184.30^{+818.63}_{-102.04}$ \\
    1101+384 & $\phn \phn 3.73^{+\phn\phn0.71}_{-\phn\phn0.66}$   & $\phn\phn3.47^{+\phn\phn0.63}_{-\phn\phn0.39}$  & $\phn \phn \phn 1.46^{+\phn\phn\phn0.52}_{-\phn\phn\phn0.42}$ & $\phn\phn1.80^{+\phn\phn0.45}_{-\phn\phn0.34}$ \\
    1127-145 & $\phn \phn 8.55^{+\phn\phn9.66}_{-\phn\phn3.11}$   & $\phn28.62^{+\phn10.90}_{-\phn\phn8.08}$        & $\phn \phn 17.48^{+\phn\phn18.07}_{-\phn\phn12.82}$           & $\phn17.76^{+\phn\phn6.38}_{-\phn\phn9.17}$ \\
    1156+295 & $\phn 94.03^{+\phn23.29}_{-\phn14.38}$             & $\phn94.90^{+\phn23.50}_{-\phn15.25}$           & $\phn \phn 26.16^{+\phn\phn26.20}_{-\phn\phn\phn9.51}$        & $\phn35.55^{+\phn17.66}_{-\phn11.41}$ \\
    1219+285 & $\phn \phn 5.06^{+\phn\phn1.93}_{-\phn\phn2.57}$   & $\phn\phn4.13^{+\phn\phn1.17}_{-\phn\phn1.64}$  & $\phn \phn 13.72^{+\phn\phn\phn9.65}_{-\phn\phn\phn7.60}$     & $\phn\phn8.59^{+\phn\phn3.47}_{-\phn\phn2.94}$ \\
    1222+216 & $\phn 14.34^{+\phn15.64}_{-\phn\phn5.62}$          & $\phn41.39^{+\phn18.46}_{-\phn13.29}$           & $\phn \phn 29.77^{+\phn\phn\phn8.15}_{-\phn\phn\phn8.56}$     & $\phn39.16^{+\phn16.69}_{-\phn\phn6.89}$ \\
    1226+023 & $\phn 36.71^{+\phn10.38}_{-\phn10.13}$             & $\phn39.89^{+\phn11.75}_{-\phn\phn5.47}$        & $\phn 140.09^{+\phn\phn56.50}_{-\phn\phn33.60}$               & $117.31^{+\phn25.06}_{-\phn15.85}$ \\
    1253-055 & $317.52^{+340.27}_{-\phn89.68}$                    & $281.66^{+\phn38.81}_{-\phn60.03}$              & $\phn 517.59^{+\phn208.74}_{-\phn111.26}$                     & $485.24^{+195.69}_{-\phn78.91}$ \\
    1308+326 & $\phn 36.04^{+\phn26.63}_{-\phn12.46}$             & $\phn33.79^{+\phn\phn9.55}_{-\phn\phn6.46}$     & $\phn 127.17^{+\phn202.28}_{-\phn\phn89.25}$                  & $485.24^{+163.52}_{-213.78}$ \\
    1406-076 & $\phn 37.05^{+\phn12.72}_{-\phn\phn9.47}$          & $\phn31.10^{+\phn\phn6.64}_{-\phn11.12}$        & $\phn \phn 44.56^{+\phn\phn95.53}_{-\phn\phn15.27}$           & $\phn42.45^{+\phn19.08}_{-\phn13.63}$ \\
    1510-089 & $\phn 84.96^{+\phn58.73}_{-\phn22.86}$             & $\phn70.65^{+\phn15.89}_{-\phn20.42}$           & $\phn \phn 73.47^{+\phn\phn62.17}_{-\phn\phn33.67}$           & $\phn84.96^{+\phn26.81}_{-\phn34.26}$ \\
    1611+343 & $\phn 22.52^{+\phn\phn8.01}_{-\phn10.93}$          & $\phn16.92^{+\phn11.97}_{-\phn\phn6.45}$        & $\phn 137.85^{+\phn156.42}_{-\phn\phn73.27}$                  & $\phn15.36^{+\phn12.10}_{-\phn\phn4.93}$ \\
    1622-297 & $\phn \phn 8.32^{+\phn26.10}_{-\phn\phn3.49}$      & $\phn\phn8.71^{+\phn\phn4.00}_{-\phn\phn3.56}$  & $\phn \phn 90.62^{+\phn136.69}_{-\phn\phn51.46}$              & $109.98^{+\phn86.61}_{-\phn36.51}$ \\
    1633+382 & $166.53^{+\phn26.47}_{-\phn34.28}$                 & $156.12^{+\phn18.26}_{-\phn22.65}$              & $\phn \phn 18.64^{+\phn\phn\phn6.69}_{-\phn\phn\phn4.47}$     & $\phn14.87^{+\phn\phn7.04}_{-\phn\phn3.19}$ \\
    1641+399 & $147.72^{+\phn50.69}_{-\phn46.50}$                 & $105.02^{+\phn26.01}_{-\phn23.90}$              & $\phn \phn 36.13^{+\phn\phn\phn7.71}_{-\phn\phn\phn7.31}$     & $\phn27.91^{+\phn\phn4.88}_{-\phn\phn3.77}$ \\
    1652+398 & $\phn \phn 0.72^{+\phn\phn0.38}_{-\phn\phn0.22}$   & $\phn\phn0.72^{+\phn\phn0.38}_{-\phn\phn0.22}$  & $\phn \phn \phn 2.69^{+\phn\phn\phn0.63}_{-\phn\phn\phn0.61}$ & $\phn\phn2.69^{+\phn\phn0.63}_{-\phn\phn0.61}$ \\
    1730-130 & $209.70^{+151.57}_{-\phn77.45}$                    & $153.27^{+\phn85.32}_{-\phn52.98}$              & $\phn \phn 60.54^{+\phn\phn24.42}_{-\phn\phn18.09}$           & $\phn61.53^{+\phn30.56}_{-\phn14.76}$ \\
    1749+096 & $221.62^{+\phn92.98}_{-\phn64.06}$                 & $247.55^{+\phn69.97}_{-\phn74.76}$              & $\phn \phn 46.02^{+\phn\phn69.41}_{-\phn\phn17.20}$           & $\phn55.85^{+111.45}_{-\phn18.54}$ \\
    1959+650 & $\phn \phn 1.64^{+\phn\phn\infty}_{-\phn\phn0.00}$ & $\phn\phn1.64^{+\phn\phn\infty}_{-\phn\phn0.00}$& $\phn \phn \phn 1.58^{+\phn\phn\phn3.47}_{-\phn\phn\phn1.27}$ & $\phn\phn1.58^{+\phn\phn3.47}_{-\phn\phn1.27}$ \\
    2200+420 & $329.45^{+270.41}_{-138.22}$                       & $259.23^{+\phn85.76}_{-\phn49.53}$              & $\phn \phn \phn 6.64^{+\phn\phn\phn1.68}_{-\phn\phn\phn1.08}$ & $\phn\phn9.78^{+\phn\phn1.35}_{-\phn\phn0.90}$ \\
    2223-052 & $\phn 10.77^{+\phn\phn2.06}_{-\phn\phn1.81}$       & $\phn12.14^{+\phn\phn2.06}_{-\phn\phn1.67}$     & $\phn \phn 19.26^{+\phn\phn\phn8.65}_{-\phn\phn\phn5.32}$     & $\phn40.45^{+\phn22.08}_{-\phn13.43}$ \\
    2230+114 & $335.58^{+154.15}_{-\phn96.99}$                    & $145.02^{+\phn93.57}_{-\phn23.30}$              & $\phn \phn 24.53^{+\phn\phn15.27}_{-\phn\phn\phn7.88}$        & $\phn28.36^{+\phn12.74}_{-\phn\phn6.45}$ \\
    2251+158 & $149.09^{+\phn80.86}_{-\phn55.06}$                 & $256.85^{+128.50}_{-\phn94.86}$                 & $1122.85^{+\phn669.92}_{\phn-212.45}$                         & $774.75^{+212.13}_{-204.55}$ \\
  \enddata
%   \tablenotetext{a}{Antenna abbreviations can be found at the NRAO VLBA website (\url{https://science.nrao.edu/facilities/vlba/docs/manuals/oss/sites}). Individual antennas are sometimes unavailable due to weather, maintenance, or undetected fringes (in the case of the weakest sources, e.g., 1959+650).}
\end{deluxetable*}

We note that this analysis is based on a few approximations, such as non-variability of the brightness temperature of components and similar properties of sources within each subclass. It is not expected \emph{a priori} that the survival function of the component brightness temperatures for a source should be the same for different time periods (e.g., the faster decrease in the survival function for knots observed from 2013 January to 2018 December versus knots observed during the entire time period for the BL 0219+428 near brightness temperatures $\sim10^{8}$ K, as in Fig.~\ref{fig:KMEstimateSrc}). However, temporal analysis of the brightness temperatures in the jets of blazars as well as an investigation into the assumptions outlined above are beyond the scope of this paper, which is focused on the kinematics of the features. We plan to present such an analysis in a separate publication (Weaver, in prep.).

%-------------------------------------------------------------------------------
%---   END MATTER   ------------------------------------------------------------
%-------------------------------------------------------------------------------

\vspace{5mm}
\facilities{VLBA, Fermi-LAT, Mets\"ahovi Radio Observatory}

\software{\texttt{AIPS} \citep{vanMoorsel1996}, \texttt{astropy} \citep{Astropy2013, Astropy2018}, \texttt{Difmap} \citep{Shepherd1997}, \texttt{lifelines} \citep{DavidsonPilon2019}, \texttt{pwlf} \citep{Jekel2019}, \texttt{statsmodels} \citep{Seabold2010}}

\clearpage

\bibliography{JetKinematicsBibliography}{}
\bibliographystyle{aasjournal}

\end{document}